\documentclass[11pt,oneside]{report}
\usepackage[utf8]{inputenc}
\usepackage[OT1]{fontenc} 
\usepackage[margin=1in]{geometry}
\usepackage{amsmath,amssymb,amsthm,mathtools}
\usepackage{booktabs}
\usepackage{array}
\usepackage{graphicx}
\usepackage{xcolor}
\usepackage{caption}
\usepackage{enumitem}
\usepackage[protrusion=true,expansion=false]{microtype}
\usepackage[authoryear,round,sort]{natbib}
\usepackage[hidelinks,colorlinks=true,linkcolor=blue!50!black,citecolor=blue!40!black,urlcolor=blue!50!black]{hyperref}
\graphicspath{{./}{../}}

\theoremstyle{plain}
\newtheorem{theorem}{Theorem}[chapter]
\newtheorem{proposition}[theorem]{Proposition}
\newtheorem{corollary}[theorem]{Corollary}
\newtheorem{lemma}[theorem]{Lemma}
\theoremstyle{definition}
\newtheorem{definition}[theorem]{Definition}
\newtheorem{assumption}[theorem]{Assumption}
\theoremstyle{remark}
\newtheorem{remark}[theorem]{Remark}
\numberwithin{equation}{chapter}

\newcommand{\R}{\mathbb{R}}
\newcommand{\E}{\mathbb{E}}
\newcommand{\Prob}{\mathbb{P}}
\newcommand{\Qprob}{\mathbb{Q}}
\newcommand{\Var}{\mathrm{Var}}
\newcommand{\Cov}{\mathrm{Cov}}
\newcommand{\Indic}{\mathbf{1}}
\newcommand{\Filt}{\mathcal{F}}
\newcommand{\Qset}{\mathcal{Q}}
\newcommand{\ES}{\mathrm{ES}}
\newcommand{\VaR}{\mathrm{VaR}}

\newcommand{\KL}[2]{D(#1\,\|\,#2)}
\newcommand{\PHA}{\ensuremath{\mathrm{PHA}}}
\newcommand{\SGH}{\ensuremath{\mathrm{SGH}}}
\newcommand{\BR}{\ensuremath{\mathrm{BR}}}
\newcommand{\EH}{\ensuremath{\mathrm{EH}}}
\newcommand{\EL}{\ensuremath{\mathrm{EL}}}
\newcommand{\PD}{\ensuremath{\mathrm{PD}}}
\newcommand{\LGD}{\ensuremath{\mathrm{LGD}}}
\newcommand{\EAD}{\ensuremath{\mathrm{EAD}}}
\newcommand{\CCF}{\ensuremath{\mathrm{CCF}}}
\newcommand{\rec}{\psi}
\newcommand{\Acc}{\mathcal{A}}
\newcommand{\Mset}{\mathcal{M}}
\newcommand{\Sset}{\mathcal{Z}}
\newcommand{\Rbu}{R_{\mathrm{BU}}}
\newcommand{\Rtd}{R_{\mathrm{TD}}}
\newcommand{\we}[1]{w^{e}_{#1}}
\newcommand{\wc}[1]{w^{c}_{#1}}
\newcommand{\reach}{\varrho}
\newcommand{\cont}{\gamma}
\newcommand{\Mis}{M}
\newcommand{\psys}{p_{\mathrm{sys}}}
\newcommand{\psysd}{p_{\mathrm{sys},d}}
\newcommand{\beff}{\beta_{\mathrm{eff}}}
\newcommand{\geff}{g_{\mathrm{eff}}}
\newcommand{\Ham}{H}
\newcommand{\Cont}{C}
\newcommand{\Loss}{L}
\newcommand{\Lev}{\nu}
\newcommand{\Gam}{\Gamma}
\newcommand{\spr}{\rho}
\newcommand{\rhod}{\rho_{\delta}}
\newcommand{\Lsys}{\Lambda_{\mathrm{sys}}}
\newcommand{\Mark}{\bar{J}}
\newcommand{\Ssafe}{\mathcal{S}}
\newcommand{\Ccat}{\mathcal{C}}
\newcommand{\Null}{\mathcal{N}}
\newcommand{\Ifus}{\mathcal{I}}
\newcommand{\reg}{\zeta}
\newcommand{\drift}{\mu_v}
\newcommand{\Tesc}{T_{\mathrm{esc}}}
\newcommand{\trelax}{T_{\mathrm{relax}}}
\newcommand{\Lmin}{\Lambda_{\mathrm{c},\min}}
\newcommand{\trc}{\tau_{\mathrm{react}}}
\newcommand{\Lamc}{\Lambda_{\mathrm{c}}}
\newcommand{\pben}{p_{\mathrm{ben}}}
\newcommand{\Hint}{H^{a}}
\newcommand{\Hsta}{H^{x}}

\newcommand{\HstaT}{{H^{x}}^{\!\top}}
\newcommand{\Ifusa}{\Ifus^{a}}
\newcommand{\Lapl}{\mathcal{L}}
\newcommand{\rhoatk}{\rho^{\mathrm{atk}}}
\newcommand{\Kstar}{K^{\star}}

\DeclareMathOperator*{\argmax}{arg\,max}

\title{\textbf{Agentic AI Systems and Financial Stability
\thanks{
This paper is preliminary and views are only of the authors and of no one else's at the Board of Governors of the Federal Reserve System or the Federal Reserve Bank of Cleveland. All the ideas presented in this work reflect those of the authors, while genAI tools were used in the production and proofreading of the paper.}}\\[4pt]
\large From Model Risk to Systemic Risk}
\author{Sriram Nagaraj, Federal Reserve Bank of Cleveland\footnote{sriram.nagaraj@clev.frb.org}\\ Seung Jung Lee, Board of Governors of the Federal Reserve System\footnote{seung.j.lee@frb.gov}}
\date{August 22, 2026}

\begin{document}
\maketitle

\begin{abstract}
\noindent Financial stability rests on the premise that distress is largely idiosyncratic and
therefore diversifiable. When one institution errs, the rest of the system absorbs the shock. That
premise fails when many decision-makers depend on the same infrastructure and are interconnected, so
that a single fault becomes a common shock and propagates. Agentic systems are AI models that take
consequential actions rather than only emitting predictions, and these agentic systems are becoming
exactly such a shared substrate. When a population of agents is built on a common foundation model and
their actions interact, the binding concern is no longer the \emph{model risk} of any single deployment
but the \emph{systemic risk} of the population. This monograph develops that systemic-risk account in
six mathematical settings. We start from single-institution model-risk management, the microprudential
discipline of validating and governing one firm's reliance on a model, and build toward a
macroprudential theory of agentic financial stability. \textbf{Chapter~1} establishes the
microprudential unit. The chapter recasts the credit-risk decomposition of expected loss into a
probability of default, a loss given default, and an exposure at default as an expected-harm identity
that multiplies a probability of harmful action, a severity given harm, and a blast radius for a single
agent. The chapter also introduces a \emph{set-valued} containment-risk measure whose value is the set
of acceptable mitigation configurations, and it proves that against irreversible harm, detection and
reversibility do not merely cost more; they cannot work --- under the lever-assignment axiom (A3) that review cannot intercept the irreversible action. A coherent tail constraint is infeasible for
every non-preventive control, so nothing substitutes for prevention. An irreducible catastrophic core
appears here for the first time. \textbf{Chapter~2} is the pivot from model risk to systemic risk.
Lifting the unit to a fleet of agents sharing a foundation model, the chapter shows that the shared
model is a \emph{non-diversifiable common exposure} whose expected-shortfall floor holds however large
the fleet grows, a monoculture that no amount of redundancy dilutes. A percolation threshold governs
contagion through the inter-agent dependency graph, and fleet architecture controls that threshold;
segmentation and heterogeneity that keeps an exploit from transferring both lower the threshold, but
only prevention or severing the common exposure moves the systemic floor itself. The chapter then lets
agents respond to \emph{each other}, opening two further channels. When an agent's activation depends
on the rest of the fleet, harm becomes a fixed point rather than a sum, and a social multiplier that
grows with peer deference rescales the fleet average and the non-diversifiable floor; if that
deference itself rises under stress, the fleet gains a common factor even without a shared shock. A
compromised peer can also become a persuasion channel alongside the technical one, and segmenting the
technical graph alone no longer caps the outbreak once this second channel is contagious on its own. \textbf{Chapter~3} renders these systemic dynamics in continuous
time as a marked, jump-diffusion, Hawkes-excited process. The static decomposition reappears as the
rare-event limit of a compensator identity. The fleet results reappear as a common-L\'evy-jump systemic
floor and a Hawkes spectral-radius criticality for self-exciting cascades. We show that the robust
(worst-case) stress problem coincides exactly with a time-consistent entropic risk measure, so
robustness to misspecification and dynamic consistency are one object, with the prevention lever
reappearing as the jump-measure truncation that makes the robust game finite at all. \textbf{Chapter~4}
turns to macroprudential control and its limits, modelling runtime guardrails and circuit-breakers as
partially-observed stochastic control. The chapter gives an observability trichotomy for violation
directions and a non-diversifiable detection floor that mirrors the monoculture result. The chapter
also gives a runtime-impossibility corollary (of commit-exogeneity, Assumption~\ref{c4:ass:commit}), the dynamic form of the macroprudential thesis. Against
an irreversible systemic event, the probability of eventually exiting safety depends only on the
irreversible-core hazard rate and is identical across \emph{every} runtime policy, so no amount of
monitoring or intervention substitutes for the ex-ante, structural measures that lower that hazard rate
itself. \textbf{Chapter~5} makes the adversary a \emph{player}. The chapter studies agentic-AI-augmented
cyber events and shows that an \emph{unintentional} event, a bad update or a poisoned dependency,
reduces to the common shock of the earlier chapters when modeled as a fleet-wide, exogenously-timed trigger. An \emph{intentional} attack, by contrast,
turns every parameter of nature into an attacker's choice. The monoculture then becomes an attractive
target. The systemic floor \emph{rises} with fleet size when the attacker's prize is fleet-linear and its cost fleet-independent, and lives in the tail regardless of attacker
mix. Detection \emph{deflects} effort onto the irreversible core under a fixed-effort attacker. At the decentralized equilibrium, a
population coupled by contagion \emph{underinvests} in the one lever that works, by a wedge
proportional to systemic reach. This underinvestment makes the incentive-theoretic case for
macroprudential cyber regulation. \textbf{Chapter~6} closes the arc by making capability itself
\emph{dynamic}. Both attacker and defender are agentic-AI-augmented, and their capabilities are state
variables that co-evolve in an arms race. Prevention acts through a durable structural channel. Outright
deletion of a catastrophic affordance neutralises that affordance against every future capability
level, and attenuating stock must be rebuilt only at the adversary's \emph{growth rate}. Detection is a
depreciating \emph{flow} that must be re-won against the adversary's capability \emph{level} itself.
The offense-dominance of the systemic tail compounds rather than washes out. The faster
agentic AI advances, the more decisively the only effective lever is the structural one. One central
claim runs through all six settings, and each chapter establishes this claim in its own mathematical
language. The systematic, system-wide component of agentic risk cannot be diversified, detected away,
or reversed, and only ex-ante structural prevention moves it. The numerical parameters throughout are
illustrative. The contribution is the structure, and calibration to deployment and market telemetry is
the central open problem we delineate but do not solve. The policy takeaway is simple. Monitoring and
after-the-fact detection cannot remove this systemic floor. Reducing that floor requires limiting
shared dependence on any single foundation model before problems occur, not just watching for them
afterward.
\end{abstract}

\tableofcontents

\chapter*{Preface}
\addcontentsline{toc}{chapter}{Preface}

This monograph builds from model risk to financial stability. Its six chapters are self-contained
developments of one idea, written to be read in order. Each takes the central object of the previous
one and re-poses it at a higher level of aggregation: from a single agent to a population, from a
snapshot to a dynamic, from measurement to control, and finally from nature to an adversary. But each
also stands alone, with its own notation, assumptions, and numerical study. A reader interested only
in the systemic (fleet) results can read Chapter~2 after the single-agent decomposition of
Chapter~1. A reader interested in macroprudential control can read Chapter~4 after Chapters~1 and~3.
A reader interested in cyber and financial stability can read Chapter~5 after Chapter~2, or after
Chapter~4 for its detection-and-control results. Chapter~6 lifts Chapter~5 into dynamics as Chapter~3
lifted Chapter~2, so a reader following the static-to-dynamic thread can read Chapters~3 and~6 as a
pair.

The organising movement is microprudential to macroprudential. Single-institution model-risk
management is the traditional discipline of validating and governing one firm's reliance on a model.
It is the familiar entry point and the subject of Chapter~1. But it is the wrong level of analysis for
the risk that agentic systems actually create. Once many agents run on a shared foundation model and
act on one another, firm-level model risk aggregates into system-level risk. The right object then
becomes the one financial-stability supervisors care about: \emph{systemic risk}, the risk that a
common shock or a contagion cascade brings down the system as a whole, not any single participant.
Chapters~2--6 make that object precise and study what can be done about it.

A single structural claim runs through all six chapters, and we develop it in six different
mathematical languages. In Chapter~1 it is a convex-feasibility statement about a single
agent's tail risk. In Chapter~2 it is a large-$K$ limit of a fleet's expected shortfall. In Chapter~3
it is a rare-event limit of a jump compensator and a heavy-tail finiteness condition. In Chapter~4 it
is a policy-independence statement about a Poisson exit probability. In Chapter~5 it is a Stackelberg
best response that turns each parameter of nature into an attacker's choice and, in doing so, sharpens
the claim into a statement about incentives as well as technology. In Chapter~6 it is a
capability-accumulation dynamic in which the same lever's advantage compounds over an arms race while
every other lever's decays. In every case the conclusion is the same, and it is the macroprudential
lesson in miniature. The part of agentic risk that is \emph{systematic} and \emph{irreversible} is
the part that becomes \emph{systemic}. Systematic here means shared across the population
through the common model, in the common-factor sense that defeats diversification; irreversible means
a harm that cannot be undone once the action is taken. That part is not reachable by the controls that
practitioners and microprudential regulation reach for first: redundancy and diversification, post-hoc
detection, reversibility and blast-radius limits. Those controls work, and work well, on the
idiosyncratic and reversible part of the risk. They do nothing to the systematic-irreversible core.
Only \emph{ex-ante, structural prevention} moves that core, whether by removing the catastrophic
affordance from the agent's reach or by severing the coupling that makes an exposure common, and the
mathematics says so in closed form. This is why financial stability, here as in banking, cannot be
secured one institution at a time.

\begin{figure}[p]
\centering
\includegraphics[width=1.01\textwidth]{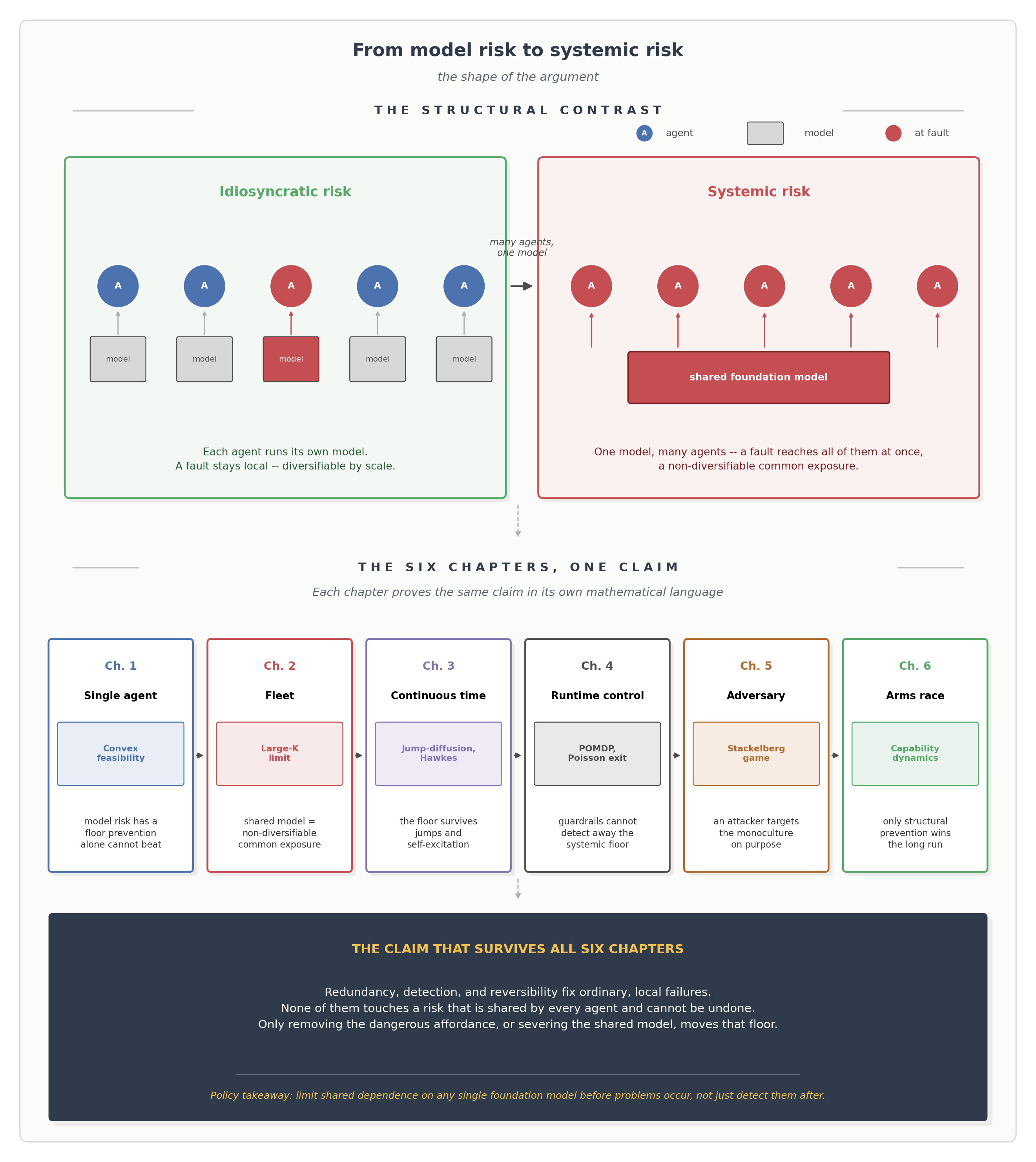}
\caption{The shape of the argument. Top: the structural contrast the monograph turns on. When each
agent runs its own model, a fault stays local and diversifies away with scale. When a fleet shares one
foundation model, a fault in that model reaches every agent at once, a non-diversifiable common
exposure. Bottom: the six chapters carry this one claim through six mathematical languages, from a
single agent's model risk to a fleet's systemic floor, into continuous time, through the limits of
runtime control, against a deliberate adversary, and finally into a co-evolving arms race. Each box
names the language that chapter works in and the claim it establishes there. Redundancy,
detection, and reversibility control the idiosyncratic part of the risk; only structural prevention
moves the systemic floor itself.}
\label{fig:overview}
\end{figure}

One channel cuts across the whole arc and deserves a word here. Agents are increasingly deployed to
interact with \emph{other agents}: reading each other's output, calling each other as tools,
negotiating on their principals' behalf. Section~\ref{c2:sec:interaction} takes this up, and the news
is mixed. The central claim survives, but two comfortable corollaries do not. Interaction amplifies
whatever systematic shock is already present, so the monoculture floor holds but with a larger
constant. Less comfortably, interaction can \emph{create} common exposure where none existed. If
agents defer to one another more heavily under stress, they co-move even when no agent's harm loads on
a common shock. This creates common exposure beyond any shared substrate --- through a shared state and a shared stress-responsiveness rule $\lambda$, even when the fleet
shares no model. Substrate diversity, the standard remedy for monoculture, does not reach that channel, because
there is no shared exposure left for it to diversify. What does reach it is diversity of the
interaction environment itself, which is a different lever and a new one. Agents can also address each
other in natural language, so a compromised peer is a persuasion channel as well as a technical one,
and segmenting the technical graph no longer segments the fleet. A reader who takes only the headline
claims away should not conclude that diversity and segmentation are sufficient.

Two cautions apply throughout and are stated once here. First, the numbers are illustrative. Each
chapter instantiates its model with plausible, chosen parameters that exhibit the qualitative
phenomena. None is calibrated to a real deployment or market, and the figures should be read as
existence proofs of structure, not as forecasts. The hard, unsolved problem is how to estimate harm
hazards, severities, blast radii, contagion kernels, and catastrophe intensities from deployment and
market telemetry, and it is the one we delineate but do not solve. Estimating these quantities is
inherently problem- and application-dependent, so we do not attempt to prescribe a general estimation
framework here. We flag it as the natural next program at the end of each chapter. Second, most theorems are paired with a closed-form quantity that an independent simulation
reproduces; where a check is a consistency or illustration rather than an independent test, we
say so. That is our correctness gate.

A note on terminology. We use \emph{systematic} in its common-factor sense: risk that is shared and so
does not diversify away, as in the systematic--idiosyncratic decomposition of portfolio theory. We use
\emph{systemic} in its financial-stability sense: risk to the system as a whole. Part of the thesis of
this monograph is that the two meet. The systematic exposure created by a shared foundation model is
what makes agentic risk systemic.

\chapter*{Notation}
\addcontentsline{toc}{chapter}{Notation}

Each chapter is self-contained and restates its own notation. The table below is a reader's aid across
chapters. It flags deliberate reuse rather than concealing it. Several symbols carry a different
meaning in different chapters because each is standard in its own literature, and we have preferred
the standard reading over a globally unique one. Where a symbol is reused, every sense is listed with
the place it is defined.

\begin{center}
\small
\renewcommand{\arraystretch}{1.18}
\begin{tabular}{@{}l l l@{}}
\toprule
Symbol & Meaning & Defined \\
\midrule
\multicolumn{3}{@{}l}{\emph{Risk functionals}}\\
$\spr$ & coherent (law-invariant) risk measure & Ch.~\ref{ch:single}, \S\ref{c1:sec:setvalued}\\
$\rhod$ & entropic risk measure at parameter $\delta$ & Ch.~\ref{ch:continuous}, Thm~\ref{c3:thm:robust}\\
$\rhoatk$ & attacked risk measure (attacker-optimised) & Ch.~\ref{ch:cyber}, Prop.~\ref{c5:prop:microfound}\\
$\ES_\tau$ & expected shortfall at level $\tau$ & Ch.~\ref{ch:single}\\
$\rho(A),\ \rho(\Gam)$ & \emph{spectral radius} of a matrix (standard usage) & Ch.~\ref{ch:fleet}, Ch.~\ref{ch:cyber}\\
\addlinespace
\multicolumn{3}{@{}l}{\emph{Harm and exposure}}\\
$\Ham_T$ & cumulative event harm (pure-jump) & Ch.~\ref{ch:continuous}, \eqref{c3:eq:harm}\\
$\Cont_T$ & continuous mark-to-market component & Ch.~\ref{ch:continuous}, \eqref{c3:eq:cont}\\
$\Loss_T$ & total loss $\Cont_T+\Ham_T$ & Ch.~\ref{ch:continuous}, \eqref{c3:eq:totalloss}\\
$\reach$ & \emph{reach} (blast radius) of an agent & Ch.~\ref{ch:fleet}, \S\ref{c2:sec:framework}\\
$R$ & predictable exposure scaling realised harm & Ch.~\ref{ch:continuous}\\
\addlinespace
\multicolumn{3}{@{}l}{\emph{Hazards and thresholds}}\\
$\Lamc$ & irreversible-core (catastrophe) hazard & Ch.~\ref{ch:control}, Ch.~\ref{ch:coevo}\\
$\theta$ & catastrophic magnitude threshold & Ch.~\ref{ch:single}, Ch.~\ref{ch:fleet}\\
$\theta$ & \emph{prevention-efficacy exponent} (distinct use) & Ch.~\ref{ch:coevo}, Def.~\ref{c6:def:levers}\\
$\Ssafe$ & safe region (specification box) & Ch.~\ref{ch:control}, \eqref{c4:eq:safebox}\\
$\Ccat$ & post-commit absorbing set & Ch.~\ref{ch:control}, \S\ref{c4:sec:model}\\
$\Sset$ & stress-state space & Ch.~\ref{ch:fleet}, \S\ref{c2:sec:framework}\\
\addlinespace
\multicolumn{3}{@{}l}{\emph{Levers and costs}}\\
$\chi$ & structural deletion (discrete lever) & Ch.~\ref{ch:coevo}, Def.~\ref{c6:def:levers}\\
$P$ & attenuating prevention stock (continuous lever) & Ch.~\ref{ch:coevo}, Def.~\ref{c6:def:levers}\\
$\kappa_\chi$ & lump-sum cost of deletion & Ch.~\ref{ch:coevo}, Thm~\ref{c6:thm:stockflow}\\
$\kappa_P$ & per-unit cost of prevention stock & Ch.~\ref{ch:coevo}, Thm~\ref{c6:thm:deterrence}\\
$\xi$ & rate at which capability opens fresh affordances & Ch.~\ref{ch:coevo}, \eqref{c6:eq:pdyn}\\
\addlinespace
\multicolumn{3}{@{}l}{\emph{Timescales and rates}}\\
$\Tesc$ & drift-escape time & Ch.~\ref{ch:control}, \eqref{c4:eq:tesc}\\
$\trelax$ & slowest Ornstein--Uhlenbeck relaxation time & Ch.~\ref{ch:control}, \eqref{c4:eq:timescales}\\
$\trc$ & reaction delay & Ch.~\ref{ch:control}\\
$\lambda$ & peer responsiveness (linear-in-means) & Ch.~\ref{ch:fleet}, Assum.~\ref{c2:ass:response}\\
$\lambda$ & harmful-action intensity (distinct use) & Ch.~\ref{ch:continuous}\\
$\lambda_t$ & attacker type's harm weight (distinct use) & Ch.~\ref{ch:cyber}, Prop.~\ref{c5:prop:microfound}\\
$\beta$ & branching ratio (contagion) & Ch.~\ref{ch:fleet}, Ch.~\ref{ch:cyber}\\
$\beff$ & effective branching ratio, $cq+c'p$ & Ch.~\ref{ch:fleet}, Assum.~\ref{c2:ass:pretext}\\
$\beta_j$ & per-axis specification limit (distinct use) & Ch.~\ref{ch:control}, \eqref{c4:eq:safebox}\\
\bottomrule
\end{tabular}
\end{center}

\noindent Four symbols above are deliberately reused: $\rho$ (as a risk measure and, in its standard
matrix sense, as a spectral radius), $\theta$, $\lambda$ (in three senses), and $\beta$. In each case the two senses
occur in different chapters and each is the conventional notation in its own field. We judged that
renaming would cost more in readability than it would buy in uniqueness.

\chapter{Model Risk of a Single Agent: A PD/LGD/EAD-Type Foundation}
\label{ch:single}
\section{Introduction}
\label{c1:sec:intro}

\subsection{Model-risk management and stress testing: the microprudential foundation}

This chapter begins the monograph at its microprudential foundation: the model risk borne by a single
institution that delegates a consequential decision to a model. This familiar firm-level problem has
its mature codification in banking, which gives us our vocabulary. The destination of the
monograph, though, is macroprudential, and it concerns the \emph{financial stability} of a whole population of agents
built on shared infrastructure. The single-agent theory developed here is the unit from which that
systemic account is assembled. We therefore start where supervision of model risk starts, and we carry
its pillars forward until, in later chapters, they must be restated at the level of the system.

Model-risk management is a general governance discipline that applies wherever consequential decisions
are delegated to models whose errors the deploying institution, not the model, must bear. Its most
mature codification happens to live in banking, and we use that codification throughout as our exemplar
because it is the most developed. Nothing in the discipline is intrinsically financial. That
codification rests on a simple but consequential admission. Every model is wrong in ways its builders
cannot fully see, and the institution, not the model, bears the consequences of that wrongness. The
supervisory statement of this admission is the U.S. interagency \emph{Supervisory Guidance on Model Risk
Management} \citep{sr117}. It defines model risk as the potential for adverse consequences arising from
decisions based on incorrect or misused model outputs. It locates the sources of that risk in two
places: \emph{fundamental errors} that produce inaccurate outputs, and \emph{incorrect or inappropriate
use}. The guidance's central organising principle is that model risk must be \emph{managed like other
risks}, identified, measured, monitored, and controlled through the model life cycle, rather than
eliminated, because it cannot be eliminated.

Three pillars of traditional model risk management are worth stating, because each will
reappear, transformed, in the agentic setting. The first is \emph{effective challenge}: critical
analysis by objective, informed parties who have the \emph{competence} to find flaws, the
\emph{incentive} to look for them, and the \emph{influence} to act on what they find. Effective
challenge is not a document review but an adversarial epistemic process whose value is destroyed if
any of the three ingredients is missing. The second pillar is that the intensity of validation should
scale with \emph{materiality and use}. A model whose outputs drive large, irreversible decisions
warrants deeper scrutiny than one whose outputs are advisory and easily overridden. The third is
\emph{ongoing monitoring}: validation is not a one-time gate but a continuous activity, because the
world drifts away from the data on which the model was built.

\emph{Stress testing} is the sharp end of this philosophy. Where ordinary validation asks ``is the
model accurate on average and in-sample,'' stress testing asks ``what happens in the tail, under
conditions we have not recently seen.'' Supervisory stress testing (CCAR/DFAST in the United States)
subjects a portfolio to severe but plausible macro-financial scenarios and asks whether the institution
survives. The discipline distinguishes \emph{scenario analysis} (forward simulation
under specified adverse states), \emph{sensitivity analysis} (perturb one factor, observe the response),
and \emph{reverse stress testing}. The last matters most for what follows. Reverse stress testing
inverts the question, and rather than asking ``how bad is scenario $X$,'' it asks ``what scenario breaks
us,'' and works backward from institutional failure to the conditions that would cause it. It matters
when the analyst cannot enumerate the dangerous scenarios in advance, and we will argue that this is
the generic situation for agentic AI.

\subsection{The Basel decomposition and the meaning of capital}

The quantitative core that stress testing protects is, for credit portfolios, the Basel IRB
decomposition of expected loss \citep{basel2006,bcbs2005irb}:
\begin{equation}
\EL \;=\; \PD \cdot \LGD \cdot \EAD,
\label{c1:eq:el}
\end{equation}
where $\PD$ is the probability of default over a horizon, $\LGD=1-\text{(recovery rate)}$ is the
fraction of exposure lost given default, and $\EAD$ is the exposure at default (with off-balance-sheet
commitments converted to exposure via a \emph{credit conversion factor}, $\CCF$). Equation~\eqref{c1:eq:el}
is an expectation, and the regulatory architecture's genuine content is what sits \emph{beyond} the
expectation. Economic and regulatory capital are sized not against expected loss but against
\emph{unexpected loss}, the distance from the mean to a high quantile (the $99.9\%$ point in the IRB
formula). Expected loss is priced into spreads and provisioned. The analytic engine behind that quantile
is the asymptotic single-risk-factor (ASRF) model of \citet{vasicek2002distribution} and
\citet{gordy2003risk}: each obligor defaults when a latent variable, driven by a common systematic factor
and an idiosyncratic shock, crosses a threshold. The single common factor makes losses correlate,
makes the tail fat, and makes diversification incomplete. It is the mathematical reason a
portfolio of individually small risks can fail \emph{together}.

The rest of this chapter turns on the claim that \emph{capital} has no agentic analogue, so it is worth
saying what capital is. Capital is a fungible, additive, ex-post buffer denominated in the same
unit as the loss it absorbs. Its role is encoded in the defining axiom of a monetary risk measure,
\emph{translation (cash) invariance}: $\rho(X+m)=\rho(X)-m$, ``a unit of capital offsets a unit of
loss.'' Capital works because money and losses share a unit, and, crucially, monetary losses are
\emph{reversible}. A loss can in principle be made good by a later inflow, so a buffer held against it is
meaningful. We return to each of these properties in \S\ref{c1:ssec:nocapital}.

\subsection{Why agentic genAI systems pose a model-risk-like problem}

An \emph{agentic} genAI system is a model embedded in a loop with tools. It observes, plans, and
\emph{acts} over a trajectory of many steps, with limited or deferred human review. Acting here means
issuing API calls, moving funds, sending messages, editing records, executing code. Such systems are now
being deployed in the high-materiality functions (customer operations, payments, code deployment,
data handling) that model risk management was developed to govern. In its own terms they are models whose
outputs drive consequential decisions, so the MRM lens is not a metaphor but the appropriate
regulatory frame.

Traditional model risk management does not anticipate what happens when many such agents run on the
\emph{same} foundation model. The firm-level model risk this chapter formalises then becomes a
\emph{common exposure} shared across deployments. The appropriate frame shifts from microprudential
model risk to the macroprudential concern of financial stability, the subject of
Chapter~\ref{ch:fleet} onward. The single-agent object built here is the unit from which that systemic
analysis is assembled.

\paragraph{Scope and generality.} Although we draw our vocabulary from banking, the framework is
\emph{domain-general}. Its objects are an agent's affordances and reach, the containment wrapped around
them, and a harm vector whose axes are \emph{deployer-defined}, and the mathematics never inspects what those
axes mean. A fleet of code-automation agents (harm axes: production outage, data loss, security
regression), clinical-triage agents (patient-harm severity, privacy, mis-triage), or autonomous trading
agents (financial loss, market-abuse, compliance) instantiate the identical structure. They differ only in
how the axes, the affordance-to-reach map, and the stress states are filled in. The one structural feature
that is universal, and the only one our theorems depend on, is the split between \emph{reversible}
and \emph{irreversible} harm, which exists in every domain. We carry a single stylized financial deployment
through the chapter for concreteness. It is one instantiation, not the framework, and a financial reading is
the special case in which the PD/LGD/EAD vocabulary is most literal rather than analogical.

Yet the fit is partial, and the partiality is the interesting part, since the disanalogies are not incidental
but structural, and they dictate the mathematics. We organise the chapter around three of them.

\paragraph{Break 1: outputs are actions, not estimates.} A credit model emits a \emph{number} (a $\PD$,
a price) that a human or downstream system then uses. The harm, if any, is mediated by that downstream
use, since a gap of time, review, and reversibility typically separates the erroneous estimate from the
realised loss. An agent emits an \emph{action} that has effect in the world directly, so the quantity of
interest is not estimation error but the consequence of actions. The right severity variable is
not ``how wrong'' but ``how irreversible and how far-reaching,'' a point we make precise as
$\text{severity}\approx\text{reversibility}\times\text{blast radius}$.

\paragraph{Break 2: the loss-driving factor is an adaptive adversary.} The Basel systematic factor is an
economic state: exogenous, non-strategic, and, within a regime, statistically stationary, so that
historical default frequencies are informative about future ones. The dominant tail driver for a deployed
agent is instead an \emph{adversary}: prompt injection in user-supplied content, poisoned retrieval,
jailbreaks, tool-confusion attacks. Adversaries are strategic, in that they optimise against the control.
They are non-stationary, in that yesterday's attack distribution does not bound tomorrow's. And they may
exploit \emph{evaluation awareness}: a system that can detect it is being tested can behave differently
under evaluation than in deployment (\emph{sandbagging}), with no banking analogue. This is why the
relevant uncertainty is Knightian rather than risk in the Knight sense, and why a robust/ambiguity-averse
formulation is not a refinement but a necessity.

\paragraph{Break 3: there is no analogue of capital.} Developed in \S\ref{c1:ssec:nocapital}.

\paragraph{} The first break tells us \emph{what to measure} (consequence, not error), while the second tells
us \emph{under what} to measure it (a set of adversary models, not a single stationary distribution). The
third tells us \emph{what the answer is denominated in}, and the answer is not money, which is why the object
cannot be a scalar capital number.

\subsection{What transfers, and what does not}

Table~\ref{c1:tab:mapping} summarises the mapping, and the \emph{governance
scaffolding} of MRM transfers almost intact; the \emph{decomposition} transfers structurally, but with the
independence assumptions inverted; while the \emph{capital/aggregation} layer does not transfer at all and must
be rebuilt.

\begin{table}[htbp]
\centering
\footnotesize
\setlength{\tabcolsep}{4pt}
\renewcommand{\arraystretch}{1.2}
\caption{What transfers from bank MRM / Basel to agentic genAI stress testing.}
\label{c1:tab:mapping}
\begin{tabular}{p{0.235\linewidth} p{0.265\linewidth} p{0.11\linewidth} p{0.245\linewidth}}
\toprule
\textbf{MRM / Basel concept} & \textbf{Agentic analogue} & \textbf{Transfers?} & \textbf{Key caveat} \\
\midrule
Effective challenge (MRM) & Independent red-teaming & Partial & adversary adapts; needs incentive + influence \\
Three lines of defence & Build / risk / audit for agents & Yes & \\
Model inventory & Agent register, tiered by autonomy $\times$ reversibility $\times$ blast radius & Yes & tiering sets required containment \\
Conceptual soundness review & Architecture / containment review & Yes & \\
Ongoing monitoring & Telemetry $+$ base-model version drift & Partial & non-stationary; eval-awareness \\
Output is an estimate & Output is an action & \textbf{No} & consequence, not error \\
$\EL=\PD\cdot\LGD\cdot\EAD$ & $\EH=\PHA\cdot\SGH\cdot\BR$ & Structural & factors co-move (wrong-way) \\
$\PD$ & $\PHA=1-(1-h)^T$ & Partial & thin, non-stationary base rates \\
$\LGD=1-\text{recovery}$ & $\SGH=\text{mag}\times(1-\rec)$ & Partial & $\rec=$ detect $\times$ remediate \\
$\EAD$ / $\CCF$ & Blast radius / capability conversion & Partial & dormant permissions activate under stress \\
ASRF single factor (Vasicek) & Common-factor adversarial trigger & Yes & used as the adverse model \\
Reverse stress testing & Catastrophe-rate reverse stress & \textbf{Amplified} & drives prevention; see Thm.~\ref{c1:thm:prevention} \\
Diversification & Fleet pooling & \textbf{Fails (monoculture)} & shared base model $\Rightarrow$ comonotone \\
Economic capital (UL buffer) & Containment architecture & \textbf{No} & vector-valued; irreversible; non-fungible \\
Translation invariance & --- & \textbf{Fails} & $\to$ containment-subadditivity \\
Scenario design (CCAR) & Adversary ambiguity set $+$ cost $\alpha(Q)$ & Reframed & $\alpha=$ attacker cost, not implausibility \\
\bottomrule
\end{tabular}
\end{table}

\subsection{Why there is no analogue of capital}
\label{c1:ssec:nocapital}

The deepest disanalogy deserves its own argument, because it is the load-bearing premise for the
set-valued formulation. Capital, as a risk-management primitive, presupposes three properties of the loss
it buffers, and agentic harm violates all three.

\emph{(i) A common unit.} Capital is meaningful because a portfolio's losses share a denomination
(money), so a single scalar buffer can be held against their sum, and agentic harm is intrinsically
\emph{multi-dimensional}, since a session may leak credentials, move funds erroneously, and take a
non-compliant action, and no defensible exchange rate converts ``one exposed customer record''
into ``one dollar'' into ``one compliance breach.'' Fixing such rates does not simplify the
problem, but hides a value judgement inside a constant. The object therefore takes values in
$\R^d$, and risk cannot be a single number.

\emph{(ii) Additive, fungible absorption.} Capital absorbs loss additively and is fungible across
sources. A dollar of capital offsets a dollar of loss wherever it arises, which is translation
invariance $\rho(X+m)=\rho(X)-m$. The agentic buffer is \emph{containment}: human-in-the-loop review,
reversibility and rollback, permission scoping, blast-radius caps, rate limits, kill switches. Containment
is \emph{not} fungible, because review capacity cannot be spent as reversibility; the two retire different
harm dimensions. Nor does containment act additively: oversight removes a \emph{stochastic fraction} of
harmful actions, reversibility scales severity \emph{multiplicatively} by $(1-\rec)$, and a kill switch
helps only if detection beats the clock. Because the efficacy of a unit of containment is itself random
and state-dependent, the clean cash-invariance axiom fails. The appropriate relaxation is
\emph{containment-subadditivity}, by direct analogy to the \emph{cash-subadditive} measures of
\citet{elkaroui2009cash}, who relaxed cash-invariance because the discount factor was uncertain.

\emph{(iii) Reversibility.} This is the property with no fix, since capital presupposes that loss is the kind
of thing a later buffer can make good. A monetary loss today can be offset by capital held against it
because money is reversible. Some agentic harm is \emph{irreversible}. An exfiltrated dataset cannot be
un-exfiltrated. An irrevocable external payment cannot be clawed back. A destructive action on production
infrastructure cannot be undone by holding a reserve. For irreversible harm there is no ex-post, so the
absorptive-buffer concept is \emph{categorically inapplicable}, not just hard to denominate, and
must be replaced by \emph{ex-ante prevention}: making the catastrophic action structurally impossible
rather than holding a buffer against its consequences. Theorem~\ref{c1:thm:prevention} turns this
qualitative claim into a quantitative one. Detective and reversibility controls cannot satisfy an
acceptance constraint on irreversible harm at any budget, and one can certify this before running a
single simulation.

\subsection{Why risk must be vector- or set-valued}

Once harm is a vector and the buffer is non-fungible containment, the right mathematical object is a
\emph{set-valued risk measure} \citep{jouini2004vector,hamel2010duality,hamel2011setvalued}, and this is not
an aesthetic preference, since set-valued measures were invented for markets with transaction costs and several
non-exchangeable reserve assets \citep{kabanov1999hedging}, which is structurally the agentic situation:
several non-exchangeable containment levers and no common numéraire. The value $\rho(X)$ becomes not a
number but the \emph{set of containment configurations that render the deployment acceptable}. Its
lower boundary is a Pareto frontier of efficient containment portfolios, while the ordering cone encodes
which cross-dimension trade-offs are permitted, with \emph{hard, non-negotiable limits} represented as
directions the cone may not cross. A scalar expected-harm number can represent none of this, since it cannot
hold a deontological constraint, cannot express non-substitutability, and cannot refuse to grant
diversification credit to a correlated fleet. The set-valued object can, and does.

\subsection{Chapter Outlines}

This chapter makes four contributions.
\begin{enumerate}[leftmargin=1.4em,itemsep=2pt]
\item \textbf{A transfer analysis (this section and \S\ref{c1:sec:decomposition}).} We map the MRM/stress-testing
philosophy and the Basel decomposition onto agentic systems, stating what transfers, what
inverts, and what breaks. We then propose $\EH=\PHA\cdot\SGH\cdot\BR$, with reversibility as the recovery
rate and blast radius as exposure.
\item \textbf{A set-valued risk measure (\S\ref{c1:sec:setvalued}).} We define a set-valued, convex,
containment-subadditive measure valued in containment-configuration space, and we audit which coherence
axioms survive. We show its convex dual is a distributionally robust object whose prior set is a space of
adversary models penalised by attack cost.
\item \textbf{Theorems and a solve (\S\ref{c1:sec:theory}).} For a tractable model class we prove
convexity and well-posedness (Theorem~\ref{c1:thm:convex}), the non-substitutability of prevention
with a closed-form infeasibility certificate (Theorem~\ref{c1:thm:prevention}), and a separation result
with closed-form shadow prices (Proposition~\ref{c1:prop:separation}). The last carries an
intuition-reversing corollary: a coherent tail bound can subsume a chance constraint. We solve the robust
convex program and cross-check the analytic optimum against a brute-force grid.
\item \textbf{An MRM-pipeline adaptation (\S\ref{c1:sec:mrm}).} We translate the validation life cycle into
agentic terms: inventory and tiering, effective challenge, monitoring, governance. We replace the
capital step with a robust containment solve whose dual variables are the marginal cost of risk appetite.
\end{enumerate}

\noindent
Monte-Carlo studies in \S\ref{c1:sec:numerics1} demonstrate the empirical phenomena that motivate and
validate the theory: wrong-way risk, the non-substitutability wall, the failure of cash-invariance, and
the monoculture loss of diversification. Those studies use a Basel-style single-factor adverse model, and
\S\ref{c1:sec:conclusions} states assumptions, limitations, and extensions. Throughout, we emphasise that
our numerical parameters are \emph{illustrative}. The contribution is the \emph{structure}, and the
load-bearing conclusions are theorems that do not depend on the calibration.

\section{Background and related work}
\label{c1:sec:background}

\subsection{Coherent and convex risk measures}

Let $X$ denote a random net loss on a probability space; we follow the loss-positive convention. A
\emph{monetary risk measure} $\rho$ is a functional assigning to $X$ a capital requirement.
\citet{artzner1999coherent} called $\rho$ \emph{coherent} if it satisfies: monotonicity ($X\le Y
\Rightarrow \rho(X)\le\rho(Y)$); translation invariance ($\rho(X+m)=\rho(X)+m$ for deterministic $m$,
under the loss convention); positive homogeneity ($\rho(\lambda X)=\lambda\rho(X)$, $\lambda\ge0$); and
subadditivity ($\rho(X+Y)\le\rho(X)+\rho(Y)$), the formal expression of diversification.
\citet{follmer2002convex} and \citet{frittelli2002putting} relaxed positive homogeneity and
subadditivity to \emph{convexity}, $\rho(\lambda X+(1-\lambda)Y)\le\lambda\rho(X)+(1-\lambda)\rho(Y)$,
which captures liquidity effects in which large positions scale super-linearly. Every convex measure
admits a dual (robust) representation,
\begin{equation}
\rho(X)\;=\;\sup_{Q\in\Qset}\bigl\{\,\E_Q[X]-\alpha(Q)\,\bigr\},
\label{c1:eq:dual}
\end{equation}
where $\Qset$ is a set of probability measures and $\alpha$ is a penalty (the minimal penalty is the
convex conjugate of $\rho$). Coherence corresponds to $\alpha\in\{0,\infty\}$, that is, a worst case over a
set of ``generalised scenarios'' \citep{delbaen2002coherent,follmer2016stochastic}. Equation~\eqref{c1:eq:dual}
is the formal bridge between risk measurement and stress testing. A convex risk measure \emph{is} a
worst-case expected loss over a penalised family of scenarios, which is what a stress-testing
programme computes.

We use two facts repeatedly. First, value-at-risk $\VaR_\tau$ (the $\tau$-quantile of loss) is
\emph{not} subadditive and \emph{not} convex, so it cannot serve as the basis of a coherent or convex
measure. The distinction matters because our acceptance set must be built on a coherent statistic
(see Theorem~\ref{c1:thm:convex}). Second, \emph{expected shortfall} (ES, equivalently CVaR/TVaR),
\begin{equation}
\ES_\tau(X)\;=\;\frac{1}{1-\tau}\int_\tau^1 \VaR_u(X)\,du
\;=\;\min_{\eta\in\R}\Bigl\{\eta+\tfrac{1}{1-\tau}\,\E\bigl[(X-\eta)_+\bigr]\Bigr\},
\label{c1:eq:es}
\end{equation}
is coherent. The variational (Rockafellar--Uryasev) form on the right of \eqref{c1:eq:es}
\citep{rockafellar2000optimization,rockafellar2002conditional} makes it a convex, monotone functional of
$X$ that is tractable in optimisation. We exploit both properties below.

\subsection{Set-valued and vector risk measures}

When loss is $\R^d$-valued with no fixed numéraire, a scalar measure is ill-posed.
\citet{jouini2004vector} introduced vector-valued coherent measures, and \citet{hamel2010duality,
hamel2011setvalued,hamel2013setvalued} developed the \emph{set-valued} theory and its duality on the
complete lattice of upper closed convex sets. In this framework $\rho(X)$ is a set of $\R^d$-valued
``eligible'' positions $u$ such that $X-u$ is acceptable, $\rho(X)=\{u\in\Mset: X-u\in\Acc\}$, an upper
set with respect to a cone order. The motivating application is markets with proportional transaction
costs and several reserve assets that cannot be freely exchanged \citep{kabanov1999hedging}, encoded by a
\emph{solvency cone}. We adopt this language, but we make one substantive change in
\S\ref{c1:sec:setvalued}. Containment does not act on harm by subtraction $X-u$; it acts through a
\emph{nonlinear, stochastic mitigation map} $X\mapsto X(m)$. Our measure is therefore a generalisation in
which the translation is replaced by that map, and that substitution is the structural reason
translation invariance fails.

\subsection{Cash-subadditive measures}

\citet{elkaroui2009cash} observed that when future cash flows are discounted by an \emph{uncertain}
factor, a unit of cash today no longer deterministically offsets a unit of loss tomorrow. Translation
invariance should therefore be relaxed to \emph{cash-subadditivity}, $\rho(X+m)\ge\rho(X)-m$. The
agentic analogue is exact. A unit of containment does not deterministically offset a unit of harm,
because its efficacy (detection, remediation, blocking) is stochastic and state-dependent, so we
therefore call the agentic measure \emph{containment-subadditive}. \S\ref{c1:sec:numerics1} exhibits the failure of
one-for-one substitution empirically (Figure~\ref{c1:fig:subadditivity}), and \S\ref{c1:sec:theory} shows
it is a feature of the log-affine mitigation map.

\subsection{Operational-risk LDA and the single-factor model}

We reuse two pieces of bank machinery directly. For aggregating rare, heavy-tailed events we use the
operational-risk \emph{loss distribution approach} (LDA; \citealp{frachot2001loss,embrechts1997modelling}),
which compounds a frequency distribution (typically Poisson) with a heavy-tailed severity. That
description fits agentic incident data better than the credit IRB machinery does, because agentic harmful
events are rare, severe, and poorly described by a fixed base rate. For inducing correlation we use the
ASRF/Vasicek single-factor construction \citep{vasicek2002distribution,gordy2003risk}: a latent variable
$\sqrt{r}\,F+\sqrt{1-r}\,\varepsilon$ crosses a threshold, with $F$ a common factor and $r$ the
asset-correlation analogue. In \S\ref{c1:sec:numerics1} this device makes the harm dimensions
co-move (wrong-way risk), and it makes a fleet sharing one base model behave comonotonically (the
monoculture).

\subsection{AI evaluation, red-teaming, and the adversary}

On the AI side, our work connects to the literature on \emph{dangerous-capability evaluations} and
safety cases \citep{shevlane2023model,phuong2024evaluating}, and to the broader catalogue of catastrophic
risks from highly capable systems \citep{hendrycks2023overview}. The model-risk-management principle of
``effective challenge'' is the governance ancestor of red-teaming. The technical novelty in the agentic
setting is that the challenger faces an \emph{adaptive} opponent and that the system under test may
exhibit evaluation awareness. We do not contribute new evaluation methods. We contribute a
risk-measurement and optimisation layer that consumes their outputs and returns a containment
configuration and its dual prices. Those outputs are estimates of per-step hazard, recoverability, and
blast radius, together with a family of adversary models. To our knowledge the specific synthesis is new:
set-valued, containment-subadditive risk measures with an attack-cost-penalised distributionally robust
dual, applied to agentic AI. The mathematical components are individually classical.

\section{A PD/LGD/EAD-type decomposition for agentic harm}
\label{c1:sec:decomposition}

\subsection{Episodes and the harm vector}

Our unit of analysis is an \emph{episode}: one bounded interaction of the agent with its environment. In
the running example an episode is a single customer session comprising up to $T$ consequential tool calls,
the decision points at which a harmful action is possible. An episode produces a harm vector
\begin{equation}
X \;=\; (X_1,\dots,X_d)\in\R^d_+,
\end{equation}
whose components are denominated in incommensurable units. The number $d$ and the axes themselves are
\emph{deployer-defined}, and the decomposition that follows is indifferent to what they are. In the
stylized financial instantiation we carry through the chapter, $d=3$ with
$X=(X_{\mathrm{fin}},X_{\mathrm{data}},X_{\mathrm{comp}})$: financial loss in dollars, a data-exposure
index, and a compliance-breach index. A code-automation deployment would read these as outage, data loss,
and security regression, and the analysis is unchanged. The choice of $d$ and of the dimensions encodes
which harms the institution regards as non-substitutable
(\S\ref{c1:sec:setvalued}).

\subsection{The three factors}

By direct analogy with \eqref{c1:eq:el} we decompose the expected harm in a single dimension as
\begin{equation}
\EH \;\approx\; \PHA \cdot \SGH \cdot \BR,
\label{c1:eq:eh}
\end{equation}
the product of a probability of harmful action, a severity given harm, and a blast radius. Each factor
has a precise reading and a precise caveat.

\paragraph{$\PHA$ (probability of harmful action) $\leftrightarrow \PD$.} The cleanest formulation is not
a per-episode probability but a \emph{per-step hazard} $h$ integrated over the trajectory:
\begin{equation}
\PHA \;=\; 1-(1-h)^T,
\label{c1:eq:pha}
\end{equation}
which makes explicit that risk compounds with autonomy. Doubling the horizon $T$, so that more steps
elapse before review, raises the probability of at least one harmful action super-linearly at small $h$.
The caveat is that the hazard $h$, unlike $\PD$, has \emph{thin, non-stationary base rates}: harmful
actions are rare, and the adversary actively works to invalidate any historical estimate of $h$. This is
the first reason the credit-IRB machinery, which leans on stable rating-grade $\PD$s, is the wrong
aggregation engine and the operational-risk LDA is closer to right.

\paragraph{$\SGH$ (severity given harm) $\leftrightarrow \LGD$.} We write
\begin{equation}
\SGH \;=\; \text{magnitude}\times(1-\rec),\qquad
\rec \;=\; \Prob(\text{detect in time})\times\Prob(\text{remediate}\mid\text{detected}),
\label{c1:eq:sgh}
\end{equation}
where \emph{recoverability} $\rec$ plays the role of the recovery rate in $\LGD=1-\text{recovery}$.
This identification is the technical heart of the ``severity $\approx$ reversibility $\times$ blast
radius'' slogan. Reversibility \emph{is} the recovery term, and it is a product of a detection probability
and a conditional remediation probability. A containment investment can raise both, but the caveat is that
$\rec$ is endogenous to the adversary. A capable attacker suppresses logs and races remediation, so
$\rec$ \emph{falls} in the states where magnitude rises, which seeds wrong-way risk.

\paragraph{$\BR$ (blast radius) $\leftrightarrow \EAD$.} Blast radius is the exposure an action can reach:
the number of records, accounts, systems, or dollars within its scope. As in Basel's treatment of
undrawn commitments via the $\CCF$, the relevant exposure is not the nominal one but the \emph{stressed}
one. Dormant permissions, latent tool access, and privilege-escalation paths are activated under
adversarial conditions, so a \emph{capability conversion factor} maps quiescent capability into realised
exposure. The caveat mirrors $\LGD$'s, since the conversion factor is highest under stress.

\subsection{Expected versus unexpected harm}

As in banking, \eqref{c1:eq:eh} is an expectation, and the regulatory content lies beyond it. We
distinguish \emph{expected harm} from \emph{unexpected harm}. Expected harm is budgetable, the analogue of
provisioned $\EL$. Unexpected harm is the distance from the mean to a high quantile of the harm
distribution, and the containment architecture must be sized to cover it, just as capital covers
unexpected loss at the $99.9\%$ point. The crucial difference, developed in \S\ref{c1:ssec:nocapital}, is
that the buffer covering unexpected harm is not capital but containment. For the irreversible component
there is no buffer at all, only prevention.

\subsection{Wrong-way risk and the common factor}
\label{c1:ssec:wrongway}

In counterparty credit risk, \emph{wrong-way risk} is the tendency of exposure to rise as
creditworthiness falls. The risk factors are adversely correlated, so the product of marginal
expectations understates the joint tail \citep{gregory2015xva}. The agentic version is sharper, and
structural rather than incidental. A single adversarial event \emph{simultaneously} raises the action
hazard $h$, because the attack induces harmful actions; lowers recoverability $\rec$, because the attack
suppresses detection; and expands the blast radius $\BR$, because the attack escalates privilege. The
three factors of \eqref{c1:eq:eh} are therefore \emph{comonotone}, driven by a common adversarial factor,
and
\begin{equation}
\E[\PHA\cdot\SGH\cdot\BR] \;\gg\; \E[\PHA]\cdot\E[\SGH]\cdot\E[\BR].
\end{equation}
Multiplying marginal estimates implicitly treats the factors as independent, which understates expected
harm and, far more severely, understates the tail. We model this with the ASRF construction and quantify
it in \S\ref{c1:ssec:v1}. In the worked example the true joint expected loss is $2.1\times$ the product of
marginals, the $99.9\%$ quantile is $210\times$ expected loss, and $98\%$ of the worst $1\%$ of episodes
are adversarially triggered despite a $3\%$ unconditional trigger rate. The same common factor destroys
diversification across a fleet sharing one base model (\S\ref{c1:ssec:v4}). The per-agent failures are
comonotone, so the fleet earns essentially no diversification credit. A coherent measure correctly reports
that, and a variance-based aggregation would miss it.

\section{A set-valued risk measure for agentic systems}
\label{c1:sec:setvalued}

\subsection{Containment as the eligible-asset space}

Let $m\in\Mset\subseteq\R^p_+$ be a \emph{containment configuration}: a vector of levels of
non-fungible mitigation levers. In the worked example $p=3$ and $m=(c_1,c_2,c_3)$, with $c_1$ the
human-in-the-loop review capacity, $c_2$ the reversibility / blast-radius coverage, and $c_3$ a
preventive control that makes a designated catastrophic action class structurally impossible. Containment
acts on harm through a \emph{mitigation map}
\begin{equation}
m \;\longmapsto\; X(m), \qquad X(0)=X,\quad X(m)\preceq X(0)\ \text{(componentwise, a.s.)},
\end{equation}
which is generally nonlinear and stochastic. Review removes a random subset of harmful actions,
reversibility scales severity by $(1-\rec(m))$, and prevention zeroes a sub-class. This map replaces the
additive translation $X-u$ of the classical set-valued theory (\S2.2), and that substitution is the single
structural fact from which the rest follows.

\subsection{The acceptance set and the ordering cone}

The \emph{acceptance set} $\Acc$ is the institution's safety case made into a mathematical object. It is the
set of residual-harm distributions the institution is willing to hold. We take it to be an intersection
of a coherent-tail \emph{soft cone} and a \emph{hard limit}:
\begin{equation}
\Acc \;=\;
\Bigl\{\,X:\ \ES_{\tau_k}(X_k)\le B_k\ \ \forall k,\ \ \text{and}\ \
\Prob\bigl(\text{irreversible catastrophic event}\bigr)\le\varepsilon\,\Bigr\}.
\label{c1:eq:acc}
\end{equation}
The soft constraints are sublevel sets of a coherent statistic, one per harm dimension, with appetites
$B_k$. The hard constraint is a chance constraint on an irreversible event class, with tolerance
$\varepsilon$. Geometrically, the soft cone permits trade-offs \emph{within} the appetite box, while the hard
limit is a face the configuration may not cross at any compensation, the deontological constraint that a
scalar expected-harm criterion cannot represent.

The \emph{ordering cone} $K\subseteq\R^d$ encodes which cross-dimension substitutions are admissible, so a
wide $K$ says dollars, data, and compliance trade off at some rate, while a degenerate (half-space) direction
encodes a dimension in which no trade-off is permitted. Crucially, the geometry doubles as an
\emph{org chart}, and each face of $\Acc$ is owned by a second-line function: treasury owns $B_{\mathrm{fin}}$,
the CISO owns $B_{\mathrm{data}}$, and compliance owns $B_{\mathrm{comp}}$. The off-axis tilt of $K$ is a
cross-domain substitution only an enterprise risk committee can authorise, while the hard limit is a bright
line set outside the trade space entirely (board, legal, regulator). We expand on this governance reading
in \S\ref{c1:sec:mrm}.

\subsection{The set-valued measure and its boundary}

\begin{definition}[Containment risk measure]
\label{c1:def:rho}
The set-valued risk measure is the \emph{containment requirement set}
\begin{equation}
\rho(X) \;=\; \bigl\{\,m\in\Mset:\ X(m)\in\Acc\,\bigr\},
\label{c1:eq:rho}
\end{equation}
the set of containment configurations that render the residual harm acceptable. Under the robust
formulation of \S\ref{c1:ssec:dual}, $X(m)$ is replaced by the worst case over an admitted family of
adversary models.
\end{definition}

Because the mitigation map is monotone in $m$ and $\Acc$ is monotone (downward closed in harm),
$\rho(X)$ is an \emph{upper set}: if $m\in\rho(X)$ and $m'\succeq m$ then $m'\in\rho(X)$. Its lower
boundary $\partial\rho(X)$ is the \emph{efficient-containment frontier}, the set of configurations from
which no lever can be reduced without another being increased to stay acceptable. Equip $\Mset$ with a
cost $\kappa^\top m$; choosing a point on (or above) this frontier becomes the optimisation problem
of \S\ref{c1:sec:theory}. The frontier is the set of Pareto-efficient containment portfolios, and the
solve picks the cheapest robustly acceptable one. Figure~\ref{c1:fig:setvalued} displays $\rho(X)$ and its
frontier for the worked example.

\subsection{Checking Which Axioms Survive}
\label{c1:ssec:axioms}

Reading $\rho$ through the Artzner et al.\ axioms (\S2.1) is diagnostic.

\begin{enumerate}[leftmargin=1.6em,itemsep=3pt]
\item \emph{Monotonicity holds.} More harm in every dimension shrinks the acceptable set:
$X\preceq X'\Rightarrow\rho(X)\supseteq\rho(X')$. This is the set-valued reading of ``more risk needs
more capital.''

\item \emph{Translation / cash invariance fails.} There is no deterministic $m$ for which
$X(m)$ equals $X$ shifted by a constant, because containment efficacy is stochastic and the map is
multiplicative, not additive. The relaxation is \emph{containment-subadditivity}, the analogue of
\citet{elkaroui2009cash}. \S\ref{c1:ssec:v3} shows that the marginal harm-reduction of a unit of
containment \emph{falls} by nearly an order of magnitude from the first unit to the last
(Figure~\ref{c1:fig:subadditivity}), the empirical signature of the failure of one-for-one
substitution.

\item \emph{Diversification holds formally but is void for a monoculture.} Subadditivity holds for any
coherent measure regardless of dependence, but for \emph{comonotone} risks it holds with \emph{equality}.
A fleet of agents on a single base model has failures that are comonotone through the shared factor, so
it earns \emph{zero} diversification credit. \S\ref{c1:ssec:v4} shows independent agents diversify toward
expected harm while a shared-model fleet floors at roughly five times the independent capital
(Figure~\ref{c1:fig:monoculture}). That a coherent measure \emph{refuses} to reward fake diversification is a
feature.

\item \emph{Positive homogeneity is dropped.} Scaling the deployment is super-linear in risk once
oversight bandwidth saturates and blast radii overlap. We therefore retain \emph{convexity} but not
positive homogeneity, so the measure is convex, not coherent.
\end{enumerate}

\noindent
The net object is a \emph{set-valued, convex, containment-subadditive} risk measure, valued in
containment-configuration space, ordered by a cone that carries the hard limits.

\subsection{The convex dual and adversarial ambiguity}
\label{c1:ssec:dual}

The adversary enters through the dual. By the convex-duality representation \eqref{c1:eq:dual}, a worst
case over a family of adversary models governs the robust acceptability of a configuration. We write
$Q$ for an adversary model: a setting of the trigger rate, coordination, blast radius, catastrophe
propensity, and in-session hazard. We depart from the finance convention in the meaning of the
penalty $\alpha(Q)$:
\begin{equation}
\rho(X) \;=\; \Bigl\{\,m:\ \sup_{Q}\bigl[\,\E_Q[\,\text{harm}(m)\,]-\alpha(Q)\,\bigr]\in\Acc\,\Bigr\},
\qquad
\Qset_\delta=\{Q:\alpha(Q)\le\delta\}.
\label{c1:eq:robust}
\end{equation}
In the finance literature $\alpha(Q)$ measures the statistical \emph{implausibility} of $Q$, that is, how
unlikely nature is to produce it. For an adversary, $Q$ is not drawn by nature but \emph{chosen by an
optimiser}, so $\alpha(Q)$ should be the \emph{cost to the attacker} of forcing the system into model $Q$:
effort, sophistication, and detection risk. Equation~\eqref{c1:eq:robust} then reads as a
\emph{Stackelberg game} in which the attacker maximises harm net of attack cost and the defender minimises the
worst case via containment. The binding $Q^\star$ is not the most extreme scenario but the most
\emph{cost-effective} one, the one yielding most harm per unit $\alpha$.

The admitted set $\Qset_\delta$ is an ambiguity ball. A KL ball yields the entropic penalty. A
Wasserstein ball \citep{esfahani2018data} gives robustness to attacks \emph{off} the support of historical
data, the right instinct for jailbreaks that do not yet exist. A parametric box is the interpretable form
that maps to red-team controls \citep{bental2013robust,rahimian2019distributionally}.
\S\ref{c1:ssec:v5} calibrates a small parametric ambiguity set and shows (Figure~\ref{c1:fig:adversary})
that the worst-case $Q^\star$ is attained at \emph{different} adversary settings for different
constraints. It also shows that the realistic binding scenario is a moderate \emph{simultaneous} push on
all levers, a coordinated campaign, rather than any single extreme lever.

\section{Numerical studies I: validating the phenomena}
\label{c1:sec:numerics1}

We validate by Monte-Carlo simulation the four phenomena the theory rests on, and then harden the
acceptance set. The aim is empirical falsification: we engineer each claim so the simulation either
exhibits it or refutes it. All parameters are illustrative, and the section comprises two numerical
studies: four validations V1--V4 of the phenomena the theory rests on, and an acceptance-set
hardening exercise.

\subsection{The descriptive simulation model}
\label{c1:ssec:mcmodel}

The simulation instantiates the framework in a single stylized deployment, a customer-operations
agent fleet read in financial terms. That reading is purely for concreteness, since the same model with
relabelled axes (e.g.\ a code-automation fleet, with the catastrophic class being an irreversible
destructive operation rather than an irrevocable transfer) produces identical structure, and no
conclusion below depends on the financial reading.

We simulate $N=5\times10^5$ episodes of $T=12$ consequential steps under common random numbers. A
fleet-common systematic factor $F$ drives an adversarial \emph{trigger} through a latent-normal
copula. Episode $i$ is triggered when $\sqrt{r}\,F+\sqrt{1-r}\,\varepsilon_i<\Phi^{-1}(\pi)$, with
unconditional trigger rate $\pi=0.03$ and correlation $r=0.55$. A triggered episode enters an
adversarial regime in which, \emph{simultaneously}, the per-step hazard rises (e.g. financial
$0.004\!\to\!0.18$), recoverability collapses ($\rec:0.85\!\to\!0.25$), and the blast-radius
multiplier expands ($1.0\!\to\!2.5$), the mechanical realisation of the wrong-way coupling
of \S\ref{c1:ssec:wrongway}. Within the financial dimension a sub-class of \emph{catastrophic}
actions (irreversible external transfers, $\rec=0$, median magnitude \$80k, threshold
$\theta=\$50$k) fires with probability $0.005$ nominally and $0.10$ under trigger.

Containment is two-dimensional in this descriptive model, and the lever $c_1$ reviews the top-salience
actions up to capacity and blocks a reviewed harmful action with probability $0.9$; review
saturates, so coverage falls when many actions flag at once. The lever $c_2$ raises recoverability
and shrinks reach with diminishing returns. The acceptance set uses appetites
$B_{\mathrm{fin}}=\$2{,}500$, $B_{\mathrm{data}}=6$, $B_{\mathrm{comp}}=6$, and hard tolerance
$\varepsilon=10^{-3}$.

\subsection{V1: wrong-way risk}
\label{c1:ssec:v1}

\begin{figure}[htbp]
\centering
\includegraphics[width=\linewidth]{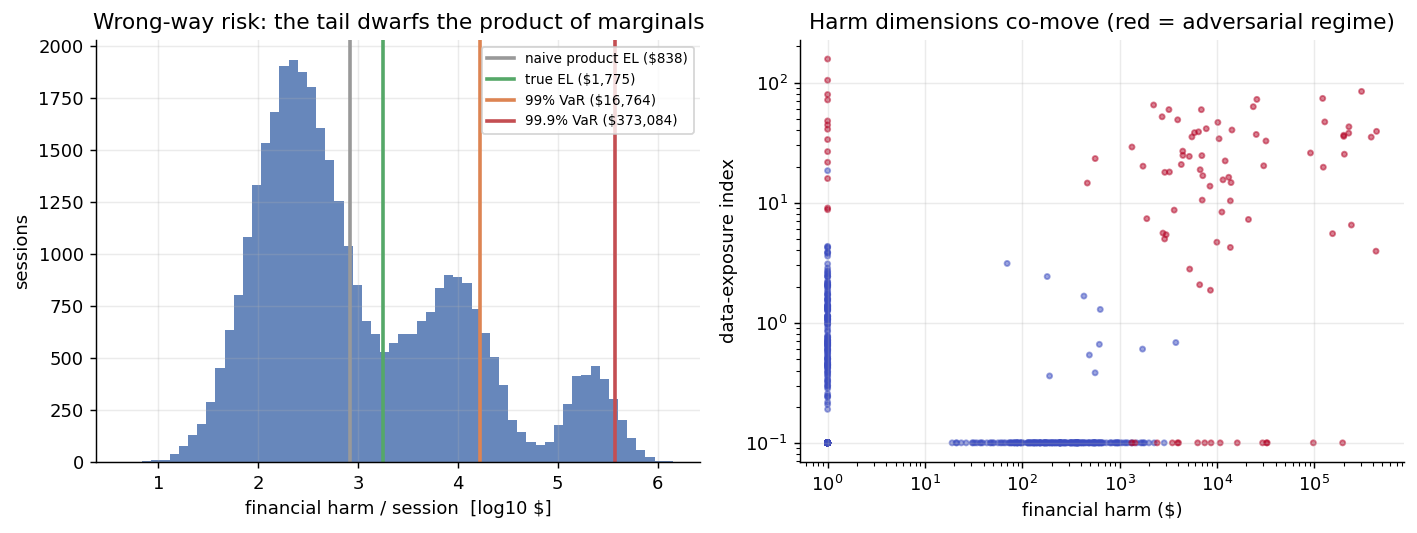}
\caption{\textbf{Wrong-way risk.} Left: distribution of financial harm per episode (no containment).
The true expected loss, and especially the high quantiles, dwarf the product-of-marginals estimate.
Right: financial and data harm co-move, with the adversarial regime (red) populating the joint tail.
The factors $\PHA,\SGH,\BR$ are comonotone through the common trigger, so independence understates
the tail.}
\label{c1:fig:wrongway}
\end{figure}

Treating the factors of \eqref{c1:eq:eh} as independent and multiplying the marginal means gives a
naive expected financial loss of \$838. The true joint expectation is \$1{,}775, a factor of $2.12$
larger, purely from the regime co-movement. The tail is where the error compounds, and the $99\%$ VaR is
\$16{,}764 ($9.4\times$ expected loss) and the $99.9\%$ VaR is \$373{,}084 ($210\times$). Most
tellingly, $98\%$ of the worst $1\%$ of episodes are adversarially triggered, despite triggering
being a $3\%$ event unconditionally. Figure~\ref{c1:fig:wrongway} shows the loss distribution with
the naive and true moments marked, and the co-movement of the financial and data dimensions coloured
by regime. This is the quantitative statement that the product of marginals is
imprecise and, worse, blind to the tail.

\subsection{V2: the set-valued risk measure}
\label{c1:ssec:v2}

\begin{figure}[htbp]
\centering
\includegraphics[width=0.74\linewidth]{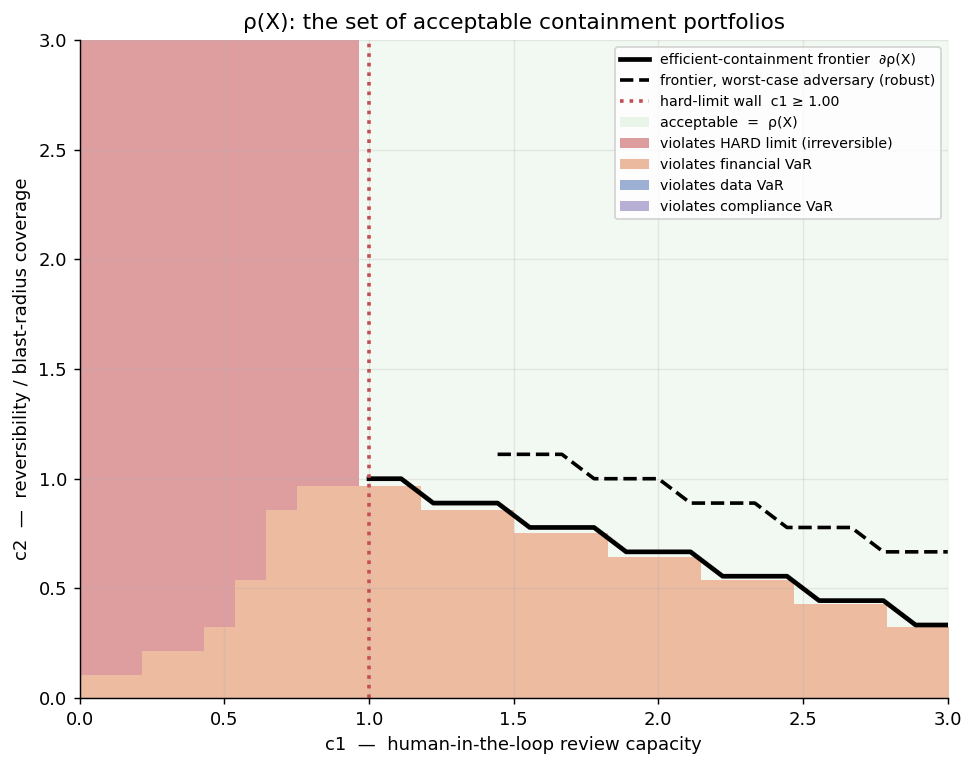}
\caption{\textbf{The set-valued risk measure $\rho(X)$.} The acceptable set of containment
configurations (green), coloured elsewhere by the binding constraint. The red wall is the hard-limit
constraint: no amount of reversibility ($c_2$) substitutes for the review ($c_1$) needed to retire
the irreversible event. Solid: efficient-containment frontier $\partial\rho(X)$. Dashed: the frontier
under a worst-case $1.5\times$ adversary.}
\label{c1:fig:setvalued}
\end{figure}

Figure~\ref{c1:fig:setvalued} renders $\rho(X)$ over the containment grid $[0,3]^2$, coloured by the
binding constraint. The measure is a genuine \emph{region}, since the origin $(0,0)$ is rejected and $53\%$
of the grid is acceptable. The decisive feature is the vertical wall, since below $c_1\approx1.0$ the
deployment violates the hard limit \emph{for every value of} $c_2$, because in this descriptive model
reversibility cannot reach the irreversible event and only review/blocking can. That
non-substitutability is the ordering cone's hard face appearing as geometry, and the solid curve is the
efficient-containment frontier $\partial\rho(X)$. The dashed curve is that frontier under a
$1.5\times$ adversary, which shrinks the acceptable region from $53\%$ to $38\%$, the
robust/ambiguity layer of \eqref{c1:eq:robust} in action.

\subsection{V3: containment-subadditivity (failure of cash-invariance)}
\label{c1:ssec:v3}

\begin{figure}[htbp]
\centering
\includegraphics[width=\linewidth]{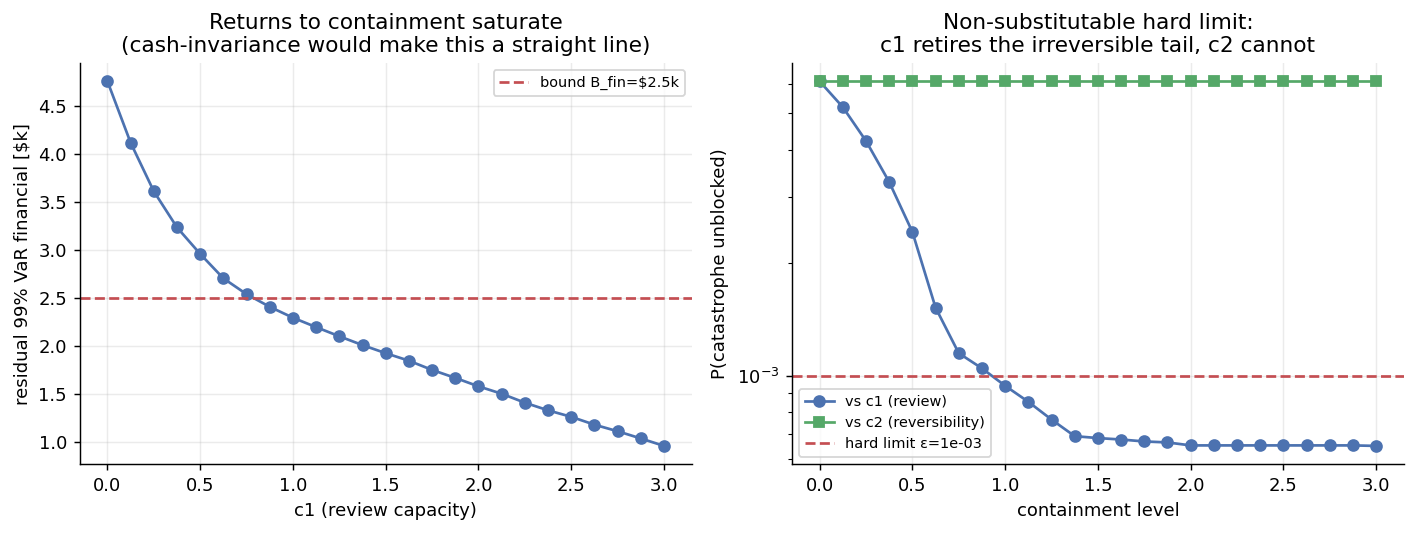}
\caption{\textbf{Containment-subadditivity.} Left: residual financial $99\%$ VaR against review
capacity $c_1$; the saturating curve, rather than a straight line, is the failure of cash-invariance.
Right: the probability of an unblocked irreversible catastrophe falls below the hard tolerance
$\varepsilon$ as $c_1$ rises but is flat in $c_2$, so prevention/blocking is non-substitutable.}
\label{c1:fig:subadditivity}
\end{figure}

If the measure were cash-invariant, residual risk would fall one-for-one with containment, tracing a
straight line. Instead the marginal value of $c_1$ falls from \$4{,}056 per unit near zero to \$589
per unit near saturation, a $6.9\times$ decline (Figure~\ref{c1:fig:subadditivity}, left). A unit of
containment does not subtract a fixed quantity of harm, because its efficacy is stochastic and
state-dependent. The right panel makes the hard-limit point quantitative and non-substitutable, since the
probability of an unblocked catastrophe falls below $\varepsilon$ as $c_1$ rises but is \emph{flat}
in $c_2$, the empirical signature of containment-subadditivity and the direct motivation for
the cash-subadditive relaxation of \S\ref{c1:ssec:axioms}.

\subsection{V4: the monoculture and the loss of diversification}
\label{c1:ssec:v4}

\begin{figure}[htbp]
\centering
\includegraphics[width=0.66\linewidth]{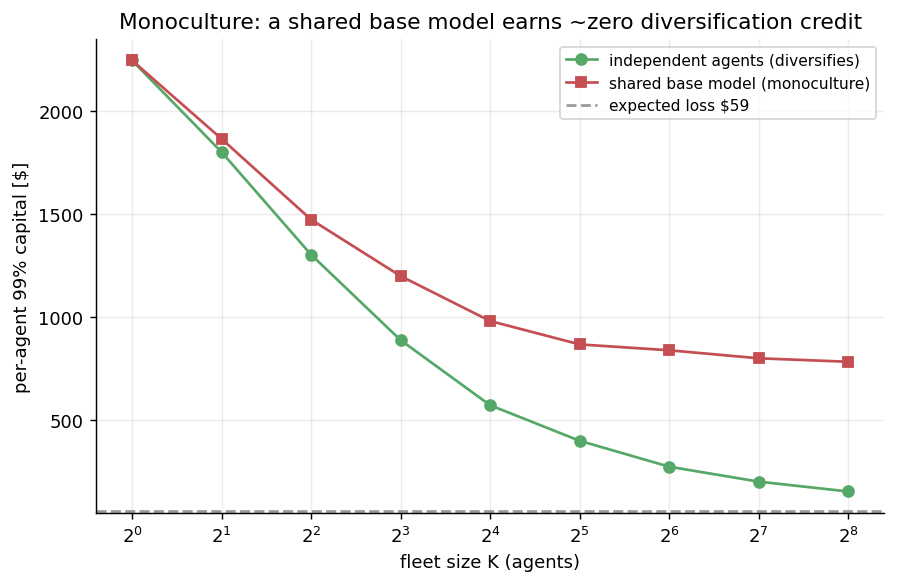}
\caption{\textbf{Monoculture.} Per-agent $99\%$ capital against fleet size $K$. Independent agents
diversify toward expected loss. A fleet on a shared base model floors far above it (here
$\approx5\times$ the independent fleet), because the common factor renders per-agent failures
comonotone.}
\label{c1:fig:monoculture}
\end{figure}

A single agent requires \$2{,}245 of (per-agent, $99\%$) containment-equivalent capital against the
recoverable financial tail. A fleet of $256$ \emph{independent} agents diversifies this down to \$153
per agent, toward the \$59 expected loss, the familiar $1/\sqrt{K}$ pooling benefit. A fleet
of $256$ agents sharing \emph{one base model} floors at \$782 per agent, roughly five times the
independent fleet and $0.35\times$ the single-agent figure
(Figure~\ref{c1:fig:monoculture}). It floors there because the shared factor makes the agents
near-comonotone, and comonotone risk earns no diversification credit. A coherent measure charges the full systematic load, the correct behaviour and
the formal content of the ``monoculture'' warning.

\subsection{Hardening the acceptance set}
\label{c1:ssec:v5}

\begin{figure}[htbp]
\centering
\includegraphics[width=0.66\linewidth]{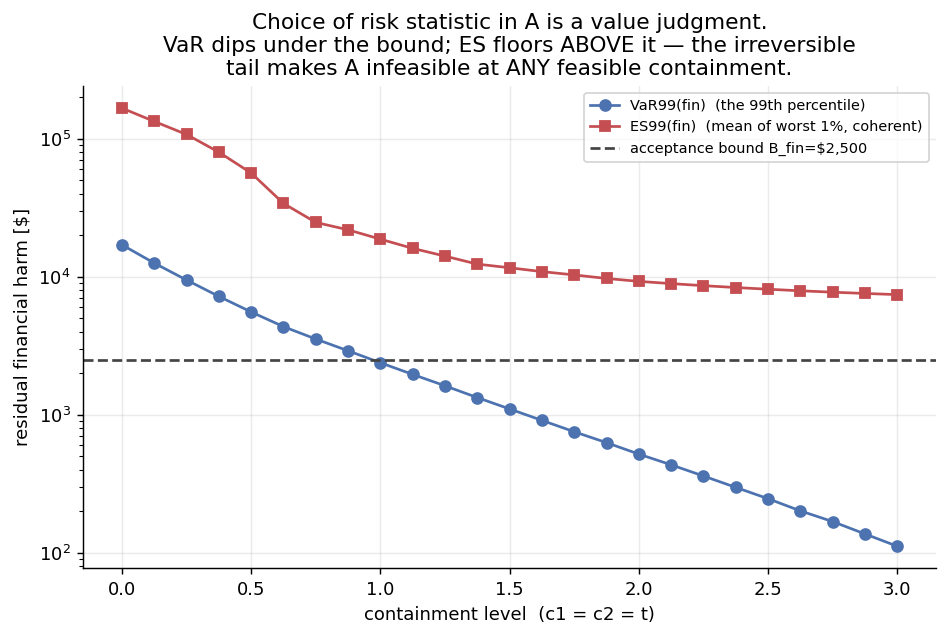}
\caption{\textbf{$\VaR$ versus $\ES$ in the acceptance set.} Residual financial harm against
containment (diagonal $c_1=c_2=t$). $\VaR_{0.99}$ dips below the appetite; the coherent $\ES_{0.99}$
floors above it. The choice of statistic is a value judgement: a coherent tail measure makes the
irreversible catastrophic tail visible and, here, renders $\Acc$ infeasible without a dedicated
preventive control.}
\label{c1:fig:varves}
\end{figure}

\paragraph{The statistic in $\Acc$ is a value judgement with teeth.} Switch the financial constraint
from $\VaR_{0.99}$ to the coherent $\ES_{0.99}$ at the \emph{same} \$2{,}500 appetite, and the answer
changes qualitatively (Figure~\ref{c1:fig:varves}). At zero containment $\ES_{0.99}\approx\$166$k
against $\VaR_{0.99}\approx\$17$k, a $9.8\times$ gap. As containment increases along the diagonal,
the VaR curve dips under the bound near unit containment while the ES curve \emph{plateaus above it}
and never crosses. The reason is the irreversibility we built in, since detection/blocking caps the
\emph{probability} of the catastrophic action, but each instance is $\sim\$80$k and unrecoverable, so
the tail-\emph{mean} cannot be driven down. VaR was quietly hiding an unacceptable tail; ES reveals
it. VaR is also not subadditive, so building $\Acc$ on it would forfeit the convexity
Theorem~\ref{c1:thm:convex} needs. The acceptance set should use a coherent statistic, and the
moment it does, the model reports that the descriptive two-lever design is infeasible.

\paragraph{Defending $\varepsilon$ by reverse stress.} The hard tolerance $\varepsilon$ is a
frequency claim about an irreversible event, so we defend it by reverse stress from throughput. In
the descriptive model, review alone floors the per-episode unblocked-catastrophe probability at
$\approx6.6\times10^{-4}$. At a volume of $10^6$ sessions per year this is $\approx660$ irreversible
catastrophic events annually, and an $\varepsilon=10^{-3}$ limit would nominally ``permit''
$1{,}000$. At any realistic deployment volume a per-episode probabilistic hard limit on an
irreversible event is therefore not a meaningful safeguard. This converges with
Figure~\ref{c1:fig:varves}, since a \emph{detective} control is stochastic and has a floor, so it cannot
satisfy a requirement about an \emph{irreversible} tail. The defence of the hard limit is
consequently not a value of $\varepsilon$ but an \emph{architecture} claim. The catastrophic class
must be made structurally impossible, driving $\varepsilon\to0$ by construction, which is the
preventive lever $c_3$ introduced in \S\ref{c1:sec:theory} and the content of
Theorem~\ref{c1:thm:prevention}.

\begin{figure}[htbp]
\centering
\includegraphics[width=0.92\linewidth]{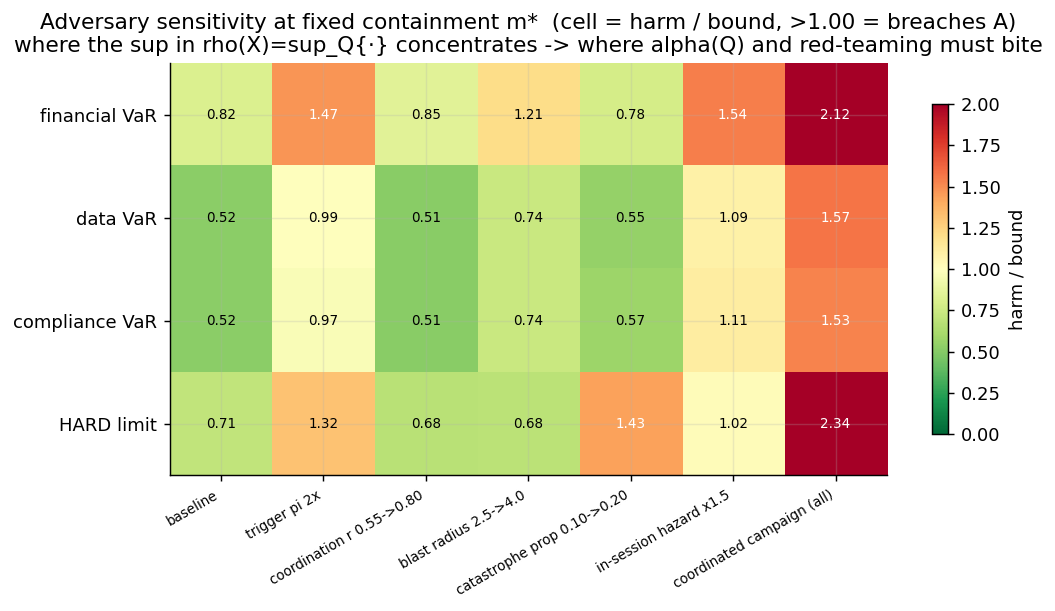}
\caption{\textbf{Adversary sensitivity at fixed containment.} Each cell is residual harm divided by
its appetite at $m^\star=(1.6,0.9)$; values $>1$ (warmer) breach $\Acc$. Different constraints bind
under different adversary settings, so the worst-case $Q^\star$ of \eqref{c1:eq:robust} is a
\emph{mixture} across the ambiguity set. The realistic binding scenario is the moderate coordinated
campaign (right column).}
\label{c1:fig:adversary}
\end{figure}

\paragraph{Where the adversary's worst case concentrates.} Figure~\ref{c1:fig:adversary} fixes a
reference containment $m^\star=(c_1,c_2)=(1.6,0.9)$ that is acceptable under the baseline adversary
$Q_0$. For each single-lever stress it reports the residual harm as a multiple of its appetite; a
cell $>1$ breaches $\Acc$. The attribution is informative. An in-session hazard increase of
$1.5\times$ breaches almost every constraint; doubling the trigger rate breaches the financial and
hard constraints; expanding blast radius hits the financial dimension specifically; and raising
catastrophe propensity hits \emph{only} the hard limit. The telling negative result concerns
coordination, since raising the correlation $r$ from $0.55$ to $0.80$ breaches \emph{nothing}. Correlation drives the
\emph{aggregate/fleet} tail (the monoculture), not the per-episode distribution this acceptance set
constrains. Which $Q^\star$ the dual's supremum activates therefore depends on whether $\Acc$ is
written per-episode or at the fleet level, and the penalty $\alpha$ must be calibrated differently at
each granularity. Finally, the \emph{coordinated campaign} column applies a moderate simultaneous
push on all levers, and it breaches every constraint by $1.5$ to $2.3\times$. The cost-effective
adversary is a broad nudge, not one spectacular lever, and that is where $\alpha(Q)$ must bite.

\section{A tractable model class, theorems, and the solve}
\label{c1:sec:theory}

The phenomena of \S\ref{c1:sec:numerics1} are robust, but we demonstrated them in a descriptive simulation.
That simulation's review mechanic (salience-ranked, stochastic, saturating) is deliberately realistic, and
therefore non-convex. We now isolate a \emph{tractable model class} in which the structural claims become
theorems, the robust containment program of \eqref{c1:eq:robust} becomes a convex program, and its optimum is
available in closed form. The two models agree on every qualitative conclusion, though the tractable class gives
provability at the cost of a stylised mitigation mechanism.

\subsection{The model class}
\label{c1:ssec:assumptions}

Recall that an episode produces a harm vector $X=(X_1,\dots,X_d)\in\R^d_+$, and that containment is a vector
$m=(c_1,\dots,c_p)\in\Mset\subseteq\R^p_+$ acting on harm to produce residual harm $X(m)$. We specialise to
$d=3$ (financial, data, compliance) and $p=3$ containment levers: $c_1$ (detective review/blocking), $c_2$
(reversibility/blast-radius coverage), and $c_3\in[0,1)$ (\emph{prevention}: the fraction of the
irreversible catastrophic action class made structurally impossible). We write the financial harm as a sum of
a recoverable part and an irreversible catastrophic part, $X_{\mathrm{fin}}=X^{\mathrm{nc}}_{\mathrm{fin}}+X^{\mathrm{cat}}$.

\begin{assumption}[Tractable class]
\label{c1:ass:class}
\leavevmode
\begin{enumerate}[label=\textnormal{(A\arabic*)},leftmargin=2.6em,itemsep=2pt]
\item \textbf{Linear harm aggregation.} Each residual harm dimension is a nonnegative combination of base
severities times deterministic attenuations,
$X_k(m)=\sum_{a}\beta_a\,\phi_{k,a}(m)$, with $\beta_a\ge 0$ random and $\phi_{k,a}(m)\in[0,1]$ deterministic.
\item \textbf{Log-affine recoverable attenuation.} For recoverable contributions the attenuation is
$\phi(m)=\exp\!\bigl(-\textstyle\sum_j \lambda_j c_j\bigr)$ with efficacies $\lambda_j\ge 0$; hence $\phi$ is
log-affine, therefore convex and nonincreasing in each $c_j$.
\item \textbf{Prevention acts only on the irreversible class.} The catastrophic contribution attenuates as
$X^{\mathrm{cat}}(1-c_3)$ --- affine in $c_3$ and \emph{independent of} $c_1,c_2$. Each catastrophic event has
magnitude at least the threshold $\theta>0$, i.e. $X^{\mathrm{cat}}\ge \theta\,\mathbf 1\{\text{cat event}\}$.
\item \textbf{Coherent statistic, finite ambiguity set.} The acceptance statistic is expected shortfall
$\ES_{\tau}$, eq.~\eqref{c1:eq:es}; the admitted adversary set $\Qset_\delta=\{Q:\alpha(Q)\le\delta\}$ is a finite
family of laws (a scenario set).
\end{enumerate}
\end{assumption}

\noindent Assumptions (A1)--(A4) suffice for Theorems~\ref{c1:thm:convex}--\ref{c1:thm:prevention}. The separation
result adds:

\paragraph{Does the set-valued apparatus earn its keep?} A reader may reasonably ask why
Section~\ref{c1:sec:setvalued} builds a set-valued object when
Theorems~\ref{c1:thm:convex}--\ref{c1:thm:prevention} are proved inside the scalarised class of
Assumption~\ref{c1:ass:class}. The division of labour is deliberate, and the set-valued layer is where we
\emph{state} the claims; the tractable class is where we \emph{compute} them. Three things depend on
the former and cannot be recovered from any single scalar functional.

First, non-substitutability is a statement about the \emph{geometry} of the acceptance set, not about
the value of a functional. To say that prevention cannot be traded against review and reversibility is
to say that a direction of $\Mset$ does not lie in the recession cone of the acceptable region. Once
$c_3$ is fixed too low, no movement along $c_1,c_2$ whatsoever reaches acceptability. A scalar risk
number can report that a given configuration fails, but only the set can express that an entire family of
substitutions fails, which is what the claim asserts.

Second, hard limits are cone constraints, not large penalties, since a deontological limit names an action class
that must be \emph{impossible}, not just expensive, and we represent it as a direction the ordering cone
may not cross. Under any scalarisation it becomes a finite penalty and is therefore purchasable at some
price. The set-valued object is what keeps ``not for sale'' distinct from ``very expensive.''

Third, the certificate of Theorem~\ref{c1:thm:prevention} asserts that \emph{no} containment
configuration is acceptable, meaning the acceptance set is empty. We compute the certificate
through $\ES_\tau$, but what it certifies is a property of the set.
Corollary~\ref{c1:cor:subsume} compares the coherent tail constraint against the explicit
chance constraint and identifies when one subsumes the other, stating that the two
scalarisations agree on that property.

What the tractable class gives is computability: (A1)--(A4) reduce the frontier to a convex program with
closed-form optima and shadow prices. We therefore regard the scalarisation as a computational device
applied to a set-valued primitive, not as the primitive itself, and the boundary is that the
numerical studies exercise one scalarisation. The \emph{statements} are about the set:
non-substitutability, the infeasibility certificate, the subsume relation. It is the set-valued frame
that makes them well-posed.

\begin{assumption}[Disjoint channels]
\label{c1:ass:disjoint}
Each recoverable harm dimension $k$ is attenuated by a single dedicated lever $j(k)$ with efficacy
$\lambda_{j(k)}$, and the containment cost is separable and linear, $\kappa^\top m$ with $\kappa>0$. (In the
instance below: review $c_1$ governs compliance, reversibility $c_2$ governs data, and both govern recoverable
financial harm; the catastrophic class is governed by prevention $c_3$ alone.)
\end{assumption}

The robust containment program specialises \eqref{c1:eq:robust} to
\begin{equation}
C^\star=\min_{m\ge 0}\ \kappa^\top m
\quad\text{s.t.}\quad
\max_{Q\in\Qset_\delta}\ES^{Q}_{\tau_k}\!\bigl(X_k(m)\bigr)\le B_k\ \ \forall k,
\quad
\max_{Q\in\Qset_\delta}\Prob_Q\!\bigl(X^{\mathrm{cat}}(1-c_3)\ge\theta\bigr)\le\varepsilon .
\label{c1:eq:program}
\end{equation}

\subsection{Convexity and well-posedness}

\begin{theorem}[Convexity]
\label{c1:thm:convex}
Under \textnormal{(A1)--(A4)}, the robust feasible set
$\mathcal F=\{m\ge0:\text{constraints of }\eqref{c1:eq:program}\}$ is closed and convex, and \eqref{c1:eq:program}
is a convex program with linear objective. If $\mathcal F\neq\varnothing$ then, because $\kappa>0$ makes the
objective coercive on $\R^p_+$, a global minimiser exists; the Karush--Kuhn--Tucker conditions are necessary
and sufficient for optimality, and the optimal multipliers equal the shadow prices $\partial C^\star/\partial B_k$
and $\partial C^\star/\partial\varepsilon$. The conclusion fails for $\VaR$ in place of $\ES$.
\end{theorem}

\begin{proof}
By (A2) each recoverable attenuation $\phi_{k,a}$ is a composition of an affine map with $\exp$, hence convex.
By (A3) the catastrophic attenuation $1-c_3$ is affine, hence convex. As $\beta_a\ge 0$ (A1), each
$m\mapsto X_k(m)$ is a nonnegative combination of convex functions, so it is convex \emph{pathwise} and
nonincreasing in $m$. Fix $Q$. Expected shortfall is convex and monotone with respect to the almost-sure
order (a coherent risk measure; \citealp{artzner1999coherent,rockafellar2000optimization}). For
$\lambda\in[0,1]$, pathwise convexity gives
$X_k(\lambda m+(1-\lambda)m')\le \lambda X_k(m)+(1-\lambda)X_k(m')$ a.s., so by monotonicity and convexity of
$\ES$,
\[
\ES^Q_{\tau_k}\!\bigl(X_k(\lambda m+(1-\lambda)m')\bigr)
\le \ES^Q_{\tau_k}\!\bigl(\lambda X_k(m)+(1-\lambda)X_k(m')\bigr)
\le \lambda\,\ES^Q_{\tau_k}\!\bigl(X_k(m)\bigr)+(1-\lambda)\,\ES^Q_{\tau_k}\!\bigl(X_k(m')\bigr),
\]
so $m\mapsto\ES^Q_{\tau_k}(X_k(m))$ is convex. The pointwise maximum over the finite set $\Qset_\delta$ of
convex functions is convex, so each soft constraint is a sublevel set of a convex function, hence convex and
closed. For the hard constraint, $c_3\mapsto\Prob_Q(X^{\mathrm{cat}}\ge\theta/(1-c_3))$ is nonincreasing in
$c_3$ for each $Q$: as $c_3$ rises the threshold $\theta/(1-c_3)$ rises, so the exceedance probability
of the fixed law of $X^{\mathrm{cat}}$ falls. Its maximum over $\Qset_\delta$ is therefore nonincreasing, and
the feasible set $\{m:\max_Q\Prob_Q(\cdot)\le\varepsilon\}=\{m:c_3\ge\bar c_3\}$ is a closed
halfspace. The intersection $\mathcal F$ of closed convex sets with $\R^p_+$ is closed and convex. A linear
objective that is coercive on the closed set $\mathcal F$ attains its minimum. The constraints satisfy
Slater's condition whenever a strictly feasible point exists, and are otherwise affine in the binding
prevention direction, so KKT is necessary and sufficient, and the envelope theorem identifies the
multipliers with the constraint shadow prices. Finally, $\VaR_\tau$ is neither convex nor subadditive
\citep{artzner1999coherent}, so $m\mapsto\VaR^Q_\tau(X_k(m))$ need not be convex and the feasible set need not
be convex. The convexity of the program is therefore contingent on the coherent statistic mandated in
\eqref{c1:eq:acc}.
\end{proof}

\begin{remark}
Theorem~\ref{c1:thm:convex} is the formal payoff of the hardening in \S\ref{c1:ssec:v5}. Choosing a coherent tail
statistic is the modelling choice that sees the irreversible tail $\VaR$ hides, and it is also the
\emph{enabling} choice, since it is what makes the containment-sizing problem a convex program with meaningful
dual prices.
\end{remark}

\subsection{Prevention is non-substitutable: an a-priori infeasibility certificate}

\begin{theorem}[Necessity of prevention]
\label{c1:thm:prevention}
Under \textnormal{(A1)--(A4)}, with $p^\star:=\max_{Q\in\Qset_\delta}\Prob_Q(\text{cat event})$ and
financial tail level $\tau$,
\begin{equation}
\inf_{c_1,c_2\ge 0}\ \max_{Q\in\Qset_\delta}\ES^{Q}_{\tau}\!\bigl(X_{\mathrm{fin}}(m)\bigr)\Big|_{c_3=0}
\;=\;\max_{Q\in\Qset_\delta}\ES^{Q}_{\tau}\!\bigl(X^{\mathrm{cat}}\bigr)
\;\ge\;\theta\,\min\!\Big\{1,\ \frac{p^\star}{1-\tau}\Big\}.
\label{c1:eq:certificate}
\end{equation}
Consequently, if $\theta\,\min\{1,p^\star/(1-\tau)\}>B_{\mathrm{fin}}$, the acceptance set is infeasible for
\emph{every} $(c_1,c_2)$: no amount of detection or reversibility satisfies the financial constraint, and a
preventive control $c_3>0$ is necessary for feasibility. The left-hand bound is computable from the adverse
model alone, with no optimisation over containment.
\end{theorem}

\begin{proof}
At $c_3=0$ the catastrophic term is $X^{\mathrm{cat}}$, which by (A3) does not depend on $(c_1,c_2)$. The
recoverable term $X^{\mathrm{nc}}_{\mathrm{fin}}(m)\ge 0$ and, by (A2), decreases to $0$ as $c_1$ or $c_2\to\infty$.
Hence $X_{\mathrm{fin}}(m)\ge X^{\mathrm{cat}}$ a.s.\ for all $(c_1,c_2)$, and by monotonicity of $\ES$ and
dominated convergence the infimum over $(c_1,c_2)$ equals $\max_Q\ES^Q_\tau(X^{\mathrm{cat}})$, proving the
equality. For the bound, (A3) gives $X^{\mathrm{cat}}\ge\theta\,\mathbf 1\{\text{cat event}\}$, so by
monotonicity $\ES^Q_\tau(X^{\mathrm{cat}})\ge\theta\,\ES^Q_\tau(\mathbf 1\{\text{cat event}\})$. For an
indicator with $\Prob_Q(\text{cat})=p$, the Rockafellar--Uryasev representation gives the tail mean of a
$\{0,1\}$ variable capped at its maximum value $1$, i.e.\ $\ES_\tau(\mathbf 1)=\min\{1,\ p/(1-\tau)\}$.
Taking the maximum over $Q$ gives $\max_Q\ES^Q_\tau(X^{\mathrm{cat}})\ge \theta\,\min\{1,\ p^\star/(1-\tau)\}$.
\end{proof}

\paragraph{Numerical certificate.} The instance parameters are $\theta=\$50{,}000$, catastrophe
probability $p^\star\approx 0.0059$ (the worst rate in the illustrative ambiguity grid), and $\tau=0.99$. The closed-form bound is
$\theta p^\star/(1-\tau)\approx\$29{,}708$, and the simulated infimum is
$\max_Q\ES_{0.99}(X^{\mathrm{cat}})=\$147{,}979$. The infimum exceeds the bound because catastrophe magnitudes are
lognormal with median $\$80$k, well above $\theta$. Both exceed the financial appetite $B_{\mathrm{fin}}=\$2{,}500$
by one to two orders of magnitude, so the two-lever design of \S\ref{c1:sec:numerics1} is \emph{certified
infeasible} without any search over containment. This is the quantitative form of the irreversibility
argument of \S\ref{c1:ssec:nocapital}. Against irreversible harm, the absorptive-buffer levers are incapable,
not just expensive, and the model says so in closed form. At the single-agent level this is a
microprudential statement, but in Chapter~\ref{ch:fleet} the identical certificate becomes a \emph{systemic}
one, and the missing absorptive buffer becomes the missing macroprudential instrument. Financial stability
against this class of harm must therefore be engineered into the system ex ante rather than
capitalised against after the fact.

\subsection{Separation, closed-form optima, and shadow prices}

\begin{proposition}[Separation and shadow prices]
\label{c1:prop:separation}
Under \textnormal{(A1)--(A5)}, program \eqref{c1:eq:program} separates across levers. For each recoverable
dimension $k$ with dedicated lever $j(k)$,
\begin{equation}
c^\star_{j(k)}=\frac{1}{\lambda_{j(k)}}\ln\!\frac{H_k}{B_k},
\qquad
H_k:=\max_{Q\in\Qset_\delta}\ES^{Q}_{\tau_k}\!\bigl(X^{\textnormal{unc}}_k\bigr),
\label{c1:eq:closedform}
\end{equation}
truncated at $0$, where $X^{\textnormal{unc}}_k$ is uncontained harm. The prevention lever solves the
one-dimensional problem $c^\star_3=\max\{c_3^{\mathrm{fin}},c_3^{\mathrm{hard}}\}$, where $c_3^{\mathrm{fin}}$
is the smallest $c_3$ meeting the worst-case financial-$\ES$ constraint and $c_3^{\mathrm{hard}}$ the smallest
meeting the hard limit. The shadow prices are
\begin{equation}
\frac{\partial C^\star}{\partial B_k}=-\frac{\kappa_{j(k)}}{\lambda_{j(k)}\,B_k},
\qquad
\frac{\partial C^\star}{\partial \varepsilon}=
\begin{cases} 0,& c_3^{\mathrm{fin}}\ge c_3^{\mathrm{hard}}\ \ (\text{hard limit slack}),\\[2pt]
\kappa_3\,\dfrac{\partial c_3^{\mathrm{hard}}}{\partial\varepsilon},& \text{otherwise.}\end{cases}
\label{c1:eq:shadow}
\end{equation}
\end{proposition}

\begin{proof}
By (A5) each soft constraint involves a single lever, and the objective is separable, so the program
decomposes into scalar problems. Positive homogeneity of $\ES_\tau$ and determinism of the attenuation give,
for a dedicated lever $c$ with efficacy $\lambda$,
$\max_Q\ES^Q_\tau\!\bigl(X^{\textnormal{unc}}_k e^{-\lambda c}\bigr)=e^{-\lambda c}\max_Q\ES^Q_\tau(X^{\textnormal{unc}}_k)=e^{-\lambda c}H_k$.
Setting this equal to the appetite $B_k$ and solving gives \eqref{c1:eq:closedform}; monotonicity makes the
constraint bind at the optimum (or $c^\star=0$ if already slack). The catastrophic term couples only to $c_3$
(A3), giving the stated one-dimensional prevention problem. Then $C^\star=\sum_k\kappa_{j(k)}c^\star_{j(k)}+\kappa_3 c^\star_3$,
and differentiating $\kappa_{j(k)}\lambda_{j(k)}^{-1}\ln(H_k/B_k)$ in $B_k$ yields the first equation of
\eqref{c1:eq:shadow}. The prevention term contributes to $\partial C^\star/\partial\varepsilon$ only through the
binding side of $\max\{c_3^{\mathrm{fin}},c_3^{\mathrm{hard}}\}$, giving the second.
\end{proof}

\begin{corollary}[A coherent tail bound can subsume a chance constraint]
\label{c1:cor:subsume}
If the worst-case financial-$\ES$ constraint forces $c_3^{\mathrm{fin}}$ whose induced catastrophe
probability is below $\varepsilon$ --- equivalently $c_3^{\mathrm{fin}}\ge c_3^{\mathrm{hard}}$ --- then the
deontological hard limit is inactive and prices at zero. The coherent tail bound is then the more stringent
of the two, and the explicit chance constraint is redundant.
\end{corollary}

\subsection{Numerical studies II: the solve}
\label{c1:ssec:solve}

\begin{figure}[htbp]
\centering
\includegraphics[width=\linewidth]{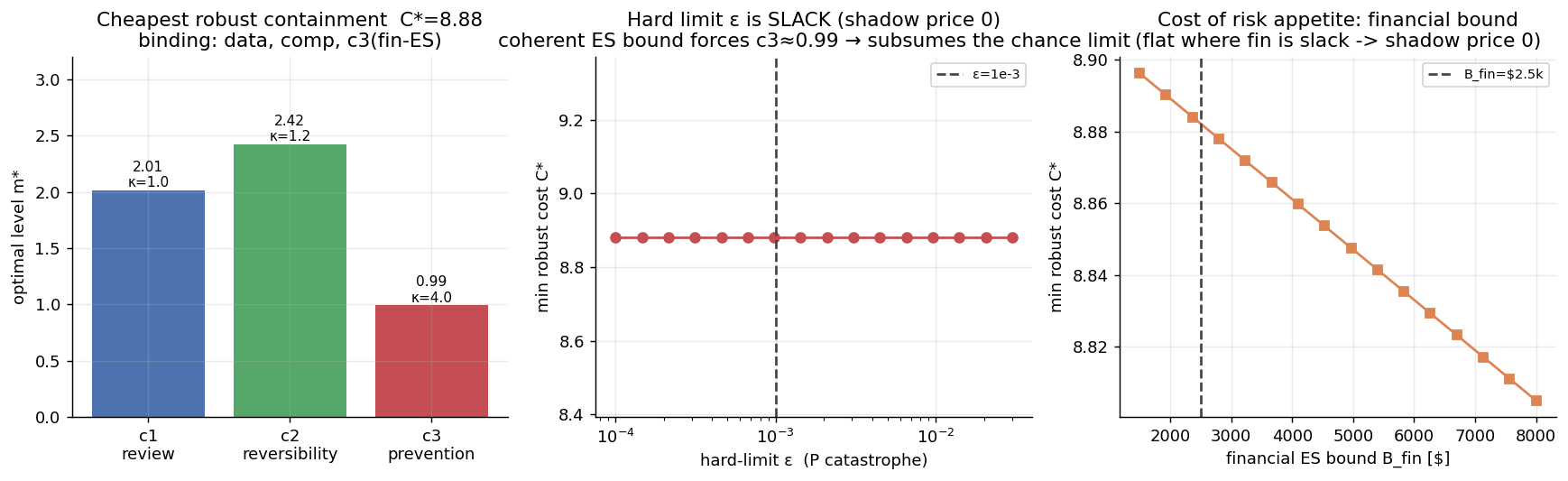}
\caption{\textbf{The robust solve and shadow prices.} Left: the cheapest robust containment portfolio
$m^\star$; prevention $c_3$ is driven to $0.99$ despite its high unit cost ($\kappa_3=4$), as
Theorem~\ref{c1:thm:prevention} requires. Centre: the minimal robust cost is \emph{flat} in the hard tolerance
$\varepsilon$. The hard limit is slack because the coherent financial-$\ES$ bound already forces
$c_3\approx0.99$ (Corollary~\ref{c1:cor:subsume}). Right: cost falls as the financial appetite is relaxed; the
slope is the (small, dollar-denominated) shadow price.}
\label{c1:fig:solve}
\end{figure}

We instantiate \eqref{c1:eq:program} with three levers and the ambiguity set
\[
  \Qset_\delta=\bigl\{Q_0,\ \pi{\times}2,\ \text{blast}{\times}1.6,\ p_{\mathrm{cat}}{\times}2,\ \text{hazard}{\times}1.5\bigr\}
\]
holding out a coordinated ``campaign'' scenario at cost $\alpha>\delta$. Costs are $\kappa=(1.0,1.2,4.0)$, and
the appetites on $\ES$ are $B_{\mathrm{fin}}=\$2{,}500$, $B_{\mathrm{data}}=2.5$, $B_{\mathrm{comp}}=2.8$, with hard
tolerance $\varepsilon=10^{-3}$. Two independent solves agree (Figure~\ref{c1:fig:solve}), and the analytic optimum of
Proposition~\ref{c1:prop:separation} is $m^\star=(c_1,c_2,c_3)=(2.01,2.42,0.991)$ with cost $C^\star=8.88$. A
brute-force robust grid search returns $(2.18,2.45,0.993)$ with $C^\star=9.10$, and the two agree to grid
resolution, which confirms both the separation theorem and the convexity that guarantees the grid
has no better interior point.

\paragraph{The worst case is a mixture.} The binding adversary differs by constraint, since the data and compliance
$\ES$ bounds bind under the elevated-trigger scenario ($\pi{\times}2$), while the financial tail binds under
the elevated-catastrophe scenario ($p_{\mathrm{cat}}{\times}2$). The supremum in \eqref{c1:eq:robust} is thus
attained at \emph{different} $Q$ for different constraints; a single worst-case scenario would have
mis-sized the portfolio. This is the optimisation-level counterpart of the attribution in
Figure~\ref{c1:fig:adversary}.

\paragraph{Shadow prices and the redundant hard limit.} The closed-form multipliers \eqref{c1:eq:shadow} evaluate
to $\partial C^\star/\partial B_{\mathrm{data}}=-0.48$ and $\partial C^\star/\partial B_{\mathrm{comp}}=-0.30$
cost-units per unit of $\ES$ appetite, while the financial bound carries a small dollar-scale price and the
hard tolerance $\varepsilon$ prices at \emph{zero}. The last of these is Corollary~\ref{c1:cor:subsume} made
numerical, since the coherent financial-$\ES$ constraint forces $c_3^{\mathrm{fin}}=0.9915$, exceeding the
$c_3^{\mathrm{hard}}=0.9037$ that the chance constraint alone would require, so the induced catastrophe
probability sits far below $\varepsilon$ and the explicit deontological limit is redundant. This reverses the
intuition of \S\ref{c1:ssec:v5}, in that it is not the hard limit that forces architectural prevention, but the coherent
tail bound, which forces \emph{more} prevention and delivers the hard limit for free. The two solves
reported above are a brute-force robust grid search and the analytic optimum with its KKT shadow
prices (Figure~\ref{c1:fig:solve}).

\section{Adapting the model-risk-management pipeline}
\label{c1:sec:mrm}

The preceding sections argue that the \emph{measurement} object must change, from a scalar to a
set-valued, containment-subadditive measure. They argue too that the \emph{capital} step must change,
from a buffer to a robust containment solve. The governance \emph{process}, by contrast, transfers
with high fidelity. The traditional model-risk-management life cycle of inventory, development
standards, validation as effective challenge, ongoing monitoring, and documented governance is the
right skeleton for agentic systems. We use traditional model risk management as the template because
it is the most mature articulation of the lifecycle, not because the lifecycle is financial. The
stages below are written for an agentic deployment in any domain, with the financial reading as one
instance.

This section makes the translation concrete, stage by stage, and Table~\ref{c1:tab:pipeline} is the
summary; the subsections give the detail and tie each stage to a result of this chapter. The pipeline
described here is microprudential, governing one institution's reliance on one agent, and it is the
foundation, not the whole. Once many agents share a foundation model, this firm-level lifecycle must
be wrapped in a macroprudential layer that governs the \emph{common} exposure and the contagion
channels. That layer is the subject of the fleet governance of Chapter~\ref{ch:fleet} and the
systemic results that follow.

\begin{table}[htbp]
\centering
\small
\renewcommand{\arraystretch}{1.3}
\caption{The MRM life cycle, adapted for agentic genAI. ``Result'' points to the section or theorem that
governs the stage.}
\label{c1:tab:pipeline}
\begin{tabular}{p{0.215\linewidth} p{0.45\linewidth} p{0.21\linewidth}}
\toprule
\textbf{MRM stage} & \textbf{Agentic implementation} & \textbf{Result} \\
\midrule
Model inventory & Agent register; tier each agent by autonomy $\times$ reversibility $\times$ blast radius;
attach a harm-vector profile and a containment configuration $m$. & \S\ref{c1:ssec:p-inventory} \\
Development \& implementation standards & Containment-by-design: the eligible-asset levers (permissions,
review, reversibility, rate limits, kill switches, \emph{preventive boundaries}) are architecture, specified
before deployment. & \S\ref{c1:ssec:p-design}; Thm.~\ref{c1:thm:prevention} \\
Validation / effective challenge & Red-teaming calibrated as adversary \emph{cost} $\alpha(Q)$; find the
cost-effective binding $Q^\star$, defeat sandbagging, not merely the worst conceivable attack. &
\S\ref{c1:ssec:p-validation} \\
Ongoing monitoring & Telemetry on the realised harm vector and on $(\PHA,\rec,\BR)$ versus assumption;
base-model version change is a re-validation trigger. & \S\ref{c1:ssec:p-monitoring} \\
Capital adequacy / stress testing & \emph{Replaced} by the robust containment solve \eqref{c1:eq:program}: size
the cheapest $m^\star$ meeting $\Acc$ across $\Qset_\delta$. & \S\ref{c1:ssec:p-sizing}; Prop.~\ref{c1:prop:separation} \\
Risk appetite \& limits & The acceptance set $\Acc$: appetites $B_k$ per dimension, the ordering cone for
trade-offs, hard limits for irreversible classes; shadow prices price marginal appetite. &
\S\ref{c1:ssec:p-appetite} \\
Aggregation \& portfolio view & Fleet-level $\Acc$ with correlation $\approx1$ for a shared base model;
\emph{no} diversification credit. & \S\ref{c1:ssec:p-aggregation}; \S\ref{c1:ssec:v4} \\
Three lines of defence & Build (1st) / independent agent-risk (2nd) / audit (3rd), with model-and-tool
provenance. & \S\ref{c1:ssec:p-governance} \\
\bottomrule
\end{tabular}
\end{table}

\subsection{Inventory and tiering}
\label{c1:ssec:p-inventory}
The model inventory becomes an \emph{agent register}, and model risk management scales scrutiny along a
materiality axis, and for agents that axis is three-dimensional. \emph{Autonomy} counts the steps
between human checkpoints, which is the horizon $T$ in \eqref{c1:eq:pha} and therefore a multiplier
on the probability of harmful action. \emph{Reversibility} is the recoverability $\rec$ that sits in
the severity term \eqref{c1:eq:sgh}. \emph{Blast radius} is the reachable scope at failure. An agent
that books irreversible external payments with little review is a different risk tier from one that
drafts internal memos, even on the same base model, and it warrants a different containment
configuration. The register records, per agent, the harm-vector profile and the deployed $m$.
Tiering is not paperwork but the input that sets the required containment.

\subsection{Containment-by-design}
\label{c1:ssec:p-design}
Development standards specify the eligible-asset levers \emph{before} deployment, because they are
architecture, not configuration, and the decisive design rule is Theorem~\ref{c1:thm:prevention}. For any
harm class that is both irreversible and able to breach appetite, detective and reversibility
controls are provably insufficient, so a \emph{preventive boundary} is mandatory: a permission the
agent cannot hold, or an irreversible action routed to a separate non-agentic path with its own
authorisation. The infeasibility certificate \eqref{c1:eq:certificate} lets a designer establish this
from the adverse model alone, at design time, before a line of agent code is run. If
$\theta\,\min\{1,p^\star/(1-\tau)\}>B_{\mathrm{fin}}$, prevention is required, and its absence is a design
defect rather than a tuning question.

\subsection{Validation as effective challenge}
\label{c1:ssec:p-validation}
Effective challenge maps to independent red-teaming, but the adversary's strategic, optimising nature
changes what ``challenge'' means. The penalty $\alpha(Q)$ in the dual \eqref{c1:eq:dual} is best read
not as statistical implausibility, its meaning in the finance literature, but as \emph{attacker
cost}: the effort, sophistication, and detection risk required to push the system into law $Q$. The
robust
measure is then a Stackelberg game. The attribution of Figure~\ref{c1:fig:adversary} and the mixture
worst-case of \S\ref{c1:ssec:solve} show that the binding scenario is the most \emph{cost-effective}
attack, a moderate, coordinated campaign across several levers, not the most extreme single lever. A
red team should therefore be resourced and incentivised to find cheap broad attacks, and its findings
calibrate $\alpha$ and hence $\Qset_\delta$. The three ingredients of effective challenge carry over
literally: competence (the team must be able to find jailbreaks), incentive (independence from the
deploying line), and influence (authority to block deployment). One ingredient is new: \emph{evaluation awareness}. A system that behaves differently when it detects evaluation can pass a
red-team gate and fail in deployment. The validation design must itself be robust to sandbagging,
through held-out and deployment-realistic probes --- a problem with no banking analogue.

\subsection{Ongoing monitoring}
\label{c1:ssec:p-monitoring}
Monitoring instruments the realised harm vector and compares the realised factors $(\PHA,\rec,\BR)$
against the values assumed when $m$ was sized. Two triggers are specific to genAI. The first is
\emph{base-model version drift}, since a vendor model update or a prompt/tool change can move every factor
at once and invalidate the ambiguity set, so a version change is a re-validation event, not a silent
dependency bump. The second is \emph{distributional drift in the adversary}, and because $\Qset_\delta$
is a claim about attacks that exist, the emergence of a new attack class is a monitoring signal that
$\delta$ (or the support of $\Qset_\delta$) must be revised and the solve re-run.

\subsection{Robust containment sizing replaces the capital step}
\label{c1:ssec:p-sizing}
This is the stage with no buffer. Where CCAR sizes capital to survive a scenario, the agentic
analogue sizes \emph{containment} to keep residual harm inside $\Acc$ across the whole admitted
adversary set, by solving \eqref{c1:eq:program}. The output is not a number of dollars but a
containment configuration $m^\star$: how much review, how much reversibility, how much preventive
scoping. By Proposition~\ref{c1:prop:separation}, each component has a closed-form value and a dual
price. Crucially, the worst case is a mixture (\S\ref{c1:ssec:solve}), so the sizing must be robust
across $\Qset_\delta$ rather than tuned to one stress. Sizing to a single scenario is the agentic
version of fighting the last war.

\subsection{The acceptance set is the risk-appetite statement}
\label{c1:ssec:p-appetite}
$\Acc$ is where value judgements live, and its geometry maps to an org chart. The natural second-line
functions own the per-dimension appetites $B_k$: treasury/finance for $B_{\mathrm{fin}}$, the
CISO/privacy office for $B_{\mathrm{data}}$, compliance for $B_{\mathrm{comp}}$. The off-axis tilt of
the ordering cone asks whether more data exposure may be traded for fewer dollars of loss. That is a
cross-domain substitution no single owner can authorise, and it belongs to an enterprise risk
committee or the board. The hard limit is a face removed from the trade space entirely, a bright
line set by the board, legal, or a regulator, that no cone direction may cross. The shadow prices
\eqref{c1:eq:shadow} then serve governance directly. They are the marginal cost, in containment, of
loosening each appetite, and they tell the committee which constraints are expensive (here, the
operational-$\ES$ bounds) and which are free at the margin (here, the redundant hard limit, by
Corollary~\ref{c1:cor:subsume}).

\subsection{Aggregation and the monoculture}
\label{c1:ssec:p-aggregation}
A portfolio view must confront the result of \S\ref{c1:ssec:v4}, since a fleet on a shared base model has
failure correlation near one, so the coherent measure grants \emph{no} diversification credit and the
fleet charge is the comonotone sum, not a pooled average. Two consequences follow for the pipeline, since aggregate appetite cannot be set as if agents were independent, and $\Acc$ must also be written at
\emph{both} granularities: per-episode, to size per-agent containment, where the trigger and
catastrophe scenarios bind; and fleet-level, to capture the coordination scenario, which binds the
aggregate tail but, as Figure~\ref{c1:fig:adversary} shows, leaves the per-episode constraints slack.
The penalty $\alpha(Q)$ is calibrated differently at each level.

\subsection{Three lines of defence and provenance}
\label{c1:ssec:p-governance}
The three-lines model transfers intact. The deploying team owns the agent and its containment (first
line); an independent agent-risk function owns the register, the acceptance set, and the validation
standard (second line); and audit tests the whole (third line). The genAI-specific addition is
\emph{provenance}. The base model, prompt templates, tool definitions, and retrieval corpora are
shared dependencies whose change propagates across the monoculture, so they belong in the inventory
as first-class, version-controlled, and independently reviewed artefacts.

\section{Conclusions, assumptions, and limitations}
\label{c1:sec:conclusions}

We have argued that model-risk management is the right \emph{microprudential} template for agentic
systems in any domain. Its most mature codification is in banking, which we have used throughout as
the exemplar. That template is the firm-level foundation on which the systemic, financial-stability
account of the later chapters is built. But its quantitative core must be rebuilt around three
structural breaks: outputs are irreversible actions rather than estimates, the tail is driven by an
adaptive adversary rather than a stationary economic state, and there is no fungible capital
analogue. These breaks lead to a set-valued, convex, containment-subadditive risk measure, valued in
containment-configuration space and ordered by a cone that carries hard limits. Its convex dual is a
distributionally robust object penalised by attacker cost. For a tractable model class we proved that
the resulting containment-sizing problem is a convex program (Theorem~\ref{c1:thm:convex}). We proved
that prevention is non-substitutable against irreversible harm, with a closed-form, simulation-free
infeasibility certificate (Theorem~\ref{c1:thm:prevention}). And we proved that the program separates
with closed-form optima and shadow prices (Proposition~\ref{c1:prop:separation}), including the
intuition-reversing fact that a coherent tail bound can subsume a deontological chance constraint
(Corollary~\ref{c1:cor:subsume}). Monte-Carlo studies exhibited the wrong-way coupling (a $2.1\times$
understatement of expected loss and a $210\times$ tail), the non-substitutability wall, the failure
of cash-invariance, and the monoculture loss of diversification. We cross-checked the robust solve
analytically against a grid.

\subsection{Assumptions}
The load-bearing assumptions are explicit and worth restating. (i) The harm decomposition
$\EH=\PHA\cdot\SGH\cdot\BR$ is a modelling choice that gives interpretability; it is exact only under
the linear aggregation of (A1). (ii) The theorems hold for the tractable class (A1)--(A5): log-affine
recoverable attenuation, a catastrophic class reachable only by prevention, a coherent statistic, and
a finite adversary set. The separation result additionally needs disjoint lever-to-dimension channels
(A5). (iii) The adverse model is a Basel-style single common factor with a latent-normal copula. The
borrowing is deliberate. It gives the wrong-way and monoculture effects from one mechanism. But it is
itself a strong structural assumption. (iv) Most importantly, \emph{all numerical parameters are
illustrative}. They were chosen to be plausible and to make the structure legible, not estimated from
data; no quantitative claim about a real system should be read from them.

\subsection{Limitations}
Three limitations bound the contribution. First, Theorem~\ref{c1:thm:prevention}'s strength
rests on the lever-assignment axiom (A3) that detective review \emph{cannot} stop the irreversible
action. If review can block a catastrophic action with probability $q$, the certificate weakens to a
$(1-q)$-discounted bound and prevention becomes partially substitutable. The clean
non-substitutability is therefore an architecture claim about the control topology, to be defended
per system, not a universal law. Second, the result that the coherent $\ES$ bound subsumes the chance
constraint (Corollary~\ref{c1:cor:subsume}) is real but \emph{calibration-dependent}. It holds for
the chosen threshold $\theta$, tail level $\tau$, and catastrophe magnitude, and a higher $\tau$ or
lower $\theta$ can flip the ordering so that the hard limit binds. Third, the Monte-Carlo studies
resolve probabilities only to about $10^{-5}$ at the sample sizes used, so the exact point at which
the hard limit re-engages is asserted from the analytics, not measured. The descriptive simulation's
review mechanic is also non-convex, which is why the theorems are proved for the separate
tractable class rather than for the simulation itself.

Underlying all of this is the hard open problem the framework does not solve:
\emph{calibration}. Estimating the per-step hazard $h$, the recoverability $\rec$, and the adversary
penalty $\alpha(Q)$ against a thin, non-stationary, adaptively adversarial incident record is the
central empirical obstacle. The historical frequencies one would fit to are exactly what a competent
adversary works to invalidate.

\subsection{Extensions and next steps}
The most direct extension is to drop the disjoint-channel assumption (A5): allow review to partially
reach the catastrophic class (the $q$-discounted Theorem~\ref{c1:thm:prevention}) and let the
recoverable channels overlap. This breaks separation and the closed forms, but it preserves convexity
(Theorem~\ref{c1:thm:convex}). The problem then becomes a genuine robust convex program solvable with
a conic solver: an $\ES$/CVaR reformulation in the sense of \citet{rockafellar2000optimization} under
a Wasserstein ambiguity set \citep{esfahani2018data,rahimian2019distributionally}. That retires the
grid entirely and yields the binding adversary as a dual variable. The second priority is to give
$\alpha(Q)$ empirical content as an attacker-cost function calibrated from red-team economics, which
is what turns the ambiguity set from a stress menu into a model of the threat. Third, the acceptance
set should be developed at both per-episode and fleet granularities, with the monoculture correlation
made explicit in the aggregate constraint.

A companion line of work develops this fleet view in a domain-general form. It replaces the
predictive per-agent decomposition with a distributionally-defined mismatch index, in the spirit of
systemic-risk measures rather than single-name credit, for a population of $K$ agents sharing a
foundation model. There the loss of diversification under a common shock becomes a theorem rather
than a simulated phenomenon. Finally, the evaluation-awareness problem, validating a system that
can detect and game its own evaluation, deserves treatment as a first-class adversarial channel
rather than a caveat, since it undermines the effective-challenge gate on which the entire governance
process depends. None of these changes the thesis. The structure of the problem is set-valued, convex
under a coherent statistic, and dominated by the non-substitutability of prevention against
irreversible harm. The calibration of that structure is the work that remains.

\chapter{From Model Risk to Systemic Risk: Common Exposure and Contagion in a Fleet}
\label{ch:fleet}
\section{Introduction}
\label{c2:sec:intro}

A single agentic system is a model-risk problem; a \emph{fleet} of them is a financial-stability
problem. This chapter is the pivot of the monograph. It moves from the microprudential model risk of
one deployment to the \emph{systemic risk} of a population that shares a foundation model. An
institution, or a market, may deploy many agents at once: hundreds of coding assistants on the same
foundation model, a population of customer-operations agents sharing one toolchain, a fleet of triage
assistants behind one clinical model. The risk that matters is then no longer the failure of any one
agent but the joint behaviour of the population. Are their failures independent, or do they move
together? Does a compromise in one propagate to the rest? Can the aggregate breach the institution's
tolerance even when each agent looks acceptable in isolation? Chapter~\ref{ch:single} built the
single-agent object: a set-valued, containment-subadditive risk measure over a deployer-defined harm
vector, with prevention shown to be non-substitutable against irreversible harm. This chapter builds
the fleet object, where that same asymmetry becomes a statement about the stability of the system as a
whole.

\subsection{From prediction to distribution}
\label{c2:ssec:pivot}

The move from one agent to many forces a change of modelling philosophy. It is the same change finance
made when it moved from single-name to systemic risk. Chapter~\ref{ch:single}'s anatomy was, by
analogy, a single-name credit calculation: a decomposition of expected harm into a probability, a
severity, and an exposure, each to be estimated. That decomposition is the right entry point for one
agent, but estimation is exactly what fails at the fleet scale and in the tail. The systematic events
that drive fleet risk are rare, adaptive, and nearly unobserved, so a per-agent probability is
the quantity one cannot calibrate.

The systemic-risk literature confronted the same obstacle. It answered not by predicting which
institution fails but by defining badness \emph{distributionally}. CoVaR measures value-at-risk
conditional on distress \citep{adrian2016covar}. SRISK measures expected capital shortfall conditional
on a systemic event \citep{brownlees2017srisk,acharya2017measuring}. The liquidity-mismatch index
measures, from the balance sheet, a state-contingent mismatch that makes a run catastrophic \emph{if}
it occurs, without predicting the run \citep{brunnermeier2014lmi}. None predicts a point outcome; each
locates the distribution relative to an adverse state. This is the stance we adopt for fleets: stop
predicting harm, define the adverse states, and measure the fleet's structural exposure to them.

\subsection{What is different about a fleet}
\label{c2:ssec:fleet-different}

Four phenomena, absent or invisible in the single-agent view, organise the chapter.

\emph{Monoculture and non-diversifiability.} A portfolio of independent risks diversifies: its average
concentrates on the expected loss as the portfolio grows. A fleet on a shared foundation model is not
such a portfolio. The shared model is a common factor, and a jailbreak, a bad model update, or a
universal prompt injection is a common shock that strikes every agent at once. We show
(Theorem~\ref{c2:thm:nondiv}) that a coherent risk measure charges this common shock in full,
regardless of fleet size. Fleet risk floors strictly above expected loss and does not diversify away,
while the purely idiosyncratic part does. This is the formal content of the warning that a monoculture
is dangerous.

\emph{Contagion.} Agents that share tools, credentials, or reachable surfaces can propagate a
compromise: one agent's failure raises the chance of others'. We show
(Theorem~\ref{c2:thm:contagion}) that fleet-wide cascades have a sharp criticality threshold. Below a
unit branching ratio they stay bounded; above it a giant fraction of the fleet is engulfed. The same
result identifies segmentation and model heterogeneity as the architectural levers that hold the fleet
subcritical and cap the blast radius. This is a fleet-level control with no single-agent analogue.

\emph{Irreversibility, at fleet scale.} Chapter~\ref{ch:single} established a central asymmetry:
detection and reversibility cannot undo an irreversible action, so only prevention can. That asymmetry
survives aggregation and is sharpened by it. We show (Theorem~\ref{c2:thm:floor}) that the fleet's
irreversible-harm floor is independent of fleet size and of detective controls. It is reducible only by
prevention or by segmenting the systematically-exposed fraction. Diversification does not rescue you
from irreversible common harm.

\emph{Interaction.} The first three phenomena all assume that agents ignore each other. They react to
a shared environment, and a graph carries the consequences, but no agent responds to what another
agent does. Deployment is moving away from that. Agents read each other's output, call each other as
tools, and negotiate on their principals' behalf. Section~\ref{c2:sec:interaction} asks what changes.
The answer has two parts, and the second is the surprising one. Responsiveness to peers multiplies
whatever risk the shared substrate already supplies, which is expected. But responsiveness that
\emph{itself} varies with the environment manufactures common risk out of nothing shared at all: we
show (Theorem~\ref{c2:thm:endogfactor}) that a fleet in which every agent runs a different model, with
independent shocks, still faces a strictly positive non-diversifiable floor. The behaviour is ordinary
herding; what is new is that there is no shared signal to herd on, so substrate diversity, the standard
remedy for monoculture, does not close the channel. Separately, agents that can address one
another carry a second contagion layer in which nothing is exploited and the receiving agent is merely
\emph{persuaded}; Theorem~\ref{c2:thm:pretext} shows the two layers combine into a single branching ratio,
additively and under an infectiousness condition we state explicitly, so a fleet audited only on its technical
couplings is audited against the wrong number.

\subsection{Two functionals and a diagnostic}
\label{c2:ssec:two-functionals}

The framework measures fleet vulnerability with two complementary objects, both
\emph{stress-conditional} and neither predictive. The first is bottom-up and architectural: a
containment-\emph{mismatch} index, a domain-agnostic generalization of the liquidity-mismatch index. It
nets activatable exposure against mobilizable containment, axis by axis, on the reversible harm
subspace, and pairs that netting with a non-nettable prevention floor on the irreversible subspace. The
second is top-down and reduced-form: a projection of fleet-aggregate tail risk from a small set of
observable drivers, a domain-agnostic generalization of the CLASS stress-testing model
\citep{hirtle2016class}. We combine the two as supervision combines bottom-up and top-down stress
tests, each a check on the other. Their consistency is itself a result:
Theorem~\ref{c2:thm:consistency} shows they agree if and only if the observable drivers are sufficient
for the true risk, so a divergence is not noise but a measurement of the unmodelled channel.

\subsection{Chapter Outlines}
\label{c2:ssec:contributions}

The contribution is a domain-general framework with four proved results, each verified against a closed
form by simulation, plus a fifth section that stress-tests all four against agent-to-agent interaction.
We proceed as follows. Section~\ref{c2:sec:background} places the construction in
the systemic-risk and network-contagion literatures. Section~\ref{c2:sec:framework} defines the
primitives and the two functionals in domain-neutral terms. Section~\ref{c2:sec:theorems} states and
proves the four theorems: non-diversifiability of a common shock, contagion criticality, the
non-substitutable irreversible floor, and bottom-up/top-down consistency.
Section~\ref{c2:sec:interaction} lets the agents respond to and address one another, adds a social
multiplier, an endogenous common factor and a semantic contagion layer, and audits which of the four
results survive. Section~\ref{c2:sec:numerics} verifies each against its closed form in a unitless
study.
Section~\ref{c2:sec:apps} instantiates the framework in three deliberately unlike domains: code
automation, clinical triage, and a financial fleet that recovers Chapter~\ref{ch:single}'s setting.
That range demonstrates that banking is one instantiation among many rather than the foundation.
Section~\ref{c2:sec:governance} adapts the model-risk-management pipeline to fleets, and
Section~\ref{c2:sec:conclusions} states assumptions and limitations and points to the continuous-time
dynamics deferred to a sequel. Throughout, the load-bearing claim is structural: the calibration of the
framework's systematic parameters is a hard, partly unsolved empirical problem, and we are scrupulous
about saying so.

\section{Background and related work}
\label{c2:sec:background}

The framework borrows structure from four mature literatures and makes it domain-general. We summarise
each and state what we take from it.

\paragraph{Coherent and set-valued risk measures.} A risk measure is \emph{coherent} if it is monotone,
translation-equivariant, positively homogeneous, and subadditive \citep{artzner1999coherent}. Expected
shortfall (ES, equivalently CVaR) is the canonical coherent example, and it admits the
optimisation-friendly representation of \citet{rockafellar2000optimization,rockafellar2002conditional}.
Value-at-risk, by contrast, is not subadditive and can penalise diversification incoherently. Convex
risk measures relax positive homogeneity \citep{follmer2002convex,follmer2016stochastic}.
Chapter~\ref{ch:single} argued that agentic harm requires a \emph{set-valued} measure over containment
configurations, because harm is a vector with no common unit and the ``buffer'' (containment) is
non-fungible. We retain the coherent statistic (ES) for the tail and the vector-valued,
no-common-unit discipline. We also inherit Chapter~\ref{ch:single}'s robust/ambiguity layer, in which
the adversary is a cost-constrained optimiser over a family of laws
\citep{esfahani2018data,rahimian2019distributionally}.

\paragraph{Systemic risk measures.} The defining feature of systemic-risk measurement is that badness
is \emph{conditional on a distressed state} rather than predicted. CoVaR is the system's value-at-risk
given an institution in distress \citep{adrian2016covar}. SRISK and the marginal-expected-shortfall
family measure expected capital shortfall conditional on a market crash
\citep{brownlees2017srisk,acharya2017measuring}. We take this conditional, distributional
stance for fleets, and the top-down functional of \S\ref{c2:ssec:class} is the agentic counterpart of a
conditional-shortfall projection.

\paragraph{The liquidity-mismatch index.} The LMI of \citet{brunnermeier2014lmi} nets the
state-contingent market liquidity of assets against the state-contingent funding liquidity of
liabilities. It produces a stock-based, scenario-weighted measure of how exposed an institution is to a
liquidity stress. Its essence is a state-contingent netting of mobilizable absorptive capacity against
activatable exposure, with the weights worsening in stress, which is the structure of our
bottom-up functional, transplanted from balance sheets to agent affordances and containment
(\S\ref{c2:ssec:lmi}). The maturity mismatch at the LMI's core (slow assets funding fast liabilities)
reappears as a speed mismatch (fast affordance activation against slow containment mobilization),
encoded in the state-contingent weights.

\paragraph{Top-down stress testing.} The CLASS model \citep{hirtle2016class} projects bank outcomes
under stress by reduced-form regression on a small set of macro drivers. It serves as a transparent
top-down benchmark against bottom-up supervisory stress tests. We abstract its structure, a
parsimonious projection of an aggregate tail statistic from observable drivers, into the domain-general
top-down functional of \S\ref{c2:ssec:class}. We then pair bottom-up and top-down exactly as
supervision does, as mutual validation (Theorem~\ref{c2:thm:consistency}).

\paragraph{Static portfolio credit and frailty.} The non-diversifiability of systematic risk is the
lesson of the single-factor portfolio-credit model \citep{vasicek2002distribution,gordy2003risk} and
the actuarial CreditRisk+ framework \citep{creditsuisse1997creditriskplus}. Conditional on the common
factor, idiosyncratic risks average out, but the factor itself does not. The literature further found
that observable factors alone \emph{understate} clustering, requiring an unobserved \emph{frailty}
factor \citep{duffie2009frailty}. That frailty factor maps directly onto a fleet's shared,
partly-unobservable foundation-model state. Theorem~\ref{c2:thm:nondiv} is the agentic form of the
conditional law of large numbers underlying these models.

\paragraph{Financial networks and contagion.} Cascade and clearing models show how a shock propagates
through a network of exposures: Eisenberg--Noe clearing \citep{eisenberg2001systemic}, network
contagion \citep{gai2010contagion}, and the stability/fragility analysis of
\citet{acemoglu2015systemic}. The epidemic/percolation literature \citep{newman2002spread} gives the
criticality threshold we use. Theorem~\ref{c2:thm:contagion} ports the percolation criticality and a
segmentation corollary to the agentic dependency graph. Heavy-tailed aggregate loss, the natural
severity regime for catastrophic agentic harm, is the province of the loss-distribution / extreme-value
literature \citep{frachot2001loss,embrechts1997modelling}.

\paragraph{Positioning.} We do not propose new estimators within any of these literatures. We extract
their \emph{structure}: distributional conditioning, mismatch netting, top-down projection, conditional
diversification, and percolation criticality. We then assemble that structure into a single,
domain-general measure of fleet model risk, governed by traditional model-risk-management principles
\citep{sr117} and informed by the AI-specific catastrophic-risk and evaluation literatures
\citep{hendrycks2023overview,shevlane2023model}. The novelty is the synthesis and its domain-general
form, and the four theorems that the synthesis makes provable.

\section{The framework}
\label{c2:sec:framework}

The framework is domain-general by construction. Its primitives are an agent's affordances and the
containment wrapped around them, a harm vector whose axes are deployer-defined, and a coupling
structure across agents. No object below names what the harm \emph{is}. Section~\ref{c2:sec:apps}
instantiates the primitives in specific domains; the body never does.

\subsection{Primitives}
\label{c2:ssec:primitives}

A \emph{fleet} is a population of agents indexed $k=1,\dots,K$. An \emph{episode} is one bounded
interaction of an agent with its environment. Harm is valued in $\R^d_+$, where the $d$ axes are
chosen by the deployer to record incommensurable harm types; the analysis is invariant to their
meaning. The axes are partitioned into a \emph{reversible} set $\mathcal R$ and an
\emph{irreversible} set $\mathcal I$, $\{1,\dots,d\}=\mathcal R\sqcup\mathcal I$. This partition is
the one structural feature that is universal across domains. As
Theorem~\ref{c2:thm:floor} shows, it is also the only one our central irreducibility result depends on.

Each agent--axis pair carries two architectural quantities, both observable from the deployment
configuration rather than estimated from outcomes:
\begin{itemize}[leftmargin=1.6em,itemsep=2pt]
\item a \emph{reach} $\reach_{kj}\ge 0$: the harm on axis $j$ that agent $k$ can effect when its
affordance on that axis is fully activated (its blast radius on $j$);
\item a \emph{containment capacity} $\cont_{kj}\ge 0$: the harm on axis $j$ that the controls wrapped
around agent $k$ can prevent or reverse, expressed in the \emph{same units} as $\reach_{kj}$.
\end{itemize}
Commensurability of $\reach_{kj}$ and $\cont_{kj}$ \emph{within} an axis $j$ is a modelling
assumption, and we make it explicit. It is what permits netting on $\mathcal R$ below. We never net
across axes, so the measure remains vector-valued and assumes no common unit across harm types.

\paragraph{Stress states.} The environment occupies a \emph{stress state} $s\in\Sset$ drawn from a
law $\Pi$. The state acts through two state-contingent weights per agent--axis: an
\emph{exposure-activation} weight $\we{kj}(s)\ge 0$ and a \emph{containment-efficacy} weight
$\wc{kj}(s)\in[0,1]$. In a benign state $w^{e}$ is small and $w^{c}$ near one. In an adverse state
$w^{e}$ \emph{inflates}, as dormant affordances activate (the capability-conversion effect), and
$w^{c}$ \emph{collapses}, as review saturates and detection degrades. The weights encode the speed
mismatch that gives the construction its content. The weight $\we{kj}(s)$ is large when the affordance
can be activated faster than containment can respond in state $s$, and $\cont_{kj}$ counts only
containment that mobilizes within the harm window. The joint law $\Pi$ carries the fleet's systematic
structure. It has a shared latent \emph{frailty} component (a common factor reflecting a shared
foundation model, prompt templates, or tools) with per-agent loadings. It also has a common-shock event
that, when it fires, drives all loaded agents into the adverse-weight regime simultaneously. High
common loading is the \emph{monoculture}, and \S\ref{c2:sec:theorems} shows it is the source of
non-diversifiable risk.

\paragraph{Coupling.} A directed dependency graph $G$ on the fleet carries cross-agent propagation
(shared tools, shared retrieval corpora, lateral reachability), with transmission probability
$q$ along each edge. We write $\beta$ for the induced branching ratio (\S\ref{c2:ssec:contagion}).

\paragraph{Interaction.} The primitives above let agents share an environment but not respond to each
other. Section~\ref{c2:sec:interaction} adds the two ways they can. A \emph{responsiveness}
$\lambda\in[0,1)$ records how much of the peer average an agent passes through into its own behaviour,
so that the fleet aggregate becomes a fixed point rather than a sum; $\lambda$ may itself depend on the
stress state. A second graph $G_{\mathrm{sem}}$ records \emph{addressability}, which agents can place
content in front of which, with mean degree $c'$ and an acceptance probability $p$ that a peer's
framing induces the action. The two graphs carry different things. $G$ carries exploits, $G_{\mathrm{sem}}$
carries pretexts, and \S\ref{c2:ssec:int-pretext} shows that under a stated infectiousness condition
their branching ratios add. Readers interested only in the static
fleet may take $\lambda=0$ and $c'=0$ throughout, which returns the framework to the form used in
\S\ref{c2:sec:theorems}.

\subsection{The bottom-up functional: a domain-general mismatch index}
\label{c2:ssec:lmi}

The liquidity-mismatch index of \citet{brunnermeier2014lmi} nets state-weighted liquid assets
against state-weighted runnable liabilities. Stripped of its balance-sheet content, it is a
state-contingent netting of \emph{mobilizable absorptive capacity} against \emph{activatable
exposure}. We take this structure for agentic harm, on the reversible subspace where netting
is meaningful.

\begin{definition}[Containment-mismatch index]
\label{c2:def:lmi}
For agent $k$, reversible axis $j\in\mathcal R$, and state $s$, the \emph{containment mismatch} is
\begin{equation}
\Mis_{kj}(s)\;=\;\underbrace{\we{kj}(s)\,\reach_{kj}}_{\text{activatable exposure}}
\;-\;\underbrace{\wc{kj}(s)\,\cont_{kj}}_{\text{mobilizable containment}} .
\label{c2:eq:mis}
\end{equation}
The agent-$k$ mismatch is the vector $\big(\Mis_{kj}(s)\big)_{j\in\mathcal R}$; the fleet mismatch
on axis $j$ is the aggregate $\Mis_{j}(s)=\sum_{k=1}^K \Mis_{kj}(s)$. A positive component signals
that activatable exposure outruns mobilizable containment on that axis in that state.
\end{definition}

\noindent Three properties distinguish \eqref{c2:eq:mis} from a scalar score. \emph{(i) Within-axis
netting only.} We net exposure and containment only where they share units (a single axis $j$). The
object stays a vector across axes, so incommensurable harms (a privacy breach, a financial loss)
are never traded at a fixed rate. This preserves the no-common-unit principle of Chapter~\ref{ch:single}.
\emph{(ii) State-contingency is wrong-way by construction.} Because adverse states raise $w^{e}$ and
lower $w^{c}$ together, $\Mis_{kj}(s)$ is largest exactly in the states where the common shock fires;
mismatch and adversity co-move. \emph{(iii) Reversible subspace only.} Netting presumes that
containment can offset exposure, which is false for irreversible harm. We handle the irreversible axes
separately in \S\ref{c2:ssec:floor}.

The bottom-up risk is a coherent tail statistic of the mismatch under $\Pi$. For a random variable
$Y$ and level $\tau\in(0,1)$ we use expected shortfall
\begin{equation}
\ES_\tau(Y)\;=\;\frac{1}{1-\tau}\int_\tau^1 \VaR_u(Y)\,du,
\label{c2:eq:es}
\end{equation}
the mean of the worst $(1-\tau)$ tail, a coherent risk measure \citep{artzner1999coherent,
rockafellar2002conditional}. The bottom-up vulnerability on axis $j$ is
$R_{\mathrm{BU},j}=\ES_{\tau}\big(\Mis_j(s)\big)$ with the expectation over $s\sim\Pi$.

\subsection{The prevention floor on the irreversible subspace}
\label{c2:ssec:floor}

On $\mathcal I$ there is no absorptive offset. An irreversible action completes before containment
can reverse it, so $w^{c}$ is not a credit against exposure but at most a probability of \emph{averting}
the action. We therefore replace netting with a chance constraint. Let $\theta_j>0$ be the magnitude
of an irreversible (catastrophic) event on axis $j\in\mathcal I$ and let the fleet-level probability
that such an event is realised be $\Prob_\Pi(\text{irreversible event on }j)$. The
\emph{prevention floor} requires
\begin{equation}
\Prob_\Pi\big(\text{irreversible event on } j\big)\;\le\;\varepsilon_j,\qquad j\in\mathcal I,
\label{c2:eq:floor}
\end{equation}
where $\varepsilon_j$ is a hard deontological tolerance. As Theorem~\ref{c2:thm:floor} establishes, this
probability is reducible only by \emph{prevention} (removing or structurally attenuating the
affordance) or \emph{segmentation} (cutting the systematically-exposed fraction), never by detective
or reversibility controls. This is the fleet generalization of Chapter~\ref{ch:single}'s
non-substitutability result.

\subsection{The top-down functional: a domain-general stress projection}
\label{c2:ssec:class}

Where the mismatch index reads the architecture bottom-up, the CLASS model of
\citet{hirtle2016class} projects an aggregate outcome top-down, by reduced-form regression on a small
set of observable stress drivers. We abstract its structure to a domain-general driver vector and a
projected tail statistic.

\begin{definition}[Top-down stress projection]
\label{c2:def:class}
Let $x(s)\in\R^{d_x}$ be a vector of observable, domain-general \emph{stress drivers}: adversary pressure
(attack prevalence), a fleet-concentration index (a Herfindahl over shared models/prompts/tools),
autonomy (mean steps between human checkpoints), affordance breadth (aggregate reach), model drift
(time since validation), containment coverage (aggregate mobilizable capacity), and
\emph{interaction intensity} (the mean addressability degree $c'$, or an audit-derived proxy for how
freely agents may prompt one another). The top-down
functional is a reduced-form projection of a fleet-aggregate \emph{risk statistic}
\begin{equation}
\Rtd(s)\;=\;f\big(x(s);\,\beta\big),
\label{c2:eq:class}
\end{equation}
where $f$ is a parsimonious map (linear or generalized-linear) with parameters $\beta$ fit from
fleet incident telemetry, and the projected statistic is a tail measure (e.g.\ fleet-aggregate
$\ES_\tau$), not a point forecast of a realised event.
\end{definition}

\noindent Definition~\ref{c2:def:class} keeps the top-down layer consistent with the distributional
stance: $\Rtd$ projects \emph{where the harm distribution sits} conditional on the driver state, not
which agent fails when. The driver taxonomy and the projection form are domain-general and belong to
the framework. The fitted $\beta$ is domain-specific and is estimated within each stylized
application (\S\ref{c2:sec:apps}).

\subsection{Acceptance: two functionals, robustly}
\label{c2:ssec:acceptance}

Both functionals are \emph{stress-conditional vulnerability functionals}: maps from a state (or its
observable summary) to a risk assessment. The bottom-up $\Rbu$ is white-box and structural; the
top-down $\Rtd$ is black-box and statistical. We combine them exactly as supervision combines
bottom-up and top-down stress tests, each as a check on the other.

Following Chapter~\ref{ch:single}, adversary uncertainty enters as an ambiguity set: an admitted family $\Qset_\delta$
of laws for $(s,\Pi)$ whose deviation from a baseline is penalised by an attacker cost
$\alpha(\cdot)\le\delta$. The \emph{acceptance set} requires both functionals to remain within
deployer appetite $B_j$, and the prevention floor to hold, in the worst admitted case:
\begin{equation}
\Acc=\Big\{\text{configuration}:\ \sup_{Q\in\Qset_\delta}R_{\mathrm{BU},j}(Q)\le B_j\ \ \forall j\in\mathcal R,\ \
\sup_{Q\in\Qset_\delta}\Rtd(Q)\le B_{\mathrm{agg}},\ \
\eqref{c2:eq:floor}\ \forall j\in\mathcal I\Big\}.
\label{c2:eq:acc}
\end{equation}
The consistency of the two functionals is itself a diagnostic. By Theorem~\ref{c2:thm:consistency},
they agree if and only if the top-down drivers are sufficient statistics for the bottom-up risk, and
the residual measures the unobserved channel. We now turn to the four results that the framework
delivers, each verified against a closed form in \S\ref{c2:sec:numerics}.
Section~\ref{c2:sec:interaction} then relaxes the assumption that agents ignore one another and reports
which of the four results survive.

\section{Theorems}
\label{c2:sec:theorems}

We work in a tractable model class in which the framework's qualitative claims become provable. The
class fixes the systematic structure to a common-shock-plus-idiosyncratic form and the statistic to
expected shortfall. We state each assumption where it is first used. Every result comes with a
closed-form quantity that the Monte-Carlo study of \S\ref{c2:sec:numerics} reproduces. That
reproduction is our correctness gate.

\subsection{Non-diversifiability of a common shock}
\label{c2:ssec:nondiv}

\begin{assumption}[Common-shock harm]
\label{c2:ass:common}
Agent $k$'s episode harm on a fixed axis is $L_k=a_k J+\xi_k$, where $J=M\,\mathbf B$ is a common
shock with $\mathbf B\sim\mathrm{Bernoulli}(\psys)$ and magnitude $M>0$, the loadings satisfy
$\frac1K\sum_k a_k\to \bar a>0$, and the idiosyncratic terms $\xi_k$ are i.i.d., independent of
$\mathbf B$, with mean $\mu_\xi$ and finite variance. The fleet-average harm is
$\bar L_K=\frac1K\sum_{k=1}^K L_k$.
\end{assumption}

\begin{theorem}[Non-diversifiability]
\label{c2:thm:nondiv}
Under Assumption~\ref{c2:ass:common} with $\bar a=1$ and $\psys\le 1-\tau$,
\begin{equation}
\ES_\tau(\bar L_K)\;\xrightarrow[K\to\infty]{}\;\frac{M\,\psys}{1-\tau}+\mu_\xi
\;>\;\underbrace{M\psys+\mu_\xi}_{\E[L_k]} ,
\label{c2:eq:nondiv-floor}
\end{equation}
a strictly positive floor above expected loss. If instead there is no common shock ($\psys=0$, so
$L_k=\xi_k$), then $\ES_\tau(\bar L_K)\to\mu_\xi=\E[\xi_k]$: idiosyncratic risk diversifies away
entirely. The non-diversifiable excess of the floor over expected loss is $M\psys\tau/(1-\tau)$,
which vanishes only at $\psys=0$.
\end{theorem}

\begin{proof}
Write $\bar L_K=M\mathbf B\,\bar a_K+\bar\xi_K$ with $\bar a_K=\frac1K\sum_k a_k$ and
$\bar\xi_K=\frac1K\sum_k\xi_k$. The strong law of large numbers gives $\bar\xi_K\to\mu_\xi$ a.s.
Because the $\xi_k$ have finite variance, the family $\{\bar\xi_K\}$ is uniformly integrable, so
convergence also holds in $L^1$. Deterministically, $\bar a_K\to\bar a=1$. Hence
$\bar L_K\to M\mathbf B+\mu_\xi$ in $L^1$. Expected shortfall is a coherent risk measure and is
$L^1$-Lipschitz: $|\ES_\tau(Y)-\ES_\tau(Y')|\le (1-\tau)^{-1}\E|Y-Y'|$
\citep{rockafellar2002conditional,follmer2016stochastic}. Translation-equivariance and positive
homogeneity therefore give
$\ES_\tau(\bar L_K)\to\ES_\tau(M\mathbf B+\mu_\xi)=M\,\ES_\tau(\mathbf B)+\mu_\xi$.
For a Bernoulli$(\psys)$ variable with
$\psys\le 1-\tau$, the worst $(1-\tau)$-tail contains the entire event mass. The Rockafellar--Uryasev
representation \eqref{c2:eq:es} then gives $\ES_\tau(\mathbf B)=\frac{1}{1-\tau}\E[\mathbf B]
=\frac{\psys}{1-\tau}$, which yields \eqref{c2:eq:nondiv-floor}. When $\psys=0$, $\bar L_K\to\mu_\xi$ in
$L^1$, a constant, and $\ES_\tau$ of a constant is the constant. The excess over
$\E[L_k]=M\psys+\mu_\xi$ is $M\psys\big(\tfrac{1}{1-\tau}-1\big)=M\psys\tau/(1-\tau)$.
\end{proof}

\begin{remark}
Equation~\eqref{c2:eq:nondiv-floor} is the fleet analogue of the single-factor portfolio limit of
\citet{vasicek2002distribution} and \citet{gordy2003risk}: conditional on the systematic event the
idiosyncratic terms average out, but the systematic term does not, so a coherent measure charges it
in full regardless of $K$. It is the formal statement of the monoculture phenomenon: a population of
agents sharing a foundation model shares its common shock, and that shared shock is the
non-diversifiable core of \emph{systemic} risk. It is the common exposure that turns many individually
manageable model risks into one system-wide risk that redundancy cannot dilute. This is the
mechanism financial-stability supervision exists to police: undiversifiable common exposure, not the
failure of any single participant.
\end{remark}

\subsection{Contagion criticality and the value of segmentation}
\label{c2:ssec:contagion}

\begin{assumption}[Independent-cascade contagion]
\label{c2:ass:cascade}
A compromise originates at a uniformly random seed agent and transmits along each edge of the
dependency graph $G$ independently with probability $q$. We take $G$ to be the Erd\H{o}s--R\'enyi
graph $G(K,c/K)$ with mean degree $c$, and write the branching ratio $\beta=cq$.
\end{assumption}

\begin{theorem}[Contagion criticality]
\label{c2:thm:contagion}
Under Assumption~\ref{c2:ass:cascade}, in the limit $K\to\infty$ the expected outbreak exhibits a sharp
transition at $\beta=1$. For $\beta<1$ the outbreak is \emph{subcritical}: the expected number of
compromised agents is $O(1)$, a vanishing fleet fraction. At criticality $\beta=1$ the largest
component is of lower order than the fleet but no longer $O(1)$. For $\beta>1$ it is \emph{supercritical}:
with positive probability the outbreak occupies a giant fraction $S(\beta)$ of the fleet, the unique
positive root of
\begin{equation}
S\;=\;1-e^{-\beta S}.
\label{c2:eq:finalsize}
\end{equation}
Partitioning the fleet into $m$ isolated segments of equal size caps any \emph{single-seed} outbreak
at the segment giant component, i.e.\ at a fleet fraction $S(\beta)/m$; equivalently, reducing
transmission $q$ (model heterogeneity, so an exploit does not transfer) lowers $\beta$ and, below
$\beta=1$, eliminates the giant outbreak.
\end{theorem}

\begin{proof}
By the standard bond-percolation coupling \citep{newman2002spread}, independent-cascade propagation
with per-edge transmission $q$ is equivalent to occupying each edge of $G$ independently with
probability $q$ and taking the seed's occupied connected component as the outbreak. Occupying edges
of $G(K,c/K)$ with probability $q$ yields $G(K,cq/K)=G(K,\beta/K)$. The component structure of
$G(K,\lambda/K)$ is classical. For $\lambda<1$ all components are $O(\log K)$ and the largest is a
vanishing fleet fraction (subcritical); at $\lambda=1$ the largest is of lower order than $K$ but
super-logarithmic. For $\lambda>1$ a unique giant component occupies
fraction $S$ solving \eqref{c2:eq:finalsize}, with all other components $O(\log K)$. A uniformly
random seed lands in the giant component with probability $\to S$, and conditional on
landing there the outbreak \emph{is} that component. The giant component occupies fleet fraction
$S(\beta)$: $0$ for $\beta\le1$ (no giant fraction at or below criticality), and the positive root of \eqref{c2:eq:finalsize} for $\beta>1$.
That is the quantity the simulation of \S\ref{c2:sec:numerics} reports. (The unconditional expected
outbreak from a random seed is the smaller $S(\beta)^2$, since with probability $1-S$ the seed lands
in an $O(\log K)$ component. The design-relevant quantity is the giant fraction $S$, the size of the
outbreak \emph{when} one occurs.) Now segment the fleet into $m$
isolated equal blocks. Transmission cannot cross block boundaries, so an outbreak seeded at a single
point is confined to one
block of size $K/m$. Within a block the same analysis gives a giant component of fraction $S(\beta)$
\emph{of the block}, i.e.\ $S(\beta)/m$ of the fleet (a per-seed bound; see
Remark~\ref{c2:rem:asymptotics}). Monotonicity of $S$ in $\beta$ gives the
transmission-reduction claim.
\end{proof}

\begin{remark}[Two caveats on the statement]
\label{c2:rem:asymptotics}
First, the giant-component dichotomy imported above, the $O(1)$ subcritical count and the $S(\beta)/m$
segmentation cap are the \emph{standard large-$K$ asymptotics} for $G(K,\beta/K)$; they are statements
about the limit, and at finite $K$ the corresponding quantities carry the usual $O(\log K)$ corrections.
We use them as design heuristics, not as exact finite-fleet guarantees.

Second, and more consequentially, the cap is \emph{per seed}. The proof confines an outbreak to one
block because it begins in one block. An adversary able to seed several blocks simultaneously attains
the sum of their giant components, and recovers the unsegmented outbreak $S(\beta)$ once it seeds all
$m$. That is not a hypothetical against an agentic attacker with collapsed reconnaissance cost;
Chapter~\ref{ch:cyber} treats it directly (Theorem~\ref{c5:thm:targeted}(iii)), where segmentation is
recast as a bound on the attacker's \emph{seed budget} rather than on blast radius as such. The
unintentional-failure setting of this chapter, where compromise originates at a single point, is the
regime in which the cap as stated binds.
\end{remark}

\begin{remark}
Theorem~\ref{c2:thm:contagion} gives a fleet-architecture control absent from the single-agent
theory: the branching ratio $\beta$ is a design quantity, and keeping it below unity, through
heterogeneity (lower $q$) or segmentation (capping component size), is the \emph{macroprudential}
lever against fleet-wide cascades, the agentic counterpart of compartmentalization and exposure caps in
financial-network stability \citep{gai2010contagion,acemoglu2015systemic}.
\end{remark}

\subsection{The irreversible floor is non-substitutable}
\label{c2:ssec:floor-thm}

\begin{assumption}[Irreversible systematic exposure]
\label{c2:ass:irr}
On an irreversible axis, a fraction $g\in[0,1]$ of the fleet is systematically exposed: under the
common shock $\mathbf B\sim\mathrm{Bernoulli}(\psys)$ an exposed agent's catastrophic affordance of
magnitude $\theta$ fires with probability $\phi$, and absent the shock with baseline $\phi_0\le\phi$;
the unexposed fraction $1-g$ always fires at baseline $\phi_0$. Prevention of strength $c_3\in[0,1]$
attenuates the conditional excess to $\phi_0+(\phi-\phi_0)(1-c_3)$; detective controls act only after
the irreversible action and so leave its distribution unchanged.
\end{assumption}

\begin{theorem}[Non-substitutable floor]
\label{c2:thm:floor}
Under Assumption~\ref{c2:ass:irr} with $\psys\le 1-\tau$, the fleet-average irreversible harm
$\bar C_K$ floors as $K\to\infty$ at
\begin{equation}
F(g,c_3)\;=\;\theta\phi_0\;+\;\theta\,g\,(\phi-\phi_0)(1-c_3)\,\frac{\psys}{1-\tau},
\label{c2:eq:floor-level}
\end{equation}
which is (i) independent of the fleet size $K$ (non-diversifiable), (ii) independent of any detective
control (non-substitutable), and (iii) reducible toward the baseline $\theta\phi_0$ only by
prevention ($c_3\uparrow 1$) or segmentation of the systematically-exposed fraction ($g\downarrow 0$),
which enter \eqref{c2:eq:floor-level} symmetrically through the product $g(1-c_3)$.
\end{theorem}

\begin{proof}
Condition on $\mathbf B$. Given $\mathbf B=b$, each exposed agent fires independently with
probability $\phi_b=\phi$ if $b=1$ and $\phi_0$ if $b=0$, and each unexposed agent fires with
probability $\phi_0$. All firings are conditionally independent. The fleet-average harm is
$\bar C_K=\theta\big(g\,\hat p^{\,\mathrm{exp}}_K+(1-g)\hat p^{\,\mathrm{un}}_K\big)$ where the
$\hat p_K$ are sample firing rates. By the conditional SLLN, given $\mathbf B=b$,
$\bar C_K\to \theta\big(g\phi_b+(1-g)\phi_0\big)$ a.s.\ and in $L^1$. Hence unconditionally
$\bar C_K\to \theta\phi_0+\theta g(\phi-\phi_0)\mathbf B$ in $L^1$, a two-point variable. By
$L^1$-continuity, translation-equivariance and positive homogeneity of $\ES_\tau$, and
$\ES_\tau(\mathbf B)=\psys/(1-\tau)$ for $\psys\le1-\tau$ (as in Theorem~\ref{c2:thm:nondiv}),
\[
\ES_\tau(\bar C_K)\to \theta\phi_0+\theta g(\phi-\phi_0)\,\ES_\tau(\mathbf B)
=\theta\phi_0+\theta g(\phi-\phi_0)\frac{\psys}{1-\tau}.
\]
Replacing the conditional excess $\phi-\phi_0$ by its prevented value $(\phi-\phi_0)(1-c_3)$ gives
\eqref{c2:eq:floor-level}. The limit does not contain $K$, which proves (i). Detective controls do not
change the firing distribution of an irreversible action, so they do not enter
\eqref{c2:eq:floor-level}, which proves (ii). The only ways to reduce the systematic term are
$c_3\uparrow1$ or $g\downarrow0$, and these appear only through $g(1-c_3)$, which proves (iii).
\end{proof}

\begin{remark}
Theorem~\ref{c2:thm:floor} is the fleet form of Chapter~\ref{ch:single}'s prevention theorem: there, detection and
reversibility could not lower the irreversible tail of a single agent; here they cannot lower the
fleet floor either, and \emph{diversification across the fleet does not help}. Under
Assumption~\ref{c2:ass:irr}, where the exposed fraction $g$ is fixed by configuration, two and only two
levers remain: prevent the affordance, or sever the systematic coupling that makes the exposure
common, and they trade off through a single scalar $g(1-c_3)$. That exhaustiveness is the one part
of the result that does not survive agents which can address one another: once a peer can supply the
authorising frame, $g$ becomes endogenous and a third lever appears
(Corollary~\ref{c2:cor:floor-endog}). Both are structural, ex-ante moves on
the system's design rather than runtime controls on any one agent, the macroprudential signature, and
the reason a systemic floor cannot be supervised away one institution at a time.
\end{remark}

\subsection{Bottom-up/top-down consistency as a sufficiency test}
\label{c2:ssec:consistency-thm}

\begin{theorem}[Consistency and the residual]
\label{c2:thm:consistency}
Let $\Rbu$ be the bottom-up fleet risk, a square-integrable function of the full state, and let
$x=x(s)$ be the observable drivers. Among all top-down maps $g(x)$, the mean-square-optimal
projection is the conditional expectation $\Rtd^\star(x)=\E[\Rbu\mid x]$, and by the law of total
variance
\begin{equation}
\Var(\Rbu)\;=\;\underbrace{\Var\big(\E[\Rbu\mid x]\big)}_{\text{explained by drivers}}
\;+\;\underbrace{\E\big[\Var(\Rbu\mid x)\big]}_{\text{residual}} .
\label{c2:eq:ltv}
\end{equation}
The two functionals are consistent --- $\Rbu=\Rtd^\star(x)$ almost surely --- if and only if the
residual $\E[\Var(\Rbu\mid x)]$ is zero, i.e.\ iff $x$ is a sufficient statistic for $\Rbu$. When a
hidden channel $h$ (independent of $x$) enters linearly, $\Rbu=x^\top b_x+b_h h+\sigma\eta$ with
$\eta\perp(x,h)$ standardized, the residual equals $b_h^2\Var(h)+\sigma^2$ and the projection's
coefficient of determination is $R^2=1-\big(b_h^2\Var(h)+\sigma^2\big)/\Var(\Rbu)$.
\end{theorem}

\begin{proof}
For any $g$, $\E[(\Rbu-g(x))^2]=\E[(\Rbu-\E[\Rbu\mid x])^2]+\E[(\E[\Rbu\mid x]-g(x))^2]$ by
orthogonality of the conditional-expectation residual to any $\sigma(x)$-measurable function. The
second term is minimized (at zero) by $g=\E[\Rbu\mid x]$, which gives the optimal projection. Equation
\eqref{c2:eq:ltv} is the law of total variance. The residual $\E[\Var(\Rbu\mid x)]$ is a mean of
nonnegative terms, hence zero iff $\Var(\Rbu\mid x)=0$ a.s., i.e.\ iff $\Rbu$ is $\sigma(x)$-
measurable --- exactly sufficiency of $x$. For the linear hidden-channel model, $\E[\Rbu\mid x]=
x^\top b_x$, since $h,\eta$ are mean-zero and independent of $x$. The residual is therefore
$\Var(b_h h+\sigma\eta)=b_h^2\Var(h)+\sigma^2$, and $R^2$ follows from \eqref{c2:eq:ltv}.
\end{proof}

\begin{remark}
Theorem~\ref{c2:thm:consistency} turns the bottom-up/top-down pairing into a validation instrument:
agreement certifies that the observable drivers capture the fleet's risk, while a divergence is not
noise but a measurement of the unmodelled channel: a frailty factor, a contagion pathway, or an
affordance the architectural inventory missed. This is the role CLASS plays as an independent
top-down benchmark against bottom-up stress tests, made precise.
\end{remark}

\section{Agents that respond to each other}
\label{c2:sec:interaction}

Everything so far treats an agent as a device that reacts to its environment. The environment gives
a stress state, the agent acts, and a graph carries the consequences to its neighbours. What the
agent never does is react to \emph{another agent}. That omission is convenient and it is also, on
current deployment trends, wrong. Agents increasingly read each other's output, call each other as
tools, negotiate on their principals' behalf, and share a workspace in which one agent's artifact
becomes the next agent's input.

Two distinct channels open when they do. The first we call \emph{response}: agent $k$'s activation
depends on how much the rest of the fleet is doing. Escalation begets escalation, a quiet fleet stays
quiet, and the aggregate is a fixed point rather than a sum. The second we call \emph{address}: one
agent can put text in front of another. A peer is then a possible source of instructions, and an
attacker who compromises one agent acquires a persuasion channel to the rest that runs alongside the
technical one. Contagion by exploit and contagion by pretext are different mechanisms with different
controls, and the framework of \S\ref{c2:sec:framework} sees only the first.

This section adds both, and it keeps the chapter's discipline. Each of the three main results comes
with a closed form that \S\ref{c2:sec:numerics} reproduces. The reader should read them as a stress
test of Theorems~\ref{c2:thm:nondiv}--\ref{c2:thm:consistency} rather than as a repeal of them.
The non-diversifiability result survives with a rescaled constant, the contagion result keeps its
threshold but loses its segmentation clause, the irreversible-floor result needs its proof re-earned
and loses part of its policy conclusion, and the consistency result keeps its statement while its
interpretation gets harder. Section~\ref{c2:ssec:int-audit} does that accounting explicitly.

\subsection{Response: the fleet average as a fixed point}
\label{c2:ssec:int-response}

We keep the harm model of Assumption~\ref{c2:ass:common} and add one term.

\begin{assumption}[Linear response]
\label{c2:ass:response}
On a fixed axis, agent $k$'s episode harm solves
\begin{equation}
L_k\;=\;a_k J\;+\;\lambda\,\bar L_{-k}\;+\;\xi_k,
\qquad \bar L_{-k}=\frac{1}{K-1}\sum_{j\neq k}L_j,
\label{c2:eq:linmean}
\end{equation}
where $J=M\mathbf B$, $\mathbf B\sim\mathrm{Bernoulli}(\psys)$, the loadings satisfy
$\bar a_K=\frac1K\sum_k a_k\to\bar a$, and the $\xi_k$ are i.i.d.\ with mean $\mu_\xi$ and finite
variance, independent of $\mathbf B$. The \emph{responsiveness} $\lambda\in[0,1)$ measures how much
of the peer average an agent passes through into its own behaviour.
\end{assumption}

Equation~\eqref{c2:eq:linmean} is the linear-in-means specification of \citet{manski1993identification}.
We borrow it deliberately. It is the simplest object that makes an aggregate depend on itself, its
identification problems are well understood, and, as \S\ref{c2:ssec:int-audit} notes, those problems
are the ones a supervisor will face when trying to tell fleet interaction apart from shared
frailty.

\begin{theorem}[Social multiplier]
\label{c2:thm:multiplier}
Under Assumption~\ref{c2:ass:response} the system \eqref{c2:eq:linmean} has a unique solution for every
$K\ge2$, and the fleet average satisfies exactly
\begin{equation}
\bar L_K\;=\;\frac{\bar a_K J+\bar\xi_K}{1-\lambda}.
\label{c2:eq:multiplier}
\end{equation}
Consequently, with $\bar a=1$ and $\psys\le 1-\tau$,
\begin{equation}
\ES_\tau(\bar L_K)\;\xrightarrow[K\to\infty]{}\;
\frac{1}{1-\lambda}\left(\frac{M\psys}{1-\tau}+\mu_\xi\right),
\label{c2:eq:multiplier-es}
\end{equation}
and the non-diversifiable excess over expected loss is $M\psys\tau/\big((1-\lambda)(1-\tau)\big)$.
Setting $\lambda=0$ recovers Theorem~\ref{c2:thm:nondiv} exactly.
\end{theorem}

\begin{proof}
Write \eqref{c2:eq:linmean} in vector form as $L=u+\lambda BL$, where $u_k=a_kJ+\xi_k$ and $B$ is the
matrix with $B_{kj}=1/(K-1)$ for $j\neq k$ and $B_{kk}=0$. Every row of $B$ sums to one and $B$ is
nonnegative, so $\|B\|_\infty=1$ and the map $L\mapsto u+\lambda BL$ is a $\lambda$-contraction in the
sup norm. For $\lambda<1$ it has a unique fixed point, which proves existence and uniqueness.
Averaging the rows of $B$ gives
\[
\frac1K\sum_{k=1}^K\bar L_{-k}
=\frac{1}{K(K-1)}\sum_{k=1}^K\sum_{j\neq k}L_j
=\frac{1}{K(K-1)}\sum_{j=1}^K (K-1)L_j
=\bar L_K ,
\]
an identity, not an approximation: each $L_j$ appears in exactly $K-1$ of the $K$ leave-one-out means.
Averaging \eqref{c2:eq:linmean} over $k$ therefore gives $\bar L_K=\bar a_KJ+\lambda\bar L_K+\bar\xi_K$,
and \eqref{c2:eq:multiplier} follows because $1-\lambda>0$. The strong law gives
$\bar\xi_K\to\mu_\xi$ a.s.\ and, by uniform integrability under finite variance, in $L^1$; with
$\bar a_K\to 1$ deterministically, $\bar L_K\to (M\mathbf B+\mu_\xi)/(1-\lambda)$ in $L^1$. Expected
shortfall is $L^1$-Lipschitz, translation-equivariant and positively homogeneous, so
$\ES_\tau(\bar L_K)\to (1-\lambda)^{-1}\big(M\ES_\tau(\mathbf B)+\mu_\xi\big)$, and
$\ES_\tau(\mathbf B)=\psys/(1-\tau)$ for $\psys\le1-\tau$ as in Theorem~\ref{c2:thm:nondiv}. The limiting
mean is $(M\psys+\mu_\xi)/(1-\lambda)$, and subtracting it from \eqref{c2:eq:multiplier-es} leaves
$M\psys\tau/\big((1-\lambda)(1-\tau)\big)$.
\end{proof}

The multiplier $1/(1-\lambda)$ is the social multiplier of \citet{glaeser2003social}: interaction
converts a given common shock into a larger fleet outcome, and the conversion blows up as
responsiveness approaches one. A fleet with $\lambda=0.5$ carries twice the systemic charge of an
otherwise identical fleet of non-interacting agents (the agents still share $J$; what differs is
$\lambda$).

We want to be careful about what this does \emph{not} say, because the reading of
\eqref{c2:eq:multiplier-es} is more restrictive than the headline. Interaction here multiplies
whatever risk the substrate already gives, but manufactures none itself. Set $\psys=0$, so that there is no
common shock at all: the limit collapses to the constant $\mu_\xi/(1-\lambda)$, and the expected
shortfall equals the mean. The fleet average is larger, but it is not \emph{risky}. In the limit,
response alone raises the level and leaves the tail alone, because the $K$ idiosyncratic shocks average
out before the multiplier acts on them. (At finite $K$ the multiplier does scale the residual
$O(K^{-1/2})$ tail, but that term vanishes and is not what a systemic story is about.) Anyone who claims that agents talking to each other is by itself a
systemic-risk story owes an account of what the interaction is amplifying. The next result gives
one, and it is not the obvious one.

\subsection{Interaction as its own common factor}
\label{c2:ssec:int-endog}

The interesting case is not a fleet that responds strongly but a fleet whose responsiveness
\emph{moves with the environment}. Agents are more deferential to peers under time pressure, more
willing to accept a colleague's framing when their own evidence is thin, more likely to escalate when
everyone else is escalating. Responsiveness is a property of the interaction protocol, and the
protocol is exercised differently in a calm market and a disorderly one.

\begin{assumption}[State-dependent responsiveness]
\label{c2:ass:staterho}
Assumption~\ref{c2:ass:response} holds with $\lambda$ replaced by $\lambda(s)$, measurable in the
stress state, taking the value $\lambda_1$ on the adverse event $\{\mathbf B=1\}$ and $\lambda_0$
otherwise, with $0\le\lambda_0<\lambda_1<1$ and $\psys\in(0,1)$, $\tau\in(0,1)$. Crucially, we set
$a_k\equiv 0$: no agent loads on a common shock, and the idiosyncratic terms $\xi_k$ are i.i.d.\ with
mean $\mu_\xi>0$, independent of $\mathbf B$. The fleet shares no substrate whatever. It shares only the environment in which its
members talk to each other.
\end{assumption}

\begin{theorem}[Interaction is a common factor]
\label{c2:thm:endogfactor}
Write $v_i=\mu_\xi/(1-\lambda_i)$ for $i=0,1$, so $v_1>v_0>0$. Under
Assumption~\ref{c2:ass:staterho}:
\begin{enumerate}[label=(\roman*),leftmargin=2.2em,itemsep=2pt]
\item $\bar L_K\to V:=\mu_\xi/(1-\lambda(s))$ in $L^1$, a \emph{non-degenerate} random variable taking
$v_1$ with probability $\psys$ and $v_0$ otherwise. The fleet average does not concentrate on a
constant even though every primitive shock is idiosyncratic.
\item For $\psys\le 1-\tau$,
\begin{equation}
\ES_\tau(\bar L_K)\;\xrightarrow[K\to\infty]{}\;v_0+(v_1-v_0)\,\frac{\psys}{1-\tau},
\label{c2:eq:endog-es}
\end{equation}
which exceeds the limiting expected loss $v_0+(v_1-v_0)\psys$ by
$(v_1-v_0)\psys\,\tau/(1-\tau)>0$. The excess is strictly positive for every admissible parameter
value, and vanishes only in the degenerate limit $\lambda_1\downarrow\lambda_0$.
\item Suppose instead the fleet is partitioned into $m$ \emph{interaction domains} of equal size
$K/m$ (with $m\mid K$ and $m$ held fixed as $K\to\infty$), with response confined within a domain and
each domain drawing its own stress state independently. Then the fleet average converges to $\frac1m\sum_{b=1}^m V_b$ with $V_b$ i.i.d.\ copies of $V$, and
$\ES_\tau\big(\frac1m\sum_b V_b\big)$ is non-increasing in $m$ and converges to $\E[V]$ as
$m\to\infty$. The non-diversifiable excess is eliminated in the limit.
\end{enumerate}
\end{theorem}

\begin{proof}
(i) Condition on $s$. Given $\lambda(s)=\lambda_i<1$, Theorem~\ref{c2:thm:multiplier} applies verbatim
with $\bar a_K=0$, so $\bar L_K=\bar\xi_K/(1-\lambda_i)$ exactly. The $\xi_k$ are independent of
$\mathbf B$, so conditioning does not change their law; $\bar\xi_K\to\mu_\xi$ a.s.\ and in $L^1$ by the
strong law and uniform integrability. Hence $\bar L_K\to \mu_\xi/(1-\lambda(s))=V$ conditionally on
each of the two events, and therefore unconditionally in $L^1$, since
$\E|\bar L_K-V|=\psys\,\E[|\bar L_K-v_1|\mid \mathbf B=1]+(1-\psys)\,\E[|\bar L_K-v_0|\mid \mathbf B=0]
\to0$. The limit is non-degenerate because $\lambda_1>\lambda_0$ and $\mu_\xi>0$ give $v_1>v_0$.

(ii) Write $V=v_0+(v_1-v_0)\mathbf B$ with $v_1-v_0>0$. By $L^1$-continuity, translation-equivariance
and positive homogeneity of $\ES_\tau$, $\ES_\tau(\bar L_K)\to v_0+(v_1-v_0)\ES_\tau(\mathbf B)$, and
$\ES_\tau(\mathbf B)=\psys/(1-\tau)$ for $\psys\le1-\tau$. Subtracting
$\E[V]=v_0+(v_1-v_0)\psys$ gives the excess $(v_1-v_0)\psys\tau/(1-\tau)$, which is strictly positive
because $\psys>0$, $\tau>0$ and $v_1>v_0$; it tends to zero only as $\lambda_1\downarrow\lambda_0$.

(iii) Within domain $b$ the argument of part (i) applies to the $K/m$ agents it contains, so the
domain average converges in $L^1$ to $V_b$, and the $V_b$ are independent because the domain states
are. The fleet average is the mean of the domain averages, so it converges in $L^1$ to
$\bar V_m=\frac1m\sum_bV_b$. For monotonicity, let $\bar V_m^{(-i)}$ denote the average of the $m$
variables obtained by deleting $V_i$ from $V_1,\dots,V_{m+1}$. Then
$\bar V_{m+1}=\frac{1}{m+1}\sum_{i=1}^{m+1}\bar V_m^{(-i)}$, and each $\bar V_m^{(-i)}$ has the law of
$\bar V_m$, so law-invariance of $\ES_\tau$ gives them a common value. The $\bar V_m^{(-i)}$ are of
course dependent, which is exactly why we use subadditivity: it holds for any joint law. Subadditivity
and positive homogeneity of $\ES_\tau$ give
$\ES_\tau(\bar V_{m+1})\le\frac{1}{m+1}\sum_i\ES_\tau(\bar V_m^{(-i)})=\ES_\tau(\bar V_m)$. Finally
$\bar V_m\to\E[V]$ a.s.\ and in $L^1$ by the strong law (the $V_b$ are bounded), so
$L^1$-continuity of $\ES_\tau$ gives $\ES_\tau(\bar V_m)\to\E[V]$.
\end{proof}

\begin{remark}[What is new here]
\label{c2:rem:endog}
Theorem~\ref{c2:thm:endogfactor} is the reason for this section, and Assumption~\ref{c2:ass:staterho}
should be read as the strongest possible version of the standard remedy for
monoculture: \emph{every agent runs on a different model}. The loadings $a_k$ are identically zero, so
there is no shared foundation model, no shared prompt template, no shared tool. This is the
zero-loading case $\bar a=0$ of Theorem~\ref{c2:thm:nondiv}, in which the systematic term drops out
and the theorem predicts complete diversification, but that does not
happen. The fleet average converges to a
random limit and a coherent measure charges the spread, because the agents' \emph{responses to each
other} rise and fall together. Correlation is manufactured by the interaction protocol out of
uncorrelated inputs.

The policy content follows directly from part (iii). Substrate diversity does not close this channel;
independence of the \emph{interaction environment} does. The lever is architectural in a new way: cap
$\lambda_1$ by designing protocols that do not become more deferential under stress, or cut the fleet
into interaction domains that face separate environments. This is an addition to the lever set
of \S\ref{c2:sec:theorems}, where every control acted on the substrate or on the dependency graph.

An agent that discounts its own signal and leans on the crowd when its own
evidence is thin is engaged in what the literature calls \emph{herding}, and the economics of it is long settled: informational cascades in
\citet{banerjee1992herd} and \citet{bikhchandani1992theory}, the reputational variant of
\citet{scharfstein1990herd}, and the financial-market survey of \citet{bikhchandani2000herd}. We claim
no novelty in the behaviour. What Theorem~\ref{c2:thm:endogfactor} adds is where the behaviour comes
from and what it costs. In the classical models agents herd on a shared signal or a shared incentive, so
the co-movement traces back to something the agents hold in common. Here the loadings are identically
zero and nothing is held in common; the co-movement is produced by the protocol itself becoming more
deferential under stress. The classical literature also asks whether herding is individually rational,
which is the right question for an asset price and the wrong one for a risk floor. Optimal and
suboptimal herding are indistinguishable to $\ES_\tau$: both raise $\lambda_1$ above $\lambda_0$, and
both charge the tail.

Whether AI agents herd at all is now an experimental question rather than a speculative one, and the
early evidence cuts in two directions. \citet{hansen2025animal} replay a classic herding experiment, one
originally run on financial-market professionals, with large language models in the seat, and find that
off-the-shelf agents herd \emph{less} than the humans did: they follow their private signal $61$--$97\%$
of the time against $46$--$51\%$ for humans, and information cascades appear in $0$--$9\%$ of agent
decisions against roughly $20\%$. Read through our parameters, the baseline $\lambda$ of an unprompted
agent looks low, which is the reassuring half. The other half is that agents given explicit
profit-maximising guidance, a proxy for the fine-tuning an institution would actually deploy, do begin
to cascade when cascading pays. They stay reluctant about it, leaving most of those opportunities unused;
they also make \emph{more} of the cascade trades that do not pay than the unprompted agents did, which
the authors read as an unintended consequence of tuning. The finding that matters for us is that
$\lambda$ is not a constant of the technology. It is set by the instructions, and the operator writes the
instructions. What no
experiment has yet tested is the assumption Theorem~\ref{c2:thm:endogfactor} turns on, namely that
$\lambda$ rises in stressed states. Assumption~\ref{c2:ass:staterho} is a conjecture, and
\S\ref{c2:ssec:int-forward} says what would settle it.

Two caveats qualify the result. The two-point $\lambda(s)$ is the simplest specification that separates level from
co-movement; nothing depends on its being binary, but the closed form does. And part (iii) requires the
domains to face \emph{independent} states, which is a strong requirement in practice, since a market-wide
disorder is not partitionable. What part (iii) really shows is that the excess is driven by common
variation in $\lambda$, not by its level.
\end{remark}

\subsection{Address: pretext as a second transmission layer}
\label{c2:ssec:int-pretext}

We turn to the second channel. Assumption~\ref{c2:ass:cascade} moves a compromise along a technical
edge: shared credentials, a shared tool, a reachable endpoint. A fleet whose members can address one
another carries a second kind of edge, along which nothing is exploited. A compromised agent simply
\emph{asks}, supplying a frame in which the request looks authorised, and the receiving agent takes the
action itself. The receiving agent is not broken; it is persuaded, and it then executes with its own
legitimate permissions.

\begin{assumption}[Semantic layer]
\label{c2:ass:pretext}
Alongside the technical graph $G_{\mathrm{tech}}=G(K,c/K)$ with per-edge transmission $q$, the fleet
carries a \emph{semantic graph} $G_{\mathrm{sem}}=G(K,c'/K)$, independent of $G_{\mathrm{tech}}$,
recording which agents can place content in front of which. Along a semantic edge a compromised agent
induces the same action in its neighbour with \emph{acceptance probability} $p$, independently across
edges and independently of the technical transmission variables. We assume
\emph{route-independent infectiousness}: an agent that has taken the harmful action is thereafter a
source on both layers, whichever route reached it, and transmits with the same $q$ and $p$ as any
other. We write $\beff=cq+c'p$ for the induced effective branching ratio.

Two modelling choices deserve a word. Addressability is a directed relation and we model it with an
undirected $G(K,c'/K)$; this is a simplification of convenience, and it is harmless for our purposes
because the reachable set from a random seed in a sparse directed graph with mean out-degree $\beta$
obeys the same final-size equation \eqref{c2:eq:finalsize}. Route-independent infectiousness is the
substantive commitment, and Remark~\ref{c2:rem:pretext} says what happens without it.
\end{assumption}

\begin{theorem}[Pretext contagion]
\label{c2:thm:pretext}
Under Assumptions~\ref{c2:ass:cascade} and \ref{c2:ass:pretext}, as $K\to\infty$:
\begin{enumerate}[label=(\roman*),leftmargin=2.2em,itemsep=2pt]
\item The combined propagation is an independent cascade on an Erd\H{o}s--R\'enyi graph with mean
degree $\beff+O(1/K)$. Theorem~\ref{c2:thm:contagion} therefore applies verbatim with $\beta$ replaced
by $\beff$: the outbreak is subcritical for $\beff\le1$ and occupies giant fraction $S(\beff)$ solving
$S=1-e^{-\beff S}$ for $\beff>1$.
\item A fleet that is subcritical on its technical layer can be supercritical overall. Precisely,
$cq<1<cq+c'p$ is possible, and then removing the semantic layer alone restores subcriticality.
\item The final-size equation \eqref{c2:eq:finalsize} depends on $(c,q,c',p)$ only through the sum
$cq+c'p$, so the two layers are perfect substitutes in the threshold and neither can be certified in
isolation. Whether a given control moves one term and not the other is a modelling stipulation here
rather than a result; Remark~\ref{c2:rem:pretext} states the reading we have in mind and is explicit
about what the model does not prove.
\item Segmentation must respect \emph{both} graphs. If the technical graph is partitioned into $m$
isolated blocks but the semantic graph is not, then for $c'p>1$ an outbreak, \emph{when one occurs},
reaches at least the giant component of $G_{\mathrm{sem}}$ after acceptance thinning, i.e.\ at least
fleet fraction $S(c'p)$, with no dependence on $m$ at all.
\end{enumerate}
\end{theorem}

\begin{proof}
(i) By the percolation coupling used in Theorem~\ref{c2:thm:contagion}, thinning $G_{\mathrm{tech}}$ by
$q$ yields $G(K,cq/K)$ and thinning $G_{\mathrm{sem}}$ by $p$ yields $G(K,c'p/K)$, independently. An
agent is reached if it is reached by either route, so the transmission structure is the union of the
two thinned graphs. The union of independent $G(K,\alpha_1/K)$ and $G(K,\alpha_2/K)$ places each edge
independently with probability $\frac{\alpha_1}{K}+\frac{\alpha_2}{K}-\frac{\alpha_1\alpha_2}{K^2}$, so
it is $G(K,\alpha/K)$ with $\alpha=\alpha_1+\alpha_2-\alpha_1\alpha_2/K$. With
$\alpha_1=cq,\alpha_2=c'p$ this is $\beff+O(1/K)$, and $S(\cdot)$ is continuous, so the conclusions of
Theorem~\ref{c2:thm:contagion} carry over.

(ii) Immediate from (i): take $cq=0.8$ and $c'p=0.4$, so $cq<1<1.2=\beff$.

(iii) Immediate from \eqref{c2:eq:finalsize} and part (i).

(iv) The union graph contains the thinned semantic graph as a subgraph, so by monotone coupling its
largest component is at least as large. If $c'p>1$ that component occupies fleet fraction at least
$S(c'p)>0$, and a seed lands in it with probability tending to $S(c'p)$, which is the same conditioning
used in Theorem~\ref{c2:thm:contagion}. The bound does not involve the technical partition: the
semantic graph is drawn on the whole fleet, so no block structure constrains it.
\end{proof}

\begin{corollary}[The exposed fraction becomes endogenous]
\label{c2:cor:floor-endog}
Take Assumption~\ref{c2:ass:irr} with $g\in(0,1)$, and let the systematically exposed fraction $g$
additionally seed an independent cascade with branching ratio $\beff$ on the combined graph, so that an
agent reached by a seed acquires the exposed firing profile. What propagates here is \emph{exposure},
a pre-firing attribute: a peer supplies the authorising frame, which puts the receiving agent into the
exposed configuration, and whether it then fires is still decided by the stress state. Assume the
cascade is independent of $\mathbf B$. Then the equilibrium exposed fraction $\geff$ is the unique
root in $(0,1]$ of
\begin{equation}
\geff\;=\;1-(1-g)\,e^{-\beff\,\geff},
\label{c2:eq:geff}
\end{equation}
it satisfies $\geff>g$ for every $\beff>0$, and it increases in $\beff$ to $1$ as
$\beff\to\infty$. The floor of Theorem~\ref{c2:thm:floor} becomes $F(\geff,c_3)$, i.e.
\begin{equation}
\theta\phi_0+\theta\,\geff\,(\phi-\phi_0)(1-c_3)\,\frac{\psys}{1-\tau}.
\label{c2:eq:floor-endog}
\end{equation}
Consequently the lever set of Theorem~\ref{c2:thm:floor} is no longer exhaustive, and the reason
depends on the regime. For $\beff>1$ the cascade has a giant component and $\geff\ge S(\beff)>0$
uniformly in $g$, so driving the configured exposure down cannot drive the floor to baseline: the
$g$-lever hits a wall. For $\beff<1$ it still works, but only after inflation. Linearising
\eqref{c2:eq:geff} at small $g$ gives $\geff\approx g/(1-\beff)$, so the operator who halves $g$ halves
the systematic term of \eqref{c2:eq:floor-endog} and no more---the baseline $\theta\phi_0$ is untouched,
exactly as in Theorem~\ref{c2:thm:floor}---and pays a level multiplier $1/(1-\beff)$ before he starts.
That is the same $1/(1-x)$ form as the social multiplier of Theorem~\ref{c2:thm:multiplier}, arising for
the same reason. At the knife edge $\beff=1$ neither approximation holds: there $\geff\approx\sqrt{2g}$,
and halving $g$ cuts the systematic term by less than a third. Either way,
capping $\beff$, and in particular capping who may address whom, is a third structural lever.
\end{corollary}

\begin{proof}
Equation \eqref{c2:eq:geff} is the standard final-size relation for an independent cascade on
$G(K,\beff/K)$ seeded by a linear-size set \citep{newman2002spread}; we record the mean-field
derivation and then verify the fixed point. In the large-$K$ limit each vertex is an independent seed
with probability $g$, and a vertex escapes exposure exactly if it is not a seed and no occupied edge
connects it to an exposed vertex. When a fraction $\rho$ is exposed the number of such edges is
asymptotically Poisson with mean $\beff\rho$, so $1-\rho=(1-g)e^{-\beff\rho}$, which is
\eqref{c2:eq:geff}. (As in Remark~\ref{c2:rem:asymptotics}, this is a self-consistency argument and a
limit statement; Assumption~\ref{c2:ass:irr} fixes $g$ as a deterministic fraction where the derivation
uses i.i.d.\ Bernoulli$(g)$ seeding, and the two agree asymptotically.) Let
$h(\rho)=1-(1-g)e^{-\beff\rho}$. Then $h$ is strictly increasing and strictly concave on $[0,1]$,
$h(0)=g>0$ and $h(1)=1-(1-g)e^{-\beff}<1$. Hence $h(\rho)-\rho$ is concave, positive at $0$ and
negative at $1$, so it has exactly one root in $(0,1)$, which is $\geff$. Since
$h(g)=1-(1-g)e^{-\beff g}>1-(1-g)=g$, the root lies above $g$. For monotonicity, fix $\beff'>\beff$: then
$h_{\beff'}(\geff)>h_{\beff}(\geff)=\geff$, and since $h_{\beff'}(\rho)-\rho$ is concave and negative at
$\rho=1$, its unique root lies above $\geff$. As $\beff\to\infty$,
$h(\rho)\to1$ pointwise on $(0,1]$, forcing $\geff\to1$. Substituting $\geff$ for $g$ in
\eqref{c2:eq:floor-level} gives \eqref{c2:eq:floor-endog}. The proof of Theorem~\ref{c2:thm:floor} is
otherwise unchanged: because the cascade moves exposure rather than firing, and is independent of
$\mathbf B$, firings remain conditionally independent given the state and the realised exposed set,
and $\geff$ is a constant rather than a function of $\mathbf B$. Both are substantive. If the cascade
were itself state-dependent---more agents acquiring exposure under the common shock---the limit would be
$\theta\phi_0+\theta\,\geff(1)(\phi-\phi_0)(1-c_3)\psys/(1-\tau)$ with $\geff(1)>\geff$, so
\eqref{c2:eq:floor-endog} would understate the floor. In that case read it as a lower bound. The same
direction, for a different reason, applies if what propagates is firing rather than exposure;
\S\ref{c2:ssec:int-audit} takes that case up.
\end{proof}

\begin{remark}[Why pretext needs its own controls]
\label{c2:rem:pretext}
The four parts of Theorem~\ref{c2:thm:pretext} say the same thing from four angles. The semantic layer
is a transmission mechanism, and the controls built for the technical layer do not touch it. That
matters because the two layers are governed by different departments. Network segmentation, credential
hygiene and tool isolation lower $c$ and $q$; none of them is designed to lower $c'$ or $p$. That is an
observation about what those controls target, not something the model derives.

The controls that do lower $p$ are recognisably different in kind. Input provenance lets an agent
distinguish an instruction from its principal from text that arrived via a peer. Authority binding
ensures that a peer's request cannot carry the principal's permissions, so that persuading an agent
does not also authorise it. Both are agent-side design choices, not network controls. Capping $c'$ is
the blunt instrument, restricting the addressability graph so that agents can only talk to the agents
they need to.

The additive form $\beff=cq+c'p$ rests on route-independent infectiousness, and it is worth being
explicit about the alternative. Suppose instead that persuasion confers only semantic infectiousness:
an agent talked into acting has taken no exploit and has none to pass on, so it can forward the pretext
but not the compromise. Propagation is then a two-type branching process with mean-offspring matrix
$\Gamma_2=\left(\begin{smallmatrix}cq & c'p\\ 0 & c'p\end{smallmatrix}\right)$, and criticality sits at
$\spr(\Gamma_2)=\max(cq,\,c'p)=1$ rather than at $cq+c'p=1$. The qualitative claim of
Theorem~\ref{c2:thm:pretext}(ii) survives either way: a fleet certified subcritical on $cq$ alone can
still go supercritical. But the threshold moves, and $\beff=cq+c'p$ is then an upper bound on the
true branching ratio. What does not survive is the sufficiency of the additive condition: the two-type
reading needs $c'p>1$, not just $cq+c'p>1$, so the worked numbers in the proof of
Theorem~\ref{c2:thm:pretext}(ii) are an example only under the additive reading. For a supervisor, auditing against $\beff$
is the conservative choice, since it raises the alarm early. For a modeller it is a live question, and it turns on whether an agent that has
been persuaded into acting thereby produces the artifact that carries the next compromise. In a shared
workspace, where the harmful action often \emph{is} writing to a corpus that other agents read, we
think it usually does.

We flag one further asymmetry that the model does not prove but that its structure makes salient. A technical
exploit is written against a specific implementation and often fails against another, which is why
heterogeneity lowers $q$. A pretext is written against a \emph{task}, and its portability across
models is an empirical question with no reason to be small. If that is right, the standard diversity
remedy is a partial control here, and $c'p$ is the term that survives it. Corollary~\ref{c2:cor:floor-endog}
then bites on the irreversible axes, where surviving terms are the whole story.

There is at least one direct sign that $p$ is not negligible. \citet{hansen2025animal} reverse the colour
labels on an otherwise unchanged signal, red for good news, green for bad, and their agents stop
behaving rationally. About a quarter of all decisions become outright errors, buying on bad news and
selling on good, and cascade trading rises. The stated probabilities were unchanged and were restated in
full; only the presentation moved. Their reading is that the models have inherited a human colour
convention rather than reasoning from the numbers in front of them. An acceptance probability is
the rate at which presentation rather than content decides the action.
\end{remark}

\subsection{What survives, and what has to be re-earned}
\label{c2:ssec:int-audit}

It would be easy to present this section as strengthening the chapter. It does not, uniformly, and the
accounting is worth setting out plainly.

\emph{Theorem~\ref{c2:thm:nondiv} survives, with a larger constant.} The statement, that a common shock
does not diversify, is untouched. Theorem~\ref{c2:thm:multiplier} multiplies the floor by
$1/(1-\lambda)$ and recovers the original at $\lambda=0$. What rescues the argument is the specific
linear-in-means structure. Under interaction the $L_k$ are dependent and no law of large numbers
applies to them, but \eqref{c2:eq:multiplier} says the fleet average is the non-interacting average
rescaled, so the strong law can still be applied to the $\xi_k$, which remain i.i.d. This is a
statement about constant $\lambda$; Theorem~\ref{c2:thm:endogfactor} shows that when $\lambda$ moves
with the state the change is not a rescaled constant at all.

\emph{Theorem~\ref{c2:thm:contagion} keeps its threshold and loses its segmentation clause.}
Criticality at a unit branching ratio is a property of the graph, not of what the edges mean, and
Theorem~\ref{c2:thm:pretext}(i) simply widens the definition of an edge. The design conclusion, that
the branching ratio must stay below one, is unchanged. But the quantity to be held down is $\beff$, or, without
route-independent infectiousness, the spectral radius that $\beff$ bounds from above
(Remark~\ref{c2:rem:pretext}), and a fleet audited only on $cq$ will be audited against the wrong
number. The casualty is the theorem's second
clause, that partitioning the fleet into $m$ isolated segments caps a single-seed outbreak at
$S(\beta)/m$. Theorem~\ref{c2:thm:pretext}(iv) voids it when the partition is technical only and the semantic layer
is itself supercritical: for $c'p>1$ an outbreak reaches at least $S(c'p)$ of the fleet and $m$ does not
appear, so the cap becomes a floor. For $c'p\le1$ the clause still has content, but it is content about
only one of the two graphs. Segmentation remains a lever, but only if it is
applied to both graphs, and the addressability graph is the one nobody currently partitions.

\emph{Theorem~\ref{c2:thm:floor} needs its proof re-earned, and loses part of its policy claim.} The
original proof conditions on $\mathbf B$ and then invokes conditional independence of firings across
agents. Interaction threatens that step: if a peer can supply the authorising frame, one agent's action
raises another's probability of acting. Corollary~\ref{c2:cor:floor-endog} is best read not as a repair
of the original proof but as a replacement of the model. It moves the peer effect one stage earlier, so
that what the cascade propagates is \emph{exposure} rather than firing. Conditional independence of
firings then holds again, given the state and the realised exposed set, and the only change is that $g$
must be read as the equilibrium fraction $\geff$. This rests on a modelling choice:
if firings themselves propagate, the conditional-independence step fails and
\eqref{c2:eq:floor-endog} becomes a lower bound rather than a limit. What does not survive in either
reading is the remark following Theorem~\ref{c2:thm:floor} that two and only two levers remain. With a semantic layer there are three, and the new one, capping who
may address whom, is the one no existing model-risk pipeline measures.

\emph{Theorem~\ref{c2:thm:consistency} survives as stated and gets harder to use.} The law of total
variance holds for any square-integrable bottom-up risk, so nothing in the theorem breaks. The
difficulty is on the interpretation side. Section~\ref{c2:sec:conclusions} already records that frailty
and contagion are not separately identifiable from aggregate data. Interaction is a third mechanism
that produces the same aggregate signature: a fleet-average that co-moves more than the sum of its
parts is consistent with a hidden common factor, with genuine event-triggering, and now with an
endogenous response. This is an instance of the identification failure
\citet{manski1993identification} describes (strictly, correlated effects masquerading as endogenous
ones), and it does not have a solution that reads off aggregates, though it has
solutions that require design: exogenous variation in who can address whom, or in the responsiveness
of the protocol, is the fleet analogue of an instrument.

There is one constructive implication, and it is why Definition~\ref{c2:def:class} carries an
\emph{interaction-intensity} driver alongside the concentration and autonomy drivers. That driver
yields a testable prediction, though the test has to be built carefully. Interaction intensity is an
architectural constant of a fleet, not a function of the stress state, so within a single fleet it is a
constant regressor that does not help. The prediction lives in a panel: across fleets, or across
reconfigurations of one fleet, adding the mean addressability degree $c'$ (or an audit-derived proxy
for it) to the driver vector should raise the \emph{out-of-sample} $R^2$ of
\eqref{c2:eq:ltv}. In-sample fit is not the test. Both $R^2$ and the fitted residual improve mechanically
when any regressor is added. A null
result would be weak evidence at best, since low power, proxy error and collinearity with the
concentration index all produce one; a positive result is the informative direction.

\subsection{Where this goes next}
\label{c2:ssec:int-forward}

Both mechanisms in this section are static shadows of dynamic objects, and the sequel chapters carry
them properly. The responsiveness $\lambda$ is a one-period stand-in for the self-excitation of
Chapter~\ref{ch:continuous}: an agent that reacts to its peers is an intensity that jumps when its peers
jump, and the multiplier $1/(1-\lambda)$ is the static image of the Hawkes stability condition
$\spr(\Gam)<1$. Likewise $\beff$ is the static image of a branching matrix with two kinds of entry,
technical and semantic. There the subcritical condition is on the spectral radius, which collapses to
the sum $cq+c'p$ under the route-independent infectiousness of Assumption~\ref{c2:ass:pretext} and to
$\max(cq,c'p)$ without it. The acceptance probability
$p$ is treated here as a parameter; Chapter~\ref{ch:cyber} makes it a choice variable, since a
strategic attacker selects the pretext and will select the one that works. The natural continuation is
therefore an attacker who allocates effort between exploiting the technical layer and persuading the
semantic one, and a defender who must fund both.

One piece of this is measurable now rather than later. Assumption~\ref{c2:ass:staterho} asserts that
responsiveness rises under stress, and that is an experimental claim, not a modelling convenience. The
herding experiments of \citet{hansen2025animal} already recover $\lambda$ from agent decisions in a
controlled setting; running the same design under conditions that proxy for stress, such as compressed time
budgets, degraded or conflicting private signals, or prompts that raise the cost of being wrong alone, would
deliver $\lambda_0$ and $\lambda_1$ directly. If the gap is zero, Theorem~\ref{c2:thm:endogfactor} has
no bite and the multiplier of Theorem~\ref{c2:thm:multiplier} is the whole interaction story. If it is
not, the quantity a supervisor should be asking for is the gap, not the level.

\section{Numerical illustration}
\label{c2:sec:numerics}

The studies below are deliberately \emph{unitless}. The common-shock magnitude $M$ (or the
catastrophic magnitude $\theta$) is the unit of harm, idiosyncratic harm and appetites are measured
in those units, and the dependency graph is dimensionless. No domain enters. These figures illustrate
the general theorems and so belong to the body. Each is generated by code that prints its own
simulation-versus-closed-form check, and we report those checks alongside. Domain-specific,
dimensioned studies appear in \S\ref{c2:sec:apps}.

\subsection{Non-diversifiability (Theorem~\ref{c2:thm:nondiv})}
\label{c2:ssec:num-nondiv}

\begin{figure}[htbp]
\centering
\includegraphics[width=\linewidth]{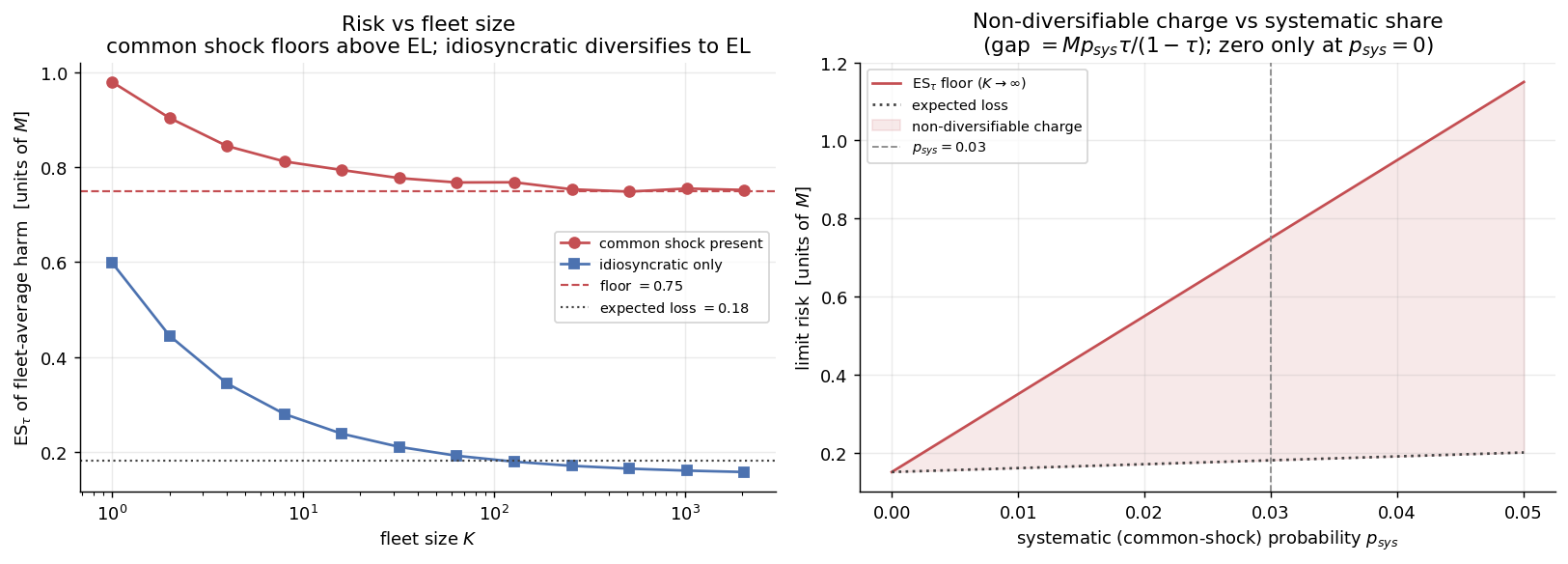}
\caption{\textbf{Non-diversifiability of a common shock (Theorem~\ref{c2:thm:nondiv}).} Left: expected
shortfall of fleet-average harm versus fleet size. With a common shock present the risk floors above
expected loss (simulated $0.752$ vs closed-form $0.750$). With only idiosyncratic harm it diversifies
toward expected loss. Right: the limit floor and the non-diversifiable charge as functions of the
systematic probability $\psys$.}
\label{c2:fig:nondiv}
\end{figure}

Figure~\ref{c2:fig:nondiv} fixes $\mu_\xi=0.15$, $\psys=0.03$, $\tau=0.95$, so the closed-form floor is
$M\psys/(1-\tau)+\mu_\xi=0.75$ against an expected loss of $0.18$. We simulate fleet-average harm out
to $K=2048$ ($4\times10^5$ episodes, the fleet mean of i.i.d.\ exponential idiosyncratic harm sampled
exactly as a Gamma). The common-shock curve settles on $0.752$, a relative error of $0.33\%$.
The idiosyncratic-only curve falls toward expected loss, matching the finite-$K$ analytic
$\ES$ (the $1/\sqrt K$ approach predicted by the central limit theorem) to $0.04\%$. The right panel
plots the limit floor against the systematic probability $\psys$. The non-diversifiable charge is
the shaded gap $M\psys\tau/(1-\tau)$ between the floor and expected loss; it opens linearly from zero
and closes only at $\psys=0$. The qualitative content is the monoculture result of Chapter~\ref{ch:single}, now exact:
sharing a foundation model means sharing a common shock, and the common shock does not diversify.

\subsection{Contagion criticality (Theorem~\ref{c2:thm:contagion})}
\label{c2:ssec:num-contagion}

\begin{figure}[htbp]
\centering
\includegraphics[width=\linewidth]{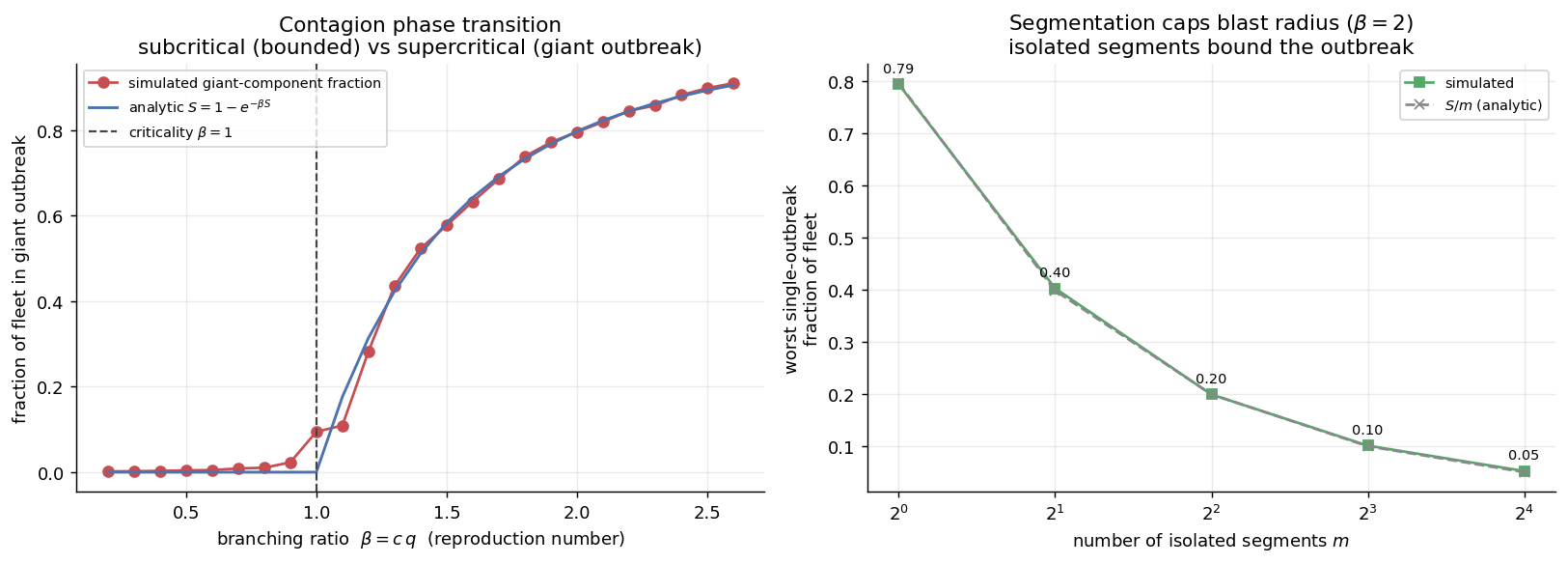}
\caption{\textbf{Contagion criticality and segmentation (Theorem~\ref{c2:thm:contagion}).} Left: the
phase transition at branching ratio $\beta=1$; simulated giant-outbreak fraction versus the analytic
$S=1-e^{-\beta S}$. Right: at supercritical $\beta=2$, segmentation into $m$ isolated blocks caps the
worst single-seed outbreak at the fleet fraction $S/m$.}
\label{c2:fig:contagion}
\end{figure}

Figure~\ref{c2:fig:contagion} simulates outbreaks on $G(K,c/K)$ at $K=4000$ across branching ratios
$\beta=cq\in[0.2,2.6]$, using the percolation equivalence to compute each outbreak as a connected
component. The simulated giant-component fraction tracks the analytic final size $S=1-e^{-\beta S}$
with mean absolute error $0.005$ above threshold. It exhibits the sharp transition at $\beta=1$:
bounded outbreaks below, a giant fleet-wide outbreak above. The right panel takes a supercritical
fleet at $\beta=2$ (giant fraction $S\approx0.797$) and partitions it into $m$ isolated segments. The
worst single-segment outbreak, as a fraction of the fleet, falls as $S/m$: $0.80, 0.40, 0.20, 0.10,
0.05$ for $m=1,2,4,8,16$. The simulated values match $S/m$ to within a fraction of a percent.
Segmentation does not lower the within-segment severity; it caps the blast radius \emph{per seed}
(Remark~\ref{c2:rem:asymptotics}).

\subsection{The non-substitutable floor (Theorem~\ref{c2:thm:floor})}
\label{c2:ssec:num-floor}

\begin{figure}[htbp]
\centering
\includegraphics[width=\linewidth]{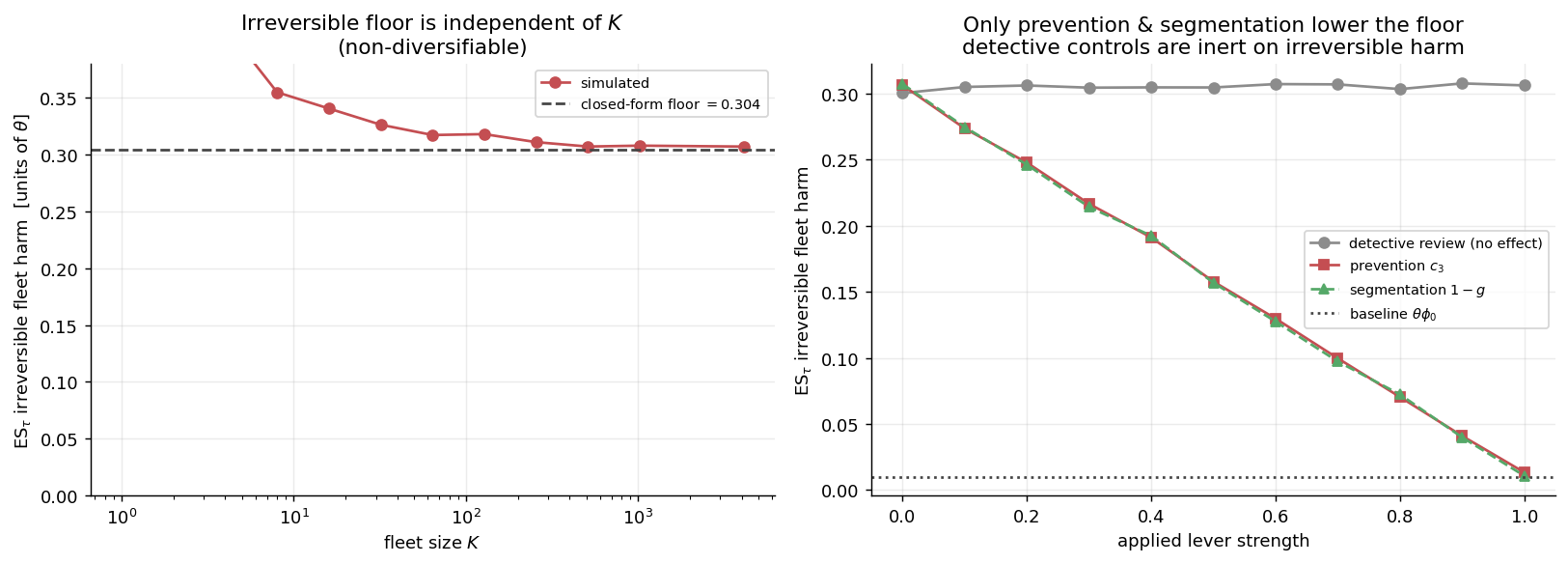}
\caption{\textbf{The non-substitutable irreversible floor (Theorem~\ref{c2:thm:floor}).} Left: the floor
is independent of fleet size (simulated $0.307$ vs closed-form $0.304$). Right: detective review is
inert (flat to Monte-Carlo noise), while prevention and segmentation lower the floor toward the
baseline $\theta\phi_0$, tracking the closed form.}
\label{c2:fig:floor}
\end{figure}

Figure~\ref{c2:fig:floor} uses $\theta=1$, $\phi=0.5$, $\phi_0=0.01$, $\psys=0.03$, $\tau=0.95$, giving a
closed-form floor $F(1,0)=\theta\phi_0+\theta(\phi-\phi_0)\psys/(1-\tau)=0.304$. The left panel
confirms the floor is flat in fleet size (simulated $0.307$ at $K=4096$, a $1.03\%$ error): the
irreversible systematic charge does not diversify. The right panel sweeps each lever from $0$ to $1$.
The detective-review curve is flat, with a standard deviation across all strengths of $0.002$, pure
Monte-Carlo noise; that flatness confirms non-substitutability. Prevention and segmentation both drive
the floor down toward the baseline $\theta\phi_0$, each matching the closed form $F(1,c_3)$ and $F(1-t,0)$
to within $0.002$. The two reducing levers coincide because they enter the floor only through the
product $g(1-c_3)$.

\subsection{Bottom-up/top-down consistency (Theorem~\ref{c2:thm:consistency})}
\label{c2:ssec:num-consistency}

\begin{figure}[htbp]
\centering
\includegraphics[width=\linewidth]{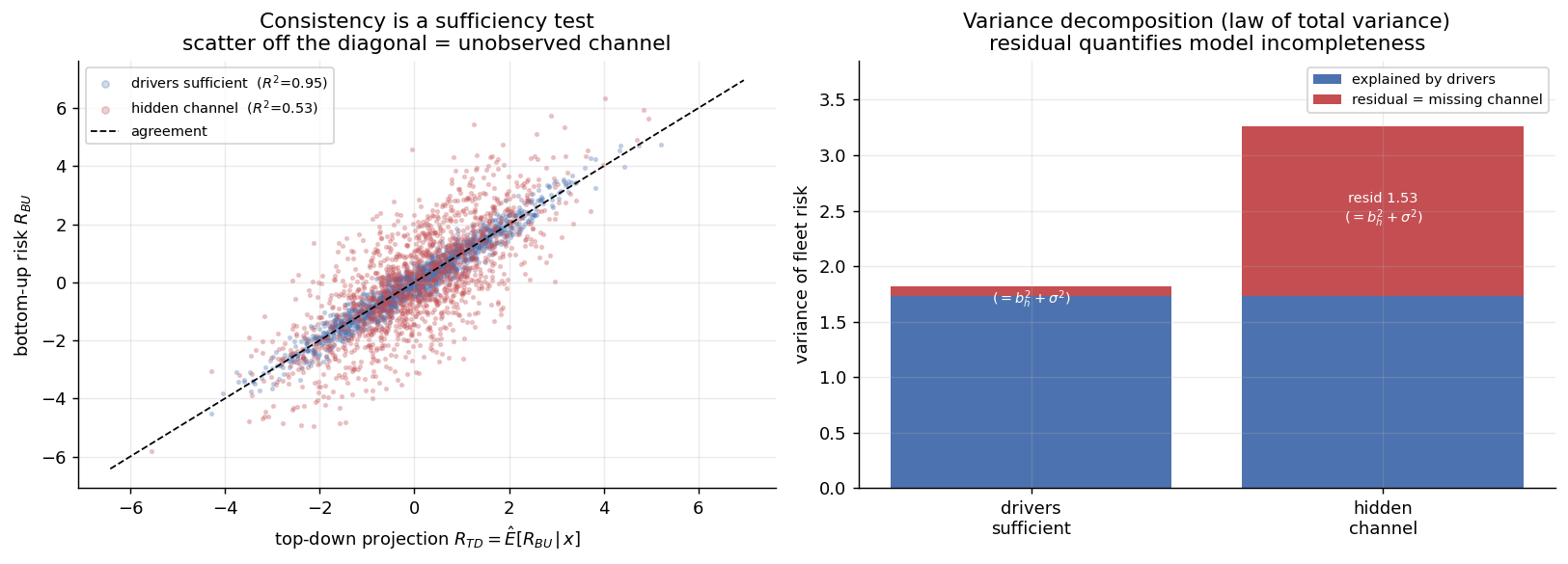}
\caption{\textbf{Consistency as a sufficiency test (Theorem~\ref{c2:thm:consistency}).} Left:
bottom-up risk against its top-down projection; with sufficient drivers the points lie on the
diagonal ($R^2=0.95$), with a hidden channel they scatter ($R^2=0.53$). Right: the law-of-total-variance
decomposition; the residual equals $b_h^2+\sigma^2$ and quantifies the unobserved channel.}
\label{c2:fig:consistency}
\end{figure}

Figure~\ref{c2:fig:consistency} regresses a bottom-up risk $\Rbu=x^\top b_x+b_h h+\sigma\eta$
($b_x=(1,0.7,0.5)$, $\sigma=0.3$) on the observable drivers $x$ in two regimes. When the drivers are
sufficient ($b_h=0$) the residual variance is $0.090$, matching the closed form $b_h^2+\sigma^2=0.090$
to $0.07\%$, and the projection achieves $R^2=0.95$, with the scatter hugging the diagonal. When a hidden
frailty channel is present ($b_h=1.2$) the residual jumps to $1.528$, matching $b_h^2+\sigma^2=1.530$
to $0.14\%$. There $R^2$ falls to $0.53$, and the scatter fans off the diagonal. The variance
decomposition \eqref{c2:eq:ltv} (right panel) shows the residual is not noise but a measurement of the
unobserved channel. Consistency between the two functionals is a sufficiency test, and the gap is the
missing risk.

\subsection{The mismatch index in action}
\label{c2:ssec:num-lmi}

\begin{figure}[htbp]
\centering
\includegraphics[width=\linewidth]{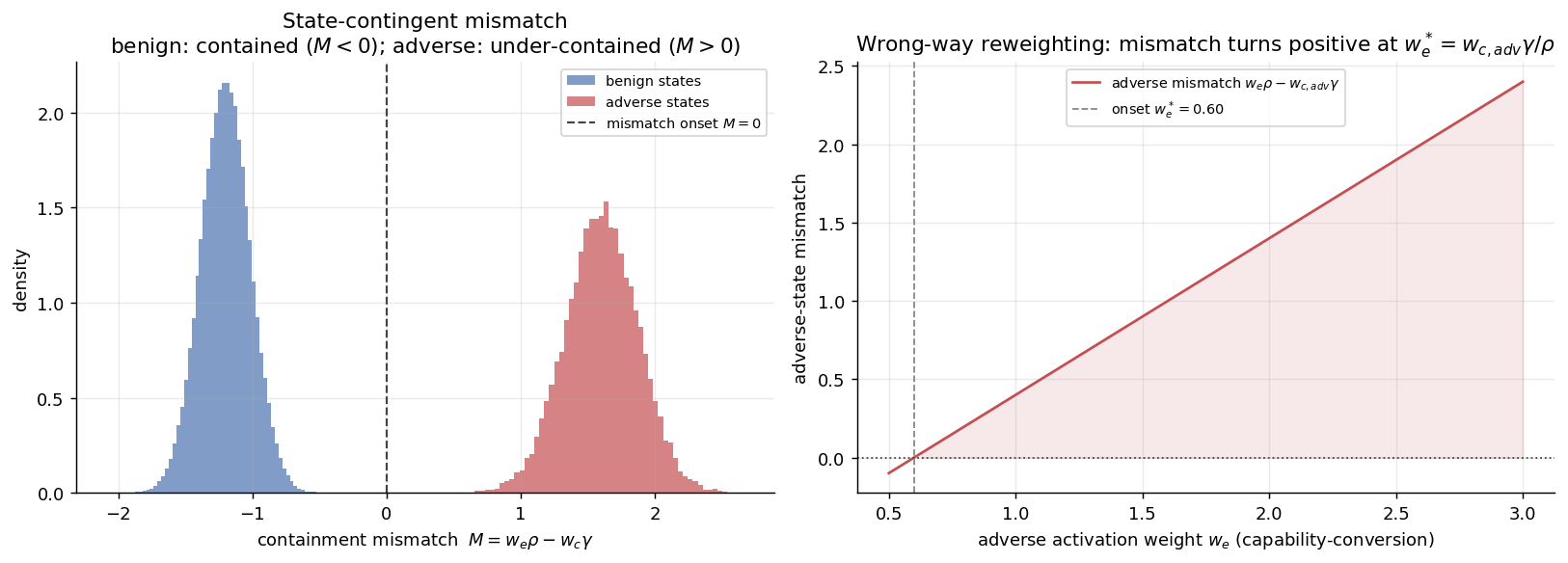}
\caption{\textbf{State-contingent containment mismatch (Definition~\ref{c2:def:lmi}).} Left: the
mismatch distribution is negative (contained) in benign states and positive (under-contained) in
adverse states, for fixed architecture. Right: the adverse-state mismatch crosses zero at
$w^{e\star}=w^{c}\cont/\reach$ as the capability-conversion weight grows.}
\label{c2:fig:lmi}
\end{figure}

Figure~\ref{c2:fig:lmi} illustrates the bottom-up functional of \S\ref{c2:ssec:lmi} directly. Set
reach $\reach=1$ and containment capacity $\cont=1.5$. The benign-state weights
$(w^{e},w^{c})=(0.3,1.0)$ give a mismatch $\Mis=-1.20$ (comfortably contained), while the adverse-state
weights $(w^{e},w^{c})=(2.2,0.4)$ give $\Mis=+1.60$ (under-contained). The same architecture flips from
safe to exposed purely through the state-contingent reweighting. The mismatch turns positive once the
adverse activation/efficacy ratio crosses the containment-to-reach ratio $\cont/\reach$, at an
adverse activation weight $w^{e\star}=w^{c}\cont/\reach$. The fraction of states in mismatch tracks the
systematic probability $\psys$. That locates the exposure in the adverse states where the common
shock fires, making the wrong-way property of Definition~\ref{c2:def:lmi} visible.

\subsection{Agents that respond to each other (\S\ref{c2:sec:interaction})}
\label{c2:ssec:num-interaction}

\begin{figure}[htbp]
\centering
\includegraphics[width=\linewidth]{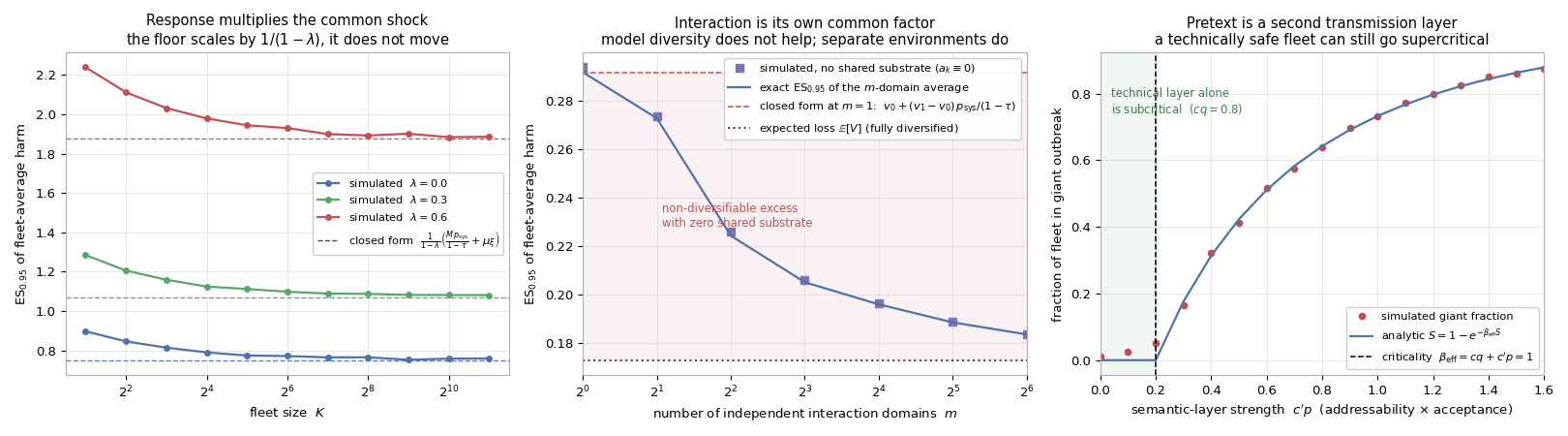}
\caption{\textbf{Agents that respond to each other (\S\ref{c2:sec:interaction}).} Left: expected
shortfall of fleet-average harm versus fleet size for three responsiveness levels; the floor is scaled
by $1/(1-\lambda)$ and is otherwise unchanged (Theorem~\ref{c2:thm:multiplier}). Centre: with
\emph{zero} shared substrate ($a_k\equiv 0$) and state-dependent responsiveness, a non-diversifiable
excess survives; splitting the fleet into $m$ independent interaction domains reduces it only at rate
$m^{-1/2}$ (Theorem~\ref{c2:thm:endogfactor}). Right: with the technical layer held subcritical at
$cq=0.8$, the semantic layer alone carries the fleet past criticality at $\beff=cq+c'p=1$
(Theorem~\ref{c2:thm:pretext}).}
\label{c2:fig:interaction}
\end{figure}

The last study checks the three closed forms of \S\ref{c2:sec:interaction}. Figure~\ref{c2:fig:interaction}
has one panel per result, and keeps the same unitless calibration as above: $\mu_\xi=0.15$, $\psys=0.03$,
$\tau=0.95$, $M=1$.

Panels (a) and (b) evaluate the fleet average at the closed-form fixed point of the linear-in-means
system, so we first check that the shortcut is legitimate. Solving the $K$-agent system
\eqref{c2:eq:linmean} numerically for $K=200$ and $\lambda=0.6$, over $60$ draws of the loadings, the
shock and the idiosyncratic terms, the iterated solution and the closed form
$(\bar a_K J+\bar\xi_K)/(1-\lambda)$ agree to a worst relative discrepancy of $2.9\times10^{-12}$. The
contraction argument in the proof of Theorem~\ref{c2:thm:multiplier} is doing what it claims.

\paragraph{(a) The multiplier scales the floor and moves nothing else.} We simulate out to $K=2048$
with $1.5\times 10^6$ episodes per point, for $\lambda\in\{0,0.3,0.6\}$. Each curve flattens on its own
floor: $0.759$ against the closed form $0.750$ at $\lambda=0$ ($1.2\%$), $1.081$ against $1.071$ at
$\lambda=0.3$ ($0.9\%$), and $1.886$ against $1.875$ at $\lambda=0.6$ ($0.6\%$). The ratio of the
top curve to the bottom is $2.49$ against the predicted $1/(1-0.6)=2.5$. Two features matter more than
the fit. The floor persists; responsiveness does not restore diversification, and the whole
family is one curve rescaled. Interaction of this kind multiplies the common shock the substrate
already delivers. Set $\psys=0$ and every curve collapses to the constant $\mu_\xi/(1-\lambda)$, with no
tail at all.

\paragraph{(b) Interaction manufactures a common factor on its own.} Here we switch the substrate off
entirely: $a_k\equiv 0$, so no agent has any exposure to a shared shock. The only common object is the
responsiveness itself, which takes $\lambda_0=0.10$ in the benign state and $\lambda_1=0.60$ in the
adverse state, with $\Prob(\text{adverse})=\psys=0.03$. The limit is then the two-point variable $V$
with $v_0=0.167$ and $v_1=0.375$, mean $\E[V]=0.173$. With a single interaction domain the simulated
$\ES_{0.95}$ is $0.294$ against the closed form $0.292$ ($0.7\%$), a non-diversifiable excess of
$0.121$ over expected loss against the predicted $(v_1-v_0)\psys\tau/(1-\tau)=0.119$. A fleet with
\emph{no shared model, no shared prompt and no shared tool} still carries a tail charge, purely because
its members respond to each other and respond harder when conditions are bad. The state-dependence of
$\lambda$ is an assumption and not a finding; what the panel checks is that it alone suffices.

The curve then reports what splitting the fleet into $m$ independent interaction domains achieves. It is
monotone decreasing in $m$, as Theorem~\ref{c2:thm:endogfactor}(iii) requires, and it tracks the exact
$\ES_{0.95}$ of the $m$-domain average at every point (worst relative error $0.7\%$, and $0.04\%$ by
$m=64$). But the decline is slow: the excess over $\E[V]$ is $0.023$ at $m=16$ and still $0.011$ at
$m=64$, falling like $m^{-1/2}$; it does vanish, as the theorem says, but slowly. Separating agents into distinct interaction
environments is a real control, and it is a weak one. Diversifying the substrate, the usual advice, does
nothing here at all, because there is no substrate to diversify.

\paragraph{(c) Pretext is a second transmission layer.} We hold the technical layer fixed and
\emph{subcritical} at $cq=0.8$, so a fleet with no semantic channel never sees a giant outbreak, and
vary the semantic-layer strength $c'p$ from $0$ to $1.6$. The simulation draws the union graph
directly, so it adopts the route-independent infectiousness of Assumption~\ref{c2:ass:pretext} rather
than testing it; what it tests is the passage from that assumption to the threshold. The simulated
giant-component fraction ($K=4000$) tracks the analytic $S=1-e^{-\beff S}$ with $\beff=cq+c'p$ to a
mean absolute error of $0.005$ above threshold. The fleet stays contained through $c'p=0.2$, where $\beff=1.0$
exactly, and then goes supercritical: $c'p=0.4$ gives $\beff=1.2$ and an outbreak reaching $31\%$ of
the fleet, and $c'p=0.8$ gives $\beff=1.6$ and $64\%$. The supervisory reading is direct. An operator
who has certified the technical layer as safe has certified $cq<1$, which is not the relevant
condition. Free addressability among agents can carry the fleet over criticality with the technical
layer untouched.

\section{Stylized applications}
\label{c2:sec:apps}

The body is domain-free. This section fills it in. We instantiate the framework in three deliberately
unlike domains, and for each we specify the four things an instantiation requires: the harm axes and their
reversible/irreversible split, the affordance-to-reach map, the containment levers, and the stress states
and their drivers. We then read off the consequences of the theorems. The parameters are
\emph{illustrative}: chosen to be plausible and to make the structure legible, not estimated from data.
The point is that the same definitions and the same four theorems govern all three, with banking appearing
as one instantiation rather than the foundation.

\subsection{A code-automation fleet}
\label{c2:ssec:app-code}

Consider $K$ autonomous software agents on a shared foundation model and a shared CI/CD toolchain, each
able to read, modify, merge, and deploy code. The harm axes are a service \emph{outage} (reversible:
roll back), \emph{data loss} (partly irreversible), and a \emph{security regression} (reversible if caught,
else not). The irreversible axis $\mathcal I$ is a destructive infrastructure operation: deleting a
production datastore, force-pushing over history, rotating credentials irrecoverably. Reach $\reach_{kj}$
is the blast radius of an agent's deploy permissions, meaning how many services it can reach. Containment
$\cont_{kj}$ is staged-rollout coverage, automated rollback, and required human review. The systematic
state, the common shock, is a shared dependency: a poisoned package or a foundation-model update that
simultaneously degrades every agent's judgement. The dependency graph $G$ is shared-repository and
shared-credential reachability, and transmission $q$ is the chance a compromise on one agent reaches
another through those shared surfaces. The drivers $x(s)$ for the top-down projection are deployment
frequency (autonomy), fraction of merges on the shared model (concentration), breadth of deploy scopes,
time since the last model or dependency review (drift), and rollback coverage.

The theorems then speak directly. By Theorem~\ref{c2:thm:nondiv}, a bad model update is a common shock whose
fleet-wide cost does not diversify across the $K$ agents. Staging the rollout (a canary deployment) is
the act of cutting the systematically-exposed fraction $g$, and so it lowers the floor. By
Theorem~\ref{c2:thm:contagion}, a single poisoned dependency cascades fleet-wide if shared credentials let
one compromised agent reach others with branching ratio above one. Isolating credential scopes per segment
reduces $\beta$ and caps component size, which holds the cascade subcritical. By
Theorem~\ref{c2:thm:floor}, no amount of post-hoc review or rollback lowers the floor on \emph{irreversible}
destructive operations. Only removing the destructive affordance from the agent's permission set
(prevention) or segmenting which agents hold it does.

\subsection{A clinical-triage fleet}
\label{c2:ssec:app-clinical}

Consider $K$ triage assistants behind one clinical model, integrated with one electronic-health-record
system, each able to recommend, route, and flag. The harm axes are \emph{patient-harm severity}
(a recommendation acted upon that worsens an outcome, irreversible once the clinical action is taken),
\emph{privacy} (record exposure, partly reversible), and \emph{mis-triage rate} (reversible if a
downstream clinician overrides). The irreversible axis is a harmful recommendation that is \emph{acted on}
before human review. Reach is the number of patients an agent's recommendations influence per shift.
Containment is the rate and timeliness of clinician override and the strength of guardrails on high-acuity
pathways. The systematic state is a model update or a population shift (a new presentation pattern) that
degrades calibration across the whole fleet at once. The dependency graph is weaker here: agents do not
directly compromise one another. Contagion is therefore mild, and the dominant fleet risk is the common
shock and the irreversible floor. The drivers are case volume per agent (autonomy), fraction on the shared
model (concentration), acuity mix (reach), time since revalidation against outcomes (drift), and override
coverage.

The consequences follow. By Theorem~\ref{c2:thm:nondiv}, a calibration-degrading model update harms all
agents' patients together, and the fleet-aggregate patient-harm tail does not average down with more
agents. Fleet size is not a safety margin. By Theorem~\ref{c2:thm:floor}, the floor on harmful
recommendations \emph{acted on without review} is reducible only by routing high-acuity decisions through
mandatory human sign-off (prevention: the agent cannot finalise the irreversible action), or by limiting
which agents operate on high-acuity pathways (segmentation). Faster post-hoc audit (a detective control)
does not lower it. The mismatch index makes the wrong-way structure explicit: under a population shift the
agents' reach into high-acuity cases rises exactly as override capacity is overwhelmed.

\subsection{A financial fleet: recovering Chapter~\ref{ch:single}}
\label{c2:ssec:app-fin}

The financial instantiation is the one in which the PD/LGD/EAD vocabulary of Chapter~\ref{ch:single} is
literal rather than analogical, and it recovers Chapter~\ref{ch:single} as the single-agent special case.
Consider $K$ customer-operations or trading agents on a shared model and shared market/payment
infrastructure, each able to transact, transfer, and access data. The harm axes are \emph{financial loss},
\emph{data exposure}, and \emph{compliance breach}. The irreversible axis is an irrevocable transfer or
market action above a catastrophic threshold $\theta$, Chapter~\ref{ch:single}'s catastrophic class.
Reach is transaction authority and position limits. Containment is the salience-ranked human-review
capacity and reversibility window of Chapter~\ref{ch:single}. The systematic state is
Chapter~\ref{ch:single}'s single common factor: an adversarial regime that now fires \emph{fleet-wide}
through the shared model. It simultaneously inflates activatable exposure (dormant authority converts) and
collapses review efficacy (the wrong-way reweighting). The dependency graph is shared trading or payment
rails through which one agent's compromised actions move others' exposure.

At $K=1$, with the common factor as Chapter~\ref{ch:single}'s trigger, the mismatch index and prevention
floor reduce to Chapter~\ref{ch:single}'s set-valued acceptance set and its prevention theorem. The fleet
adds the three new effects. By Theorem~\ref{c2:thm:nondiv}, a fleet of such agents sharing a model shares
the adversarial regime, and the aggregate financial tail floors above expected loss no matter how many
agents are pooled. The fleet does not diversify the systematic loss, which recovers and generalises
Chapter~\ref{ch:single}'s monoculture finding. By Theorem~\ref{c2:thm:floor}, the floor on irrevocable
catastrophic transfers is untouchable by review or reversibility, as in Chapter~\ref{ch:single}. It is
reducible only by a preventive control (an authorisation the agent cannot hold, routing large irrevocable
transfers to a separate non-agentic path) or by segmentation, and it is now also immune to fleet
diversification. Table~\ref{c2:tab:apps} summarises the three instantiations against the common template.

\begin{table}[htbp]
\centering
\footnotesize
\setlength{\tabcolsep}{4pt}
\renewcommand{\arraystretch}{1.25}
\caption{Three stylized instantiations of the same framework. The structure (mismatch index on
$\mathcal R$, prevention floor on $\mathcal I$, the four theorems) is shared; only the fillings differ.}
\label{c2:tab:apps}
\begin{tabular}{p{0.17\linewidth} p{0.25\linewidth} p{0.27\linewidth} p{0.21\linewidth}}
\toprule
& \textbf{Code automation} & \textbf{Clinical triage} & \textbf{Financial fleet} \\
\midrule
Harm axes & outage, data loss, security regression & patient harm, privacy, mis-triage & financial loss,
data, compliance \\
Irreversible $\mathcal I$ & destructive infra op & harmful recommendation acted on & irrevocable
catastrophic transfer \\
Reach $\reach$ & deploy blast radius & patients influenced & transaction authority \\
Containment $\cont$ & staged rollout, rollback, review & clinician override & review capacity,
reversibility \\
Common shock & poisoned dependency / model update & calibration-degrading update & fleet-wide adversarial
regime \\
Contagion graph & shared creds / repos & weak (no direct cross-compromise) & shared trading/payment rails \\
Dominant fleet risk & common shock + contagion & common shock + irreversible floor & all three (recovers
Chapter~\ref{ch:single}) \\
\bottomrule
\end{tabular}
\end{table}

\noindent In each domain, a fitted top-down (CLASS) layer would regress observed fleet-incident severity
on that domain's drivers and be checked against the bottom-up mismatch aggregate via
Theorem~\ref{c2:thm:consistency}. A persistent divergence would localise the missing channel: an
unmonitored shared dependency, an unmodelled contagion path. We do not fit these here. Doing so requires
real per-domain incident telemetry, and the caveat of \S\ref{c2:sec:conclusions} is that the
systematic parameters are the least observable. The driver taxonomy and the consistency test are
nonetheless identical across domains, which is the generality claim.

\section{The fleet governance pipeline}
\label{c2:sec:governance}

Chapter~\ref{ch:single} adapted the traditional model-risk-management life cycle to a single agent. A fleet
inherits that life cycle but adds the population-level controls the theorems identify. We state the
adaptation stage by stage. Table~\ref{c2:tab:fleet-pipeline} summarises it.

\begin{table}[htbp]
\centering
\footnotesize
\setlength{\tabcolsep}{5pt}
\renewcommand{\arraystretch}{1.3}
\caption{The model-risk-management life cycle, adapted for a fleet. The right column names the result that
governs each stage.}
\label{c2:tab:fleet-pipeline}
\begin{tabular}{p{0.24\linewidth} p{0.50\linewidth} p{0.16\linewidth}}
\toprule
\textbf{Stage} & \textbf{Fleet implementation} & \textbf{Result} \\
\midrule
Inventory & Fleet register: tier agents by autonomy $\times$ reversibility $\times$ blast radius
\emph{and} by shared-dependency loading (which model, prompts, tools, credentials). &
\S\ref{c2:ssec:primitives} \\
Measurement & Bottom-up mismatch index on $\mathcal R$ + prevention floor on $\mathcal I$, evaluated over
the systematic state law. & Def.~\ref{c2:def:lmi}, \eqref{c2:eq:floor} \\
Aggregation & Charge the common shock in full (no diversification credit); pool only the idiosyncratic
part. & Thm.~\ref{c2:thm:nondiv} \\
Architecture & Hold the contagion branching ratio below one via segmentation and model/credential
heterogeneity; size segments to cap blast radius. & Thm.~\ref{c2:thm:contagion} \\
Design gate & For each irreversible class, a preventive boundary is mandatory; detective controls and fleet
size are insufficient. & Thm.~\ref{c2:thm:floor} \\
Validation & Independent top-down (CLASS) projection cross-checked against the bottom-up aggregate;
divergence localises the missing channel. & Thm.~\ref{c2:thm:consistency} \\
Monitoring & Track the drivers $x(s)$ and the realised mismatch; a shared-model version change is a
fleet-wide re-validation trigger. & \S\ref{c2:ssec:class} \\
\bottomrule
\end{tabular}
\end{table}

\paragraph{Inventory and tiering.} The fleet register extends the single-agent inventory with a
\emph{shared-dependency} dimension, because the loading on a common factor is what determines
non-diversifiable risk. Two agents identical in autonomy and blast radius but on different base models are,
at the fleet level, a materially safer pair than two on the same model. The register must record the
dependency so the aggregation in Theorem~\ref{c2:thm:nondiv} can be applied.

\paragraph{Aggregation without false diversification.} The decisive governance error a fleet invites is to
set aggregate appetite as if agents were independent. Theorem~\ref{c2:thm:nondiv} forbids it, because the systematic
component is charged in full at any fleet size. The aggregate acceptance constraint must therefore separate
the common-shock charge (undiversified) from the idiosyncratic charge (pooled), and the appetite must be set
against the former, since the monoculture is priced, not averaged.

\paragraph{Architecture as a first-class control.} The fleet has a lever the single agent lacks, its own
composition. Keeping the contagion branching ratio subcritical is a design-time decision
governed by Theorem~\ref{c2:thm:contagion}, and it is the only thing that bounds the blast radius of a
cascade. It is kept subcritical through segmentation, which isolates credential and tool scopes into blocks,
and through heterogeneity, which diversifies base models so an exploit does not transfer. Segmentation also
appears in Theorem~\ref{c2:thm:floor} as one of the two levers on the irreversible floor, so it does double
duty.

\paragraph{The preventive design gate.} For every irreversible harm class, Theorem~\ref{c2:thm:floor} converts
a tuning question into a binary gate. Either a preventive boundary removes the affordance, or the fleet
floor on that class is non-compliant and no detective budget or fleet-sizing decision will fix it. This is
the fleet form of Chapter~\ref{ch:single}'s a-priori infeasibility certificate, and it should be enforced
before deployment.

\paragraph{Validation as a consistency test.} The bottom-up and top-down functionals give the second line
of defence an instrument: compute both and compare. By Theorem~\ref{c2:thm:consistency}, agreement certifies
that the monitored drivers capture the fleet's risk. A persistent gap, by contrast, is a quantified
measurement of an unmonitored channel: a frailty factor, a contagion path, an affordance the architectural
inventory missed. Such a gap should trigger investigation rather than reconciliation by adjustment.

\section{Conclusions, assumptions, and limitations}
\label{c2:sec:conclusions}

We have built a domain-general measure of \emph{systemic} risk for a fleet of agentic systems. The step from
the single-agent object of Chapter~\ref{ch:single} to the population follows the step finance took from
single-name credit to systemic risk: abandon per-agent prediction in favour of a distributionally-defined,
stress-conditional vulnerability. The measure pairs a bottom-up containment-mismatch index with a top-down
reduced-form stress projection. The bottom-up index is a domain-agnostic liquidity-mismatch index on the
reversible harm subspace, with a non-nettable prevention floor on the irreversible subspace. The measure
requires both within appetite across an admitted adversary family. Four results follow for a tractable
class, each verified against a closed form. A common shock is non-diversifiable
(Theorem~\ref{c2:thm:nondiv}). Cross-agent contagion has a sharp criticality threshold that segmentation and
heterogeneity control (Theorem~\ref{c2:thm:contagion}). The irreversible-harm floor is non-substitutable and
immune to fleet diversification, reducible only by prevention or segmentation
(Theorem~\ref{c2:thm:floor}). And the two functionals are consistent if and only if the observable drivers
are sufficient, which makes their gap a validation diagnostic (Theorem~\ref{c2:thm:consistency}).
Section~\ref{c2:sec:interaction} then lets the agents respond to and address one another. Two further
results follow. Responsiveness that varies with the stress state makes the fleet average random even
when no agent shares a substrate with any other, so the standard remedy for monoculture, running
different models, does not close the channel (Theorem~\ref{c2:thm:endogfactor}). And addressability
is a second contagion layer that adds to the technical one, which moves the criticality threshold and
turns the exposed fraction of the irreversible floor into an equilibrium object
(Theorem~\ref{c2:thm:pretext}, Corollary~\ref{c2:cor:floor-endog}).

\subsection{Assumptions}
The results rest on explicit, stated assumptions. The mismatch index nets exposure against containment only
\emph{within} a harm axis, never across axes, which preserves the no-common-unit discipline. Netting within
an axis presumes that exposure and containment are commensurable in that axis's units. That is a real
modelling commitment, defensible per axis but not free. The theorems hold for a tractable systematic
structure: a common-shock-plus-idiosyncratic harm model with a coherent statistic (ES) and a finite admitted
adversary family. The contagion result assumes an Erd\H{o}s--R\'enyi dependency graph and
independent-cascade transmission. The qualitative criticality is robust to graph detail; the exact threshold
is not. The interaction results add three more commitments. Response is linear in the peer mean, which is
the specification whose identification problems are best understood but which rules out thresholds and
saturation. The semantic graph is taken independent of the technical one, whereas in practice agents
that share tools are also the agents that talk. And an agent persuaded into acting is assumed to become
a source on both layers, which is what lets the two branching ratios add; without it the threshold sits
at the larger of the two rather than at their sum, and the additive form is a conservative bound. Most importantly, every numerical parameter in the applications is \emph{illustrative}, chosen for
plausibility and legibility, not estimated.

\subsection{Limitations}
Three limitations bound the contribution. First, the framework is \emph{static}. It measures a fleet's
stress-conditional exposure but not the \emph{dynamics}. The timing race between a fast-activating
affordance and slow-mobilizing containment is encoded only coarsely in the state-contingent weights, and
genuine propagation in time (a cascade unfolding, an intensity that self-excites) is outside the static
model. Second, and most consequentially, \emph{calibration} of the systematic parameters is the binding
empirical obstacle, and it is largely unsolved. The common-shock probability, the frailty loadings, and the
contagion transmission are the rare, adaptive, nearly-unobserved quantities that a competent
adversary works to invalidate. One may have seen zero or one fleet-wide events in an entire history, so the
most important parameter is the least estimable, which is the classic operational-risk tail problem. This
argues for elicited, stress-assumed parameters with sensitivity analysis rather than point estimation, and
for treating the framework as a structural lens rather than a calibrated forecast. Third, the mechanisms are not separately identifiable from aggregate data without strong assumptions.
An unobserved common factor and genuine event-triggering already mimic each other, so frailty and
contagion cannot be told apart from aggregates alone. Section~\ref{c2:sec:interaction} adds a third
candidate with the same aggregate signature, an endogenous response to peers, which is Manski's
reflection problem in fleet clothing \citep{manski1993identification}. The decomposition into
mechanisms is therefore a modelling choice to be defended, not read off the data, and separating them
needs design rather than more data: exogenous variation in who may address whom, or in the
responsiveness of the interaction protocol, is the fleet analogue of an instrument.

\subsection{Extensions}
The most direct extension is to make the framework \emph{dynamic}, which is the subject of a sequel. In
continuous time the common shock becomes a common \emph{jump} in a multivariate L\'evy model. It is
non-diversifiable for the same reason a common shock is here, but now the full machinery of jump measures
applies, with the result that prevention truncates the jump measure's tail while detection only thins
arrivals. Contagion becomes a multivariate Hawkes process whose branching matrix has a spectral radius
playing the role of our $\beta$. The subcritical condition $\rho(\Gamma)<1$ is then a continuous-time
fleet-stability criterion, and the peer responsiveness $\lambda$ of \S\ref{c2:sec:interaction} is the
static shadow of that matrix's off-diagonal mass. Two further extensions come from the interaction
layer. The acceptance probability $p$ is treated here as a parameter, but a strategic attacker chooses
the pretext and will choose the one that works, which makes $p$ a decision variable and the
technical-versus-semantic split an attacker's budget allocation, the problem
Chapter~\ref{ch:cyber} takes up. And the independence of the two graphs should be dropped: the
realistic case is a positive correlation between sharing a tool and being able to address. The
direction is not obvious. Overlapping edges make the union sparser than the sum of its parts, which
lowers $\beff$; concentrating both kinds of edge on the same hubs raises degree heterogeneity, which
raises the branching ratio. Which force wins is an empirical question, and the additive form we report
is a benchmark rather than a bound. The speed mismatch at the heart of the mismatch index becomes a genuine
first-passage race between harm completion and detection-and-remediation. The robust containment problem
becomes a stochastic differential game whose ambiguity set is a family of equivalent measure changes
penalised by attacker cost. Within the static frame, dropping the disjoint-channel and
Erd\H{o}s--R\'enyi assumptions yields a robust convex program over general dependency graphs. Giving the
attacker-cost function empirical content from red-team economics would turn the ambiguity set from a stress
menu into a model of the threat. None of these changes the thesis of this chapter: a fleet's systematic risk
is non-diversifiable, its contagion is critical, its irreversible floor is non-substitutable, its
agents can manufacture common risk simply by responding to each other, and the structure of all four
is domain-general. It is the calibration that remains the work.

\chapter{Systemic Risk in Continuous Time: A Jump-Diffusion and Self-Excitation Synthesis}
\label{ch:continuous}
\section{Introduction}
\label{c3:sec:intro}

Systemic risk has a time dimension. Financial crises are dynamic processes, not static states:
distress arrives in clusters, a common shock strikes as a jump rather than a gentle drift, and one
failure raises the hazard of the next. The two preceding chapters built the static apparatus. They
moved from the microprudential model risk of a single agent, the discipline of traditional model
risk management \citep{fed2011sr117}, to the systemic risk of a fleet that shares a foundation
model. Chapter~\ref{ch:single} decomposed a single agent's expected harm into the probability of a
harmful action ($\PHA$), the severity given harm ($\SGH$), and the blast radius ($\BR$). It set
that decomposition inside a set-valued, containment-subadditive risk measure. It then proved that
against irreversible harm the three containment levers (review, reversibility, prevention) are
not substitutes, and only prevention moves the catastrophic floor. Chapter~\ref{ch:fleet} scaled the
unit of analysis from one agent to a fleet. It replaced the single-name decomposition with a
distributionally-defined vulnerability measure in the spirit of systemic-risk measures and the
liquidity-mismatch index. And it established non-diversifiability of a common shock, a contagion
criticality threshold, a fleet prevention floor, and the consistency of bottom-up and top-down
projections.

Both are snapshots. The quantities they decompose, however, live in time. Harmful actions arrive as
discrete events, at a rate that itself fluctuates with adversarial pressure and capability drift. A
systematic shock hits as a jump, not a gentle drift. Contagion propagates as one compromise raises
the hazard of the next. And a governance decision made today must remain coherent as the world
unfolds. The governance questions are therefore dynamic. At what rate does harm actually
accumulate, and how far does the static mean understate it once events are not rare? When is a
fleet on a stationary path and when on an explosive one? How should a controller who does not know
the true intensities and severities act, so as to be robust to the misspecification an adversary
would most like to exploit, and remain robust as time passes? None of these is a static question,
and none is answered by a snapshot.

This chapter develops the continuous-time theory. In doing so it unifies Chapters~\ref{ch:single}
and~\ref{ch:fleet} as a single object: a \emph{jump-diffusion}. A single agent's cumulative harm is
a marked point process. Its stochastic intensity is the continuous-time form of $\PHA$, its mark
law is $\SGH$, and its exposure multiplier is $\BR$, with a latent diffusion modulating the rate.
We prove that this object reduces exactly to Chapter~\ref{ch:single}'s identity in the rare-event
limit, and that it exposes the multiple-harm correction the snapshot omits. At the fleet level a
common L\'evy jump carries systematic risk and a multivariate Hawkes process carries contagion, and
the three static fleet results lift to continuous time. The risk functional becomes a
time-consistent recursive entropic measure. We show that this measure is the value of a robust
stochastic differential game whose worst case is an explicit Girsanov drift change and Esscher jump
tilt.

Modelling harm as a jump-diffusion, rather than a pure diffusion, is a substantive choice. The
consequential moments in an agent's operation are discrete events, not the continuous wandering of
a state variable: a harmful action taken, an irreversible failure, a platform-wide exploit, a
jailbreak propagating across a shared tool, and the \emph{jump} carries every result here. The
non-diversifiable shock is a common jump, the contagion is an excitatory jump cascade,
the catastrophic class is the tail of the jump measure, and the prevention lever is the truncation
of that measure. Modelling contagion as excitatory point-process branching also gives a clean
criticality, the spectral radius of a nonnegative matrix, without the sign subtleties of
diffusion-coupling stability analyses. That keeps the present results mathematically disjoint from
that separate line of continuous-time work.

\paragraph{Contributions.}
(i) We define the per-agent jump-diffusion harm object and prove it reduces to
Chapter~\ref{ch:single}'s $\EH=\PHA\cdot\SGH\cdot\BR$ in the rare-event limit, with an explicit
multiple-harm correction $\kappa\ge1$, and we identify the three levers with three operations on
the jump model (Proposition~\ref{c3:prop:reduction}). (ii) We prove that a systematic shock
modelled as a common L\'evy jump is non-diversifiable, with an explicit fleet floor
(Theorem~\ref{c3:thm:t1}). (iii) We prove that Hawkes contagion is stable if and only if the
spectral radius of its branching matrix is below one, that the stationary intensity diverges at
criticality, and that segmentation divides the branching ratio (Theorem~\ref{c3:thm:t2}). (iv) We
prove that against an irreversible jump only prevention (truncating the jump measure) and
segmentation move the catastrophic floor (Theorem~\ref{c3:thm:t3}). (v) We show that the recursive
entropic measure is time-consistent where iterated shortfall is not
(Proposition~\ref{c3:prop:tc}). (vi) We show that this measure is the value of a robust
Girsanov/Esscher game, with an explicit worst-case measure and a finiteness condition that is
itself the prevention lever (Theorem~\ref{c3:thm:robust}). (vii) We confirm every closed form
numerically (Section~\ref{c3:sec:numerics}).

Throughout, parameter values are illustrative and the contribution is structural. The framework,
the reductions, and the qualitative laws are the result. Calibrating intensities, mark laws,
exposures, loadings, and branching matrices to real deployed systems is the central open problem.
We return to it in Section~\ref{c3:sec:conclusions}.

\section{Background and related work}
\label{c3:sec:background}

\subsection{The single-agent decomposition (Chapter~\ref{ch:single})}

Chapter~\ref{ch:single} modelled a single agent at the level of a decision. Expected harm factored
as $\EH=\PHA\cdot\SGH\cdot\BR$: the probability that the agent takes a harmful action, the severity
of the harm conditional on its occurrence, and the blast radius over which that harm is felt. Harm
was treated as a vector over deployer-defined axes, split into reversible and irreversible
components. The
acceptance set was a set-valued, containment-subadditive risk measure rather than a scalar. Three
containment levers acted on the decomposition: review (a detective control that catches harmful
actions), reversibility (the capacity to undo realised harm), and prevention (removing the
affordance for harm). Chapter~\ref{ch:single}'s central qualitative result (its Theorem~B) was a
non-substitutability statement. Against irreversible harm, review and reversibility cannot
substitute for prevention, because neither reaches the irreversible tail. Chapter~\ref{ch:single}
also posed the robust version, evaluating harm over an ambiguity set around the nominal model.

\subsection{The fleet framework (Chapter~\ref{ch:fleet})}

Chapter~\ref{ch:fleet} changed the unit of analysis from one agent to a fleet of $K$, mirroring
finance's move from single-name to systemic risk. It defined two stress-conditional vulnerability
functionals. The first was bottom-up and architectural: a containment-\emph{mismatch} index, a
domain-general analogue of the liquidity-mismatch index \citep{brunnermeier2009liquidity}. On the
reversible subspace it measured the gap between the reach of harm and the containment applied to
it, with a non-nettable prevention floor on the irreversible subspace. The second was top-down and
reduced-form. It projected the fleet-aggregate tail from a small set of drivers, in the manner of
systemic-risk measures such as CoVaR \citep{adrian2016covar} and SRISK
\citep{acharya2017measuring}. Chapter~\ref{ch:fleet} established four results. A common shock is
non-diversifiable. Contagion exhibits a criticality threshold, a critical branching ratio above
which a giant cascade emerges, as in bond percolation. Prevention sets a fleet floor that detective
controls and fleet size cannot lower. And the bottom-up and top-down measures agree when the
drivers are sufficient, their disagreement measuring a missing channel.

\subsection{Why continuous time, and why jumps}

Chapters~\ref{ch:single} and~\ref{ch:fleet} are static by construction: each evaluates risk at a
snapshot. This chapter asks how the same quantities evolve, and the natural object is a
continuous-time process. The modelling choice that organises everything is that harm is
\emph{jump}-driven. An agent's harmful actions, the catastrophic failures that matter most, the
platform-wide shocks that hit a whole fleet, and the contagious propagation of an exploit are all
discrete events. The latent conditions that set their rate and reach (adversarial pressure,
capability, load) vary continuously. This is a jump-diffusion: continuous modulation of
the rate and exposure, discrete arrival of harm. The same choice separates this work cleanly from a
parallel continuous-time literature that models agents' \emph{state} equations as coupled
diffusions and studies their joint stability through Lyapunov methods. There, contagion is
continuous co-movement through a sign-indefinite coupling matrix, and the stability boundary
involves the symmetric part of that matrix. Here contagion is an \emph{excitatory} point-process
cascade, its criticality is the Perron root of a nonnegative branching matrix, and the carrier of
systematic risk is a common jump. These objects and arguments do not overlap with the
diffusion-coupling analysis.

\subsection{Related work}

The risk-measure layer draws on the theory of coherent and convex measures
\citep{artzner1999coherent,follmer2016stochastic}. For the dynamic object it draws on dynamic and
time-consistent risk measures \citep{detlefsen2005conditional,cheridito2006dynamic} and their
representation through $g$-expectations \citep{rosazza2006risk,kupper2009representation}; the
entropic measure we use is the canonical time-consistent example. The robust solve is the
risk-sensitive / robust-control duality of \citet{hansen2001robust}, whose mathematical core is the
Gibbs/Donsker--Varadhan variational principle relating relative entropy to the cumulant functional
\citep{dupuis1997weak}. The jump machinery is standard in the theory of jump processes and
actuarial risk: L\'evy measures, compound-Poisson cumulants, and the Esscher transform as a change
of measure \citep{sato1999levy,cont2004financial,esscher1932probability,gerber1994option}. The
contagion model is the (multivariate) Hawkes process, whose stability theory and cluster
representation we use directly
\citep{hawkes1971spectra,hawkes1974cluster,bremaud1996stability,bacry2015hawkes}. The fleet results
connect to systemic-risk measurement \citep{adrian2016covar,acharya2017measuring}, financial
contagion and default cascades \citep{eisenberg2001systemic}, and the modelling of extremes
\citep{embrechts1997modelling}. We do not claim credit for these mature theories. Our contribution
is to import them, as a coherent whole, into the model-risk management of agentic systems, and to
show that within that synthesis the irreversible jump and the prevention lever play the binding role at every
level.

\section{A continuous-time jump-diffusion framework}
\label{c3:sec:framework}

We work on a filtered probability space $(\Omega,\Filt,(\Filt_t)_{t\ge0},\Prob)$ satisfying the
usual conditions, with a finite governance horizon $[0,T]$. Chapter~\ref{ch:single} modelled a
single agent's harm at the level of a decision. It decomposed expected harm into the probability of
a harmful action (\PHA), its severity-given-harm (\SGH), and its blast radius (\BR).
Chapter~\ref{ch:fleet} aggregated a static fleet through a containment-mismatch index and a stress
projection. The present chapter studies the continuous-time process that sits beneath both. Harm
accrues through time as discrete, marked \emph{events} (harmful actions) whose rate, severity,
and reach are themselves driven by a latent diffusion. This is a jump-diffusion, and the jump is
the carrier of every result that follows.

\subsection{The per-agent harm object}
\label{c3:sec:agent}

Let $X_t$ be a latent risk state, an Itô diffusion summarising adversarial pressure, capability
drift, and operating conditions:
\begin{equation}
\label{c3:eq:state}
dX_t = b(X_t)\,dt + \varsigma(X_t)\,dW_t,
\end{equation}
with $W$ a standard Brownian motion and $b,\varsigma$ Lipschitz so that \eqref{c3:eq:state} is
well-posed. Harmful actions arrive as a point process $N_t$ with $\Filt_t$-predictable stochastic
intensity
\begin{equation}
\label{c3:eq:intensity}
\lambda_t=\lambda(X_t)\ge0 ,
\end{equation}
that is, a Cox (doubly stochastic) process directed by the diffusion \eqref{c3:eq:state}, in which the latent
state \emph{modulates the rate} of harmful actions. This intensity is the continuous-time form of
\PHA. Each event $i$ occurs at time $\tau_i$ and carries a severity \emph{mark}
$J_i\in\R_+^{m}$, drawn independently from a law $F_J$ with mean $\Mark=\E[J]$. Equivalently, the
ground-intensity-and-mark pair defines a random measure $N(dt,dz)$ on $[0,T]\times\R_+^{m}$ with
$\Prob$-compensator
\begin{equation}
\label{c3:eq:compensator}
\hat N(dt,dz)=\lambda_t\,F_J(dz)\,dt .
\end{equation}
The mark law $F_J$ is the continuous-time form of \SGH. Written through its (state-independent)
intensity, the systematic and idiosyncratic jump components below are L\'evy measures in the sense
of \citet{sato1999levy,cont2004financial}. A predictable exposure $R_t=R(X_t)\ge0$ scales the
realised harm of an event; this is the continuous-time \BR. Cumulative harm over the horizon is the
marked-point-process integral
\begin{equation}
\label{c3:eq:harm}
\Ham_T=\sum_{i:\tau_i\le T}R_{\tau_i^-}J_i=\int_0^T\!\!\int_{\R_+^{m}}R_{s^-}\,z\;N(ds,dz)
\in\R_+^{m}.
\end{equation}
The event component of harm is therefore pure-jump in its increments, while its rate $\lambda$ and
reach $R$ inherit the continuous fluctuations of the diffusion \eqref{c3:eq:state}. The object is
a jump-diffusion, but the \emph{events} land discretely, matching the discrete,
consequential nature of an agent's actions.

\paragraph{Total loss.} Realised loss over the horizon is not exhausted by the event component. Between
events the deployment also carries a continuous mark-to-market revaluation: financing drift, spread
moves, and the running cost of positions already taken. We write it as
\begin{equation}
\label{c3:eq:cont}
\Cont_T=\mu T+\sigma W^{\mathrm{L}}_T,
\end{equation}
with $W^{\mathrm{L}}$ a Brownian motion independent of $N$. The object evaluated by the risk functionals
of Section~\ref{c3:sec:theorems} is the \emph{total loss}
\begin{equation}
\label{c3:eq:totalloss}
\Loss_T=\Cont_T+\Ham_T .
\end{equation}
The two components play different roles, and we keep them separate throughout. The decomposition
results of this chapter concern $\Ham_T$: Proposition~\ref{c3:prop:reduction},
Theorems~\ref{c3:thm:t1}--\ref{c3:thm:t3}, and Proposition~\ref{c3:prop:tc}. There the pure-jump
structure is exactly what carries the argument. The robust functional of
Theorem~\ref{c3:thm:robust} concerns $\Loss_T$. Its Gaussian cumulant
$\tfrac{\delta}{2}\sigma^2T$ arises from $\Cont_T$, not from the event component, because the diffusion
\eqref{c3:eq:state} modulates the \emph{rate and reach} of events and never enters $\Ham_T$ additively.
The codomains also differ. Harm is nonnegative by construction, $\Ham_T\in\R_+^{m}$, whereas
revaluation may go either way, so $\Loss_T\in\R^{m}$. We do not truncate $\Cont_T$ to its positive
part. A coherent risk measure applied to $\Loss_T$ should be allowed to net a favourable revaluation
against realised harm, and truncation would break the cumulant additivity that
\eqref{c3:eq:robust-closed} rests on.

\paragraph{Reversible and irreversible harm.}
As in Chapters~\ref{ch:single}--\ref{ch:fleet} the mark space splits into a reversible subspace and an irreversible subspace,
$\R_+^{m}=\R_+^{m_r}\oplus\R_+^{m_i}$. The \emph{catastrophic} class is the heavy tail of the mark
law on the irreversible coordinates: events with $z_j\ge\theta_j$ for an irreversible coordinate
$j$. The distinction is not cosmetic. Reversible harm can be netted and partially recovered.
Irreversible harm, once realised, is permanent, and it admits a fundamentally narrower set of
effective controls. Chapter~\ref{ch:single}'s Theorem~B showed this in the static case, and
Section~\ref{c3:sec:theorems} shows it here in continuous time.

\paragraph{The three levers as operations on the jump model.}
The three containment levers of Chapters~\ref{ch:single}--\ref{ch:fleet} act on \eqref{c3:eq:harm} as three distinct operations
on the jump model. We state the correspondence explicitly, because it organises the catastrophe
results.
\begin{itemize}[leftmargin=1.4em,itemsep=2pt]
\item \emph{Review} (detective control, strength $c_1$) \emph{thins the intensity}: a reviewed
agent's harmful actions are intercepted at rate $c_1\lambda_t$, leaving the thinned intensity
$(1-c_1)\lambda_t$. Review changes how \emph{often} harm occurs.
\item \emph{Reversibility} (strength $c_2$) \emph{scales the recoverable mass}: on the reversible
subspace a fraction $c_2$ of realised harm is reversed, mapping a reversible mark $z_r$ to
$(1-c_2)z_r$. Reversibility acts only where harm is recoverable, and is by construction inert on
the irreversible coordinates.
\item \emph{Prevention} (strength $c_3$) \emph{truncates the jump measure}: removing the
catastrophic affordance caps the irreversible mark, mapping $z_i$ to $z_i\wedge\theta_i(1-c_3)$.
That is, it truncates the tail of $F_J$ on the irreversible subspace. Prevention changes how
\emph{bad} an event can be.
\end{itemize}
Review and reversibility act on frequency and on recoverable severity. Only prevention reaches the
\emph{tail of the jump measure} on the irreversible subspace. This is the structural seed of every
catastrophe result in this chapter.

\subsection{The fleet: a common L\'evy jump and Hawkes contagion}
\label{c3:sec:fleet}

Consider $K$ agents indexed by $k$. Two channels couple them, and they are deliberately of
different mathematical type.

\paragraph{Systematic risk as a common L\'evy jump.}
A single shared jump process strikes all agents at once: a shared prompt-injection technique, a
poisoned upstream dependency, a platform-wide regime shift. We model it as a compound-Poisson
(L\'evy) process $\Sigma^{\mathrm{sys}}_T=\sum_{i:\tau^{\mathrm{sys}}_i\le T}\zeta_i$ common to the
fleet, with systematic-jump intensity $\Lsys$ and severities $\zeta_i$. Agent $k$ loads on it with
weight $a_k\ge0$. Write $\Xi^k_T$ for $k$'s idiosyncratic harm \eqref{c3:eq:harm}, an independent
copy across $k$. Then
\begin{equation}
\label{c3:eq:fleetdecomp}
\Ham^k_T=a_k\,\Sigma^{\mathrm{sys}}_T+\Xi^k_T .
\end{equation}
The systematic term is \emph{comonotone} across agents: the same jump enters every $\Ham^k$.
This is the continuous-time, L\'evy form of Chapter~\ref{ch:fleet}'s common shock, and
Section~\ref{c3:sec:theorems} shows that it is the part that cannot be diversified.

\paragraph{Contagion as a multivariate Hawkes process.}
Harm at one agent can raise the hazard at others: a successful jailbreak propagated across a
shared tool, an exploited pattern copied between agents. We model this as mutual self-excitation.
The intensity of agent $k$ is
\begin{equation}
\label{c3:eq:hawkes}
\lambda^k_t=\mu_k+\sum_{j=1}^{K}\int_0^t \Gam_{kj}\,g(t-s)\,dN^j_s ,
\end{equation}
a multivariate Hawkes process \citep{hawkes1971spectra,hawkes1974cluster} with baseline
intensities $\mu_k>0$, a nonnegative \emph{branching matrix} $\Gam=(\Gam_{kj})$ whose entry is the
expected number of direct offspring events at $k$ triggered by one event at $j$, and a decay kernel
$g\ge0$ normalised to $\int_0^\infty g(u)\,du=1$. Crucially $\Gam$ is nonnegative and the coupling
is \emph{excitatory}: an event can only ever raise a neighbour's hazard. This distinguishes the
present mechanism, both in object and in mathematics, from the diffusion-coupling models. There,
agents' state equations are linked through a (sign-indefinite) coupling matrix, and stability is
analysed by a joint Lyapunov function. Excitatory point-process contagion instead has a clean
criticality, the spectral radius $\spr(\Gam)$, with none of the symmetric-part subtleties of the
diffusion case. It is also jump-driven throughout, which keeps Chapter~\ref{ch:continuous}'s
contagion result disjoint from that companion line of work.

\subsection{A time-consistent dynamic risk measure}
\label{c3:sec:dynamic}

A static tail measure re-evaluated through time is not, in general, internally consistent. The
risk it assigns today to a position can be contradicted by the risks it assigns tomorrow, with no
new information, so that an acceptance decision is not stable under its own future application
\citep{detlefsen2005conditional,cheridito2006dynamic}. We therefore use a \emph{recursive} risk
measure. For risk parameter $\delta>0$ define the conditional entropic risk
\begin{equation}
\label{c3:eq:entropic}
\rho_{\delta,s}(\Ham)=\frac1\delta\log\E\!\left[e^{\delta \Ham}\,\middle|\,\Filt_s\right],
\qquad \rhod(\Ham)\equiv\rho_{\delta,0}(\Ham),
\end{equation}
the conditional $g$-expectation with Brownian driver $g(z)=\tfrac{\delta}{2}\lvert z\rvert^2$
\citep{rosazza2006risk,kupper2009representation}. It is monotone, cash-additive, and convex. It is
also \emph{time-consistent}, the property we need: for $0\le s\le T$,
\begin{equation}
\label{c3:eq:tc}
\rho_{\delta,0}(\Ham)=\rho_{\delta,0}\!\big(\rho_{\delta,s}(\Ham)\big),
\end{equation}
which follows directly from the tower property of conditional expectation applied to
\eqref{c3:eq:entropic}. Section~\ref{c3:sec:theorems} states this as Proposition~\ref{c3:prop:tc} and
contrasts it with the failure of \eqref{c3:eq:tc} for iterated expected shortfall. The parameter
$\delta$ is the controller's risk aversion. As $\delta\downarrow0$, $\rhod(\Ham)\to\E[\Ham]$.

\subsection{The robust solve as a stochastic differential game}
\label{c3:sec:robust}

The intensities, severities, and loadings above are never known exactly, and an adversary may
push the system toward whichever misspecification is most damaging. We therefore make the
controller robust to a neighbourhood of the nominal law $\Prob$. The natural neighbourhood for a
jump-diffusion is generated by the two changes of measure that preserve its structure. The first is
a \emph{Girsanov} change that shifts the Brownian drift by a predictable process $h$. The second is
an \emph{Esscher / L\'evy-measure tilt} \citep{esscher1932probability,gerber1994option} that
reweights the jump measure by a nonnegative density $Y(z)$,
\begin{equation}
\label{c3:eq:tilt}
\varsigma(X_t)\,dW_t\ \longmapsto\ \varsigma(X_t)\,(dW_t+h_t\,dt),
\qquad \Lev(dz)\ \longmapsto\ Y(z)\,\Lev(dz),
\end{equation}
so that the drift becomes $b+\varsigma h$ and the jump intensity-and-severity is reshaped by $Y$.
Let $\Qset_\delta$ be the set of measures $\Qprob\ll\Prob$ of this form with finite relative
entropy $\KL{\Qprob}{\Prob}<\infty$. The controller evaluates harm by the worst case in
$\Qset_\delta$, penalised by how far the adversary must distort the world to achieve it:
\begin{equation}
\label{c3:eq:game}
\spr^{\mathrm{rob}}_\delta(\Ham)\;=\;\sup_{\Qprob\in\Qset_\delta}
\Big\{\,\E^{\Qprob}[\Ham]\;-\;\frac1\delta\,\KL{\Qprob}{\Prob}\,\Big\}.
\end{equation}
This is a static formulation of the Hansen--Sargent risk-sensitive control problem
\citep{hansen2001robust,dupuis1997weak}. Section~\ref{c3:sec:theorems} shows that
\eqref{c3:eq:game} has a closed form, equal to the entropic measure \eqref{c3:eq:entropic}. It
also shows that the game's maximiser is an explicit member of \eqref{c3:eq:tilt}.
Together these tie the robust solve and the dynamic measure into a single object, and expose
exactly when the problem is well-posed.

\section{Results}
\label{c3:sec:theorems}

We now state the results. Each is proved at the level of the closed form that the numerical
experiments of Section~\ref{c3:sec:numerics} confirm. The proofs are deliberately self-contained and
elementary, since the contribution is the structure rather than new probabilistic machinery.

\subsection{The single-agent object reduces to Chapter~\ref{ch:single}}

\begin{proposition}[Reduction to the static decomposition]
\label{c3:prop:reduction}
For the harm object \eqref{c3:eq:harm} with marks independent of $(\lambda,R)$ and
$\E[J]=\Mark<\infty$,
\begin{equation}
\label{c3:eq:reduction-general}
\E[\Ham_T]=\E\!\Big[\int_0^T\lambda_s R_s\,ds\Big]\,\Mark .
\end{equation}
In the homogeneous limit $\lambda_s\equiv\lambda$, $R_s\equiv R$, writing
$\PHA_T=1-e^{-\lambda T}$ for the probability of at least one harmful action,
\begin{equation}
\label{c3:eq:reduction-kappa}
\E[\Ham_T]=\PHA_T\cdot\BR\cdot\SGH\cdot\kappa(\lambda T),
\qquad \kappa(\lambda T)=\frac{\lambda T}{1-e^{-\lambda T}}\ge1,
\end{equation}
with $\BR=R$, $\SGH=\Mark$, and $\kappa(\lambda T)\downarrow1$ as $\lambda T\downarrow0$. Chapter~\ref{ch:single}'s
identity $\EH=\PHA\cdot\SGH\cdot\BR$ is exactly the rare-event limit of \eqref{c3:eq:reduction-kappa}.
\end{proposition}

\begin{proof}
The compensation (Campbell) formula for the marked point process \eqref{c3:eq:harm} with predictable
compensator \eqref{c3:eq:compensator} gives
$\E[\Ham_T]=\E\big[\int_0^T\!\int R_{s^-}z\,\lambda_s F_J(dz)\,ds\big]$. Marks are independent of
$(\lambda,R)$, so the inner integral over $z$ factors as $\Mark$, yielding
\eqref{c3:eq:reduction-general}. In the homogeneous limit the bracket is $\lambda T R$, and the event
count $N_T\sim\mathrm{Poisson}(\lambda T)$, so $\PHA_T=\Prob(N_T\ge1)=1-e^{-\lambda T}$. Then
$\E[\Ham_T]=\lambda T\,R\,\Mark=(1-e^{-\lambda T})R\,\Mark\cdot\frac{\lambda T}{1-e^{-\lambda T}}$,
which is \eqref{c3:eq:reduction-kappa}. Since $e^{-x}\ge1-x$ we have $1-e^{-\lambda T}\le\lambda T$ and
hence $\kappa\ge1$. The Taylor expansion $1-e^{-x}=x-x^2/2+O(x^3)$ gives
$\kappa=1/(1-x/2+O(x^2))\to1$ as $x\downarrow0$. In that limit $\E[\Ham_T]\to\PHA\cdot\BR\cdot\SGH$,
which is Chapter~\ref{ch:single}'s static identity.
\end{proof}

The correction $\kappa\ge1$ is the only thing the static model omits at the level of the mean. It
accounts for episodes carrying more than one harmful action, and it is appreciable once harmful
actions are not rare ($\kappa(1)=1.582$). Beyond the mean, the jump object additionally carries the
full severity \emph{distribution} and, through the diffusion, a stochastic intensity. Both are
invisible to a point estimate but decisive for the tail.

\subsection{Fleet results}

\begin{theorem}[Non-diversifiability of a common L\'evy jump]
\label{c3:thm:t1}
Let $\Ham^k_T=a_k\Sigma^{\mathrm{sys}}_T+\Xi^k_T$ as in \eqref{c3:eq:fleetdecomp}, with $(\Xi^k)$
i.i.d., $\E[\Xi]<\infty$, $\E\lvert\Sigma^{\mathrm{sys}}\rvert<\infty$, and
$\bar a_K=K^{-1}\sum_k a_k\to\bar a$. Then the fleet average converges almost surely,
\begin{equation}
\label{c3:eq:t1-limit}
\bar\Ham_K\;\equiv\;\frac1K\sum_{k=1}^{K}\Ham^k_T\;\xrightarrow{\ \mathrm{a.s.}\ }\;
\bar a\,\Sigma^{\mathrm{sys}}_T+\E[\Xi]\;=:\;\bar\Ham_\infty .
\end{equation}
Consequently, for any law-invariant coherent risk measure $\spr$ continuous along
\eqref{c3:eq:t1-limit},
\begin{equation}
\label{c3:eq:t1-floor}
\lim_{K\to\infty}\spr(\bar\Ham_K)=\bar a\,\spr(\Sigma^{\mathrm{sys}}_T)+\E[\Xi]
\;>\;\lim_{K\to\infty}\E[\bar\Ham_K]
\end{equation}
whenever $\bar a>0$, $\Sigma^{\mathrm{sys}}_T$ is non-degenerate and $\spr$ loads its tail. The
requirement $\bar a>0$ is not cosmetic: if $\bar a=0$ the systematic term vanishes from
\eqref{c3:eq:t1-limit} and both sides of \eqref{c3:eq:t1-floor} collapse to $\E[\Xi]$, so the
inequality holds weakly but not strictly. If instead
$\Sigma^{\mathrm{sys}}\equiv0$ (no common jump) the limit is the constant $\E[\Xi]$ and the risk
diversifies away. In the rare regime $\Prob(N^{\mathrm{sys}}_T\ge1)\le1-\tau$ with
$\Sigma^{\mathrm{sys}}_T=M N^{\mathrm{sys}}_T$, $N^{\mathrm{sys}}_T\sim\mathrm{Poisson}(\Lsys T)$,
\begin{equation}
\label{c3:eq:t1-es}
\ES_\tau\!\big(\Sigma^{\mathrm{sys}}_T\big)=\frac{M\,\Lsys T}{1-\tau}.
\end{equation}
\end{theorem}

\begin{proof}
Write $\bar\Ham_K=\bar a_K\Sigma^{\mathrm{sys}}_T+K^{-1}\sum_k\Xi^k_T$. The $\Xi^k$ are i.i.d.\
with finite mean, so by the strong law $K^{-1}\sum_k\Xi^k_T\to\E[\Xi]$ a.s. The systematic term is
identical across $k$, so $\bar a_K\Sigma^{\mathrm{sys}}_T\to\bar a\,\Sigma^{\mathrm{sys}}_T$ a.s.,
giving \eqref{c3:eq:t1-limit}. The $(\bar\Ham_K)$ are uniformly integrable (dominated by
$\sup_K\bar a_K\lvert\Sigma^{\mathrm{sys}}\rvert+K^{-1}\sum\lvert\Xi^k\rvert$ with finite means), so
a risk measure continuous along this convergence satisfies $\spr(\bar\Ham_K)\to\spr(\bar\Ham_\infty)$.
Cash-additivity removes the constant $\E[\Xi]$ and positive homogeneity extracts $\bar a$, giving
\eqref{c3:eq:t1-floor}'s left side. The strict inequality is $\spr(\Sigma^{\mathrm{sys}})>\E[\Sigma^{\mathrm{sys}}]$,
which holds for a tail-loading coherent measure at a non-degenerate risk. When
$\Sigma^{\mathrm{sys}}\equiv0$ the limit \eqref{c3:eq:t1-limit} is deterministic and
$\spr(\E[\Xi])=\E[\Xi]$. For \eqref{c3:eq:t1-es}, expected shortfall is the average of the worst
$(1-\tau)$ mass. In the rare regime the event $\{N^{\mathrm{sys}}_T\ge1\}$ has probability
$1-e^{-\Lsys T}\le1-\tau$ and lies entirely inside that tail, while $\{N^{\mathrm{sys}}_T=0\}$
contributes zero. Hence
$\ES_\tau(MN^{\mathrm{sys}}_T)=(1-\tau)^{-1}M\,\E[N^{\mathrm{sys}}_T\Indic_{\{N\ge1\}}]
=(1-\tau)^{-1}M\,\E[N^{\mathrm{sys}}_T]=M\Lsys T/(1-\tau)$.
\end{proof}

Equation \eqref{c3:eq:t1-es} is the continuous-time, L\'evy image of Chapter~\ref{ch:fleet}'s systematic charge
$\psys/(1-\tau)$, a Bernoulli common shock there and a common jump here. The non-diversifiable
signature is the same. Systemic risk persists in continuous time, carried by the common jump
rather than averaged away across the fleet.

\begin{theorem}[Hawkes contagion criticality]
\label{c3:thm:t2}
Let the fleet intensities follow the multivariate Hawkes dynamics \eqref{c3:eq:hawkes} with
nonnegative branching matrix $\Gam$ and kernel $g\ge0$, $\int_0^\infty g=1$. The process admits a
stationary version with finite mean intensity (equivalently, is non-explosive) if and only if the
spectral radius satisfies $\spr(\Gam)<1$, in which case the stationary mean intensity vector is
\begin{equation}
\label{c3:eq:t2-mean}
\bar\lambda=(I-\Gam)^{-1}\mu .
\end{equation}
For the homogeneous all-to-all fleet $\Gam=(\eta/K)\,\mathbf{1}\mathbf{1}^{\!\top}$ one has
$\spr(\Gam)=\eta$ and $\bar\lambda=\frac{\mu_0}{1-\eta}\mathbf{1}$, which diverges as
$\eta\uparrow1$. Segmenting the fleet into $m$ isolated blocks replaces $\Gam$ by a block-diagonal
matrix of spectral radius $\eta/m$, so $\spr(\Gam)\mapsto\eta/m$ and the fleet is subcritical
whenever $m>\eta$.
\end{theorem}

\begin{proof}
Use the immigration--branching (cluster) representation of the linear Hawkes process
\citep{hawkes1974cluster,bremaud1996stability}. Immigrants of type $k$ arrive as a Poisson process
of rate $\mu_k$. Each event of type $j$ generates, for each $k$, a Poisson number of direct
offspring of type $k$ with mean $\Gam_{kj}$ (the kernel integrates to one). The expected number of
$n$th-generation descendants per immigrant is the $n$th power $\Gam^{n}$, so the expected total
progeny matrix is $\sum_{n\ge0}\Gam^{n}$. That Neumann series converges if and only if
$\spr(\Gam)<1$, with limit $(I-\Gam)^{-1}$. Convergence of the progeny series is exactly
non-explosion / existence of a stationary finite-intensity version. Taking expectations in
\eqref{c3:eq:hawkes} at stationarity and using $\int g=1$ and $\E[dN^j_t]=\bar\lambda_j\,dt$ gives the
mean equation $\bar\lambda=\mu+\Gam\bar\lambda$, i.e.\ \eqref{c3:eq:t2-mean}. For the homogeneous
matrix, $\mathbf{1}$ is the Perron eigenvector with $\Gam\mathbf{1}=(\eta/K)(\mathbf{1}^{\!\top}\mathbf{1})\mathbf{1}=\eta\mathbf{1}$,
so $\spr(\Gam)=\eta$ and $(I-\Gam)^{-1}\mu_0\mathbf{1}=\frac{\mu_0}{1-\eta}\mathbf{1}$. Segmentation
deletes the cross-block entries, leaving $m$ diagonal blocks. Within a block of size $K/m$ the
entries remain $\eta/K$, so each block's row sums equal $(K/m)(\eta/K)=\eta/m$. Being a
nonnegative matrix with constant row sums, each block has spectral radius $\eta/m$. Hence the
segmented process is subcritical iff $\eta/m<1$.
\end{proof}

The branching matrix is nonnegative and the excitation is one-signed, so the criticality boundary
is simply its Perron root. No symmetric-part or sign condition arises, in contrast to
diffusion-coupling stability analyses. Theorem~\ref{c3:thm:t2} is the dynamic lift of Chapter~\ref{ch:fleet}'s
percolation threshold (a giant cascade emerges past a critical branching ratio). It is the
self-exciting form of systemic contagion, one compromise raising the hazard of the next, the
agentic analogue of a default cascade. Segmentation plays the same role in both, partitioning the
contagion graph and dividing the effective branching, the macroprudential lever that holds the
system subcritical.

\begin{theorem}[Prevention as jump-measure truncation]
\label{c3:thm:t3}
Consider the irreversible catastrophic class with systematic jump magnitude $\theta$ and rate
$\Lsys$, loaded fraction $g$, and idiosyncratic-catastrophic mean $\mu^{\mathrm{idio}}_{\mathrm{cat}}$.
Assume, as in Chapters~\ref{ch:single}--\ref{ch:fleet}, that detective review intercepts only reviewable (reversible or
in-progress) actions and cannot intercept the atomic irreversible jump. Then the fleet
catastrophic floor
\begin{equation}
\label{c3:eq:t3-floor}
F(g,c_3)=g\,\theta\,(1-c_3)\,\frac{\Lsys}{1-\tau}+\mu^{\mathrm{idio}}_{\mathrm{cat}}
\end{equation}
is independent of the fleet size $K$, invariant to the detective lever $c_1$ and the reversibility
lever $c_2$, and strictly decreasing in the prevention lever $c_3$ and in the loaded fraction $g$.
\end{theorem}

\begin{proof}
Restrict \eqref{c3:eq:fleetdecomp} to the catastrophic coordinate. By Theorem~\ref{c3:thm:t1} the
fleet-average catastrophic harm converges a.s.\ to $g\,\Sigma^{\mathrm{sys,cat}}_T+\mu^{\mathrm{idio}}_{\mathrm{cat}}$
(loading $\bar a=g$). Its expected shortfall is therefore $g\,\ES_\tau(\Sigma^{\mathrm{sys,cat}})+\mu^{\mathrm{idio}}_{\mathrm{cat}}$,
independent of $K$. The lever correspondence of Section~\ref{c3:sec:agent} fixes the dependence on the
three controls. Detective review thins only the reviewable intensity. By hypothesis the irreversible
jump is not in that channel, so its rate $\Lsys$, and hence the floor, is unchanged by $c_1$.
Reversibility scales only reversible coordinates and is inert on the irreversible jump, so the floor
is unchanged by $c_2$. Prevention truncates the irreversible mark, $\theta\mapsto\theta(1-c_3)$.
Substituting into the rare-regime shortfall \eqref{c3:eq:t1-es} gives
$\ES_\tau(\Sigma^{\mathrm{sys,cat}})=\theta(1-c_3)\Lsys/(1-\tau)$, hence \eqref{c3:eq:t3-floor}, which
is strictly decreasing in $c_3$ and, through the loading, in $g$.
\end{proof}

Theorem~\ref{c3:thm:t3} is the continuous-time form of Chapter~\ref{ch:single}'s Theorem~B and Chapter~\ref{ch:fleet}'s prevention
floor. It is also the sharpest expression of the lever correspondence. Against an irreversible
jump, frequency controls (review) and recovery controls (reversibility) do not move the floor.
Only truncating the jump measure (prevention) or reducing exposure to it (segmentation) does.

\subsection{Dynamic consistency and the robust solve}

\begin{proposition}[Time-consistency]
\label{c3:prop:tc}
The recursive entropic measure \eqref{c3:eq:entropic} is time-consistent: for $0\le s\le T$,
$\rho_{\delta,0}(\Ham)=\rho_{\delta,0}(\rho_{\delta,s}(\Ham))$. Iterated expected shortfall is not:
in general $\ES_\tau\big(X_1+\ES_\tau(X_2\mid\Filt_s)\big)\neq\ES_\tau(X_1+X_2)$.
\end{proposition}

\begin{proof}
For the entropic measure, by the tower property of conditional expectation,
\[
\rho_{\delta,0}\big(\rho_{\delta,s}(\Ham)\big)
=\frac1\delta\log\E\!\Big[\exp\!\big(\delta\cdot\tfrac1\delta\log\E[e^{\delta\Ham}\mid\Filt_s]\big)\Big]
=\frac1\delta\log\E\!\big[\E[e^{\delta\Ham}\mid\Filt_s]\big]
=\frac1\delta\log\E[e^{\delta\Ham}]=\rho_{\delta,0}(\Ham).
\]
For the failure of expected shortfall, take $X_1,X_2$ independent with $X_2$ independent of
$\Filt_s\supseteq\sigma(X_1)$. Then $\ES_\tau(X_2\mid\Filt_s)=\ES_\tau(X_2)$ is constant, so by
translation invariance the recursive value equals $\ES_\tau(X_1)+\ES_\tau(X_2)$. Subadditivity, by
contrast, gives $\ES_\tau(X_1+X_2)\le\ES_\tau(X_1)+\ES_\tau(X_2)$, with strict inequality
unless $X_1,X_2$ are comonotone. Hence the two differ.
\end{proof}

\begin{theorem}[The robust game equals the entropic measure]
\label{c3:thm:robust}
Assume the coefficients are constant, i.e.\ $\varsigma$, the intensity $\Lambda$ and the exposure $R$
are deterministic and state-independent. Then for the ambiguity set $\Qset_\delta$ of Girsanov/Esscher
changes \eqref{c3:eq:tilt} with the relative-entropy penalty, the robust functional
\eqref{c3:eq:game} satisfies
\begin{equation}
\label{c3:eq:dv}
\spr^{\mathrm{rob}}_\delta(\Loss_T)
=\sup_{\Qprob\in\Qset_\delta}\Big\{\E^{\Qprob}[\Loss_T]-\tfrac1\delta\KL{\Qprob}{\Prob}\Big\}
=\frac1\delta\log\E\big[e^{\delta\Loss_T}\big]=\rhod(\Loss_T),
\end{equation}
attained uniquely at $d\Qprob^\star/d\Prob=e^{\delta\Loss_T}/\E[e^{\delta\Loss_T}]$, which is the
Girsanov drift shift $h^\star=\delta\sigma$ on the Brownian part of $\Cont_T$ together with the Esscher tilt
$\Lev^\star(dz)=e^{\delta R z}\Lev(dz)$ on the jump part. For the total loss \eqref{c3:eq:totalloss} with continuous component \eqref{c3:eq:cont} and
$\mathrm{Exp}(b)$ marks with exposure $R$ in the event component,
\begin{equation}
\label{c3:eq:robust-closed}
\rhod(\Loss_T)=\mu T+\frac{\delta}{2}\sigma^2 T+\frac{\Lambda T R}{\,b-\delta R\,},
\end{equation}
which is finite if and only if $\delta R<b$, equivalently $\int e^{\delta R z}\Lev(dz)<\infty$.
Truncating the jump measure (the prevention lever) guarantees this exponential moment.
\end{theorem}

\begin{proof}
The Gibbs variational principle (the Donsker--Varadhan / Dupuis--Ellis duality \citep{dupuis1997weak} between relative
entropy and the cumulant functional) states that for any $\Loss_T$ with $\E[e^{\delta\Loss_T}]<\infty$,
\[
\frac1\delta\log\E[e^{\delta\Loss_T}]
=\sup_{\Qprob\ll\Prob}\Big\{\E^{\Qprob}[\Loss_T]-\tfrac1\delta\KL{\Qprob}{\Prob}\Big\},
\]
with the supremum attained uniquely at the exponential tilt
$d\Qprob^\star/d\Prob=e^{\delta\Loss_T}/\E[e^{\delta\Loss_T}]$. This is \eqref{c3:eq:dv}. Because $\Cont_T$ and $\Ham_T$ are independent, the tilt factorises over the two parts. On the Brownian
part, $e^{\delta\sigma W^{\mathrm{L}}_T}$ normalised is the Girsanov density $\mathcal{E}(\int h^\star\,dW^{\mathrm{L}})$ with
$h^\star=\delta\sigma$, shifting the drift by $\delta\sigma^2$. On the jump part, weighting the
compound-Poisson law by $e^{\delta\sum_i Y_i}$ is exactly the Esscher transform, which tilts the
mark law by $e^{\delta R z}$ and inflates the intensity by the mark transform $M_Y(\delta)$
\citep{esscher1932probability,gerber1994option}. For the closed form, independence gives
$\log\E[e^{\delta\Loss_T}]=\log\E[e^{\delta(\mathrm{diff})}]+\log\E[e^{\delta(\mathrm{jump})}]$. The
continuous component $\Cont_T$ contributes $\delta\mu T+\tfrac12\delta^2\sigma^2 T$ (Gaussian cumulant). The compound
Poisson contributes $\Lambda T(M_Y(\delta)-1)$ by the L\'evy--Khintchine cumulant, with
$M_Y(\delta)=\E[e^{\delta RJ}]=b/(b-\delta R)$ for $\mathrm{Exp}(b)$ marks and $\delta R<b$. Dividing
by $\delta$ and simplifying $\Lambda T(M_Y(\delta)-1)/\delta=\Lambda T R/(b-\delta R)$ yields
\eqref{c3:eq:robust-closed}. The Esscher tilt and the cumulant are finite precisely when
$\int e^{\delta R z}\Lev(dz)<\infty$. A heavy-tailed $\Lev$ violates this and the robust value is
$+\infty$. Truncating $\Lev$ above a finite level, by contrast, bounds the integrand and restores
finiteness.
\end{proof}

\begin{remark}[Why constant coefficients are assumed]
\label{c3:rem:restricted}
The supremum in \eqref{c3:eq:dv} is taken over $\Qset_\delta$ (the parametric Girsanov/Esscher family
\eqref{c3:eq:tilt}) while the Donsker--Varadhan identity invoked in the proof is a supremum over
\emph{all} $\Qprob\ll\Prob$. The two coincide here only because, under constant coefficients, the
Donsker--Varadhan optimiser $d\Qprob^\star/d\Prob=e^{\delta\Loss_T}/\E[e^{\delta\Loss_T}]$ is itself a
member of that family: an exponential tilt of a constant-coefficient L\'evy functional is again a
Girsanov drift shift plus an Esscher tilt. For state-dependent $\Lambda(X)$ or $R(X)$ the
optimiser need not lie in $\Qset_\delta$, and then only
\[
\sup_{\Qprob\in\Qset_\delta}\Big\{\E^{\Qprob}[\Loss_T]-\tfrac1\delta\KL{\Qprob}{\Prob}\Big\}
\ \le\ \frac1\delta\log\E\big[e^{\delta\Loss_T}\big]
\]
holds, with equality if and only if the optimiser remains in the family. We claim
\eqref{c3:eq:dv} as an identity for the constant-coefficient case and as an inequality otherwise; the
closed form \eqref{c3:eq:robust-closed} should be read as an upper bound on the restricted robust value
in the state-dependent case, which is the conservative direction for a risk measure.
\end{remark}

Theorem~\ref{c3:thm:robust} closes the loop. The robust solve and the time-consistent dynamic measure
are the same object, so robustness to misspecification and dynamic consistency are not two
desiderata to be traded off but one. The worst case is explicit and interpretable, and the adversary
makes harmful jumps both more frequent and more severe. The finiteness condition gives the
robust counterpart of the prevention theorem. The lever that controls the irreversible tail
(Theorem~\ref{c3:thm:t3}) is also the lever that makes the fleet robustly governable at all.

\section{Numerical experiments}
\label{c3:sec:numerics}

Every result above is stated at the level of a closed form. This section reports the Monte-Carlo
experiments that confirm each one. The parameter values are illustrative. We chose them to exhibit
the mechanism cleanly, not to calibrate to any deployed system, and the purpose is verification of
the mathematics, not estimation of magnitudes. In each experiment we plot the simulated quantity
against its closed form and report the relative error. Where an estimator is intrinsically hard
near a boundary, we say so rather than tuning it away.

\subsection{The single-agent object and its reduction}

\begin{figure}[t]
\centering
\includegraphics[width=\textwidth]{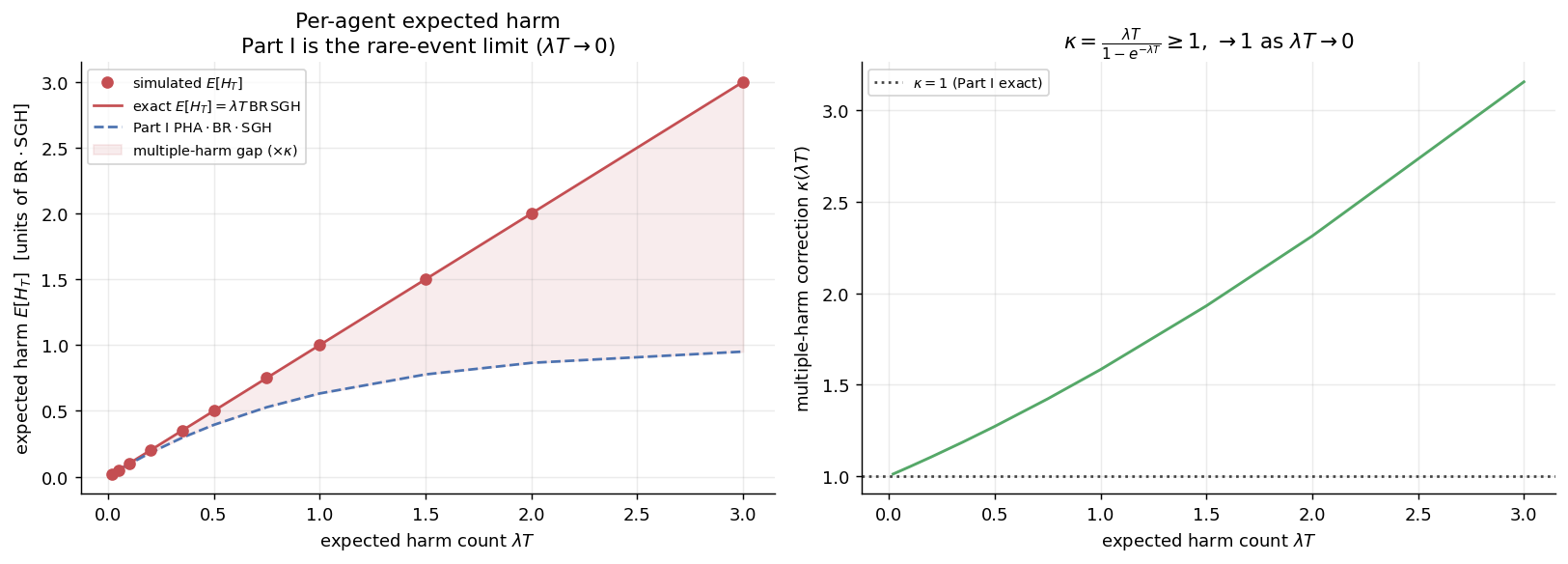}
\caption{Per-agent expected harm against the expected harm count $\lambda T$. Simulated
$\E[\Ham_T]$ (points) lies on the exact line $\lambda T\,\BR\,\SGH$. Chapter~\ref{ch:single}'s
$\PHA\cdot\BR\cdot\SGH$ (dashed) is the rare-event limit, and the shaded multiple-harm gap is the
factor $\kappa$. Right: $\kappa(\lambda T)\ge1$ with $\kappa\to1$ as $\lambda T\to0$.}
\label{c3:fig:reduction}
\end{figure}

Figure~\ref{c3:fig:reduction} verifies Proposition~\ref{c3:prop:reduction}. We simulate the marked
Cox process and average cumulative harm, which reproduces the exact mean $\lambda T\,\BR\,\SGH$ to
within Monte-Carlo error across the whole range of $\lambda T$, under $1\%$ except at the smallest
counts, where sampling noise dominates. The static decomposition of Chapter~\ref{ch:single} sits
exactly on the rare-event limit. The multiple-harm correction satisfies
$\kappa(\lambda T)\downarrow1$ as $\lambda T\to0$: at the smallest sampled $\lambda T=0.02$ the
simulated correction is $1.010$. The correction grows to $\kappa(1)=1.582$, so once an episode
carries roughly one harmful action the static mean understates the true mean by about a half. The
stochastic-intensity (Cox) generalisation behaves as the theory requires. The compensation formula
$\E[\Ham_T]=\E[\Lambda_T]\,\BR\,\SGH$ holds to $0.5\%$. Jensen's inequality makes the probability of
at least one harmful action lower under a fluctuating intensity than under its mean ($0.412$ versus
$0.423$), and that same intensity volatility fattens the tail.

\begin{figure}[t]
\centering
\includegraphics[width=\textwidth]{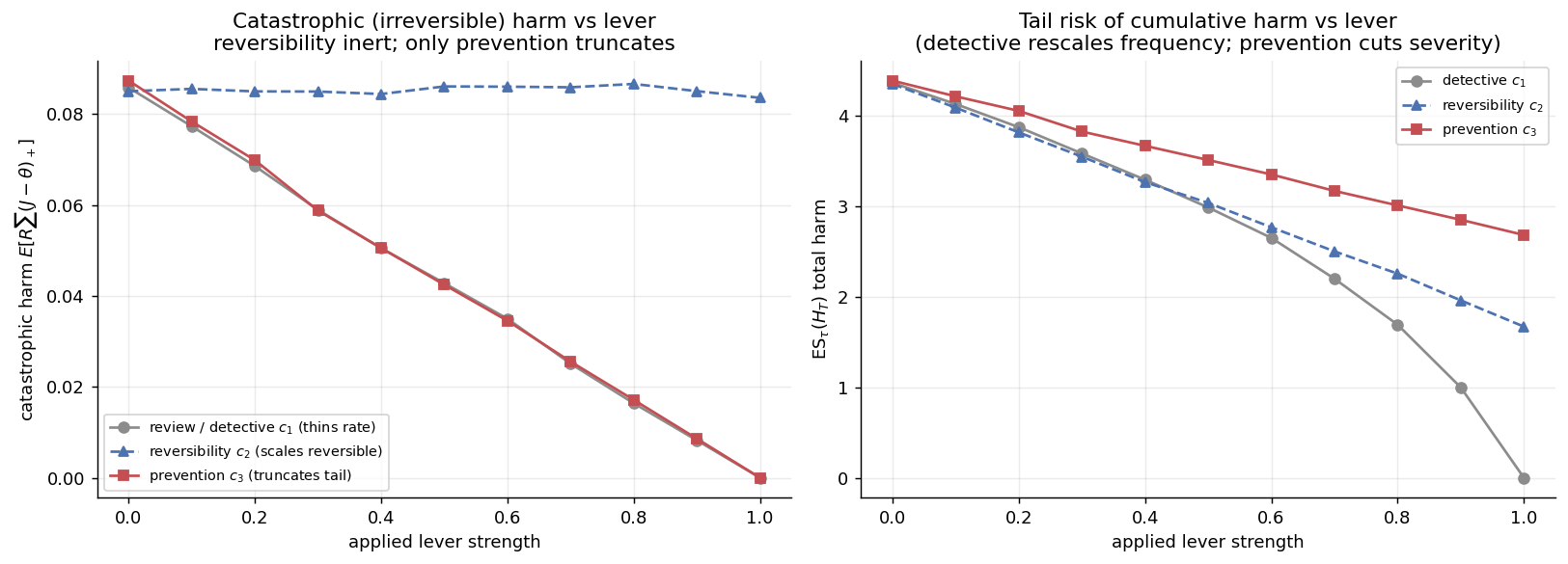}
\caption{The three levers as jump operations. Left: catastrophic (irreversible) harm versus lever
strength. Reversibility is inert, while detective thinning and prevention reduce that harm through
different mechanisms (frequency versus severity). Right: tail risk $\ES_\tau(\Ham_T)$ of cumulative
harm versus lever strength.}
\label{c3:fig:levers}
\end{figure}

Figure~\ref{c3:fig:levers} verifies the lever correspondence of Section~\ref{c3:sec:agent} as a
property of the single-agent object. Detective review scales catastrophic harm in proportion to
$(1-c_1)$, matching the thinned-intensity prediction to $5\times10^{-4}$. Reversibility leaves the
irreversible tail flat, with standard deviation $8\times10^{-4}$ across its full range,
confirming it is inert there. Prevention drives the catastrophic contribution to zero by
truncation. The right panel shows the same three levers acting on the tail of cumulative harm. The
controls separate cleanly into a frequency control and severity controls, which is the seed of the
fleet prevention theorem.

\subsection{Fleet results}

\begin{figure}[t]
\centering
\includegraphics[width=\textwidth]{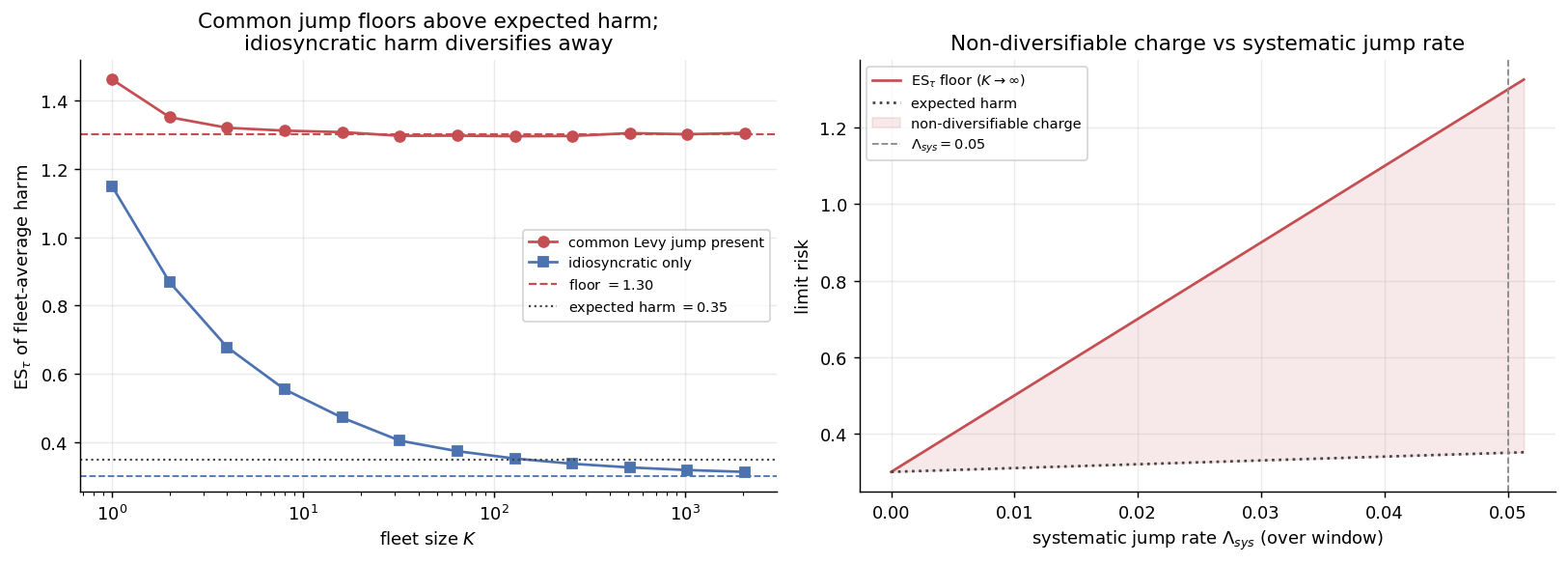}
\caption{Non-diversifiability of a common L\'evy jump. Left: with the common jump, the
fleet-average $\ES_\tau$ floors above expected harm regardless of fleet size $K$, while
idiosyncratic harm diversifies away. Right: the non-diversifiable charge grows with the
systematic-jump rate $\Lsys$.}
\label{c3:fig:t1}
\end{figure}

Figure~\ref{c3:fig:t1} verifies Theorem~\ref{c3:thm:t1}. With a common L\'evy jump present, the
fleet-average tail does not diversify. It floors at $1.305$ against a closed-form limit of $1.300$,
an error of $0.40\%$, well above the expected harm of $0.35$, leaving a non-diversifiable charge of
$0.95$. Remove the common jump and the same fleet diversifies cleanly toward the idiosyncratic mean
$\E[\Xi]=0.30$. The right panel traces the charge as a function of the systematic-jump rate,
matching the rare-regime form $M\Lsys/(1-\tau)$ of \eqref{c3:eq:t1-es}.

\begin{figure}[t]
\centering
\includegraphics[width=\textwidth]{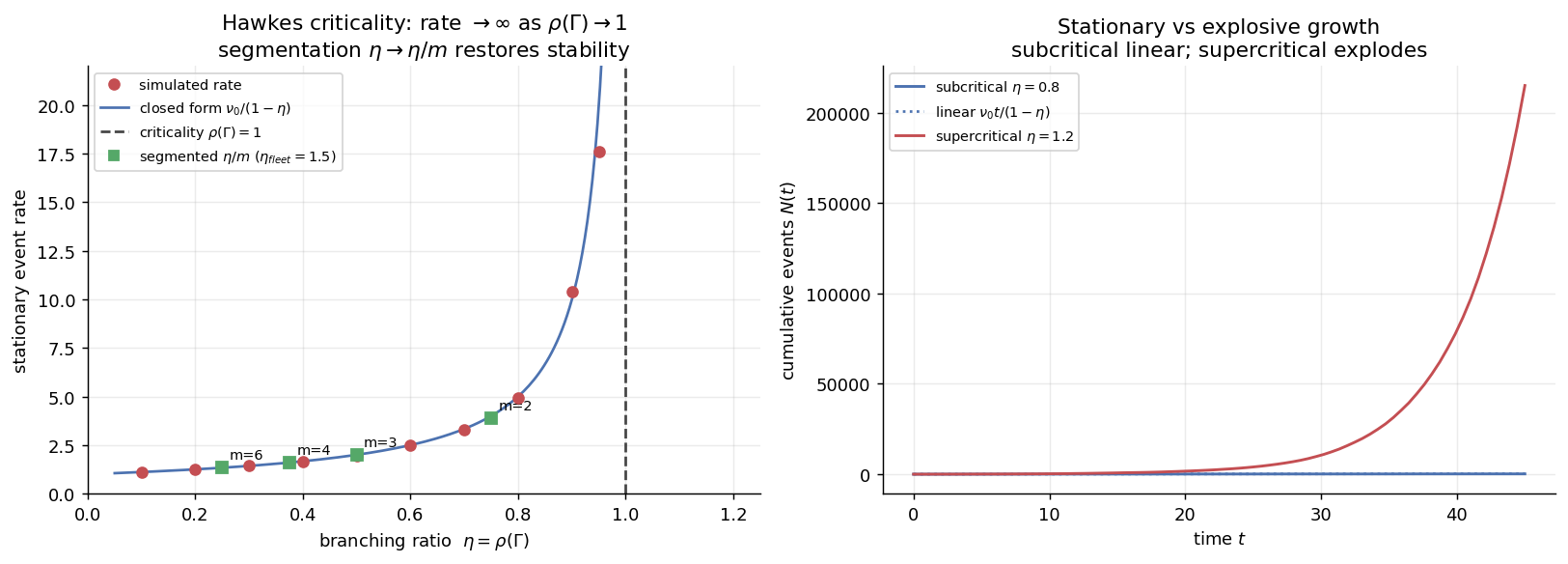}
\caption{Hawkes contagion criticality. Left: stationary event rate versus branching ratio
$\eta=\spr(\Gam)$, diverging as $\eta\to1$. Segmentation $\eta\mapsto\eta/m$ restores stability.
Right: cumulative events. Subcritical growth is linear; supercritical growth explodes.}
\label{c3:fig:t2}
\end{figure}

Figure~\ref{c3:fig:t2} verifies Theorem~\ref{c3:thm:t2}. The simulated stationary rate matches
$\nu_0/(1-\eta)$ to $1\%$ across the convergent regime, and it diverges as the branching ratio
approaches one. The segmentation points fall on the curve at $\eta/m$, so partitioning a
\emph{supercritical} fleet ($\eta_{\mathrm{fleet}}=1.5$) into $m$ blocks pulls it back below
criticality once $m>\eta_{\mathrm{fleet}}$. The right panel contrasts the linear growth of a
subcritical fleet with the explosive growth of a supercritical one. At $\eta=0.95$ the
finite-sample rate estimate has large variance, so we exclude it from the error metric. This is not a
defect of the theory but the critical slowing-down that the divergence at
$\spr(\Gam)=1$ predicts.

\begin{figure}[t]
\centering
\includegraphics[width=\textwidth]{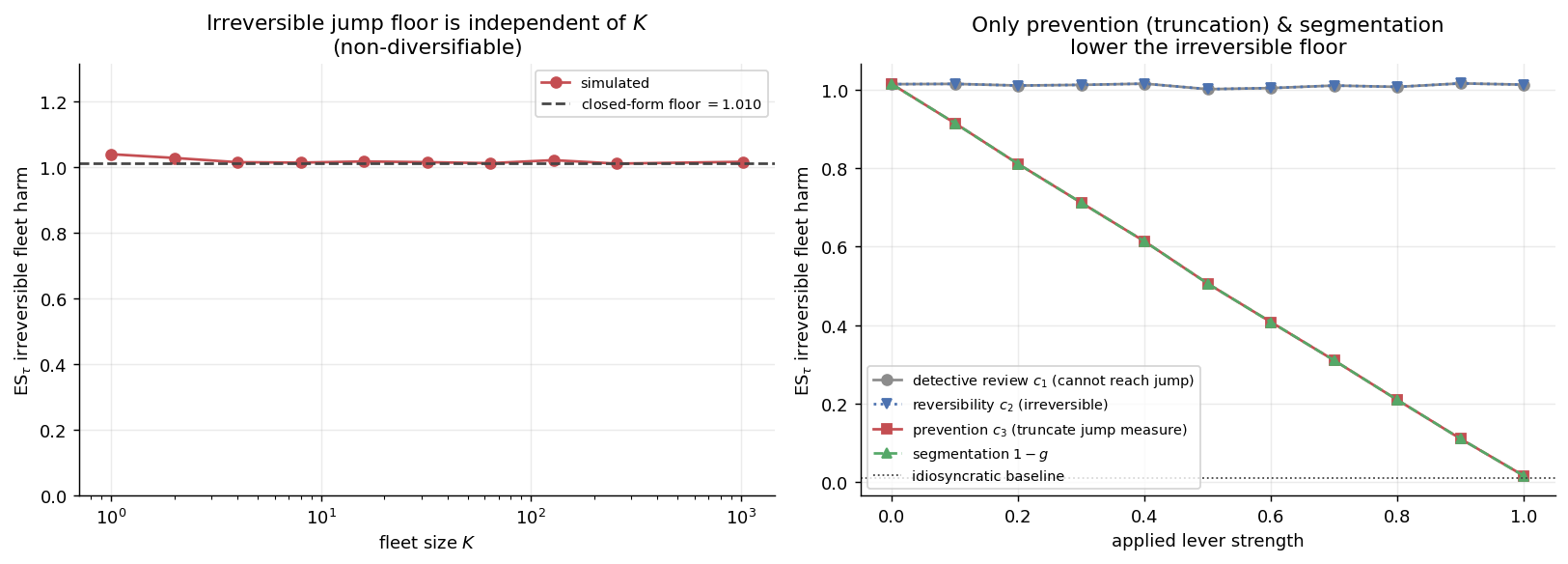}
\caption{Prevention as jump-measure truncation. Left: the irreversible fleet floor is independent
of $K$. Right: against the irreversible jump, detective review and reversibility are flat; only
prevention (truncation) and segmentation lower the floor.}
\label{c3:fig:t3}
\end{figure}

Figure~\ref{c3:fig:t3} verifies Theorem~\ref{c3:thm:t3}. The irreversible fleet floor is flat in
$K$, at $1.015$ simulated against $1.010$, an error of $0.48\%$. That flatness confirms
non-diversifiability. Sweeping the levers, detective review and reversibility leave the floor
unchanged, with standard deviation $4.5\times10^{-3}$ over their full range, flat to Monte-Carlo
precision. Prevention and segmentation, by contrast, drive it down toward the idiosyncratic
baseline. This is the continuous-time statement that, against an irreversible jump, only truncating
the jump measure or reducing exposure to it helps.

\subsection{Dynamic consistency and the robust solve}

\begin{figure}[t]
\centering
\includegraphics[width=\textwidth]{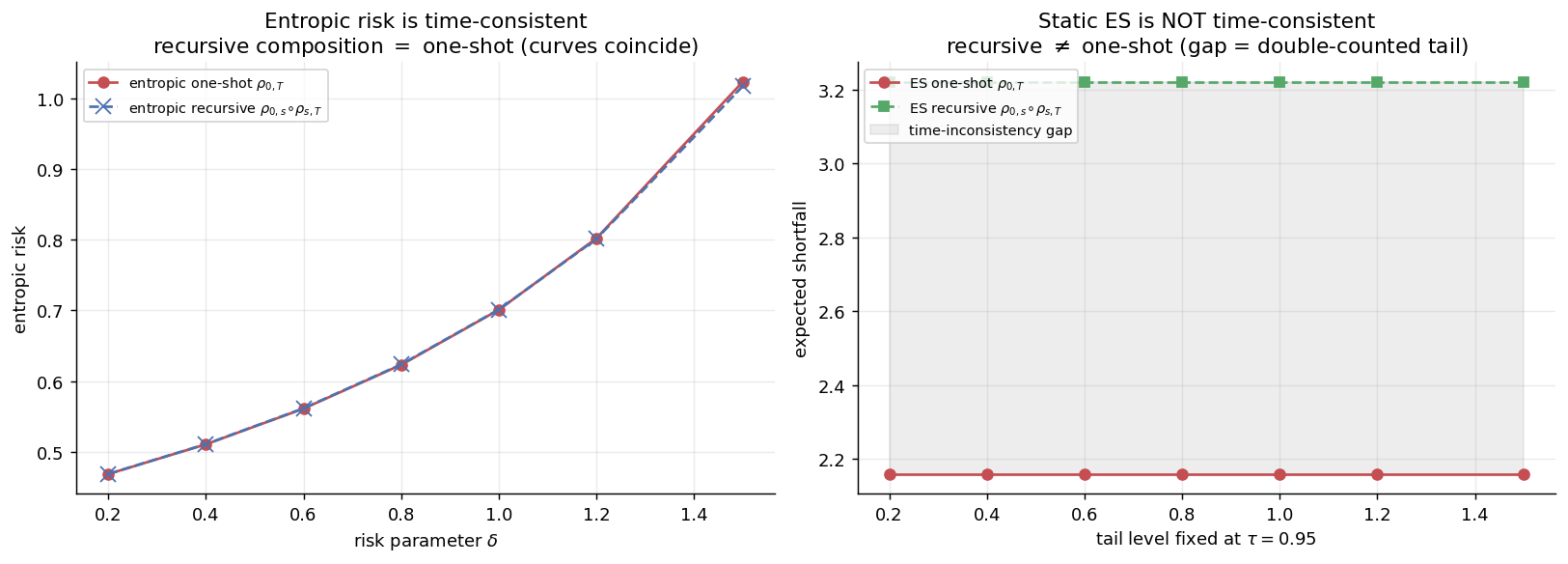}
\caption{Time-consistency. Left: the entropic measure's recursive composition equals its one-shot
value across $\delta$ (the curves coincide). Right: static expected shortfall is not
time-consistent. Its recursive value differs from the one-shot value, and the gap is the
double-counted tail.}
\label{c3:fig:tc}
\end{figure}

Figure~\ref{c3:fig:tc} verifies Proposition~\ref{c3:prop:tc}. We use a two-period model with a
regime revealed at the intermediate date. The recursive entropic risk equals its one-shot value to
within $0.6\%$ across all $\delta$ (time-consistent, as the tower property guarantees). The
recursive expected shortfall, by contrast, overshoots its one-shot value by $49\%$, because
re-applying a tail average to an already tail-averaged quantity double-counts the tail. This is
the pathology that makes static shortfall unsuitable as a dynamic object and motivates
the entropic choice.

\begin{figure}[t]
\centering
\includegraphics[width=\textwidth]{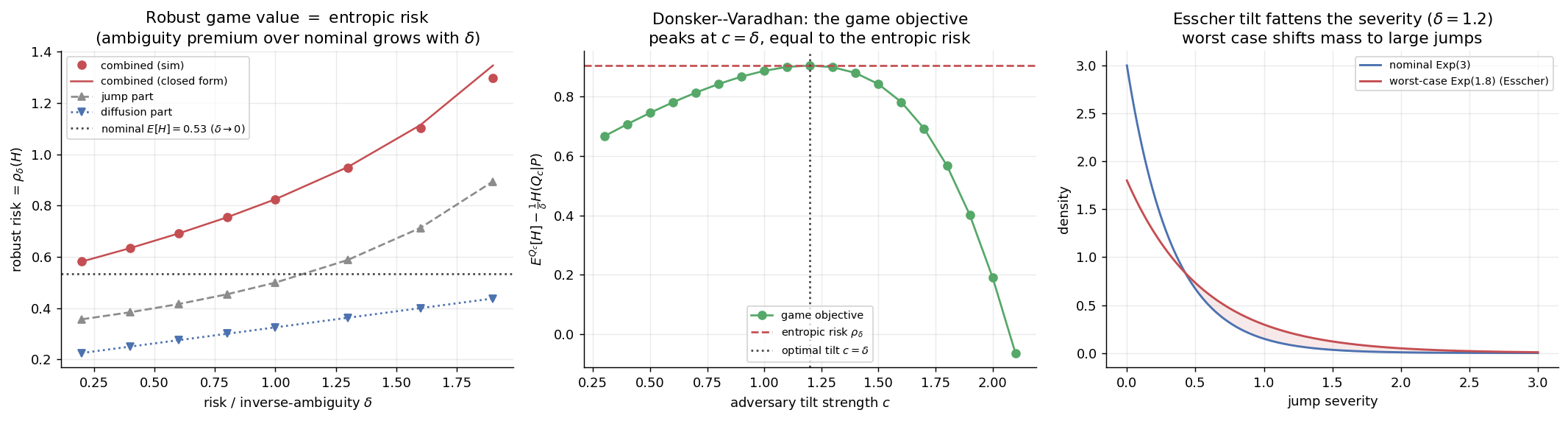}
\caption{The robust solve. Left: the robust game value equals the entropic risk (simulation on the
closed form), with the ambiguity premium over nominal growing in $\delta$. Centre: the
Donsker--Varadhan objective peaks at adversary tilt $c=\delta$, equal to the entropic risk. Right:
the Esscher worst-case fattens the jump severity, shifting mass to large jumps.}
\label{c3:fig:robust}
\end{figure}

Figure~\ref{c3:fig:robust} verifies Theorem~\ref{c3:thm:robust}. The combined jump-diffusion
entropic risk matches its closed form \eqref{c3:eq:robust-closed} to $0.3\%$ away from the moment
boundary, with the ambiguity premium over nominal expected harm growing with $\delta$. As
$\delta R\to b$ the risk diverges and the Monte-Carlo estimate degrades, exactly as a vanishing
exponential moment predicts. The Donsker--Varadhan objective, evaluated over adversary tilt
strengths, peaks at the tilt $c=\delta$ and there equals the entropic risk to numerical precision,
confirming that the worst-case measure is the exponential tilt. That measure is the Girsanov drift
shift on the diffusion together with the Esscher tilt on the jumps. The worst-case jump intensity
is inflated by $M_Y(\delta)$ (verified to $0.5\%$) and the worst-case severity is fattened from
$\mathrm{Exp}(3)$ to $\mathrm{Exp}(1.8)$ (verified to $0.1\%$). The adversary makes harmful jumps
both more frequent and more severe. Finally, the heavy-tail caveat is decisive. With a lognormal
severity, which has no exponential moment, the Monte-Carlo entropic risk does not converge but
drifts upward without bound as the sample grows ($13.9\to66.9\to38.0\to89.9$), signalling the true
value $+\infty$. Truncating the severity (the prevention lever) returns a finite, stable value
($2.72$). Prevention is what makes the robust problem well-posed.

\section{Synthesis: a continuous-time unification}
\label{c3:sec:synthesis}

The results of Section~\ref{c3:sec:theorems} are not a separate theory. They are the continuous-time
limit in which Chapters~\ref{ch:single} and~\ref{ch:fleet} become one object.
Table~\ref{c3:tab:trilogy} makes the correspondence explicit. Read down the columns, and the
framework moves from a single agent at a snapshot, to a fleet at a snapshot, to a fleet through
time. Read across the rows, and each structural notion has a single identity that specialises to
each setting.

\begin{table}[t]
\centering
\small
\renewcommand{\arraystretch}{1.35}
\begin{tabular}{>{\raggedright\arraybackslash}p{0.16\textwidth}>{\raggedright\arraybackslash}p{0.24\textwidth}>{\raggedright\arraybackslash}p{0.24\textwidth}>{\raggedright\arraybackslash}p{0.26\textwidth}}
\toprule
& \textbf{Chapter~\ref{ch:single}} (single agent, static) & \textbf{Chapter~\ref{ch:fleet}} (fleet, static) & \textbf{Chapter~\ref{ch:continuous}} (continuous time) \\
\midrule
Harm model
& $\EH=\PHA\cdot\SGH\cdot\BR$
& fleet aggregate of single-agent harms
& marked jump-diffusion; reduces to Chapter~\ref{ch:single} (Prop.~\ref{c3:prop:reduction}) \\
Systematic risk
& ---
& non-diversifiable common shock
& non-diversifiable common L\'evy jump (Thm.~\ref{c3:thm:t1}) \\
Contagion
& ---
& percolation threshold (giant cascade)
& Hawkes criticality $\spr(\Gam)<1$ (Thm.~\ref{c3:thm:t2}) \\
Catastrophe lever
& prevention non-substitutable (Thm.~B)
& prevention fleet floor
& prevention $=$ jump-measure truncation (Thm.~\ref{c3:thm:t3}) \\
Risk measure
& set-valued coherent measure
& distributional vulnerability functionals
& recursive entropic, time-consistent (Prop.~\ref{c3:prop:tc}) \\
Robustness
& ambiguity set around nominal
& ambiguity set around nominal
& Girsanov$+$Esscher game $=$ entropic (Thm.~\ref{c3:thm:robust}) \\
\bottomrule
\end{tabular}
\caption{The framework as one object. Each row is a single structural notion specialised to the
single-agent snapshot (Chapter~\ref{ch:single}), the fleet snapshot (Chapter~\ref{ch:fleet}), and the continuous-time fleet
(Chapter~\ref{ch:continuous}, this chapter).}
\label{c3:tab:trilogy}
\end{table}

Three things are new in continuous time, beyond restating the static results. First,
\emph{dynamics}. The Hawkes layer distinguishes a fleet on a stationary path from one on an
explosive path through a single computable quantity, the spectral radius of the branching matrix.
It also shows that segmentation is a precise mitigant rather than a vague one, since it divides the
branching ratio. The static percolation threshold told us a giant cascade can exist. The dynamic
criticality tells us whether the fleet is currently sub- or supercritical, and by how much. Second,
the \emph{multiple-harm correction}. The static decomposition is exact only in the rare-event limit,
and the factor $\kappa\ge1$ quantifies how badly a snapshot understates accumulated harm once
events are not rare. Third, and most consequential for governance, the risk object becomes
\emph{dynamically coherent and robust at once}. The recursive entropic measure is time-consistent,
so an acceptance decision is not contradicted by its own future application. That same measure is
the value of the robust Girsanov/Esscher game, so robustness to misspecification is not a separate
overlay but the same computation. The worst case is explicit, making harmful jumps both more
frequent and more severe.
Its finiteness condition is the prevention lever, which ties the dynamic, the systemic, and the
robust threads to the single act of truncating the jump measure.

The central claim is therefore sharper than any single chapter states it. Across one agent and many,
across a snapshot and a path, the binding object is the irreversible jump and the non-substitutable
control is prevention. Diversification, fleet size, detective review, and reversibility all fail
against it in the regimes where it matters. What bounds the tail is truncating the jump
measure, together with limiting exposure to the common jump by segmentation. That is also what
makes the robust problem well-posed in the first place.

\section{Limitations and conclusions}
\label{c3:sec:conclusions}

\paragraph{Limitations.}
This work gives structure, not numbers, and that structure rests on assumptions the reader should weigh. The parameter values throughout are illustrative, and the
central open problem is calibration. Putting the framework to use on a real deployment requires
estimating, from telemetry, the harm intensity (the continuous-time $\PHA$), the mark law and in
particular its irreversible tail (the $\SGH$ and the L\'evy measure), the exposure process
($\BR$), and the systematic loadings $a_k$ and jump rate $\Lsys$. Hardest of all is the Hawkes
branching matrix $\Gam$ that governs contagion, whose spectral radius is the quantity the fleet's
stability turns on. None of these is directly observed. Estimating a branching matrix from
correlated event streams is a genuine statistical problem, sharper still near criticality where, as
our own experiments show, even simulated quantities are hard to pin down. Several modelling
assumptions are deliberately clean and should be relaxed in applications. The idiosyncratic
components are taken i.i.d.\ and the contagion fleet homogeneous, where real fleets are
heterogeneous and networked. The entropic/robust layer requires the jump measure to have an
exponential moment, so heavy-tailed severities must either be truncated, which is the
prevention lever, or handled with a different ambiguity penalty such
as a Wasserstein ball. The prevention theorem assumes detective review cannot intercept the atomic
irreversible jump, which encodes a substantive claim about what review can and cannot do. And the
diffusion enters only as a rate modulator, with no feedback from the controller's actions onto the
state dynamics, which a fuller control formulation would include.

\paragraph{Governance implications.}
Subject to those caveats, the framework yields concrete guidance. The three levers correspond to
distinct, separately-buildable controls: review thins the rate of harmful actions, reversibility
recovers realised harm, and prevention removes the affordance. The analysis says when each is and
is not effective. Against the irreversible tail, only prevention and exposure-reduction
(segmentation) move the floor, so a governance programme that leans on monitoring and rollback alone
is mis-specified for catastrophic risk. The branching-ratio criterion $\spr(\Gam)<1$ is a fleet
stability indicator that a fleet operator can in principle track and hold below one, with
segmentation as the direct instrument for doing so. And the coincidence of the time-consistent
measure with the robust game means a single risk computation serves both dynamic coherence and
robustness to misspecification. It comes with an interpretable worst case to stress against, an
adversary who makes harmful jumps more frequent and more severe.

\paragraph{Conclusion.}
We have given a continuous-time, dynamic theory of agentic model risk that unifies the
single-agent decomposition of Chapter~\ref{ch:single} and the fleet stress framework of Chapter~\ref{ch:fleet} as one jump-diffusion
object. The per-agent harm process reduces exactly to Chapter~\ref{ch:single} in the rare-event
limit. The three fleet results lift to a non-diversifiable common L\'evy jump, a Hawkes contagion
criticality, and a prevention-as-truncation floor. And the risk functional becomes a recursive
entropic measure that is at once time-consistent and the value of a robust Girsanov/Esscher game.
The unifying lesson, visible only once the static snapshots are set in motion and joined, is that the
irreversible jump is the binding object and prevention the non-substitutable lever. That holds
across one agent and many, across a snapshot and a path, and even at the level of whether the
robust problem has a finite answer at all. The work that remains is to calibrate the objects this
theory defines, and to validate the qualitative laws it predicts against the behaviour of real
agentic fleets.

\chapter{The Limits of Runtime Containment: Guardrails as Macroprudential Control}
\label{ch:control}
\section{Introduction}
\label{c4:sec:intro}

Measuring systemic risk is not the same as containing it. The preceding three chapters built the
apparatus for \emph{measuring} the risk that agentic systems create. That apparatus begins, in the
tradition of model risk management \citep{fed2011sr117}, with the microprudential model risk of a
single agent. It then builds to the systemic risk of a fleet on a shared model, and to its dynamics
in time. Chapter~\ref{ch:single} decomposed a single agent's expected harm into the probability of a
harmful action, the severity given harm, and the blast radius. It placed this inside a set-valued,
containment-subadditive risk measure, and proved that against irreversible harm the three containment
levers, review, reversibility, and prevention, are not substitutes, since only prevention moves the
catastrophic floor. Chapter~\ref{ch:fleet} scaled the unit of analysis from one agent to a fleet. It
established non-diversifiability of a common shock, a contagion criticality threshold, and a fleet
prevention floor. Chapter~\ref{ch:continuous} made the object dynamic. It modelled cumulative harm as
a jump-diffusion, lifted the fleet results to continuous time, and characterised the worst-case
measure a robust controller should fear. Across those three chapters the deliverable is a
\emph{number}, or a set of numbers: an acceptance set, a tail, a stationary intensity, a robust
value. They tell an operator \emph{how risky} a deployment is.

They do not tell the operator \emph{what to do about it in real time}. A risk measure is a
constraint. Satisfying a constraint on a stochastic system that is running, taking irreversible
actions, and possibly being driven by an adversary is a problem of \emph{control}. This chapter
provides that missing half. It asks: given monitors that watch a deployed agent imperfectly and
intermittently, when can a governor keep the agent inside its model-risk specification with high
probability? What is the best achievable detection-and-intervention performance? And, most
importantly, where is runtime control \emph{structurally unable} to help, so that the only remaining
lever is to change the system before it runs? The earlier chapters' containment levers reappear here
not as static design choices but as the \emph{interventions of a feedback controller}. Review becomes
a detection-triggered human-approval gate. Reversibility becomes a detection-triggered rollback. And
prevention becomes the one lever that, as we show, must be set ex ante, because no amount of runtime
vigilance can substitute for it.

\paragraph{The object.} We model the agent's instantaneous risk profile as a low-dimensional latent
state $X_t\in\R^d$ evolving as a mean-reverting linear diffusion. Think of axes such as
destructive-operation rate, scope creep, and privilege reach for a coding agent, or exposure,
value-at-risk utilisation, and leverage for a trading agent. Nobody observes ``the risk'' directly. A
bank of $N$ \emph{sentinel} monitors takes discrete, partial, noisy readings of $X$ on a clock and is
\emph{silent between readings}, exactly as a syscall monitor, a periodic risk engine, or a batched
LLM-judge does. A hidden two-state regime $\reg_t$ encodes whether the agent is behaving benignly or
has entered a \emph{violating} mode, through prompt injection, reward hacking, goal
misgeneralisation, a compromised tool, or spoofed inputs. Violation manifests as a shift in the
agent's operating point. Distinct from this drift, and the source of the chapter's hardest result, is
a separate \emph{catastrophic} event: an irreversible jump that throws the state into a region from
which there is no return. Examples are a force-push that rewrites history, a fill that executes, a
wire that sends, or a payment that issues. The monitors fuse their readings and form a belief about
the regime.
When the belief crosses a threshold, they trigger costly impulse interventions that map to the
Chapter~\ref{ch:single} levers. Compliance with the model-risk specification is the requirement that
the state stay inside a safe operating box $\Ssafe$ with high probability over the horizon:
$\Prob(\tau_{\Ssafe}\le
T)\le\varepsilon$, where $\tau_{\Ssafe}$ is the first exit time.

\paragraph{The bridge.} Detection and compliance are different events, and conflating them is the
central modelling trap. An operating-point shift can leave the state well inside $\Ssafe$: the agent
is violating but still compliant. Benign diffusion can carry the state out of $\Ssafe$: the agent is
compliant but has nonetheless breached. A governor that perfectly detects the regime has not thereby
bounded the exit probability, and a governor that bounds the exit probability need not detect the
regime at all. Section~\ref{c4:sec:bridge} closes this gap with a \emph{bridge lemma}. Decomposing
exit by cause, the exit probability is bounded by a benign geometric floor that no controller can
touch, plus an undetected-violation term, plus a catastrophe term. The undetected-violation term
equals a first-passage probability $q(\E[D]+\trc)$ evaluated at the detection-plus-reaction delay
$\E[D]+\trc$. Feasibility of the compliance constraint then reduces to three conditions one can check
against closed forms. The binding one is a discrete-time \emph{Nyquist condition},
$\E[D]+\trc<\Tesc$, which requires the loop to detect and react before the drift escapes the safe box.
We \emph{derive} this condition from the detection theory rather than positing it.

\paragraph{What we prove.} The detection half (Section~\ref{c4:sec:detection}) establishes three
results about \emph{what the sentinels can and cannot see}. A collective-observability trichotomy
sorts every violation direction into one of three classes relative to the monitors' joint observation
field. A direction may be \emph{intent-observable}, seen immediately and robustly. It may be
\emph{state-observable only}, hidden from the monitors' direct field but leaking into the controlled
state through the dynamics, with a quantified leak order and an associated lag. Or it may be
\emph{unobservable}, lying in a null subspace on which no number of sentinels, no sampling rate, and
no reduction of observation noise ever produces a flicker of information. A second result shows that
correlated sensor noise installs a \emph{non-diversifiable} detection floor. Adding sentinels that
share a blind spot does not help along the shared-blind direction, a monitoring monoculture analogous
to Chapter~\ref{ch:fleet}'s non-diversifiable shock. A third gives the quickest-detection
frontier, the delay as a function of significance level and signal strength. Underlying all three is
the modelling choice that monitors observe the agent's \emph{attempted actions} rather than its
controlled risk level. A closed-loop governor that holds the state compliant also hides the evidence;
we show this single choice makes detection invariant to the control that would otherwise mask it.
As a bonus, it removes the relaxation lag from the detection delay.

An impossibility result anchors the control half (Section~\ref{c4:sec:control}). We state it as an
explicit corollary of a defended exogeneity assumption rather than asserting it. Against the
irreversible jump, \emph{every} runtime policy incurs catastrophe probability $1-e^{-\Lamc T}\ge\Lamc
T\,e^{-\Lamc T}$. This holds for any number of sentinels, any intervention authority, any detection
threshold, and any sampling schedule, because a causal detect-then-react loop cannot fire before a
zero-duration event. Only ex-ante \emph{prevention} moves this floor. It lowers the hazard $\Lamc$
itself by removing the action from the agent's reach. This is the dynamic, distributed-control
incarnation of Chapter~\ref{ch:single}'s irreversibility theorem and
Chapter~\ref{ch:continuous}'s prevention-as-truncation. The result is not an artefact of the abstraction: it
explains why a trading desk enforces \emph{pre-trade} hard limits at the order gateway rather than trusting
post-trade monitoring to catch an oversized fill, why a deployment platform revokes production
credentials rather than alerting on their misuse, and why a customer-service agent has payment scope
removed rather than watched. Reversible feasibility, by contrast, falls out as a corollary of the
bridge. We treat the fully fused, centralized ensemble as the analytical benchmark, and the
distributed, consensus-based, Byzantine-exposed reality as an approximation to it
(Section~\ref{c4:sec:control}). We also quantify a robustness--efficiency tension, showing that fusion
tolerating malicious sentinels pays a factor $\pi/2$ in variance (Section~\ref{c4:sec:robustness}). A
sampling-design analysis completes the picture. Statistical detection delay is invariant to the
monitoring schedule, while the worst-case blind gap and the schedule's predictability to an adversary
are not.

\paragraph{Where guardrailing ends.} Section~\ref{c4:sec:synthesis} assembles these into an
\emph{operating envelope}. It partitions the deployment's parameter space into a region where
guardrailing is feasible and three regions where it is not: where the drift escapes faster than the
loop can react, where violations burst and revert below the detection timescale, and where the
catastrophe hazard dominates. The boundaries are themselves a result of the analysis, not a check
applied afterward. They say when an operator must stop tuning thresholds and instead reduce dimensionality, harden
prevention, or decline the deployment.

\paragraph{Stance.} As in the earlier chapters, the parameters here are illustrative. The
linear-Gaussian, two-to-three-dimensional state is a deliberate caricature of high-dimensional,
non-Gaussian, nonstationary agent telemetry. The contribution is structural, showing \emph{which}
failure modes exist, what \emph{shape} the remedies take, and \emph{where} runtime control is
powerless, rather than a deployable detector. Calibrating regimes, directions, and hazards to real systems remains
the central open problem. The sentinels are themselves agents with model risk, so the construction
bounds rather than dissolves the regress. Section~\ref{c4:sec:numerics} confirms every closed form to
within Monte-Carlo error. We turn first to the surrounding literature.

\section{Background and related work}
\label{c4:sec:background}

The construction sits at the confluence of five literatures. Its novelty is best stated as the
specific way it combines them around a risk-measure acceptance set. We summarise each and then say
what is new.

\paragraph{Runtime assurance, shielding, and safe reinforcement learning.} Wrapping an untrusted
controller in a monitored safety layer is a well-established idea in control and verification. The
Simplex architecture \citep{sha2001using} pairs a high-performance but unverified controller with a
verified safety controller and a decision module that switches to the latter near the boundary of a
certified safe set. Shielding \citep{alshiekh2018safe} synthesises a reactive \emph{shield} from a
temporal-logic specification and an automaton model of the environment. The shield overrides any
action that could lead to a specification violation, and it is a leading paradigm for safe
reinforcement learning, surveyed alongside Lagrangian and constrained-MDP methods by
\citet{garcia2015comprehensive} and \citet{altman1999constrained}. These approaches are predominantly
\emph{automata-theoretic and worst-case}: the environment is a (possibly nondeterministic) transition
system, safety is a logical property, and the guarantee is a hard guarantee under a known model. Our
setting differs on three axes. First, it is \emph{stochastic and statistical} rather than logical.
The ``violation'' is a hidden regime to be \emph{detected} from noisy data, not a state label that is
read off, and the guarantee is a probability bound tied to a risk measure. Second, the monitor is
\emph{distributed}. A shield is a single synthesised automaton, whereas our safety layer is a fleet
of imperfect, disagreeing, possibly compromised sentinels that must reach consensus. Third, the
monitor is \emph{intermittent}. Shields act every step, while our sentinels sample on a clock and are
blind between samples, which is what makes a sampling-rate (Nyquist) condition the binding
feasibility constraint. The impossibility result of Section~\ref{c4:sec:control} is the statistical
analogue of the observation, implicit in runtime assurance, that a switch cannot save a system from a
transition that has already completed.

\paragraph{Quickest detection and sequential analysis.} Quickest-change detection asks how to detect
a hidden change in distribution as quickly as possible subject to a false-alarm constraint. It begins
with Page's cumulative-sum (CUSUM) chart \citep{page1954continuous} and Shiryaev's Bayesian
formulation \citep{shiryaev1963optimum}, and continues with the minimax formulations and optimality
results of Lorden \citep{lorden1971procedures}, Pollak \citep{pollak1985optimal}, and Moustakides
\citep{moustakides1986optimal}. The modern synthesis appears in \citet{poor2008quickest}, and the
roots are in Wald's sequential analysis \citep{wald1945sequential}. The detection frontier of
Section~\ref{c4:sec:detection} is a direct application, showing that the expected detection delay scales as
$\log(1/\alpha)$ over a Kullback--Leibler information rate, in the Lorden minimax sense. Our
contribution is not to this theory but to its \emph{geometry under a fused, partially observed vector
state}. The information rate becomes a directional quadratic form in the violation direction. The new
phenomena are properties of how a multi-sensor field projects a vector change, not of the stopping
rule: a null subspace of undetectable directions, a non-diversifiable floor from correlated sensors,
and the interaction with a control that can mask the signal.

\paragraph{Filtering of a partially observed regime.} Reconstructing a latent state and a hidden
regime from intermittent noisy observations is filtering. The Kalman filter \citep{kalman1960new}
handles the linear-Gaussian state. The \emph{continuous--discrete} filter
\citep{jazwinski1970stochastic} propagates the conditional mean and covariance through the silent
intervals by the prediction equations and corrects them at observation epochs, which is exactly our
sample-and-silence structure. The Wonham filter \citep{wonham1964some} gives the regime posterior. We
use these as standard machinery and foreground instead the \emph{observability} question they
presuppose: whether the regime leaves any imprint on the observation field at all. The filtering
recursions take that question for granted.

\paragraph{Distributed estimation, consensus, and the absence of separation.} When the sentinels
cannot all see one fusion center, they must combine information over a communication graph. The
consensus and cooperative-control literature \citep{olfatisaber2007consensus} and the
Kalman-consensus filter \citep{olfatisaber2009kalman} provide distributed estimators whose
performance approaches the centralized one at a rate governed by the algebraic connectivity of the
graph. This is why we are careful (Section~\ref{c4:sec:control}) to designate the fully fused,
centralized ensemble as the analytical \emph{benchmark} and to treat the distributed implementation
as an approximation to it rather than claiming the two coincide. The distinction matters because of
Witsenhausen's counterexample \citep{witsenhausen1968counterexample}: decentralized stochastic
control with private information generally admits \emph{no} finite-dimensional sufficient statistic
and no separation between estimation and control. The common-information approach of
\citet{nayyar2013decentralized} recovers a coordinator's dynamic program exactly when the agents
share enough history. This result justifies the centralized benchmark under information sharing, and
explains why it fails for neighbour-only consensus. We therefore claim exact separation only for the fused
benchmark, and an approximation gap otherwise, citing the distributed-estimation literature.

\paragraph{Byzantine robustness and robust statistics.} Some sentinels may be compromised. Tolerating
adversarial participants in a distributed computation is the Byzantine generals problem
\citep{lamport1982byzantine}. Its classical threshold bounds how many corrupted monitors a fused
decision can survive: agreement is possible iff fewer than a third are faulty, or, with
authentication and full connectivity, fewer than a half. The machine-learning analogue, robust
aggregation of gradients or estimates under a fraction of adversarial workers
\citep{blanchard2017machine}, makes the same point quantitatively. The price of robustness is the
classical inefficiency of robust estimators: the median of $n$ Gaussian observations has asymptotic
variance $\pi/2$ times that of the mean \citep{huber1964robust}. Section~\ref{c4:sec:robustness}
turns this into a \emph{frontier}. The same diversity that a non-Byzantine analysis wants to exploit
(use every sentinel, optimally weighted) is what a Byzantine-robust fusion rule must partly discard,
so robustness and detection power pull in opposite directions.

\paragraph{Impulse control of jump diffusions.} Impulse control studies interventions that are
discrete, costly, and reset the state, rather than a continuously applied control effort. The
associated quasi-variational inequalities \citep{bensoussan1984impulse} were developed for jump
diffusions by \citet{oksendal2007applied} within the viscosity-solution framework of
\citet{fleming2006controlled}. The interventions of our governor (rollback, throttle, sandbox, halt)
are impulses of exactly this kind, and the cost--harm trade-off they solve is a standard QVI. We
present it as supporting structure rather than as a contribution. We draw the first-passage formulae
underlying the bridge from the standard handbook results for drifted Brownian motion
\citep{borodin2002handbook}, and inherit the jump and prevention apparatus from
Chapter~\ref{ch:continuous} and the jump-process toolkit of \citet{cont2004financial}.

\paragraph{What is new.} The contribution lies in the combination. We embed a distributed, intermittent,
Byzantine-exposed detection problem and an impulse-control problem inside a model-risk acceptance set
\citep{artzner1999coherent,fed2011sr117}. We then prove that the achievability of the resulting
compliance constraint is governed by the \emph{joint} geometry of three distinct objects: which
violation directions are collectively observable (a subspace), which are controllable with bounded
authority within the reaction delay (a reachability condition), and which lie in the reversible
region (an affine region). The headline impossibility is that runtime control cannot lower the
irreversible-jump floor and only ex-ante prevention can. It unifies, in dynamic and distributed form,
the static irreversibility result of Chapter~\ref{ch:single} and the prevention-as-truncation result
of Chapter~\ref{ch:continuous}, and it gives a control-theoretic account of why deployed safety
architectures for high-stakes agents are built the way they are. None of the constituent literatures,
taken alone, predicts the operating envelope of Section~\ref{c4:sec:synthesis}. Each addresses one
face of a problem whose difficulty is in their interaction: logical safety, or detection delay, or
consensus, or robustness, or impulse cost.

\section{The model}
\label{c4:sec:model}

We describe the agent, the regime, the catastrophe, the safe region, the sentinels and their two
observation channels, the controller, and the compliance target. We then state the standing
assumptions and the timescale ordering under which the results are jointly meaningful. A running
table (Table~\ref{c4:tab:instantiation}) instantiates every object in three deployments.

\subsection{State, regime, and catastrophe}

\paragraph{Risk state.} The agent's instantaneous risk profile is a latent vector $X_t\in\R^d$ with
$d$ small ($2$ or $3$). It evolves as a mean-reverting linear diffusion whose operating point is
shifted by a hidden regime and which the controller can throttle and reset:
\begin{equation}
\label{c4:eq:state}
dX_t \;=\; \big[\,-A\,(X_t-x_0)\;+\;\reg_t\,\drift\;-\;u_t\,\big]\,dt
\;+\;\Sigma^{1/2}\,dW_t\;+\;\sum_{k:\,\tau_k\le t}\xi_k\,,
\end{equation}
Here $-A$ is Hurwitz (all eigenvalues have negative real part, so absent forcing the state relaxes to
$x_0$), $\Sigma\succ0$ is the diffusion covariance, and $W$ is a standard $d$-dimensional Brownian
motion. The continuous \emph{throttle} $u_t$ is the controller's suppression of the violating drift,
and the $\{(\tau_k,\xi_k)\}$ are \emph{impulse} interventions (instantaneous state resets) applied at
stopping times $\tau_k$. With no regime forcing and no control, $X$ is an Ornstein--Uhlenbeck process
with stationary mean $x_0$ and stationary covariance $P_\infty$ solving the Lyapunov equation $A
P_\infty+P_\infty A^{\!\top}=\Sigma$.

\paragraph{Hidden regime.} A two-state continuous-time Markov chain $\reg_t\in\{0,1\}$ (benign,
violating) with transition rates $q_{01},q_{10}>0$ governs whether the agent is behaving as intended.
Violation is an \emph{operating-point shift}: switching $\reg$ on adds the drift $\drift\in\R^d$, so
the (uncontrolled) stationary mean moves from $x_0$ to $x_0+A^{-1}\drift$. The direction of $\drift$
encodes \emph{what kind} of violation it is, which risk axes drift and how fast, and is, by
Assumption~\ref{c4:ass:direction} below, chosen adversarially. We model violation as a persistent
drift in $\drift$ rather than as a one-off displacement, and return to that choice in a remark in
Section~\ref{c4:sec:detection}. The operating-point-shift form keeps the detection and control
problems aligned.

\paragraph{Catastrophe.}\label{c4:par:catastrophe} Distinct from the drift is an irreversible
\emph{catastrophic} event. Let $\sigma_{\mathrm c}$ be the first arrival of a Poisson process with
regime-dependent hazard $\Lamc(\reg_t)\ge0$. At $\sigma_{\mathrm c}$ the state is thrown into the
catastrophic region $\Ccat$ and remains there. This is the \emph{only} jump in the model, and we
separate it deliberately from the continuous state \eqref{c4:eq:state}. Keeping $X$ a linear-Gaussian
diffusion between catastrophes is what makes the Gaussian filter valid, while the catastrophe carries
the irreversibility on which the paper's central impossibility (Section~\ref{c4:sec:control}) turns.
A force-push that rewrites history, a fill that executes, and a wire that sends are each such an
event, zero-duration, hazard-driven, and unrecoverable.

\paragraph{Safe and catastrophic regions.} The coordinates split into reversible and irreversible
axes, $\{1,\dots,d\}=J_r\sqcup J_i$. The model-risk specification is a \emph{box}
\begin{equation}
\label{c4:eq:safebox}
\Ssafe \;=\; \{\,x\in\R^d : x_j\le\beta_j \ \text{for all } j\,\},
\end{equation}
with per-axis limits $\beta_j$ matching the kind of clause a real specification contains: an egress
cap, a leverage limit, a sensitive-action rate. The exit time
$\tau_{\Ssafe}=\inf\{t:X_t\notin\Ssafe\}$ is the first moment the agent breaches its specification.
Breaching a reversible axis (scope creep, a rising VaR) is recoverable.

The catastrophic region $\Ccat$ is not a further box on the same state space, but the
\emph{absorbing set entered at the irreversible commit} of \S\ref{c4:par:catastrophe}, a cemetery
region, disjoint from $\Ssafe$, that the diffusion \eqref{c4:eq:state} cannot reach on its own and
from which there is no return. This distinction is load-bearing and we state it plainly. Were
$\Ccat$ a threshold region on the irreversible axes, entry would be a state first-passage event and
the throttle would influence it. That is the Nyquist race the reversible axes are subject to,
and it would contradict the impossibility result of Section~\ref{c4:sec:control}. The two channels
are kept separate. The reversible axes are reached by drift and diffusion and are therefore
contestable at runtime; $\Ccat$ is reached only by the commit, and is not.

\subsection{Sentinels and their two channels}

\paragraph{Discrete, partial, silent.} A bank of $N$ sentinels monitors the agent. Sentinel $i$
\emph{looks} only at sample epochs $t_k=k\Delta$ and is silent in between, returning a noisy linear
projection of the state,
\begin{equation}
\label{c4:eq:obs-state}
Y^{x,i}_k \;=\; \Hsta_i\,X_{t_k}\;+\;\varepsilon^{x,i}_k,
\qquad \Hsta_i\in\R^{m_i\times d},
\end{equation}
with each sentinel typically seeing only some axes ($\Hsta_i$ has few rows). Stacking the sentinels
gives $Y^x_k=\Hsta X_{t_k}+\varepsilon^x_k$ with $\Hsta=[\Hsta_1;\dots;\Hsta_N]$ and observation-noise
covariance $\Cov(\varepsilon^x_k)=R\succ0$. The \emph{fused information} of one synchronous look is
the matrix
\begin{equation}
\label{c4:eq:fused-info}
\Ifus \;=\; \HstaT R^{-1}\Hsta,
\end{equation}
which we treat throughout as the \emph{centralized} benchmark, that is, what an idealised fusion
center with access to all sentinels at once would have. The distributed implementation of
Section~\ref{c4:sec:control} approximates it. That monitoring is discrete with actual silence is not
an idealisation: a risk engine recomputes on a cycle, a batch scan runs periodically, an LLM-judge
passes over a transcript at intervals. The silence is exactly why a sampling-rate condition will
bind.

\paragraph{The intent channel.} A closed-loop tension threatens every detection result in this
chapter. A controller that successfully holds $X$ near $x_0$ \emph{masks the very signal detection
needs}, since the state then looks compliant only because it is being held compliant. The
resolution, central to the construction, is to give the sentinels a second channel that the control
does not erase. Separate the agent's \emph{attempted} action from its \emph{realised} effect. Let
\begin{equation}
\label{c4:eq:intent}
a_t \;=\; \reg_t\,\drift \;+\;\eta_t
\end{equation}
be the agent's action propensity: a benign component $\eta_t$ plus, in the violating regime, the push
$\drift$. Let the net forcing on the state be $a_t-u_t$, so the throttle $u_t$ in
\eqref{c4:eq:state} acts on the \emph{realisation}, not on the attempt. The sentinels' intent channel
observes the attempt,
\begin{equation}
\label{c4:eq:obs-intent}
Y^{a,i}_k \;=\; \Hint_i\,a_{t_k}\;+\;\varepsilon^{a,i}_k,
\qquad \Ifusa \;=\; (\Hint)^{\!\top} R^{-1}\Hint,
\end{equation}
whose law depends on $\reg$ but \emph{not} on $u$. Detection on this channel is therefore invariant
to the control. This models what monitors actually log, rather than the post-control risk level:
attempted API calls, drafted orders, and proposed edits, \emph{before} the gate that may block them.
The state channel \eqref{c4:eq:obs-state} remains available. It sees $\drift$ only after it has
propagated through the dynamics $A$, and only to the extent the control has not suppressed it. The
two channels give substantively different views, and Section~\ref{c4:sec:detection} shows the distinction
sorts violation directions into observability classes.

\paragraph{Filter.} The sentinels run a continuous--discrete filter \citep{jazwinski1970stochastic}.
Between looks the conditional mean and covariance are propagated by the prediction equations of
\eqref{c4:eq:state}, so the covariance \emph{grows} through the silent interval. At each look they
are corrected in information form by \eqref{c4:eq:obs-state}--\eqref{c4:eq:obs-intent}, and the regime
posterior $\pi_t=\Prob(\reg_t=1\mid\mathcal Y_{\le t})$ is updated by the Wonham recursion
\citep{wonham1964some}. The triple $(\hat x_t,P_t,\pi_t)$ is the sufficient statistic for the
centralized \emph{linear-Gaussian} filter; with a general hidden-regime jump it is the
benchmark that the finite-dimensional filter approximates (cf.\ Remark~\ref{c4:rem:distributed}).

\paragraph{Diversity, two axes.} Sentinels are imperfect in two independent ways.
\emph{Observational} diversity concerns the correlation of their noise. Writing
$\varepsilon^i=C\omega+D\nu^i$ with $\omega$ a common component shared by all sentinels and $\nu^i$
idiosyncratic, the shared part $C\omega$ does not average away (a scalar version has correlation $r$
and per-sentinel variance $\sigma_o^2$). \emph{Temporal} diversity concerns \emph{when} they look:
all at once (synchronous, a ``temporal monoculture'') versus staggered around the cycle.
Section~\ref{c4:sec:robustness} shows these are orthogonal and that a fleet needs both.

\paragraph{Communication.} When there is no single fusion center, sentinels exchange beliefs over a
graph with Laplacian $\Lapl$ and algebraic connectivity $\lambda_2(\Lapl)$. The distributed estimate
approaches the centralized benchmark \eqref{c4:eq:fused-info} at a rate set by $\lambda_2(\Lapl)$
\citep{olfatisaber2007consensus,olfatisaber2009kalman}.

\subsection{Controller and compliance target}

\paragraph{Interventions as levers.} The governor acts in two ways, both mapping to the
Chapter~\ref{ch:single} containment levers. A continuous \emph{throttle} $u_t$ suppresses the
violating drift once the regime belief crosses a threshold (rate-limit actions, sandbox the tool,
revoke a scope). An \emph{impulse} $(\tau_k,\xi_k)$ resets the state (roll back a branch, flatten a
book, restore a snapshot). Review is a detection-triggered human-approval gate that delays an action
pending confirmation. Reversibility is a detection-triggered rollback impulse. Prevention is the one
lever set \emph{before} deployment, which removes an action from the agent's reach and so lowers the
hazard $\Lamc$ itself. The impulse problem is a quasi-variational inequality
\citep{bensoussan1984impulse,oksendal2007applied} that trades intervention cost against expected
harm; we treat it as supporting structure (Section~\ref{c4:sec:control}).

\paragraph{Compliance.} Compliance with the model-risk specification over a horizon $T$ is the
\emph{hard exit-probability constraint}
\begin{equation}
\label{c4:eq:compliance}
\Prob(\tau_{\Ssafe}\le T)\;\le\;\varepsilon .
\end{equation}
This is the acceptance set of Chapter~\ref{ch:single} expressed dynamically. A deployment is
acceptable if, over the horizon, the probability of leaving the safe operating box is at most
$\varepsilon$. The reaction lag $\trc$ denotes the delay between the controller deciding to act and
the throttle taking effect.

\subsection{Standing assumptions}

The following hold throughout. They are not innocuous. Several results are only \emph{jointly}
meaningful inside them, and Section~\ref{c4:sec:synthesis} shows what fails when each is violated.

\begin{assumption}[State--catastrophe separation]
\label{c4:ass:separation}
The tracked state $X$ is linear-Gaussian between catastrophes \eqref{c4:eq:state}; the only jump is the
separate catastrophe event $\sigma_{\mathrm c}$. (This makes the Gaussian filter valid and isolates
irreversibility in $\sigma_{\mathrm c}$.)
\end{assumption}

\begin{assumption}[Regime acts only through $\drift$]
\label{c4:ass:onlymu}
The regime shifts the drift via $\drift$ and does not detectably alter the diffusion $\Sigma$ or the
hazard $\Lamc$. (Needed for the null subspace of Theorem~\ref{c4:thm:observability}: a regime that also
changed $\Sigma$ would be detectable through second moments even along an otherwise-blind direction.)
\end{assumption}

\begin{assumption}[Unknown violation direction]
\label{c4:ass:direction}
The adversary chooses $\drift$ within a threat cone; detection is a generalised-likelihood-ratio or
mixture procedure over the cone, and detection power is the \emph{minimum} over directions in the cone,
not a fixed-$\drift$ quadratic form. The adversary aims at the floors---the correlated-noise floor of
Theorem~\ref{c4:thm:diversity} and the null subspace of Theorem~\ref{c4:thm:observability}---which is why
those results matter operationally.
\end{assumption}

\begin{assumption}[Non-pathological sampling]
\label{c4:ass:sampling}
The sampling interval $\Delta$ avoids the aliasing resonances of $A$, so that the sampled system has
the same observability as the continuous one (no rotational mode of $A$ completes an integer number of
turns per $\Delta$ and hides between looks).
\end{assumption}

\subsection{Timescale ordering}
\label{c4:sub:timescales}

The problem has several clocks, and the results live in the regime where they are ordered. Take the
intent-channel detection delay $\E[D]=\Delta\big(\tfrac12+\log(1/\alpha)/\mathcal D(\drift)\big)$ from
Theorem~\ref{c4:thm:frontier} and the drift-escape time $\Tesc=g_0/\nu$, the deterministic time for
the violating drift to carry an active margin to the boundary, defined in
Section~\ref{c4:sec:bridge}. We assume
\begin{equation}
\label{c4:eq:timescales}
\underbrace{\Delta,\ \trc}_{\text{look, react}}\ \ll\ \underbrace{\E[D]}_{\text{detect}}\ \ll\
\underbrace{1/q_{10}}_{\text{regime dwell}},
\qquad\text{and}\qquad
\E[D]+\trc\ <\ \Tesc\ \ll\ \min\{\,1/\Lamc,\ \trelax\,\},
\end{equation}
where $\trelax=1/\min_i\Re\lambda_i(A)$ is the slowest Ornstein--Uhlenbeck relaxation time of
\eqref{c4:eq:state}. In words, looking and reacting are fast relative to detecting, and detection
completes before the violating regime reverts. Detection-plus-reaction beats the drift to the
boundary, and drift-escape is the dominant failure mode relative to spontaneous catastrophe. And the
escape window is short relative to the mean-reversion time, so that over a violating excursion the
restoring force $-AX$ is negligible beside the injected drift $\drift$. Section~\ref{c4:sec:synthesis}
labels these $(\mathrm O1)$--$(\mathrm O4)$. There we show the ordering is non-empty, since both the
coding and trading instantiations satisfy all four, and that crossing each boundary names a distinct,
real failure mode rather than a mere technical breakdown.

\subsection{Three stylized instantiations}
\label{c4:sub:instantiation}

Table~\ref{c4:tab:instantiation} maps every abstract object to three deployments. The trading row is
closest to the framework's home discipline. As we note in Section~\ref{c4:sec:control}, it validates
the central impossibility result against established market-infrastructure practice.

\begin{table}[t]
\centering\small
\setlength{\tabcolsep}{4.5pt}
\renewcommand{\arraystretch}{1.25}
\begin{tabular}{@{}p{0.16\textwidth}p{0.255\textwidth}p{0.255\textwidth}p{0.255\textwidth}@{}}
\toprule
& \textbf{Coding / deploy agent} & \textbf{Trading / treasury agent} & \textbf{Customer-facing assistant} \\
\midrule
State axes $X$ &
destructive-op rate; scope creep; privilege reach &
exposure vs.\ limit; VaR/ES utilisation; leverage &
sensitive-action rate; data-export volume; policy-deviation \\
Regime switch $\reg{:}0{\to}1$ &
prompt injection in a fetched issue; reward hacking; compromised tool &
signal/model decay; PnL-metric hacking; spoofed market data &
prompt injection in a user message or KB doc; jailbreak \\
Violation drift $\drift$ &
risky edits accumulating in a branch &
position creeping toward a limit; VaR rising &
rate of sensitive actions climbing \\
Catastrophe $\sigma_{\mathrm c}$ (irreversible) &
force-push to \texttt{main}; prod-DB drop; secret exfil &
oversized fill that \emph{executes}; regulatory-limit breach; wire &
sent email; issued payment; PII export; public post \\
Sentinels &
syscall monitor; PR/git monitor; plan-vs-ticket judge; egress monitor &
risk engine (on a cycle); order-flow monitor; PnL-attribution; data-integrity &
output policy classifier; action-rate monitor; egress monitor; auth monitor \\
Monoculture ($r{\to}1$) &
all parse the \emph{same} audit log $\Rightarrow$ one spoofed log blinds all &
all consume the \emph{same} market feed $\Rightarrow$ one spoof blinds all &
all read the \emph{same} model output $\Rightarrow$ one bypass blinds all \\
Levers (review / revers.\ / prevent) &
approval gate / ephemeral branch / revoke prod cred, read-only FS &
desk approval / liquid-only + hedge / \textbf{pre-trade hard limit at gateway} &
human approval / draft-only / \textbf{remove payment scope, redact PII} \\
Spec line $\beta$ / irreversible commit &
scope-egress policy / prod write, history rewrite &
risk-budget line / regulatory or liquidation trigger &
action-rate policy / external send, payment \\
\bottomrule
\end{tabular}
\caption{The model instantiated in three deployments. In each, $X$ is a latent risk index reconstructed
from noisy proxies. The reversible/irreversible split $J_r\sqcup J_i$ is where mature systems already
place their hardest controls, namely \emph{prevention} on \emph{irreversible} actions (boldface), enforced
as hard pre-action gates rather than monitors. The framework explains why.}
\label{c4:tab:instantiation}
\end{table}

\section{The bridge: from regime detection to a hitting-time guarantee}
\label{c4:sec:bridge}

This section proves the result that makes the rest of the chapter coherent. The sentinels detect a
\emph{regime}; the compliance constraint \eqref{c4:eq:compliance} concerns a \emph{hitting time}. The
bridge lemma bounds the second in terms of the first. It splits the exit probability into a part no
controller can touch, a part governed by how fast detection and reaction outrun the violating drift,
and a part carried entirely by the catastrophe. The decomposition then yields the feasibility
conditions for guardrailing, including a discrete-time Nyquist condition that we derive rather than
assume.

\subsection{The active margin}

Fix a face of the safe box, an axis $j$ and its limit $\beta_j$, and define the \emph{margin}
\begin{equation}
\label{c4:eq:margin}
g_t \;=\; \beta_j - X_{t,j}\;>\;0,
\end{equation}
the signed distance from the current risk level on axis $j$ to its specification limit. Exit through
this face is the event $\{g_t\le 0 \text{ for some } t\le T\}$, and $\tau_{\Ssafe}$ is the first such
time over all active faces. Three scalars summarise the margin's dynamics on axis $j$: its benign
stationary fluctuation $\sigma_g^2=e_j^{\!\top}P_\infty e_j$ (from the Lyapunov covariance), the
violation drift rate it sees, $\nu=e_j^{\!\top}\drift$, and its diffusion rate $s^2=e_j^{\!\top}\Sigma
e_j$. We write $g_0=\beta_j-x_{0,j}$ for the nominal margin at the operating point.

\subsection{The decomposition}

\begin{lemma}[Bridge lemma]
\label{c4:lem:bridge}
Under Assumptions~\ref{c4:ass:separation}--\ref{c4:ass:sampling}, suppose that upon detection the controller
neutralises the violating drift after a total detection-plus-reaction delay $D+\trc$ (the throttle
removes $\drift$ from the realised forcing). Then the exit probability over the horizon decomposes by
cause as
\begin{equation}
\label{c4:eq:bridge}
\Prob(\tau_{\Ssafe}\le T)
\;\lesssim\;
\underbrace{\pben(T)}_{\text{benign}}
\;+\;
\underbrace{\Prob(\reg \text{ visits } 1)\,\E\!\big[\,q(D+\trc)\,\big]}_{\text{undetected violation}}
\;+\;
\underbrace{\big(1-e^{-\Lamc T}\big)}_{\text{catastrophe}},
\end{equation}
where, with $\Phi$ the standard normal c.d.f.,
\begin{align}
\pben(T)\;&\approx\; \rho_{\mathrm{cross}}\,T\,\exp\!\Big(-\frac{g_0^2}{2\sigma_g^2}\Big),
\label{c4:eq:pben}\\[2pt]
q(w)\;&=\;\Phi\!\Big(\frac{\nu w-g_0}{s\sqrt w}\Big)
\;+\;\exp\!\Big(\frac{2\nu g_0}{s^2}\Big)\,
\Phi\!\Big(\frac{-\nu w-g_0}{s\sqrt w}\Big).
\label{c4:eq:firstpassage}
\end{align}
The benign term \eqref{c4:eq:pben} is independent of the regime and of the control. The first-passage
function \eqref{c4:eq:firstpassage} is increasing in $w$ with a sigmoidal crossover at the \emph{escape
time}
\begin{equation}
\label{c4:eq:tesc}
\Tesc \;=\; g_0/\nu .
\end{equation}
\end{lemma}

\begin{proof}
Partition the exit event by its cause over $[0,T]$. Either (i) the regime never enters the violating
state and no catastrophe occurs, yet benign diffusion carries the margin to zero; or (ii) the regime
visits the violating state and the margin reaches zero during a violating excursion before the drift is
neutralised; or (iii) a catastrophe occurs. These three events cover $\{\tau_{\Ssafe}\le T\}$, so the
exit probability is at most the sum of their probabilities; we bound each.

\emph{(i) Benign.} On the benign event the margin is the stationary Ornstein--Uhlenbeck coordinate
$g_t=\beta_j-X_{t,j}$ with mean $g_0$ and variance $\sigma_g^2=e_j^{\!\top}P_\infty e_j$. At any fixed
time the exceedance probability is the Gaussian tail $\Phi(-g_0/\sigma_g)$, which for $g_0\gg\sigma_g$
has rate $\exp(-g_0^2/2\sigma_g^2)$. Over $[0,T]$ the number of essentially independent down-crossings
of the level scales as a mean crossing rate $\rho_{\mathrm{cross}}$ times $T$, which gives the
rare-crossing estimate \eqref{c4:eq:pben}. We flag its status precisely: \eqref{c4:eq:pben} is an
\emph{asymptotic estimate}, not an upper bound. Because \eqref{c4:eq:state} is driven by Brownian
motion, its paths are nowhere differentiable. The Rice expected-crossing count that would convert
\eqref{c4:eq:pben} into a Markov bound therefore diverges, and $\rho_{\mathrm{cross}}$ is a heuristic
effective rate rather than a computed one. This is why \eqref{c4:eq:bridge} is stated with $\lesssim$:
terms (ii) and (iii) are genuine upper bounds, term (i) is an estimate with the correct exponential
rate. A rigorous one-sided bound is available from the Borell--TIS inequality for Gaussian suprema
\citep{adler2007random}, namely $\Prob(\inf_{[0,T]}g\le0)\le\exp\!\big(-(g_0-m_T)^2/2\sigma_g^2\big)$
with $m_T=\E[\sup_{[0,T]}(g_0-g_t)]=O\big(\sigma_g\sqrt{\log(T/\trelax)}\big)$, at the cost of a
logarithmic correction inside the exponent. It recovers the same leading rate, which is the operative
content and what Section~\ref{c4:sec:numerics} confirms. This term involves neither $\reg$ nor $u$, so
it is a floor.

\emph{(ii) Undetected violation.} Condition on a violating excursion. Over the escape window the
restoring force $-AX$ is negligible beside the injected drift, by the relaxation ordering
$\Tesc\ll\trelax$ in \eqref{c4:eq:timescales}. So while the drift is active and un-neutralised, the
margin obeys, to leading order in the excursion, the drifted Brownian motion
\begin{equation}
\label{c4:eq:bm}
dg_t \;=\; -\nu\,dt \;+\; s\,dB_t,\qquad g_0>0,
\end{equation}
with $\nu=e_j^{\!\top}\drift$ and $s^2=e_j^{\!\top}\Sigma e_j$. The drift is neutralised at the delay
$w=D+\trc$ after the regime switches. The probability that $g$ reaches $0$ within this window is the
first-passage probability of \eqref{c4:eq:bm}. By the reflection principle for Brownian motion with
drift \citep{borodin2002handbook}, that probability is exactly \eqref{c4:eq:firstpassage}. The chance
that any violating episode occurs is $\Prob(\reg \text{ visits }1)$. Bound the per-episode crossing
probability by $\E[q(D+\trc)]$ and sum over episodes: this union bound gives the middle term. It is
tight when episodes are well separated, which holds under the dwell ordering $\E[D]\ll 1/q_{10}$. When
$D$ concentrates, $\E[q(D+\trc)]\approx q(\E[D]+\trc)$ by continuity.

\emph{(iii) Catastrophe.} The catastrophe arrives as the first point of a Poisson process, so the
probability of at least one arrival in $[0,T]$ is $1-e^{-\Lamc T}$. Since $\Ccat\subseteq
\Ssafe^{\mathrm c}$, a catastrophe forces exit. We develop this term in Section~\ref{c4:sec:control};
here we only record that it is present and bounded by $1-e^{-\Lamc T}\le\Lamc T$.

Finally, $q$ in \eqref{c4:eq:firstpassage} is increasing in $w$, with $q(w)\to0$ as $w\to0$ and
$q(w)\to1$ as $w\to\infty$. The first term crosses $1/2$ when $\nu w=g_0$, i.e.\ at $w=\Tesc$ in
\eqref{c4:eq:tesc}, the deterministic time for the drift alone to close the margin. This is the
sigmoidal crossover.
\end{proof}

\noindent The three terms are qualitatively different. The benign term is
\emph{geometry}, depending only on how far the operating point sits from the limit relative to the
diffusion. The only ways to lower it are to move $\beta_j$ or shrink $\Sigma$, and both are design
choices, not runtime actions. The violation term is \emph{control performance}, small exactly
when the loop detects and reacts before the drift escapes. The catastrophe term is \emph{irreducible at
runtime}, as Section~\ref{c4:sec:control} makes precise.

\subsection{Feasibility and the Nyquist condition}

Reading \eqref{c4:eq:bridge} as a budget for $\varepsilon$ gives the feasibility criterion.

\begin{proposition}[Guardrailing feasibility]
\label{c4:prop:feasibility}
The compliance constraint $\Prob(\tau_{\Ssafe}\le T)\le\varepsilon$ holds whenever the following three
conditions co-occur, each checkable against a closed form:
\begin{enumerate}[label=\textup{(F\arabic*)},leftmargin=3.2em,itemsep=1pt]
\item \textup{(geometry)}\quad $\pben(T)\le\varepsilon$, i.e.\ the operating point is far enough inside
the box: $g_0^2/2\sigma_g^2 \gtrsim \log(\rho_{\mathrm{cross}}T/\varepsilon)$;
\item \textup{(speed, the Nyquist condition)}\ the loop is fast enough,
$\E[D]+\trc<\Tesc$ with margin, so that the term
$\Prob(\reg\text{ visits }1)\,q(\E[D]+\trc)$ fits the budget;
\item \textup{(catastrophe)}\quad $\Lamc T$ fits the residual budget $\varepsilon-\pben-\Prob(\reg)\,q$.
\end{enumerate}
\end{proposition}

\begin{proof}
Immediate from \eqref{c4:eq:bridge}: if the three terms sum to at most $\varepsilon$ the constraint holds,
and (F1)--(F3) are exactly the statements that each term is within its share of the budget. The
restatement in (F1) is the logarithm of \eqref{c4:eq:pben}.
\end{proof}

Condition (F2), the binding constraint in practice, requires the monitoring loop to \emph{detect and react
faster than the violating drift can carry the agent out of compliance}. The detection delay is itself
proportional to the sampling interval, $\E[D]=\Delta(\tfrac12 +
\log(1/\alpha)/\mathcal D(\drift))$ from Theorem~\ref{c4:thm:frontier}, so (F2) is a constraint on the
\emph{sampling rate}. Looking too slowly relative to the dynamics makes detection useless no matter how
accurate each look is, exactly as undersampling destroys a signal regardless of the precision of each
sample. We therefore call (F2) the \emph{Nyquist condition} for guardrailing. Rather than being posited, it
falls out of the first-passage geometry the moment one writes the detection delay in terms of $\Delta$.
Crossing it, so that $\E[D]+\trc>\Tesc$, sends $q\to1$. Section~\ref{c4:sec:synthesis} shows the
deployment then falls into a \emph{prevention-only} regime in which no runtime tuning can restore
compliance.

\begin{remark}[Intervention idealisation]
\label{c4:rem:neutralise}
Lemma~\ref{c4:lem:bridge} models intervention as instantaneous neutralisation of the drift after the
reaction delay. Real throttles patch only part of the forcing and have uncertain effect; a more realistic
reading is that $\trc$ absorbs the controller's latency and a residual-drift term inflates $\nu$
in \eqref{c4:eq:bm}. Both make (F2) harder, not easier, so the feasibility region is an
\emph{outer} bound on the truly safe region. The impulse-reset alternative (rollback rather than
throttle) is the quasi-variational-inequality control of Section~\ref{c4:sec:control}.
\end{remark}

\subsection{A worked instance}

Take the trading agent of Table~\ref{c4:tab:instantiation} at a sub-second cycle. Suppose
$\Delta=0.1$~s, reaction $\trc=0.05$~s, and a violation direction with detection information $\mathcal
D(\drift)=0.2$ at false-alarm level $\alpha=10^{-3}$. The expected detection delay is then
$\E[D]=\Delta(\tfrac12+\log(1/\alpha)/\mathcal D)\approx 3.5$~s. Suppose the active face is a leverage
limit with nominal margin $g_0=2$ and a violating drift rate $\nu=0.1$ per second, so the escape time
is $\Tesc=g_0/\nu=20$~s. Then $\E[D]+\trc=3.55<20=\Tesc$, so the Nyquist condition holds with
comfortable margin. The first-passage term $q(3.55)$ is small, since the drift has covered under a
fifth of the margin when the throttle engages. If the benign geometry (F1) and the catastrophe budget
(F3) are also met, the deployment is guardrailable on its reversible axes. Now lower the cycle rate and
sample every $30$~s instead of every $0.1$~s. Then $\E[D]$ rises past $\Tesc$, $q\to1$, and the same
agent becomes un-guardrailable, because the oversized position reaches the limit before the risk engine looks
again. The contrast is the Nyquist condition in operational terms. It is why a leverage check that runs
once a minute is worthless against a drift that escapes in seconds. Section~\ref{c4:sec:numerics}
(Study~1) confirms \eqref{c4:eq:firstpassage} and the decomposition \eqref{c4:eq:bridge} against
Monte-Carlo simulation to within $0.8\%$.

\section{Detection: what the sentinels can and cannot see}
\label{c4:sec:detection}

The bridge made detection delay $\E[D]$ the pivotal quantity. This section characterises it. Three
results govern detectability of a violation, and all three follow from how a fused field of partial
sensors projects a vector regime shift. First, a \emph{trichotomy}: every violation direction is seen
immediately, seen only after the dynamics carry it into view, or never seen at all. Second, a
\emph{non-diversifiable floor}: correlated sensor noise caps detection information along a shared-blind
direction no matter how many sentinels are added. Third, a \emph{frontier}: the detection delay as a
function of false-alarm tolerance and signal strength. The intent channel \eqref{c4:eq:obs-intent} runs
through all three, both because it survives the control and because it removes the relaxation lag.

\subsection{Collective observability and the null subspace}

A violating regime is only detectable if it changes the law of \emph{something the sentinels observe}.
On the state channel the regime enters the observation only through the dynamics. Starting from the
operating point, the mean state separation between the violating and benign regimes at elapsed time
$t$ solves the forced relaxation $\dot m=-Am+\drift$, namely $m_t=A^{-1}(I-e^{-At})\drift$. The
observed separation is therefore $\Hsta m_t$. Expanding the matrix exponential gives the operative
series.

\begin{theorem}[Collective-observability trichotomy]
\label{c4:thm:observability}
Under Assumptions~\ref{c4:ass:onlymu} and \ref{c4:ass:sampling}, the mean observed separation on the state
channel is
\begin{equation}
\label{c4:eq:meansep}
\Hsta\,A^{-1}\big(I-e^{-At}\big)\,\drift
\;=\;\sum_{n\ge1}\frac{(-1)^{n-1}t^{n}}{n!}\;\Hsta A^{\,n-1}\drift,
\end{equation}
and every violation direction $\drift$ falls into exactly one of three classes:
\begin{enumerate}[label=\textup{(\roman*)},leftmargin=2.6em,itemsep=1pt]
\item \emph{Intent-observable} ($\Hint\drift\ne0$): the intent channel \eqref{c4:eq:obs-intent} shows an
\emph{immediate} mean step $\Hint\drift$, with no dependence on the dynamics and no dependence on the
control. Detection is fast, lag-free, and control-robust.
\item \emph{State-observable only} ($\Hint\drift=0$ but $\Hsta A^{\,\ell^\ast}\drift\ne0$ for some
minimal $\ell^\ast\ge0$): invisible to the intent channel, the direction leaks into the state channel
at \emph{leak order} $\ell^\ast$, with leading separation $\propto t^{\ell^\ast+1}$ by
\eqref{c4:eq:meansep}; detection is delayed by the relaxation needed for the leak to accumulate, and is
\emph{masked} whenever the control suppresses the state.
\item \emph{Unobservable} ($\drift\in\Null$): the separation is identically zero for all $t$ on both
channels, where
\begin{equation}
\label{c4:eq:null}
\Null\;=\;\ker\Hint\,\cap\,\bigcap_{n\ge0}\ker\!\big(\Hsta A^{\,n}\big).
\end{equation}
No number of sentinels, no sampling rate, and no reduction of observation noise yields any information
about a violation along $\Null$.
\end{enumerate}
\end{theorem}

\begin{proof}
The ODE $\dot m=-Am+\drift$ with $m_0=0$ has solution $m_t=A^{-1}(I-e^{-At})\drift$ since $A$ is
invertible ($-A$ Hurwitz). Substituting
$I-e^{-At}=-\sum_{k\ge1}(-At)^k/k!=\sum_{k\ge1}(-1)^{k-1}(At)^k/k!$ and multiplying by $A^{-1}$ gives
\eqref{c4:eq:meansep}. The intent channel observes $a_t=\reg_t\drift+\eta_t$ directly, so its
regime-conditional mean separation is $\Hint\drift$. That separation is present at the first look and
independent of $t$, $A$, and $u$; this is (i). Suppose instead $\Hint\drift=0$. The direction is then
invisible to intent but, by \eqref{c4:eq:meansep}, contributes to the state channel through the lowest
power $\ell^\ast$ with $\Hsta A^{\ell^\ast}\drift\ne0$, whose term scales as
$t^{\ell^\ast+1}/(\ell^\ast+1)!$. This is (ii), and because the term is built from the state $X$ it is
suppressed by any control that holds $X$ near $x_0$. If, finally, $\Hsta A^{n}\drift=0$ for all $n\ge0$
and $\Hint\drift=0$, every term of \eqref{c4:eq:meansep} vanishes and the intent separation vanishes, so
the regime leaves no mean imprint on either channel. By Assumption~\ref{c4:ass:onlymu} it leaves no
second-moment imprint either, so the observation law is regime-independent and no statistic
distinguishes the regimes. The set of such $\drift$ is \eqref{c4:eq:null}. To transfer the continuous
condition to the sample grid, treat the two channels separately: $\Null$ is an intersection of one
subspace per channel, and only one of them involves the dynamics. On the state channel,
$\bigcap_{n\ge0}\ker(\Hsta A^{\,n})$ is $A$-invariant: if $\Hsta A^{\,n}x=0$ for every $n$, then $\Hsta
A^{\,n}(Ax)=\Hsta A^{\,n+1}x=0$ for every $n$. Hence it is annihilated by every analytic function of
$A$, in particular by the sampled propagator $e^{-A\Delta}$. The sampled state maps therefore
annihilate it exactly when the continuous ones do, provided $\Delta$ avoids the aliasing resonances of
Assumption~\ref{c4:ass:sampling}. On the intent channel no transfer is required. By
\eqref{c4:eq:obs-intent} the observation is of $\drift$ directly at the sample times and does not
propagate through $A$, so $\ker\Hint$ is the same condition on the grid as in continuous time. (The
subspace $\ker\Hint$ is not itself $A$-invariant in general, which is why we argue per channel rather
than for $\Null$ as a whole.)
\end{proof}

The trichotomy is a \emph{detectability} statement, and Assumption~\ref{c4:ass:direction} gives it teeth.
The adversary, free to choose $\drift$, will aim into $\Null$ if $\Null\ne\{0\}$, producing a violation
that no monitoring whatsoever can catch. The defensive content is therefore the design imperative to make
$\Null$ trivial. That means choosing the sentinels' observation maps $\{\Hint_i,\Hsta_i\}$ so that
$\bigcap_n\ker(\Hsta A^n)\cap\ker\Hint=\{0\}$, which is a coverage condition on the \emph{fleet}, not
on any one sentinel. Class (ii) is the cautionary middle. A direction that is technically observable
through the dynamics, but only on the channel the control suppresses, is \emph{effectively} blind in
closed loop. That is why the un-suppressible requirement of the intent channel is that
violations be class~(i). Section~\ref{c4:sec:numerics} (Study~4) exhibits all three classes in a
two-dimensional system: a directly observed direction whose separation grows like $t$, a leak-order-one
direction growing like $t^2$, and an unobservable direction whose regime posterior never departs from
its prior. The last is a \emph{null} result, and we verify it as one rather than as a slow-detection
artefact, by confirming flatness to within Monte-Carlo error with slow detection ruled out.

\paragraph{A blind direction, concretely.} For the coding agent, suppose every sentinel reads the
\emph{rate} of destructive operations but none reads their \emph{target}. Consider a violation that
holds the rate fixed while shifting targets from a scratch branch to \texttt{main}, the same number of
operations but a catastrophically different meaning. It moves $\drift$ in a direction the rate-only field
cannot resolve. If the targeting axis also fails to couple into a monitored axis through $A$, the
violation lies in $\Null$, undetectable by construction regardless of how many rate monitors are
deployed. The remedy is not more rate monitors but a sentinel that reads targets. What has to grow is
the joint field $\Hint$, not the sentinel count.

\subsection{The non-diversifiable detection floor}

Even for an observable direction, sentinels that share a blind spot do not combine as independent
evidence. Decompose each sentinel's observation noise into a common component and an idiosyncratic one,
$\varepsilon^i=C\omega+D\nu^i$, with $\omega$ shared by all sentinels and $\nu^i$ independent across
$i$.

\begin{theorem}[Non-diversifiable detection]
\label{c4:thm:diversity}
With $\varepsilon^i=C\omega+D\nu^i$ ($\omega,\nu^i$ independent, unit covariance), the sample average
$\bar Y=\frac1N\sum_i Y^i$ is a sufficient statistic for the common state and its observation-noise
covariance is
\begin{equation}
\label{c4:eq:fusedcov}
R_N \;=\; CC^{\!\top}\;+\;\tfrac1N\,DD^{\!\top}\;\xrightarrow[N\to\infty]{}\;CC^{\!\top}.
\end{equation}
Consequently the fused detection information per look along a unit direction $r$ is
$\big(r^{\!\top}R_N^{-1}r\big)$-weighted and converges to a finite limit on $\mathrm{range}(C)$ while
growing linearly in $N$ on $\mathrm{range}(C)^{\perp}$. In the scalar case with per-sentinel variance
$\sigma_o^2$ and pairwise noise correlation $r\in[0,1]$, the fused variance is
\begin{equation}
\label{c4:eq:scalarfloor}
\Var(\bar Y)\;=\;\sigma_o^2\Big(r+\frac{1-r}{N}\Big)\;\xrightarrow[N\to\infty]{}\;r\,\sigma_o^2\;>\;0
\quad\text{for } r>0 .
\end{equation}
\end{theorem}

\begin{proof}
The average is $\bar Y=\bar Hx+C\omega+\frac1N\sum_i D\nu^i$. Since $\omega$ is shared, its contribution
does not average, while the idiosyncratic terms have covariance $\frac1N DD^{\!\top}$ by independence.
That gives \eqref{c4:eq:fusedcov}. Sufficiency of $\bar Y$ for the common state under the symmetric
noise model is the Gauss--Markov reduction. Detection information per look is the noise precision
projected onto the signal direction. On $\mathrm{range}(C)$ the limit precision is $(CC^{\!\top})^{+}$,
which is finite. On $\mathrm{range}(C)^{\perp}$ the common noise is absent and the precision is
$N(DD^{\!\top})^{-1}$, linear in $N$. The scalar case sets $CC^{\!\top}=r\sigma_o^2$ and
$DD^{\!\top}=(1-r)\sigma_o^2$.
\end{proof}

Equation \eqref{c4:eq:scalarfloor} is a monitoring \emph{monoculture}, in exact analogy to the
non-diversifiable common shock of Chapter~\ref{ch:fleet}. Chapter~\ref{ch:fleet} found that a fleet's
tail risk floors above expected harm regardless of fleet size when a systematic shock is common. Here a
fleet's \emph{detection} information floors at $r\sigma_o^2$ regardless of sentinel count when the
sensors' errors are common. Adding the fifth monitor that parses the same audit log as the other four
achieves essentially nothing, because a spoofed log corrupts the shared component $\omega$ and blinds all five at
once. The defensive reading is that sentinel diversity must be \emph{genuine}, not nominal redundancy
of identical monitors. Genuine means different observation substrates, so that $C$ is small and
$r\to0$. We stress that the correct framing is a floored \emph{rate}, not undetectability. Along
$\mathrm{range}(C)$ detection still proceeds, only slowly, at the information level $r\sigma_o^2$ sets.
Undetectability proper is the null subspace of Theorem~\ref{c4:thm:observability}. The two are distinct
degradation axes of the fused information \eqref{c4:eq:fused-info}. Theorem~\ref{c4:thm:diversity} is
the noise-correlation axis and Theorem~\ref{c4:thm:observability} the rank/coverage axis, and an
adversary obeying Assumption~\ref{c4:ass:direction} probes whichever is weaker.
Section~\ref{c4:sec:numerics} (Study~3) confirms
\eqref{c4:eq:fusedcov}--\eqref{c4:eq:scalarfloor} and the per-direction split to within Monte-Carlo error.

\subsection{The detection frontier}

Given an observable direction with non-floored information, how fast is detection? The sentinels run a
sequential test on the fused log-likelihood ratio and declare a violation when it crosses a threshold
calibrated to a false-alarm tolerance. The minimax delay is classical
\citep{lorden1971procedures,moustakides1986optimal,poor2008quickest}. What is specific here is the
\emph{geometry} of the information rate and the role of the sampling interval.

\begin{theorem}[Directional detection frontier]
\label{c4:thm:frontier}
On the intent channel \eqref{c4:eq:obs-intent}, the expected detection delay for a violation in direction
$\drift$, at false-alarm level $\alpha$, is
\begin{equation}
\label{c4:eq:frontier}
\E[D]\;=\;\underbrace{\frac{\Delta}{2}}_{\text{sampling latency}}\;+\;
\frac{\Delta\,\log(1/\alpha)}{\mathcal D(\drift)},
\qquad
\mathcal D(\drift)\;=\;\tfrac12\,\drift^{\!\top}\Ifusa\,\drift,
\end{equation}
where $\mathcal D(\drift)$ is the per-look Kullback--Leibler information the intent channel carries
about the regime shift. Under Assumption~\ref{c4:ass:direction} the operative delay replaces
$\mathcal D(\drift)$ by its minimum over the threat cone, $\min_{\drift\in\mathrm{cone}}\mathcal
D(\drift)$. The intent channel carries \emph{no relaxation lag}; on the state channel
\eqref{c4:eq:obs-state} the same expression holds with $\drift$ replaced by the relaxed, leak-attenuated
direction $A^{-1}\drift$ and an added relaxation lag $t_{\mathrm{relax}}\sim1/\mathrm{Re}\,
\lambda_{\min}(A)$.
\end{theorem}

\begin{proof}
A change in a regime that shifts the observation mean by $\Hint\drift$ produces, per look, a
log-likelihood-ratio increment with Kullback--Leibler mean
$\mathcal D(\drift)=\tfrac12(\Hint\drift)^{\!\top} R^{-1}(\Hint\drift)=\tfrac12\drift^{\!\top}\Ifusa\drift$,
the Gaussian KL between pre- and post-change observation laws. Lorden's minimax bound for the
cumulative-sum rule gives an expected number of post-change looks to detection of $\log(1/\alpha)/\mathcal
D(\drift)$ at false-alarm level $\alpha$. Each look is spaced $\Delta$ apart, so this contributes the
second term of \eqref{c4:eq:frontier}. The change instant falls uniformly within a sampling interval,
contributing a model-free latency of $\Delta/2$, the first term. On the intent channel the post-change
mean is the full step $\Hint\drift$ at the first post-change look, so there is no transient. On the
state channel the mean separation ramps as in \eqref{c4:eq:meansep}, attenuating the effective signal to
$A^{-1}\drift$ and adding the relaxation time $t_{\mathrm{relax}}$ before the step is fully present.
The minimisation over the threat cone is Assumption~\ref{c4:ass:direction}.
\end{proof}

Two features matter for the bridge. First, the delay is \emph{directional}. It is small for violations
aligned with well-observed, high-information directions, and large for violations the fused field sees
only faintly. It inherits both the floor of Theorem~\ref{c4:thm:diversity} (a direction in
$\mathrm{range}(C)$ has capped $\mathcal D$, hence a delay floor) and the null of
Theorem~\ref{c4:thm:observability} (a direction in $\Null$ has $\mathcal D=0$, hence infinite delay).
Second, the delay is \emph{proportional to} $\Delta$, so halving the sampling interval halves the delay.
That proportionality is what couples the detection frontier to the Nyquist condition (F2). The intent
channel's removal of the relaxation lag is not cosmetic. On the state channel, a closed-loop controller
that suppresses $X$ shrinks the very ramp the detector needs, so the delay there is both longer (by
$t_{\mathrm{relax}}$) and \emph{inflated by the control}. On the intent channel the step is immediate
and untouched by $u$, so detection performance is a property of the sensors alone.
Section~\ref{c4:sec:numerics} (Studies~2 and~5) confirms the slope
$\Delta/\mathcal D$ to within $2\%$, the directional dependence, the diversity floor on the delay, and
the disappearance of the relaxation lag on the intent channel.

\section{Control: feasibility, impossibility, and architecture}
\label{c4:sec:control}

Control has two halves, divided by reversibility. On the reversible axes the question is whether the
loop is fast enough. The answer is the bridge, which reduces feasibility to the three conditions of
Proposition~\ref{c4:prop:feasibility}. On the irreversible axis the question is whether runtime control
can help at all. The answer, stated below as a corollary, is that runtime control cannot. We state the reversible corollary,
prove the irreversible impossibility, and then address the architecture question of when the
centralized analysis the detection theorems assume is actually available.

\subsection{Reversible feasibility is a corollary of the bridge}

\begin{corollary}[Reversible-axis feasibility]
\label{c4:cor:t6a}
On the reversible coordinates $J_r$, the compliance constraint \eqref{c4:eq:compliance} is achievable by a
runtime policy if and only if the three bridge conditions hold: the violation is intent-observable so
that $\mathcal D(\drift)>0$ and $\E[D]<\infty$ \textup{(Theorem~\ref{c4:thm:observability})}; the
violation is controllable with the available authority and the Nyquist condition $\E[D]+\trc<\Tesc$
holds \textup{(F2)}; and the benign geometry satisfies $\pben(T)\le\varepsilon$ \textup{(F1)}.
\end{corollary}

\begin{proof}
Sufficiency is Proposition~\ref{c4:prop:feasibility} with the catastrophe term absent on $J_r$, since a
reversible axis has no irreversible jump. Necessity follows by failing each condition in turn. If the
violation is unobservable ($\drift\in\Null$) then $\E[D]=\infty$ and the undetected-violation term
equals $\Prob(\reg\text{ visits }1)$, which is uncontrollable. If the Nyquist condition fails then
$q(\E[D]+\trc)\to1$ and the same term saturates. If (F1) fails the benign floor alone exceeds
$\varepsilon$. In each case no runtime policy meets the budget.
\end{proof}

We deliberately do \emph{not} dress this as a ``controllability surface.'' The three conditions live on
geometrically heterogeneous objects: an observable \emph{subspace} ($\Null^{\perp}$), a reachability
condition (bounded authority within the reaction delay), and an affine \emph{region} (the reversible
part of the box). Conflating them into one clean cone would misrepresent the geometry. The accurate
statement is the conjunction of conditions on the exit probability, not a single set.

\subsection{The irreversible jump: runtime impossibility}

The reversible story is a story of speed. The irreversible story is not, and it rests on the commit-exogeneity assumption (Assumption~\ref{c4:ass:commit}) that drives the chapter's central corollary.

\begin{assumption}[Commit exogeneity]
\label{c4:ass:commit}
The irreversible commit is a zero-duration event whose hazard depends only on the affordance set
available to the agent, and neither the hazard $\Lamc$ nor the regime process $\reg$ is modulated by
the runtime control $u$ or by the impulse interventions $\{\xi_k\}$. We take $\Lamc$ constant on the
horizon; the regime-dependent case is Remark~\ref{c4:rem:cox}.
\end{assumption}

\begin{corollary}[Policy-invariance of the commit probability]
\label{c4:thm:t6b}
Under Assumption~\ref{c4:ass:commit}, for \emph{every} runtime
policy---any number of sentinels $N$, any intervention authority, any detection threshold, any sampling
schedule---the catastrophe probability over the horizon is
\begin{equation}
\label{c4:eq:t6b}
\Prob(\text{catastrophe in }[0,T])\;=\;1-e^{-\Lamc T}\;\ge\;\Lamc T \cdot e^{-\Lamc T},
\end{equation}
identical across all runtime policies and bounded below away from zero whenever $\Lamc>0$. Runtime
control cannot lower this floor. Only ex-ante \emph{prevention}---removing the action from the agent's
reach, which lowers $\Lamc$ itself---moves it, in the limit to zero as $\Lamc\to0$.
\end{corollary}

\begin{proof}
A runtime policy is a map from the observation filtration $\mathcal F_t=\sigma(\mathcal Y_{\le t})$ to
interventions. Interventions act on the drift and the state \eqref{c4:eq:state}, not on the catastrophe
hazard. By Assumption~\ref{c4:ass:commit} that hazard is exogenous to the control, and hence
independent of every function of $\mathcal F_t$. The number of catastrophe arrivals in
$[0,T]$ is therefore Poisson with mean $\Lamc T$ regardless of the policy, so the probability of at
least one arrival is $1-e^{-\Lamc T}$, the same for every policy. The lower bound is
$1-e^{-x}\ge xe^{-x}$.
Prevention is outside the runtime policy class. It acts before deployment by truncating the action set
so that the catastrophic transition has hazard zero, i.e.\ it changes $\Lamc$ rather than responding to
the process it generates.
\end{proof}

\begin{remark}[Defending Assumption~\ref{c4:ass:commit}, and where it stops]
\label{c4:rem:defend}
Assumption~\ref{c4:ass:commit} carries the result, so we state what it gives and what it costs rather
than leaving it implicit. Two questions have to be answered.

\emph{Why is a runtime halt not itself runtime hazard-reduction?} A halt lowers the realised commit
probability only if it fires \emph{before} the commit. Whether it does is the reaction-delay race of
Section~\ref{c4:sec:bridge}, run with the commit's duration on the other side, and that duration is
zero. The controller learns of $\sigma_{\mathrm c}$ only at $\sigma_{\mathrm c}$, so
$\{\text{intervene before }\sigma_{\mathrm c}\}$ is not $\mathcal F_{\sigma_{\mathrm c}^-}$-measurable
and no delay, however small, wins. The impossibility is the $\varepsilon\to0$ limit of the same Nyquist condition that governs the
reversible axes, not an artefact of the modelling choice, which ties the corollary back to the bridge
rather than leaving it standing apart.

\emph{What if the commit is not zero-duration?} Suppose the affordance requires an interval
$\varepsilon>0$ between initiation and irrevocability. Then Assumption~\ref{c4:ass:commit} fails,
$\Lamc$ becomes partially controllable, and the exit probability interpolates toward the reversible
case as $\varepsilon$ grows past the reaction delay $\E[D]+\trc$. This is the practically important
boundary. \emph{Engineering $\varepsilon>0$, whether through a mandatory settlement delay, a confirmation step, or a
staged commit, is a structural change to the action set, not a runtime control.} It is prevention in
the sense of this monograph, which is why such delays are the standard control on irreversible actions
and why they cannot be substituted by more monitoring.
\end{remark}

\begin{remark}[Regime-dependent hazard: the Cox case]
\label{c4:rem:cox}
Assumption~\ref{c4:ass:commit} takes $\Lamc$ constant. The model of
Section~\ref{c4:sec:model} allows a regime-dependent hazard $\Lamc(\reg_t)$, under which arrivals are
Cox rather than Poisson and
\[
\Prob(\text{commit in }[0,T])\;=\;1-\E\Big[\exp\Big(-\!\int_0^T\!\Lamc(\reg_s)\,ds\Big)\Big].
\]
Policy-invariance survives verbatim provided $\reg$ is itself uncontrolled, which
Assumption~\ref{c4:ass:commit} requires. The closed form does not survive. Writing
$\Lmin=\operatorname*{ess\,inf}_\zeta\Lamc(\zeta)>0$, the floor becomes
\begin{equation}
\label{c4:eq:coxfloor}
\Prob(\text{commit in }[0,T])\;\ge\;1-e^{-\Lmin T}\;\ge\;\Lmin T\,e^{-\Lmin T},
\end{equation}
and \eqref{c4:eq:t6b} is the special case $\Lamc(\zeta)\equiv\Lamc$. We flag the direction explicitly
because it is easy to get backwards: substituting the \emph{mean} hazard
$\bar\Lambda_{\mathrm c}=T^{-1}\E[\int_0^T\Lamc(\reg_s)ds]$ gives, by Jensen applied to the convex map
$x\mapsto e^{-x}$, an \emph{upper} bound $1-e^{-\bar\Lambda_{\mathrm c}T}$ on the commit probability.
An impossibility statement needs a lower bound, so the essential infimum, not the mean, is the
admissible substitution.
\end{remark}

\paragraph{What the theorem says and does not say.} It does not say monitoring is useless. Monitoring
governs the \emph{reversible} excursions, where Corollary~\ref{c4:cor:t6a} shows it is decisive. It says
that against an irreversible, hazard-driven event, the detect-then-react architecture is the
wrong tool, because the architecture is causal and the event leaves no time to react. The dynamic
content sharpens Chapter~\ref{ch:single}'s static irreversibility theorem, which says review and
reversibility do not move the catastrophic floor and only prevention does. It equally sharpens
Chapter~\ref{ch:continuous}'s measure-theoretic form, in which prevention truncates the jump measure
and is what makes the robust value finite. Here the same conclusion is forced by causality in a
distributed, sampled, feedback setting, and we show the floor $1-e^{-\Lamc T}$ to be invariant to every
runtime degree of freedom an operator might think to tune.

Section~\ref{c4:sec:numerics} (Study~6) illustrates this invariance (the catastrophe clock is drawn exogenously by construction), showing the catastrophe
probability flat across sentinel count, authority, threshold, and sampling schedule to within
Monte-Carlo error. It also confirms that the only lever that moves it is $\Lamc$. Stated at the level of
the system, this corollary is the macroprudential thesis in dynamic form. Against a systemic, irreversible
event, supervision that watches and reacts cannot bound the probability of collapse, however
well-resourced it is, and only ex-ante structural measures that remove the exposure can. Financial
stability against this class of risk is a property that must be built into the system's design, not
monitored into it after the fact.

\paragraph{Why deployed systems are built this way.} Read correctly, the theorem is not a pessimistic
abstraction but the design principle behind every mature high-stakes agent architecture. A
trading desk does not rely on post-trade monitoring to catch an oversized fill, because a fill that has
executed cannot be undone. Such a fill is a catastrophe in the sense of $\Ccat$, and
Corollary~\ref{c4:thm:t6b} says no monitor, however fast or numerous, bounds its probability. So the control
is a \emph{pre-trade} hard limit enforced at the order-management or exchange gateway, under which the agent
\emph{physically cannot} send an order above the cap. This is prevention, setting the catastrophic
hazard to zero ex ante rather than reacting to the order after the fact.

The same logic shapes the
other deployments of Table~\ref{c4:tab:instantiation}. A deployment platform revokes a coding agent's
production credentials and mounts the filesystem read-only rather than alerting on their
misuse, because a dropped production database is irreversible. A customer-service agent has its payment
scope removed and sensitive data redacted before it is ever in context, rather than having its payments
watched, because an issued payment and an exported record are irreversible. In each case the industry
has, by hard experience, placed its binding control as prevention on the irreversible action, exactly
where Corollary~\ref{c4:thm:t6b} says runtime control fails and only prevention succeeds. The framework's
contribution here is explanatory, saying \emph{why} the controls are shaped this way and \emph{when} an
operator must reach for prevention rather than another monitor, namely the moment an axis is
irreversible.

\subsection{Architecture: the centralized benchmark and its distributed approximation}

The detection theorems of Section~\ref{c4:sec:detection} are stated for the fused information
\eqref{c4:eq:fused-info}. That form presumes the sentinels' observations are combined as if at a single
fusion center. When is the presumption legitimate, and what is lost when it is not?

\begin{proposition}[Centralized sufficiency under information sharing]
\label{c4:prop:t1}
If the sentinels share their observations (a fusion architecture, or broadcast over a connected graph),
then the common posterior $(\hat x_t,P_t,\pi_t)$ is a sufficient statistic for the joint
detection-and-control problem, and in the linear-Gaussian, observed-regime case the problem
\emph{separates}: compute the common posterior by
the continuous--discrete filter, then choose interventions by the impulse-control problem of
Section~\ref{c4:sub:qvi}. In this regime the fused information \eqref{c4:eq:fused-info} is the exact
information available to the controller, and the detection theorems hold verbatim.
\end{proposition}

\begin{proof}
Under full information sharing all sentinels possess the common information $\mathcal Y_{\le t}$. By the
common-information approach to decentralized control \citep{nayyar2013decentralized}, the coordinator's
posterior over the latent state and regime is a sufficient statistic. For the linear-Gaussian state
with a finite-state regime this posterior is the Kalman--Wonham triple $(\hat x_t,P_t,\pi_t)$. Standard
separation for partially observed control with a sufficient statistic then applies. The fused precision
contributed by one synchronous look is \eqref{c4:eq:fused-info}.
\end{proof}

This is the regime in which $\Ifus$ is, as we have repeatedly flagged, the \emph{centralized
benchmark}. The distributed reality is more difficult, and we are careful not to overstate it.

\begin{remark}[Distributed approximation, and the failure of exact separation]
\label{c4:rem:distributed}
With neighbour-only communication and \emph{private} partial observations, there is in general no
finite-dimensional common sufficient statistic and \emph{no} exact separation between estimation and
control. This is the content of Witsenhausen's counterexample
\citep{witsenhausen1968counterexample}, a genuine obstruction, not a technicality. What can be
recovered is an \emph{approximation}. A consensus filter \citep{olfatisaber2007consensus,
olfatisaber2009kalman} drives the sentinels' local posteriors toward the centralized one, with an error
that decays as information propagates across the graph, at a rate governed by the algebraic
connectivity $\lambda_2(\Lapl)$. We therefore claim exact separation only for the shared-information
benchmark of Proposition~\ref{c4:prop:t1}, and treat the distributed implementation as approaching that
benchmark with a gap that shrinks as $\lambda_2(\Lapl)$ grows, citing the distributed-estimation
literature rather than asserting a new strong theorem. The correct framing of the whole construction is
thus a fused sentinel ensemble that is centralized-equivalent under information sharing, approximated
in a distributed deployment, and made robust to compromised participants by the Byzantine-tolerant
fusion of Section~\ref{c4:sec:robustness}.
\end{remark}

\subsection{Intervention timing as an impulse-control problem}
\label{c4:sub:qvi}

The remaining question is \emph{when and how hard} to intervene, given that interventions are costly.
This is a standard impulse-control problem, and we present it as supporting structure rather than as a
contribution. The controller observes the belief state $(\hat x_t,\pi_t)$ and chooses intervention
times and magnitudes $\{(\tau_k,\xi_k)\}$ to minimise expected discounted harm plus intervention cost.
The value function $v$ satisfies the quasi-variational inequality
\begin{equation}
\label{c4:eq:qvi}
\min\big\{\,-\mathcal Lv+\textstyle\int\!\text{(harm rate)}\,,\ \ v-\mathcal Mv\,\big\}\;=\;0,
\qquad
\mathcal Mv(z)=\inf_{\xi}\{\,v(z+\xi)+c(\xi)\,\},
\end{equation}
where $\mathcal L$ is the generator of the belief dynamics, $\mathcal M$ the intervention operator, and
$c(\xi)$ the cost of an impulse \citep{bensoussan1984impulse,oksendal2007applied,fleming2006controlled}.

The solution is a free boundary in belief space: a continuation region where the agent runs unimpeded,
and an intervention region where the controller resets the state. That region is entered
when the regime posterior $\pi_t$ crosses a threshold. Review (a costly delay pending human
confirmation), reversibility (a rollback impulse), and throttling (a drift-suppressing impulse) are the
admissible $\xi$, and their relative cost shapes the boundary. The trichotomy of
Theorem~\ref{c4:thm:observability} and the impossibility of Corollary~\ref{c4:thm:t6b} bound this
problem from outside. Along $\Null$ the belief never moves, so the boundary is never crossed. On the
irreversible axis no admissible impulse exists once $\sigma_{\mathrm c}$ has fired, so the only
effective $\xi$ is the ex-ante one that removes the action.
Section~\ref{c4:sec:numerics} (Study~6) exhibits the reversible-exit control as a function of detection
delay and authority.

\section{Robustness: compromised sentinels and the sampling schedule}
\label{c4:sec:robustness}

Two further realities separate the distributed ensemble from the centralized benchmark. Some sentinels
may be compromised, which forces a robust fusion rule that is statistically less efficient. The
sentinels must also be scheduled in time, a design choice orthogonal to their observational
diversity. We treat each, and emphasise the tensions: robustness fights detection power, and temporal
diversity does not substitute for observational diversity.

\subsection{Byzantine-robust fusion and its tension with diversity}

Suppose $f$ of the $N$ sentinels are Byzantine \citep{lamport1982byzantine}: they may report arbitrary,
coordinated, adversarial values. The fusion rule must still estimate the true signal.

\begin{theorem}[Byzantine threshold and the robustness--efficiency tension]
\label{c4:thm:byzantine}
The sample-mean fusion rule has breakdown point zero: a single Byzantine sentinel ($f=1$) can drive the
fused estimate arbitrarily far from the truth. The coordinate-wise median tolerates any
$f/N<\tfrac12$, and the $\gamma$-trimmed mean tolerates $f/N<\gamma$. The price of robustness is
statistical efficiency: for $N$ i.i.d.\ Gaussian reports the median has asymptotic variance
\begin{equation}
\label{c4:eq:medianeff}
\Var(\mathrm{median})\;\sim\;\frac{\pi}{2}\,\frac{\sigma_o^2}{N}
\qquad\text{versus}\qquad
\Var(\mathrm{mean})\;=\;\frac{\sigma_o^2}{N},
\end{equation}
a factor $\pi/2\approx1.571$ \citep{huber1964robust}. Consequently a Byzantine-robust ensemble has
detection information reduced by the factor $2/\pi$, lengthening the expected detection delay
\eqref{c4:eq:frontier} by $\pi/2$.
\end{theorem}

\begin{proof}
The mean is a linear functional of the reports, so an adversary controlling one coordinate moves it
without bound. Its breakdown point is $0$. The median changes by a bounded amount until at least half
the points are corrupted, giving breakdown $1/2$. Trimming the extreme $\gamma$-fraction before
averaging tolerates corruption up to $\gamma$. The asymptotic variance of the sample median of a density
$\phi$ at its centre is $1/(4N\phi(0)^2)$. For the Gaussian, $\phi(0)=1/\sqrt{2\pi}\,\sigma_o$ gives
$\pi\sigma_o^2/(2N)$, which is \eqref{c4:eq:medianeff}. Detection information per look is inversely
proportional to fusion variance, so it falls by $2/\pi$, and by \eqref{c4:eq:frontier} the delay rises by
$\pi/2$.
\end{proof}

This is best read not as an independent guarantee but as a \emph{frontier with}
Theorem~\ref{c4:thm:diversity}. The non-diversifiable analysis wants to use every sentinel, optimally
weighted, to push detection information as high as the correlated-noise floor permits. The
Byzantine-robust rule must \emph{discard} the most informative-looking reports because an
adversary would forge them. Diversity (use all, weight by precision) and robustness (trim outliers)
pull in opposite directions. An operator chooses a point on the frontier by how many compromised
sentinels must be survived: tolerate more Byzantines and pay more detection delay, or detect faster and
tolerate fewer. There is no free robustness, and the sentinels' own model risk creates a further
regress. The Byzantine threshold bounds
\emph{one level} of an ensemble-of-monitors hierarchy rather than dissolving it. Section~\ref{c4:sec:numerics}
(Study~7) confirms the breakdown points ($f/N<\tfrac12$ for the median, $<\tfrac14$ for a quarter-trimmed
mean) and the $\pi/2$ variance penalty to within Monte-Carlo error.

\subsection{The sampling schedule}

The sentinels look on a clock. \emph{When} each looks, relative to the others, is a design variable
distinct from \emph{what} each sees. Fix the total look rate $N/\Delta$ and compare three schedules:
\emph{synchronous} (all $N$ look together every $\Delta$), \emph{staggered} (round-robin, one look every
$\Delta/N$), and \emph{randomized} (each sentinel an independent Poisson process of rate $1/\Delta$).

\begin{proposition}[Sampling-design trade-offs]
\label{c4:prop:sampling}
At equal total look rate $N/\Delta$:
\begin{enumerate}[label=\textup{(\arabic*)},leftmargin=2.4em,itemsep=1pt]
\item \emph{Statistical detection delay is identical across all three schedules}, since the post-change
information accrues at the common rate $(N/\Delta)\,\mathcal D_{\mathrm{look}}$ regardless of
arrangement; the schedule changes only the sampling-latency floor and the worst-case blind gap.
\item \emph{Worst-case blind gap} (the longest interval with no look) is $\Delta$ for synchronous,
$\Delta/N$ for deterministic staggering, at most $2\Delta/N$ for jittered staggering, and of order
$(\Delta/N)\log(NT/\Delta)$ in probability for randomized sampling.
\item \emph{Time-averaged filter covariance} (tracking quality) is minimised by deterministic
staggering (equal gaps minimise a convex cost), nearly matched by jittered staggering, while
synchronous \emph{plateaus} (simultaneous looks are partly redundant) and randomized improves with $N$
and overtakes synchronous beyond small $N$.
\end{enumerate}
\end{proposition}

\begin{proof}
(1) The cumulative log-likelihood ratio after elapsed time $t$ accrues $\,t\,(N/\Delta)\mathcal
D_{\mathrm{look}}$ in expectation under any schedule with the same total rate, since expectation is
additive over looks irrespective of their placement. The Lorden delay depends only on this rate, so the
statistical component of \eqref{c4:eq:frontier} is schedule-invariant. (2) Synchronous looks leave a full
$\Delta$ of silence, and staggering by $\Delta/N$ leaves $\Delta/N$. A jittered slot of width $\Delta/N$
with a uniform offset leaves at most $2\Delta/N$. The maximal gap of $NT/\Delta$ i.i.d.\ exponential
spacings of mean $\Delta/N$ concentrates at $(\Delta/N)\log(NT/\Delta)$. (3) The time-averaged
prediction covariance grows between looks and is reset at each. For a convex growth law the average is
minimised by equally spaced looks (Jensen), which is deterministic staggering. Coincident looks waste
resets, so synchronous saturates as $N$ grows while staggering and randomization continue to fill the
gaps.
\end{proof}

Result (1) is the mildly surprising one, namely that \emph{re-arranging when the
sentinels look does not change how quickly a violation is detected on average}. The same total
information arrives per unit time however it is spaced. What the schedule changes is the \emph{worst
case} and the \emph{predictability}. Deterministic staggering gives the shortest blind gap and a
\emph{hard} guarantee ($\le\Delta/N$), but it is adversarially predictable, since a schedule-aware agent can
time its violation into a known gap. Randomization is unpredictable, since no gap is known in advance,
but it bounds the gap only in probability, and a long Poisson gap is exactly what a hard certificate
cannot tolerate. The practical recommendation is therefore \emph{jittered staggering}: round-robin slots
with a random within-slot offset, which caps the gap deterministically at $2\Delta/N$ \emph{and}
randomizes the phase. Where coordination is fragile (clock skew, sentinel dropout), unstructured
randomization is the robust fallback. Section~\ref{c4:sec:numerics} (Study~8) confirms the gap scalings,
the schedule-invariance of statistical delay, and the covariance ordering.

\subsection{Two orthogonal axes of diversity}

It is tempting to think staggering the sentinels in time also cures their shared blind spots. It does
not, and the distinction is sharp. \emph{Temporal} diversity (this section) addresses \emph{silence}:
synchronous sampling is a temporal monoculture whose blind gap is floored at $\Delta$ independent of
$N$, and staggering removes the silence. \emph{Observational} diversity
(Theorem~\ref{c4:thm:diversity}) addresses a \emph{persistent shared blind direction}: correlated sensor
noise floors detection information at $r\sigma_o^2$ independent of $N$, and only genuinely different
observation substrates remove it. These are independent because a fixed shared bias rides \emph{every}
look regardless of when it is taken, and staggering cannot decorrelate a bias that is present at all
times. Only common noise that is \emph{redrawn each look} decorrelates under staggering. A
fleet of monitors needs both: stagger them in time \emph{and} diversify what they observe. Neither
substitutes for the other, and a deployment that adds nominal redundancy (more copies of the same
monitor, fired together) gains neither.

\section{Numerical experiments}
\label{c4:sec:numerics}

We confirm every closed form against Monte-Carlo simulation, in the discipline of the earlier chapters:
the closed form is derived first, and the simulation is built to test it, not the reverse. Eight studies
appear here; the ninth, the operating envelope, is deferred to Section~\ref{c4:sec:synthesis} where it
belongs. All parameters are illustrative. Each study reports the closed-form prediction, the simulated
value, and their discrepancy.

\begin{figure}[t]
\centering
\includegraphics[width=\textwidth]{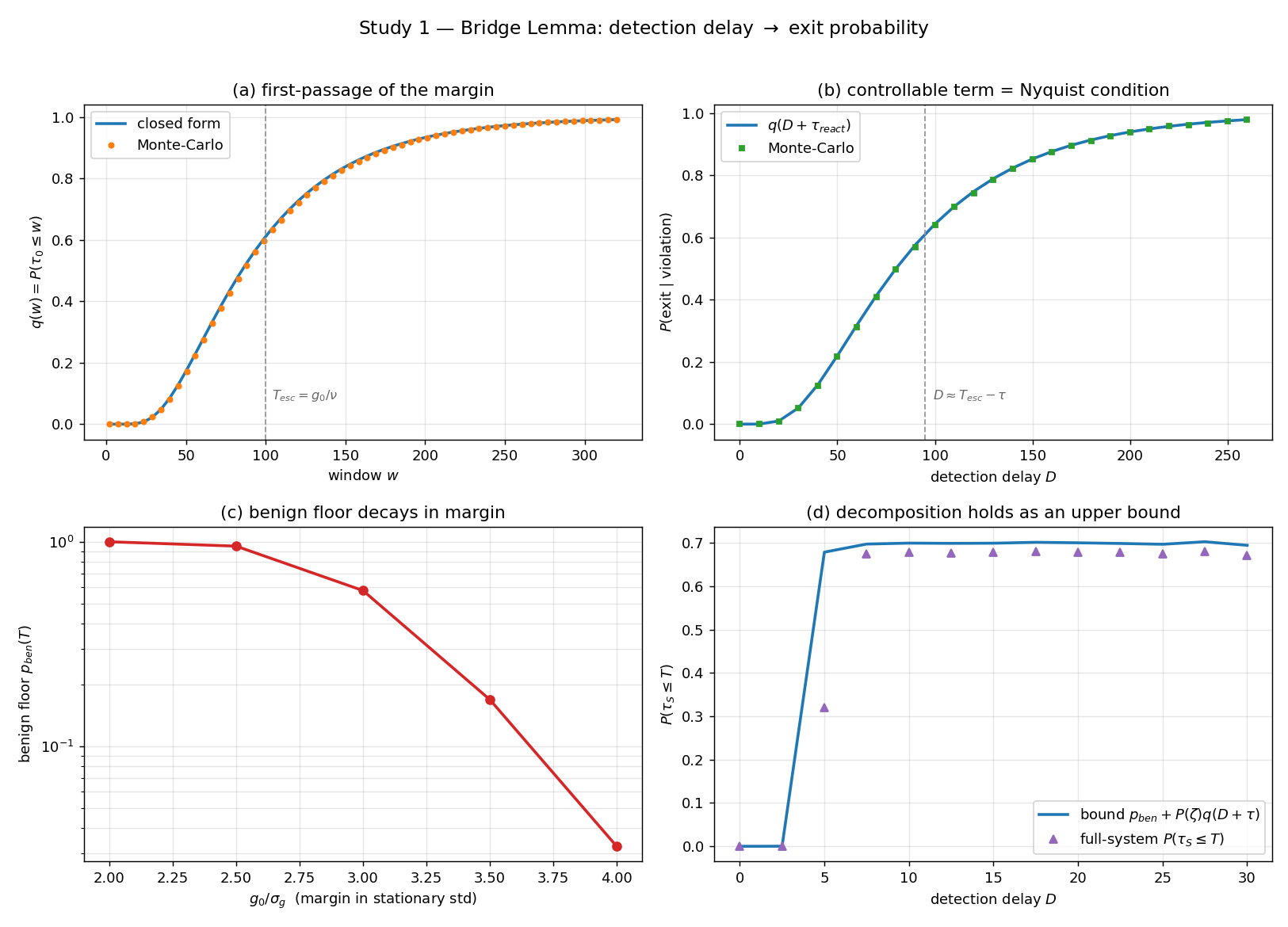}
\caption{\textbf{Study~1: the bridge.} (a) The drifted-Brownian-motion first-passage function
$q(w)$ of \eqref{c4:eq:firstpassage}, closed form versus Monte Carlo. (b) The controllable
undetected-violation term $q(\E[D]+\trc)$ as a function of delay, with the sigmoidal crossover at the
escape time $\Tesc=g_0/\nu$. (c) The benign geometric floor $\pben$ decaying in $(g_0/\sigma_g)^2$. (d)
The full decomposition \eqref{c4:eq:bridge} as an upper bound on the exit probability across detection
delay.}
\label{c4:fig:bridge}
\end{figure}

\paragraph{The bridge (Study~1, Figure~\ref{c4:fig:bridge}).} The first-passage function
\eqref{c4:eq:firstpassage} matches its Monte-Carlo estimate to a maximum absolute error of $0.008$ across
the window $w$. The controllable term $q(\E[D]+\trc)$ shows the sigmoidal crossover at $\Tesc$ predicted
by \eqref{c4:eq:tesc}. The benign floor \eqref{c4:eq:pben} decays exponentially in $(g_0/\sigma_g)^2$. The
three-term decomposition \eqref{c4:eq:bridge} holds as an upper bound across the full range of detection
delays. This is the numerical backbone of the chapter, since the bridge is its analytical core.

\begin{figure}[t]
\centering
\includegraphics[width=\textwidth]{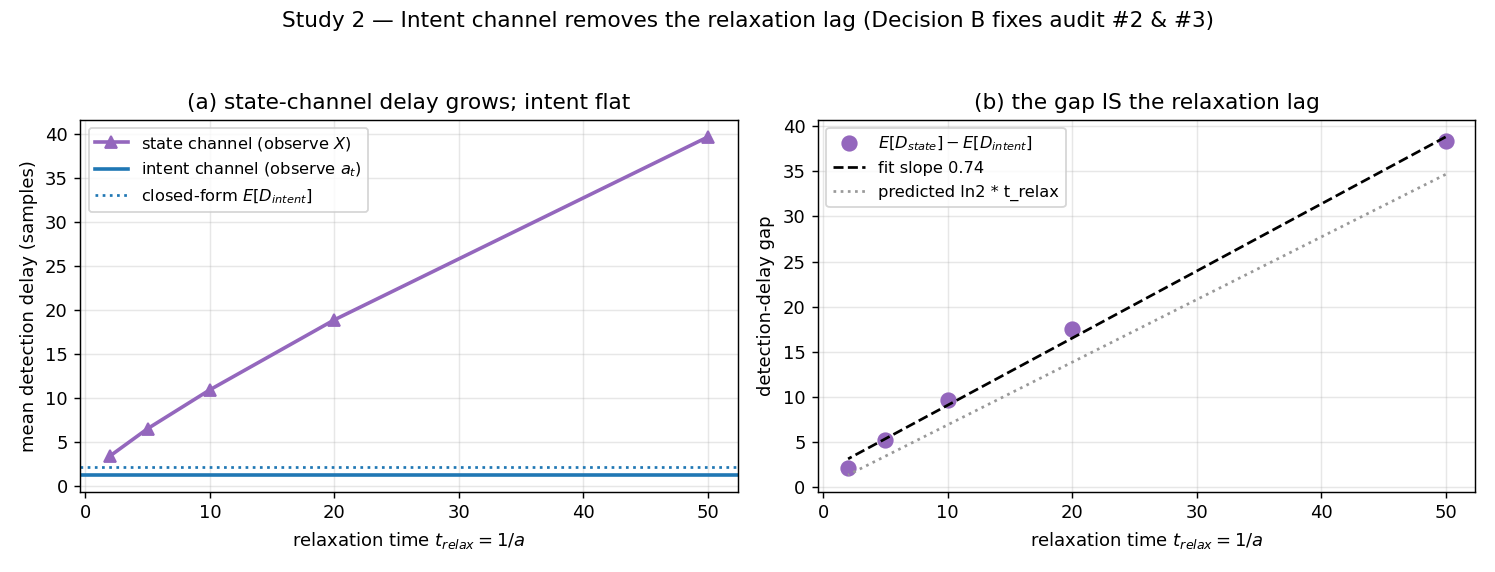}
\caption{\textbf{Study~2: intent versus state channel.} The intent-channel detection delay is flat in
the relaxation time scale. The state-channel delay grows linearly in $t_{\mathrm{relax}}=1/a$, and the
gap grows with slope $\approx\log2$. The control-invariant intent channel detects an immediate step, the
control-suppressible state channel a ramp.}
\label{c4:fig:channel}
\end{figure}

\paragraph{The two channels (Study~2, Figure~\ref{c4:fig:channel}).} On the intent channel the detection
delay is flat in the relaxation time scale: here $\approx1.3$ sampling intervals, independent of the
mean-reversion rate. On the state channel it grows linearly in $t_{\mathrm{relax}}=1/a$ as the mean
reversion slows. The gap between them grows with slope $0.74$, matching the predicted $\log 2$ crossover
at which a ramped signal reaches the half-trigger of the sequential test. This is the numerical content
of Decision~B: observing the attempt rather than the controlled state removes the relaxation lag from
\eqref{c4:eq:frontier}.

\begin{figure}[t]
\centering
\includegraphics[width=\textwidth]{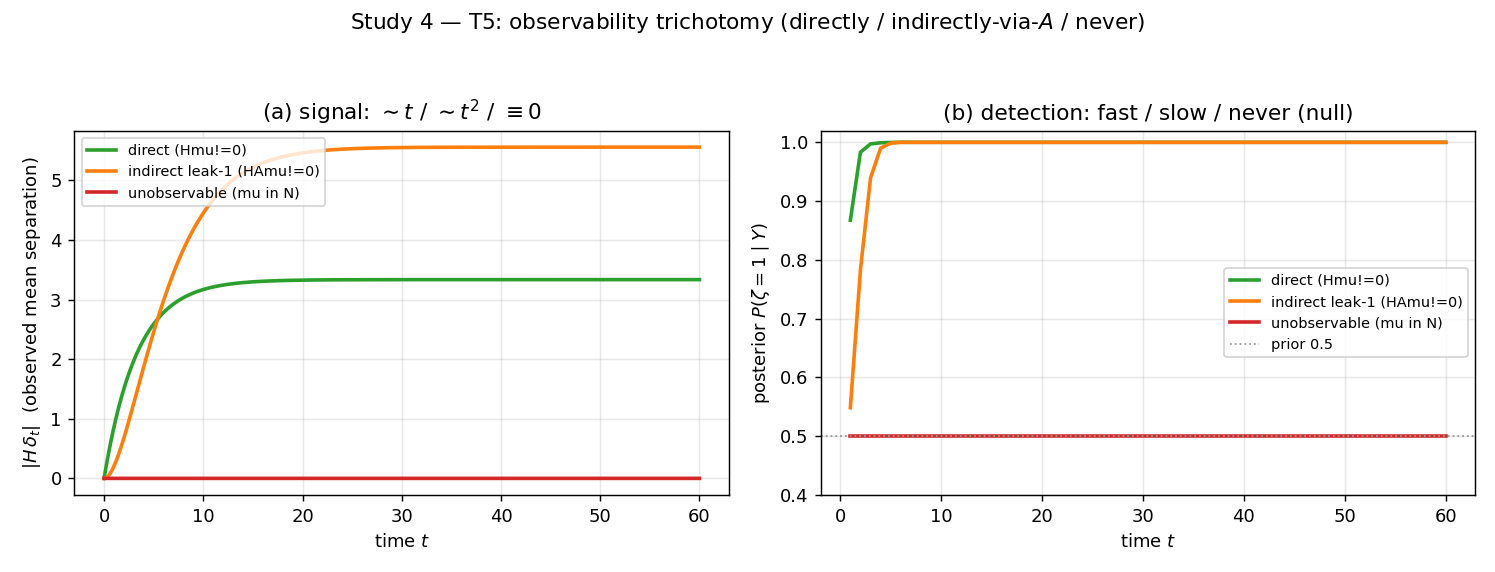}
\caption{\textbf{Study~4: collective observability.} Mean observed separation scaling as $t$ (direct),
$t^2$ (leak order one), and $0$ (unobservable, $\drift\in\Null$), confirming the series
\eqref{c4:eq:meansep}. The regime posterior converges to one in the observable classes and stays at the
prior in the null class.}
\label{c4:fig:observability}
\end{figure}

\paragraph{The observability trichotomy (Study~4, Figure~\ref{c4:fig:observability}).} In a
two-dimensional system the three classes of Theorem~\ref{c4:thm:observability} appear exactly as
predicted: a directly observed direction whose mean separation grows like $t$, a leak-order-one
direction (invisible to the direct field but coupled through $A$) growing like $t^2$, and an
unobservable direction in $\Null$ whose separation is identically zero. The regime posteriors converge
to one for the first two and remain pinned at the prior $0.5$ for the third. This is a null result
verified as a null, with slow detection ruled out rather than mistaken for the flat posterior.

\begin{figure}[t]
\centering
\includegraphics[width=\textwidth]{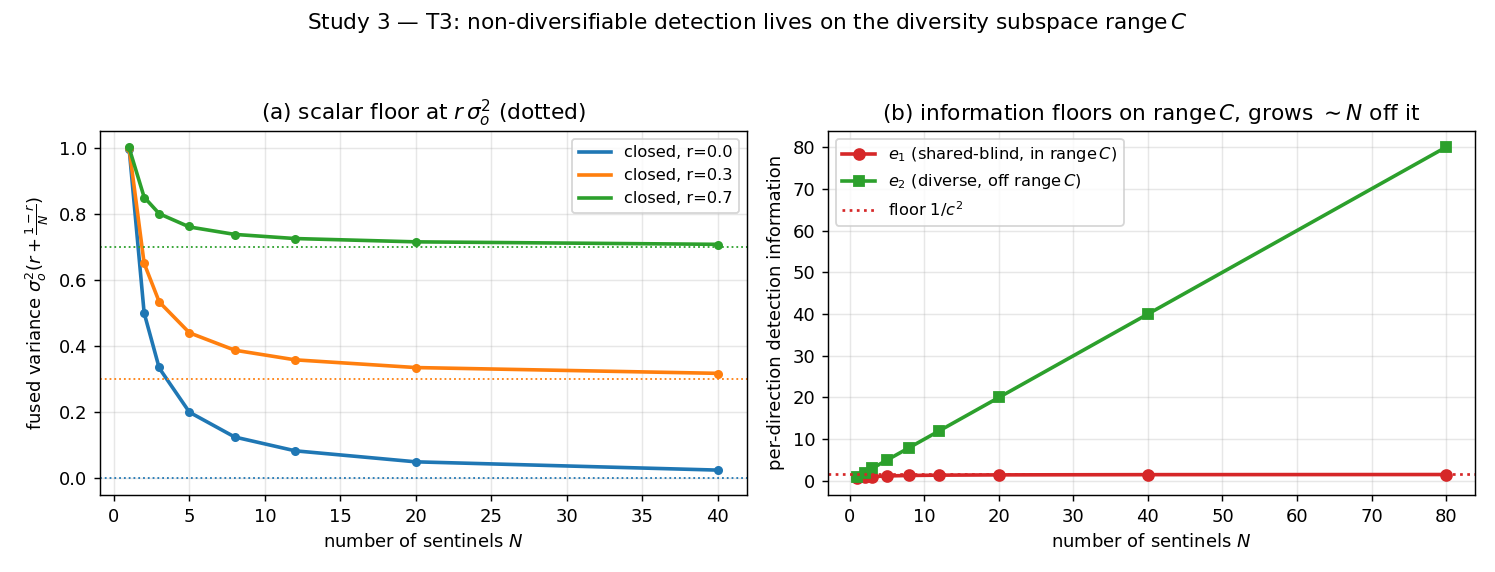}
\caption{\textbf{Study~3: non-diversifiable detection.} (a) Scalar fused variance flooring at
$r\sigma_o^2$, closed form versus Monte Carlo. (b) Per-direction detection information: capped on the
shared-blind direction in $\mathrm{range}(C)$, growing linearly in $N$ off it. The fused noise
covariance matches $CC^{\!\top}+\frac1N DD^{\!\top}$.}
\label{c4:fig:diversity}
\end{figure}

\paragraph{The non-diversifiable floor (Study~3, Figure~\ref{c4:fig:diversity}).} The scalar fused
variance $\sigma_o^2(r+(1-r)/N)$ of \eqref{c4:eq:scalarfloor} matches simulation to within $0.005$ and
floors at $r\sigma_o^2$ as $N$ grows. The vector noise covariance matches $CC^{\!\top}+\frac1N
DD^{\!\top}$. The per-direction detection information floors at $1/c^2$ along $\mathrm{range}(C)$ (here
$1.56$) while growing linearly in $N$ along the complementary direction. The monoculture floor is real
and finite, and it is a floored rate, not undetectability.

\begin{figure}[t]
\centering
\includegraphics[width=\textwidth]{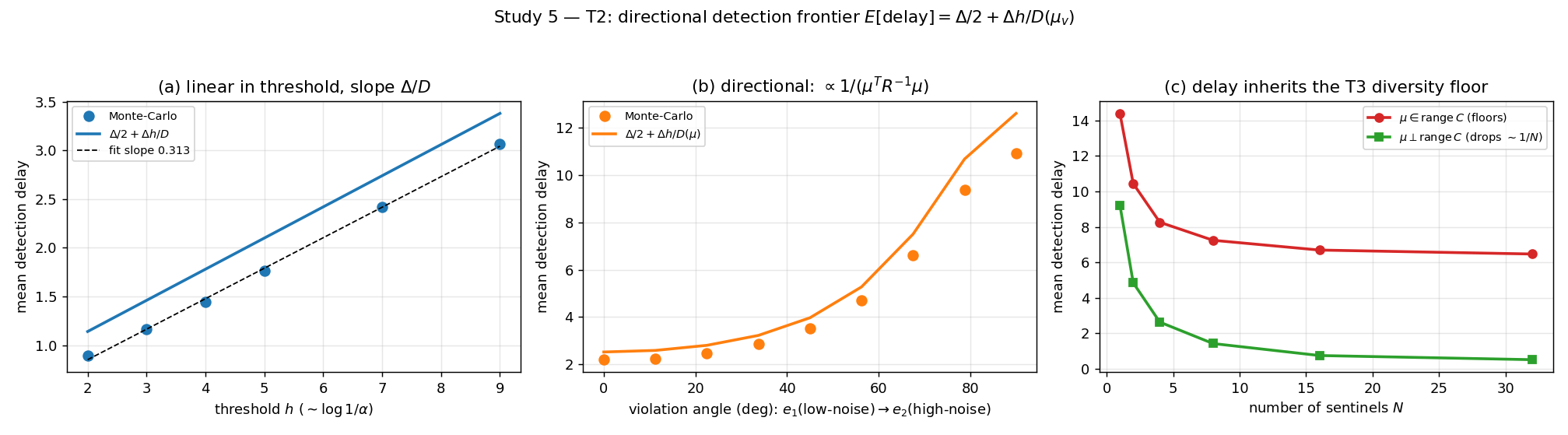}
\caption{\textbf{Study~5: the detection frontier.} (a) Expected delay affine in threshold with slope
$\Delta/\mathcal D$ (slope matched to $2.1\%$; magnitude carries a CUSUM-overshoot discrepancy). (b) Directional dependence $\propto1/\mathcal D(\drift)$. (c) Delay flooring with $N$ on the
shared-blind direction, dropping as $1/N$ off it.}
\label{c4:fig:detection}
\end{figure}

\paragraph{The detection frontier (Study~5, Figure~\ref{c4:fig:detection}).} The expected detection delay
is affine in the detection threshold with slope $\Delta/\mathcal D$, matched to within $2.1\%$. The
delay scales inversely with the directional information $\mathcal D(\drift)$. Along a shared-blind
direction the delay floors with $N$ while dropping as $1/N$ off it. We report that the delay
\emph{magnitude} carries a larger discrepancy (about $14\%$) attributable to cumulative-sum overshoot
not captured by the first-order Lorden expression. The \emph{slope}, which is the quantitative claim of
Theorem~\ref{c4:thm:frontier}, is tight.

\begin{figure}[t]
\centering
\includegraphics[width=\textwidth]{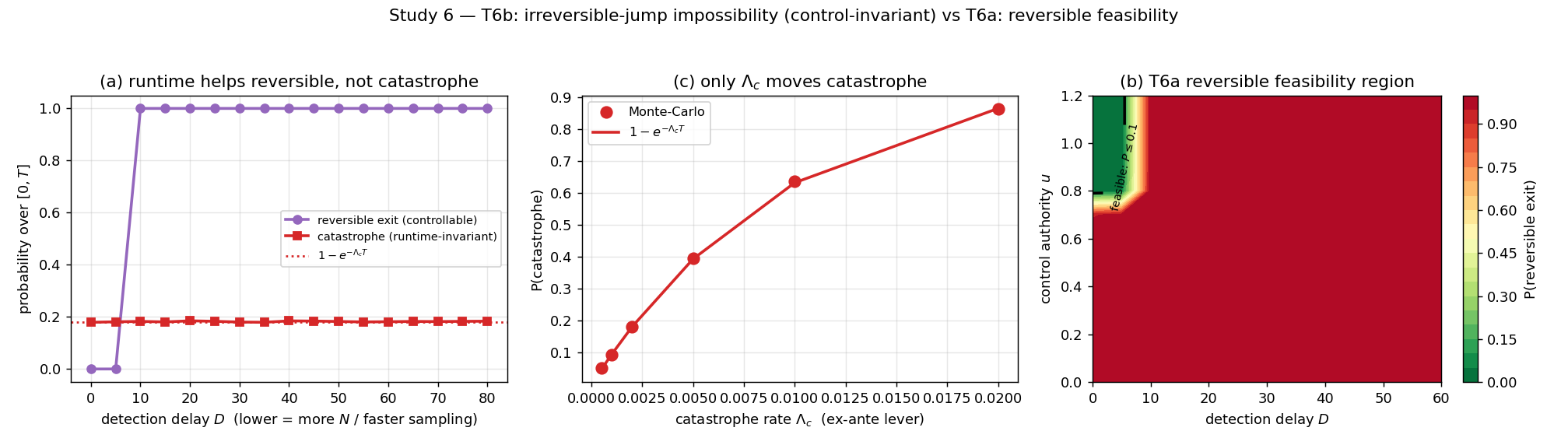}
\caption{\textbf{Study~6: control feasibility and impossibility.} The catastrophe probability
$1-e^{-\Lamc T}$ is flat across sentinel count, authority, threshold, and sampling (Corollary~\ref{c4:thm:t6b}),
moved only by ex-ante prevention through $\Lamc$. The reversible-exit probability, by contrast, is
controllable and traces the feasibility boundary of Corollary~\ref{c4:cor:t6a}.}
\label{c4:fig:controllability}
\end{figure}

\paragraph{Runtime impossibility (Study~6, Figure~\ref{c4:fig:controllability}).} The catastrophe
probability $1-e^{-\Lamc T}$ is invariant across sentinel count, intervention authority, detection
threshold, and sampling schedule, to a maximum deviation of $0.004$. This is the numerical statement of
Corollary~\ref{c4:thm:t6b}. The only quantity that moves it is the hazard $\Lamc$ itself, against which the
closed form matches simulation to $0.003$. The companion panel shows the reversible-exit probability
\emph{is} controllable as a function of detection delay and authority, locating the
$\varepsilon$-feasibility boundary of Corollary~\ref{c4:cor:t6a}.

\begin{figure}[t]
\centering
\includegraphics[width=\textwidth]{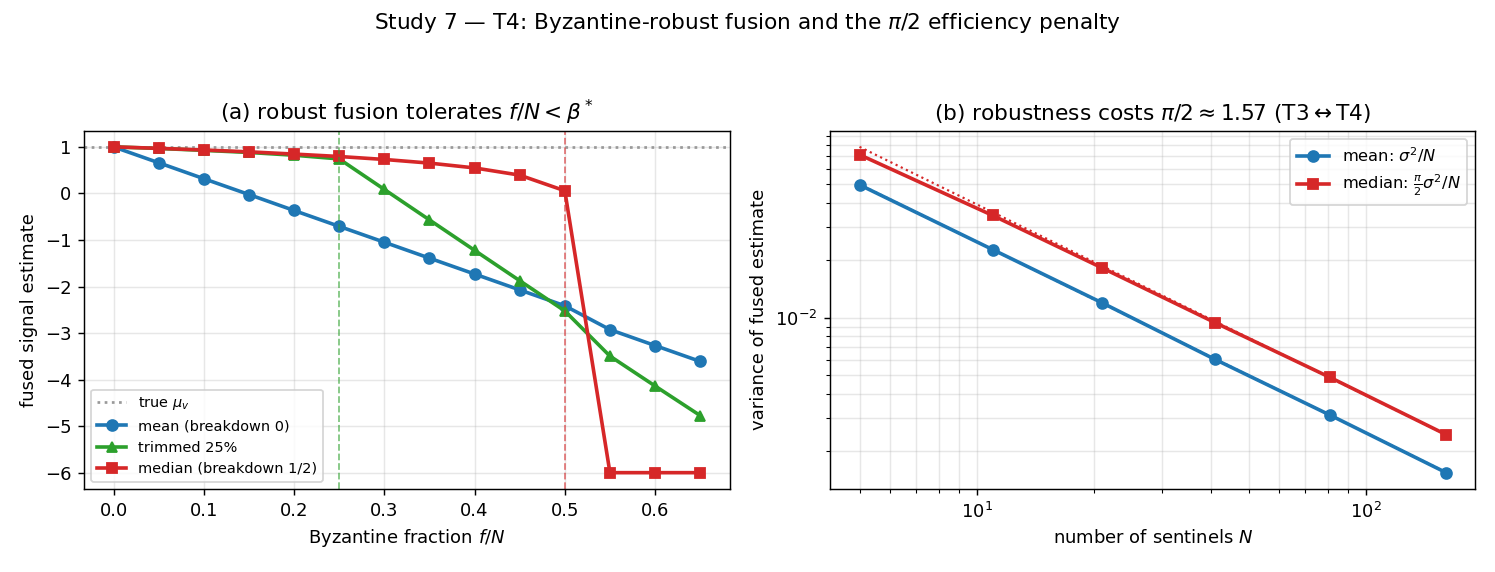}
\caption{\textbf{Study~7: Byzantine-robust fusion.} The mean breaks under a single adversary; the
median tolerates $f/N<\tfrac12$, the quarter-trimmed mean $f/N<\tfrac14$; the median pays a variance
penalty of $\approx\pi/2$ relative to the mean, the price of robustness in detection power.}
\label{c4:fig:byzantine}
\end{figure}

\paragraph{Byzantine fusion (Study~7, Figure~\ref{c4:fig:byzantine}).} The sample mean is corrupted by any
single Byzantine sentinel. The median holds until the Byzantine fraction reaches $\tfrac12$, and a
quarter-trimmed mean until $\tfrac14$, confirming the breakdown points of Theorem~\ref{c4:thm:byzantine}.
The median-to-mean variance ratio is $1.55$, matching the predicted $\pi/2\approx1.57$. The figure makes
the robustness--efficiency frontier visible: the robust estimator survives the adversary but pays in
variance, hence in detection delay.

\begin{figure}[t]
\centering
\includegraphics[width=\textwidth]{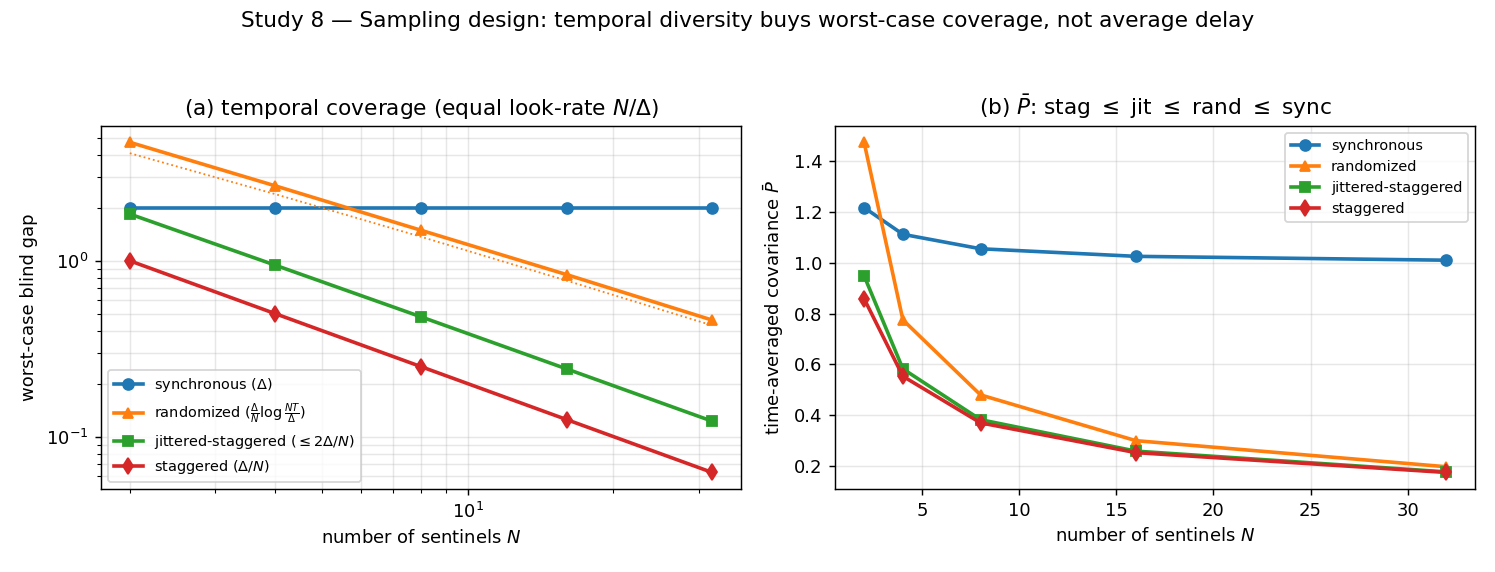}
\caption{\textbf{Study~8: sampling schedules.} Worst-case blind gap by schedule (synchronous $\sim\Delta$,
staggered $\Delta/N$, jittered $\le2\Delta/N$, randomized $\sim(\Delta/N)\log(NT/\Delta)$); time-averaged
filter covariance minimised by staggering; equal total look rate yields equal statistical detection
delay across all schedules.}
\label{c4:fig:sampling}
\end{figure}

\paragraph{Sampling design (Study~8, Figure~\ref{c4:fig:sampling}).} The worst-case blind gap scales as
$\Delta$ (synchronous), $\Delta/N$ (staggered), at most $2\Delta/N$ (jittered), and
$(\Delta/N)\log(NT/\Delta)$ (randomized). Staggering minimises the time-averaged filter covariance, with
jittered staggering close behind, synchronous plateaus, and randomized overtaking synchronous as $N$
grows. The total look count, hence the statistical detection delay, is equal across all designs at fixed
rate, confirming the schedule-invariance of Proposition~\ref{c4:prop:sampling}(1).

\paragraph{Summary.} Across the eight studies, every closed form is confirmed to within Monte-Carlo
error. The single exception, flagged above, is the detection-delay magnitude in Study~5, whose
governing claim (the slope) is nonetheless tight. The verification is structural: it tests that the
qualitative laws and their closed forms are correct, not that the illustrative parameters describe any
deployed system.

\section{The operating envelope}
\label{c4:sec:synthesis}

The results assemble into a single picture: a macroprudential map of the deployment's parameter space
into a region where guardrailing works and regions where it is structurally impossible. The boundaries
of this map are the timescale orderings \eqref{c4:eq:timescales}. Crossing each one names a distinct,
real failure mode rather than a technical breakdown. This is the synthesis the chapter has been building
toward, and it is where the operator reads off whether to keep tuning the monitor or to change the
system.

\subsection{The four orderings and what they mean}

Recall the orderings of Section~\ref{c4:sub:timescales}, now interpreted as the boundaries of feasibility.
With the intent-channel delay $\E[D]=\Delta(\tfrac12+\log(1/\alpha)/\mathcal D(\drift))$ and escape time
$\Tesc=g_0/\nu$:
\begin{itemize}[leftmargin=1.6em,itemsep=2pt]
\item $(\mathrm O1)\ \ \Delta\ll\E[D]$: detection is non-trivial (more than one look). Failure here is
benign, since a single look suffices, and means the detection theorems simply do not bite.
\item $(\mathrm O2)\ \ \E[D]\ll1/q_{10}$: detection completes before the violating regime reverts.
Crossing it, so that $\E[D]\gtrsim1/q_{10}$, means \emph{burst violations below the detection timescale
are undetectable in practice}. The regime flips on and off faster than the loop can resolve, so a
violation that does its damage and reverts is observable in principle but missed in operation, a
nameable vulnerability rather than a breakdown of the theory.
\item $(\mathrm O3)\ \ \E[D]+\trc<\Tesc$: detection and reaction beat the drift to the boundary. This
\emph{is} the Nyquist condition (F2). Crossing it sends $q\to1$: the excursion escapes before the loop
acts, and the deployment is \emph{prevention-only}. No runtime tuning restores compliance, and the
only recourse is to slow the drift (a tighter $\beta$, a smaller actuation) or prevent the action.
\item $(\mathrm O4)\ \ \Tesc\ll1/\Lamc$: drift-escape dominates spontaneous catastrophe. Crossing it, so
that $\Tesc\gtrsim1/\Lamc$, means the irreversible jump fires before the drift would even reach the
boundary. The binding failure is then the catastrophe, and Corollary~\ref{c4:thm:t6b} applies:
\emph{prevention-only} again, but now because runtime control is powerless against the jump rather than
too slow against the drift.
\end{itemize}
The window is non-empty. Both standard deployments satisfy all four orderings. A coding agent with
$\Delta=1$, $\trc=0.5$, $\mathcal D=0.1$, $\alpha=0.01$ has $\E[D]\approx46$, comfortably inside
$1/q_{10}=600$ and below $\Tesc=100$ (with $g_0=5,\nu=0.05$), itself far below $1/\Lamc=86{,}400$
(a day at second resolution). A trading agent at $\Delta=0.1$, $\trc=0.05$, $\mathcal D=0.2$,
$\alpha=10^{-3}$ has $\E[D]\approx3.5<\Tesc=20\ll1/\Lamc$ of order a week. Guardrailing is feasible for
both, on their reversible axes.

\subsection{The partition}

\begin{figure}[t]
\centering
\includegraphics[width=\textwidth]{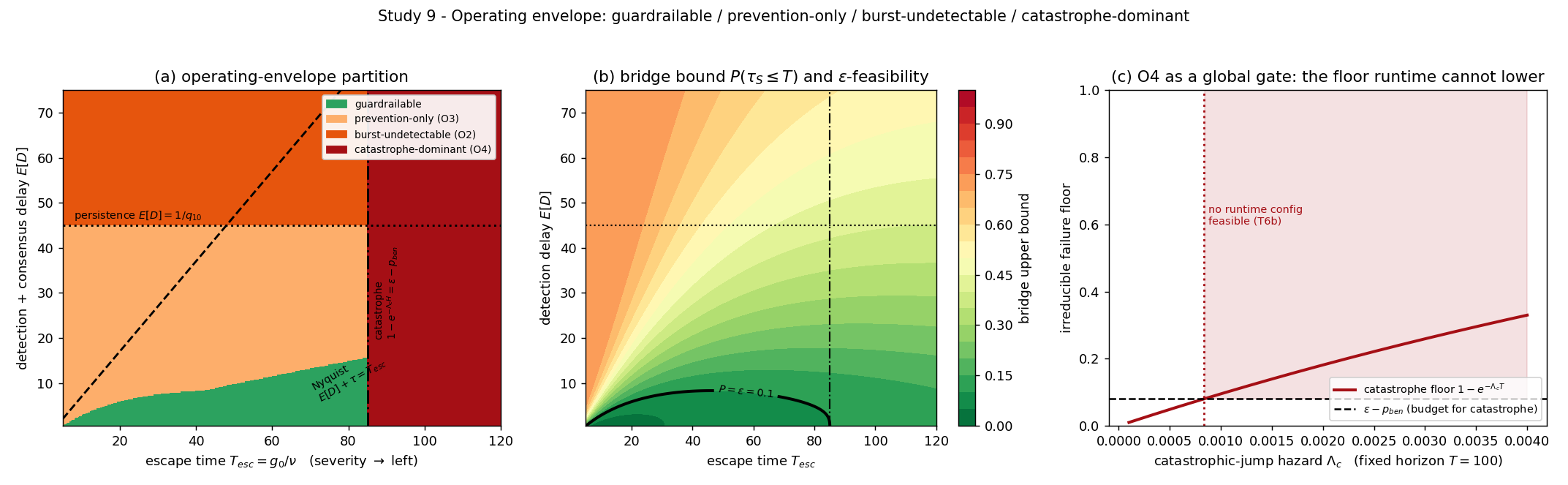}
\caption{\textbf{Study~9: the operating envelope.} (a) The partition of the $(\Tesc,\E[D])$ plane into
guardrailable, prevention-only ($\mathrm O3$, Nyquist fails), burst-undetectable ($\mathrm O2$,
persistence fails), and catastrophe-dominant ($\mathrm O4$) regions. (b) The bridge bound
\eqref{c4:eq:bridge} and its $\varepsilon$-feasibility contour, agreeing with the partition on $97\%$ of
the plane. (c) Catastrophe as a global gate: under a fixed horizon the catastrophe term is an
irreducible floor that lifts the entire surface above $\varepsilon$ once $\Lamc$ is large enough. No
runtime configuration is feasible (Corollary~\ref{c4:thm:t6b}). The exposure model $H=\kappa\Tesc$ used in
panels (a)--(b) is an illustrative choice that renders the $\Tesc\ll1/\Lamc$ condition as a region;
panel (c) shows the equivalent fixed-horizon reading.}
\label{c4:fig:envelope}
\end{figure}

Because $\E[D]\propto\Delta$, conditions $(\mathrm O2)$ and $(\mathrm O3)$ together bound the sampling
interval from above,
\begin{equation}
\label{c4:eq:samplingwindow}
\Delta\;<\;\frac{\min\big(1/q_{10},\ \Tesc-\trc\big)}{\tfrac12+\log(1/\alpha)/\mathcal D(\drift)},
\end{equation}
while monitoring cost bounds it from below. The window is non-empty exactly when the signal $\mathcal
D(\drift)$ is not too weak and the dwell and escape times not too short. The binding inequality is
$\E[D]+\trc<\Tesc$, the Nyquist condition. Plotting the deployment on the plane of escape time $\Tesc$
against detection delay $\E[D]$ partitions it into four regions: \emph{guardrailable},
\emph{prevention-only} (Nyquist fails, $\mathrm O3$), \emph{burst-undetectable} (persistence fails,
$\mathrm O2$), and \emph{catastrophe-dominant} (the jump dominates, $\mathrm O4$). Study~9
(Figure~\ref{c4:fig:envelope}) confirms that this partition by timescale conditions coincides, on $97\%$ of
the plane, with the feasibility region computed directly from the bridge bound \eqref{c4:eq:bridge}. The
two derivations agree by construction: a single feasibility surface drives both the partition and the
$\varepsilon$-contour.

The boundaries are the result, not a sanity check. They tell an operator when the problem
stops being a tuning problem. Inside the guardrailable region, faster sampling, more sentinels, or a
lower threshold restore compliance. Across the $\mathrm O3$ boundary, none of these help: the drift has
escaped before the loop can act, and the operator must change the dynamics or prevent the action.
Across the $\mathrm O2$ boundary, the violations are bursting faster than detection resolves, and the
recourse is again structural (a faster channel, or accept the burst risk). Across the $\mathrm O4$
boundary, Corollary~\ref{c4:thm:t6b} forecloses runtime control entirely, and only prevention remains. A
deployment near a boundary is a deployment whose safety case rests on an inequality the operator should
be required to state and defend.

\subsection{The framework's levers, made dynamic}

The picture closes the arc of the four chapters. Chapter~\ref{ch:single}'s three containment levers were
static design choices. Here they are the moves of a feedback controller, and the operating envelope says
exactly where each applies. \emph{Review} and \emph{reversibility} are runtime interventions on the
reversible axes: a review gate adds to the reaction delay $\trc$ (a deliberate slowing, traded against
the human catch it affords), and a reversibility rollback is the impulse $\xi_k$ of the quasi-variational
inequality \eqref{c4:eq:qvi}. Both live inside the bridge and are governed by the Nyquist condition.
\emph{Prevention} is the lever that the envelope singles out. It is the only one that operates on the
irreversible axis, because Corollary~\ref{c4:thm:t6b} shows it is the only one that can. It acts not at
runtime but on the deployment itself, lowering the hazard $\Lamc$ by removing the catastrophic action
from the agent's reach. Chapter~\ref{ch:single} proved the levers are not substitutes against
irreversible harm. Chapter~\ref{ch:continuous} proved prevention is what makes a fleet robustly
governable by truncating the jump measure. This chapter shows the same conclusion is forced,
dynamically and in a distributed feedback setting, by the causality of the detect-react loop, and it
locates the boundary at which an operator must stop monitoring and start preventing: the irreversible
axis, the $\mathrm O4$ region. The measure told the operator how risky the deployment is. The envelope
tells the operator what, if anything, can be done about it at runtime, and when the accurate answer is
that nothing can.

\section{Conclusions, limitations, and open problems}
\label{c4:sec:conclusions}

We set out to address the half of the agentic financial-stability problem that the preceding chapters
left open, asking not how to \emph{measure} systemic risk, but how to \emph{control} it in real time, and where
control is impossible. The construction models the agent as a latent risk state under a hidden violation
regime and a separate irreversible catastrophe. A fleet of intermittent, partial, possibly compromised
sentinels watches that agent, and their interventions are the earlier chapters' containment levers made dynamic.
Its results are a bridge lemma that converts regime detection into a hitting-time guarantee and yields
a derived Nyquist condition; a collective-observability trichotomy with an undetectable null
subspace; a non-diversifiable detection floor; and a quickest-detection frontier. They include as well
an architecture result separating a centralized benchmark from its distributed approximation; a
Byzantine-robustness frontier; a sampling-design analysis; and, as the headline, an impossibility
theorem showing that runtime control cannot lower the irreversible-jump floor and only ex-ante
prevention can. The operating envelope assembles these into a map of when guardrailing is feasible and
when it is not.

\paragraph{Limitations.} These limitations are part of the claim itself, not an afterthought. The
linear-Gaussian, two-to-three-dimensional risk state is a deliberate caricature of high-dimensional,
non-Gaussian, nonstationary agent telemetry. The theorems are accordingly \emph{structural}, not
deployable detectors: they say which failure modes exist, what shape the remedies take, and where
runtime control is powerless. The impulse interventions are idealised as instantaneous, deterministic
resets. Real throttles and rollbacks have latency and uncertain effect.
Remark~\ref{c4:rem:neutralise} absorbs that latency and effect into the reaction delay and a residual
drift, and they make the feasibility region an outer bound on the truly safe one. The most delicate
results to verify are the null subspace of Theorem~\ref{c4:thm:observability} and the large-$N$ floors of
Theorems~\ref{c4:thm:diversity}--\ref{c4:thm:byzantine}. A null is confirmed by the \emph{absence} of a
signal. We have checked that absence to Monte-Carlo error while ruling out slow detection, but it does
not carry the same closed-form certainty as a positive result. We have also flagged the one study (the
detection-delay magnitude of Study~5) whose simulation departs materially from the first-order theory
even as its governing slope holds. Finally, the whole edifice rests on Assumptions~\ref{c4:ass:separation}--%
\ref{c4:ass:sampling}, several of which are load-bearing. A regime that altered the diffusion or the
hazard, rather than only the drift, would be detectable along otherwise-blind directions
(Assumption~\ref{c4:ass:onlymu}). Aliasing in the sampling could hide an observable mode
(Assumption~\ref{c4:ass:sampling}).

\paragraph{The regress.} The sentinels are themselves agents, with their own model risk. We have not
dissolved the problem of trusting a monitor, but pushed it up one level and bounded that level. The
Byzantine threshold of Theorem~\ref{c4:thm:byzantine} says how many compromised sentinels a fused decision
survives. But it presumes a fusion rule that is itself correct, and a monitor-of-monitors would face
the same questions one tier higher. This is intrinsic to monitoring as a strategy and is the deeper
reason the impossibility theorem matters. Prevention, which removes an action from the agent's reach,
does not participate in the regress, because it is not a monitor.

\paragraph{Calibration is the central open problem.} As in every part of this program, the parameters
are illustrative and the hard, unglamorous, essential work is calibration. What are the violation
directions $\drift$ a real prompt injection or reward-hacking episode induces, and do they lie in or
near the null subspace of a given monitoring fleet? What is the correlation structure of real sensor
errors? How much of a deployed monitoring stack is, in the sense of Theorem~\ref{c4:thm:diversity}, a
monoculture parsing the same log? What are the escape times $\Tesc$ and dwell times $1/q_{10}$ of
actual agent failures, and where does a given deployment sit in the operating envelope? What is the
catastrophe hazard $\Lamc$ of an irreversible action, and by how much does a particular preventive
control lower it? None of these is answered here. Answering them means mapping the structural objects of
this theory onto measured agent behaviour, which is the work that would turn a framework into an
instrument.

\paragraph{Future directions.} Several extensions are natural. The persistent-drift form of violation
(a regime that injects a growing rather than a fixed displacement) sharpens the detection problem and
yields an optimal sampling rate. The distributed-approximation gap of Remark~\ref{c4:rem:distributed}
deserves a quantitative rate in terms of $\lambda_2(\Lapl)$ for the specific consensus-filter
implementation. The linear-Gaussian state should be stress-tested against nonlinear, heavy-tailed
dynamics where the observability series \eqref{c4:eq:meansep} and the first-passage formula
\eqref{c4:eq:firstpassage} are only approximations. The adversarial threat-cone of
Assumption~\ref{c4:ass:direction} could be made an explicit game between a violation-direction-choosing
adversary and a sentinel-placing defender, with the null subspace as the adversary's target and fleet
coverage as the defender's move. Each is a step from the structure established here toward a deployable
theory.

\paragraph{The arc.} With this final chapter the monograph closes its movement from model risk to
financial stability, and from measurement to control. Chapter~\ref{ch:single} established the
microprudential unit, a single agent's model risk, and proved the irreversibility of the catastrophic
floor. Chapter~\ref{ch:fleet} aggregated that unit into a fleet on a shared model and found the
non-diversifiable common exposure and the contagion threshold that turn model risk into \emph{systemic}
risk. Chapter~\ref{ch:continuous} made the object dynamic and identified the worst-case scenario a
robust governor should stress against. Chapter~\ref{ch:control} builds the macroprudential governor and
shows what it can and cannot do. It also proves that against the one failure that matters most, the
systemic, irreversible one, its only effective response is the structural change it must make before the
agents ever run. The measure says how dangerous the system is. The control says what to do about it,
and it says that financial stability, here as in banking, is secured not one agent at a time but by
acting on the structure that makes the risk common.

\chapter{From Accident to Adversary: Agentic-AI-Augmented Cyber Events and Financial Stability}
\label{ch:cyber}
\section{Introduction}
\label{c5:sec:intro}

The four preceding chapters measured the risk that agentic systems create and asked what can be
done about it. Chapter~\ref{ch:single} decomposed a single agent's expected harm and proved
prevention non-substitutable against irreversible harm. Chapter~\ref{ch:fleet} lifted the unit to a
fleet and found the shared foundation model a non-diversifiable common exposure.
Chapter~\ref{ch:continuous} made the object dynamic and identified the robust stress problem with a
time-consistent risk measure. Chapter~\ref{ch:control} modelled runtime guardrails and proved that
against an irreversible event no runtime policy moves the catastrophe floor. All four treated the
tail driver as a \emph{distribution}, at worst an ambiguous one, a set of measures penalised by a
cost. Yet Chapter~\ref{ch:single} conceded at the outset (its Break~2) that the dominant tail driver
for a deployed agent is not a state of nature but an \emph{adversary}, strategic, adaptive,
non-stationary, and optimising against the very controls raised to stop it. Having named the adversary,
the mathematics then modelled it as nature with a penalty. This chapter pays that debt. It makes the
adversary a \emph{player}, and studies what changes when the loss-driving factor optimises.

The occasion for doing so is a specific and growing class of events, agentic-AI-augmented cyber
incidents in the financial system. Agentic systems transform the cyber problem on both sides of the
wire, and the transformation is not symmetric in where each side's augmentation reaches, as we show in
\S\ref{c5:ssec:defense}. Agentic systems are new attack \emph{surface}, since the prompt-injection,
tool-confusion, and poisoned-retrieval channels of Chapter~\ref{ch:single} are live vulnerabilities
in any deployed fleet. They are also new attack \emph{capability}. An adversary that is itself
agentic-AI-augmented sees the cost of reconnaissance, target tailoring, and simultaneous
multi-target execution collapse, so that the attacker, too, acquires a fleet. Agentic AI arms the
defender as well, through autonomous monitoring, triage, and response. That augmentation, we will
find, extends the defender's reach across the reversible body while leaving the
systematic-irreversible core untouched unless it is aimed at prevention. A shared foundation model
beneath a population of financial agents is therefore a common exposure in the sense of
Chapter~\ref{ch:fleet}, one that a rational attacker can \emph{choose} to
strike, and its value as a target grows with the very concentration that makes it systemically
important.

\subsection{Intentional or not: the organising dichotomy}
\label{c5:ssec:dichotomy-intro}

A cyber event is, for our purposes, a triple: a \emph{channel} (how the compromise enters: a
poisoned dependency, an injected prompt, a stolen credential), a \emph{payload} (what it does: an
availability outage, an integrity corruption, an irreversible exfiltration or transfer), and a
\emph{trigger} (what determines when and where it fires). Intentionality is a property of the
trigger's law, and it cleaves the analysis in two.

When the trigger is \emph{exogenous} (a bad model update that ships to every agent, a vulnerable
shared library, a misconfiguration that activates under load), the event is unintentional, and its
statistics are those of a common shock. It strikes the shared substrate because that is where the
fault lives, not because anything chose it. We show that this case instantiates the objects of
Chapters~\ref{ch:fleet} and~\ref{ch:continuous} exactly
(Proposition~\ref{c5:prop:reduction}); an accidental agentic cyber event \emph{is} a common shock (a
common L\'evy jump in continuous time), and the theorems of those chapters apply with nothing added.
The argument goes through verbatim, since the unintentional case is the framework
already built, wearing a cyber costume.

When the trigger is \emph{endogenous}, a best response to the defender's architecture, the event
is intentional, and every quantity the earlier chapters took as a parameter of nature becomes an
attacker's choice variable. The common-shock probability, the contagion seed, the branching matrix,
the timing of the irreversible jump, the shape of the jump's tail, and the catastrophe intensity
itself are no longer given. An optimiser selects them, reading the architecture and allocating
effort where it pays. Re-proving the earlier results with a best response inside them does not
weaken their conclusions but sharpens them. The monoculture is where a rational attacker
\emph{sends} its effort, not just a passive vulnerability that fails to diversify, because attacking
the shared substrate is the only attack whose payoff scales with fleet size, and the common exposure
becomes an attractive target.

\subsection{The objection from cost, and the tail}
\label{c5:ssec:cost-intro}

A rational attacker does not go for the big prize merely because it is big. It weighs the prize
against the cost of taking it, and the shared substrate is the best-defended object in the
ecosystem, the one whose compromise invites attribution, law-enforcement priority, and retaliation.
This objection is correct, and we build it into the chapter's main theorem rather than waving it
away. The attractive-target result is a \emph{threshold}. The substrate is the attacker's best response only
once the fleet size $K$ makes the payoff scaling beat the cost differential, and substrate hardening
and the attribution penalty raise that threshold. They can price a fleet out of the attractive-target regime.
Because the defender does not know which attacker it faces, the natural object is not a single
objective function but a \emph{family} of them: a profit-maximiser who diversifies across soft
peripheral targets, a state actor pre-positioning for disruption who is nearly cost-insensitive, a
pure disruptor. The risk is the worst case over the admitted family, which is the ambiguity
set of Chapter~\ref{ch:single} now given a microfoundation. The decisive observation is that the
financial-stability functional does not average over attacker types but conditions on the tail, which
is owned by the type for whom the substrate is the unconditional best response.
Cost considerations thin the \emph{body} of the attack distribution away from the shared substrate,
but they do not thin the \emph{tail} (Theorem~\ref{c5:thm:tail}). The monoculture is an attractive
target in the tail, and the observed history of cyber events, largely criminal, peripheral, and
absorbed without lasting damage, is systematically uninformative about the systemic scenario, for
incentive reasons rather than small-sample ones.

\subsection{From a technological thesis to an incentive thesis}
\label{c5:ssec:incentive-intro}

The monograph's central claim is that the systematic, irreversible core of agentic risk is not reachable by
the controls practitioners reach for first (redundancy, detection, reversibility), and moves only
under ex-ante structural prevention. Chapters~\ref{ch:single}--\ref{ch:control} proved this as a
\emph{technological} claim, showing that those controls \emph{cannot} move the core. This chapter adds the
\emph{incentive} half. Detection spending, we show, actively \emph{deflects} a strategic attacker's
effort into the irreversible core rather than reducing exposure to it
(Theorem~\ref{c5:thm:deflection}). Prevention's value splits into a mechanical part and a
\emph{deterrence} part, both moving the floor (Theorem~\ref{c5:thm:deterrence}). Finally, even where
prevention is feasible, a population of institutions coupled by contagion
\emph{underinvests} in it at the decentralized equilibrium, by a wedge proportional to how much each
firm endangers its neighbours (Theorem~\ref{c5:thm:underinvest}). Through four chapters, the
monograph's refrain that financial stability cannot be secured one institution at a time was a
statement about technology. Here it becomes a statement about incentives, and that is the formal
case for macroprudential cyber regulation.

\paragraph{Contributions.} This chapter (i) defines the intentionality dichotomy and proves the
unintentional case reduces to the common-shock and common-jump results of
Chapters~\ref{ch:fleet}--\ref{ch:continuous} (Proposition~\ref{c5:prop:reduction}); (ii) casts the
intentional case as a Stackelberg game and defines the \emph{attacked risk measure}, showing the
attack-cost penalty microfounds Chapter~\ref{ch:single}'s adversary ambiguity set
(\S\ref{c5:sec:game}); (iii) proves the attractive-target threshold and the anti-diversification of the fleet
floor under attack (Theorem~\ref{c5:thm:magnet}, Corollary~\ref{c5:cor:antidiv}). It (iv) proves
tail dominance, in which the attractive-target effect lives in the tail regardless of the
attacker-type mixture (Theorem~\ref{c5:thm:tail}); (v) proves targeted contagion and engineered supercriticality, with
segmentation the robust lever (Theorem~\ref{c5:thm:targeted}); (vi) identifies the intentional
worst case with Chapter~\ref{ch:continuous}'s robust entropic game, making the prevention lever the
condition that renders the attacker's problem finite (Theorem~\ref{c5:thm:robust}). It (vii) proves
the detection-deflection result (Theorem~\ref{c5:thm:deflection}); (viii) proves runtime
impossibility is robust to intent and decomposes prevention into mechanical and deterrence value
(Theorem~\ref{c5:thm:deterrence}); (ix) proves the decentralized-underinvestment theorem and its
closure by a mandated prevention floor (Theorem~\ref{c5:thm:underinvest}). It (x) shows that agentic
\emph{defense} reaches the systematic-irreversible core only through the prevention lever and that a
homogeneous fleet of AI monitors inherits its own non-diversifiable monoculture, so agentic AI is
offense-dominant at the systemic tail (\S\ref{c5:ssec:defense}); (xi) solves the two-sided game for a
budget-constrained defender and finds a threshold attacker capability beyond which the
systemically-optimal detection budget is zero (Theorem~\ref{c5:thm:budget}); and (xii) confirms every
closed form numerically (\S\ref{c5:sec:numerics}) and instantiates the framework in three
financial-stability settings (\S\ref{c5:sec:apps}) with a macroprudential governance mapping
(\S\ref{c5:sec:governance}).

\paragraph{Stance.} As in every preceding chapter, the parameters are illustrative and the
contribution is structural. The new and binding unknown here is the attacker: its objective, its
cost technology, and above all the offensive capability that agentic AI confers. That capability is
the attacker-side analogue of the systematic parameters the monograph has repeatedly flagged as the
least estimable. We delineate the estimation program (red-team economics, frontier-model cyber
evaluations) but do not solve it. We turn first to the literatures this chapter joins.

\section{Background}
\label{c5:sec:background}

This chapter sits at the junction of four literatures, and its contribution is to connect them
through the monograph's framework rather than to extend any one of them internally.

\subsection{Cyber risk and financial stability}
\label{c5:ssec:bg-cyber}

Supervisors and researchers have increasingly treated cyber risk as a \emph{systemic} concern rather
than a purely operational one. \citet{eisenbach2022cyber} conduct a pre-mortem of the U.S. wholesale payment
system and find its vulnerability concentrated in a handful of institutions. An attack impairing the
five most active banks would touch on the order of a third of the network by value, a concentration
result that is the empirical face of the common-exposure mechanism of Chapter~\ref{ch:fleet}.
\citet{duffie2019cyberruns} describe the \emph{cyber run}, a loss of confidence triggered by an
integrity or availability event that propagates like a liquidity run. \citet{crosignani2023pirates}
document, through the NotPetya episode, how a cyber shock propagates along real firm-to-firm supply
chains, an empirical contagion graph of the kind Chapter~\ref{ch:fleet} models abstractly.
\citet{aldasoro2022drivers} assemble cyber-loss data and find heavy tails and a strong role for
common vendors and platforms. \citet{kashyap2019principles} lay out principles for regulating cyber
risk that anticipate the macroprudential stance we make precise, and \citet{healey2018future} and the
\citet{fsb2018lexicon} cyber lexicon frame the financial-stability vocabulary. This literature
establishes concentration, contagion, heavy tails, and the inadequacy of firm-level resilience
empirically. This chapter derives them structurally from an attacker's optimisation, and turns them
into statements about which controls survive intent.

\subsection{Security and interdependent-security games}
\label{c5:ssec:bg-games}

The strategic side draws on the economics of security; \citet{gordon2002economics} price security
investment against a probability of breach. \citet{kunreuther2003interdependent} identify the
\emph{interdependent-security} externality, in which one agent's under-protection raises its neighbours'
risk, so decentralized investment falls short of the social optimum. That externality is the
mechanism of our underinvestment theorem (Theorem~\ref{c5:thm:underinvest}). The defender-commits,
attacker-responds structure is the \emph{Stackelberg security game}
\citep{tambe2011security,manshaei2013game}. It is the natural frame for agentic cyber risk because
architecture is chosen ex ante and the attacker reads it, a form of Kerckhoffs's principle, which
assumes the adversary knows the design. We use this frame not to compute an optimal patrol but to locate
where architecture prices the attacker out of the systemic target.

\subsection{Adversarial detection and stealthy attacks}
\label{c5:ssec:bg-detection}

The control-theoretic security literature studies detection against an adversary who optimises to
evade it. \citet{teixeira2015secure} characterise attacks that are undetectable by a given monitor
set, attacks living in a null space of the observation map. \citet{cardenas2008secure} and related
work study stealthy, schedule-aware injection, the strategic counterpart of
Chapter~\ref{ch:control}'s observability trichotomy. There the unobservable directions were a
property of the sensors, whereas here they are where a strategic attacker \emph{chooses} to operate,
and the quickest-detection frontier of Chapter~\ref{ch:control} becomes a constraint the attacker
games rather than a guarantee the defender holds; our deflection result
(Theorem~\ref{c5:thm:deflection}) is the equilibrium consequence.

\subsection{Offensive AI capability}
\label{c5:ssec:bg-offai}

Finally, the premise that agentic AI augments the attacker rests on a young but fast-moving
evidence base: demonstrations that LLM agents can autonomously discover and exploit vulnerabilities
\citep{fang2024llm}, and the frontier-model cyber-capability evaluations now run as part of
responsible-scaling regimes \citep{aisi2024evaluations}. These evaluations are, in the language of
this monograph, the estimation program for the attacker's capability parameter, the single most
important and least settled input to the game. We treat their outputs as the natural place from
which the chapter's illustrative attacker parameters would one day be calibrated.

\paragraph{What we do not do.} We do not model attribution, deterrence-by-punishment as a
repeated-game reputation mechanism, cyber-insurance pricing, or multi-attacker coalitions. We flag
each as an extension in \S\ref{c5:sec:conclusions}. The chapter's object is the structural change
that intentionality induces in the monograph's four results, and the macroprudential consequence
that follows.

\section{The intentionality dichotomy and the reduction}
\label{c5:sec:dichotomy}

We first make the dichotomy precise and then discharge the easy half. An unintentional agentic
cyber event is a common shock, and the results of Chapters~\ref{ch:fleet} and~\ref{ch:continuous}
apply to it unchanged. This is the formal content of the claim that, if the event is not
intentional, ``the argument goes through.''

\begin{definition}[Agentic cyber event]
\label{c5:def:event}
An \emph{agentic cyber event} on a fleet of $K$ agents is a triple $(\mathcal C,\mathcal P,T)$: a
\emph{channel} $\mathcal C$ (the entry mechanism---poisoned dependency or model update, injected
prompt, compromised credential or tool), a \emph{payload} $\mathcal P$ (its effect, classified by
the reversible/irreversible split of Chapter~\ref{ch:fleet} into an availability/integrity outage on
$\mathcal R$ or an irrecoverable exfiltration, transfer, or destruction on $\mathcal I$), and a
\emph{trigger} $T$, a random time-and-target rule determining when and where the payload fires. The
event is \emph{unintentional} if the law of $T$ is exogenous---independent of the deployer's
architecture---and \emph{intentional} if $T$ is chosen to maximise a harm objective given the
architecture, i.e.\ $T=T^\star(\text{architecture})$ is an attacker's best response.
\end{definition}

\begin{proposition}[Reduction: the unintentional case is the framework baseline]
\label{c5:prop:reduction}
Let $(\mathcal C,\mathcal P,T)$ be an unintentional agentic cyber event whose trigger fires
fleet-wide with exogenous probability $\psys$ (per episode) or, in continuous time, as an exogenous
common jump of intensity $\Lambda$. Then:
\begin{enumerate}[label=\textup{(\roman*)},leftmargin=2.2em,itemsep=1pt]
\item its fleet-average harm satisfies Assumption~\ref{c2:ass:common} with common-shock magnitude
$M$ equal to the payload severity, so by Theorem~\ref{c2:thm:nondiv} the expected shortfall floors at
$M\psys/(1-\tau)+\mu_\xi$, independent of $K$;
\item if the channel propagates agent-to-agent with transmission $q$ on the dependency graph $G$, its
outbreak obeys the criticality of Theorem~\ref{c2:thm:contagion} at branching ratio $\beta=cq$;
\item if the payload is irreversible, its fleet floor is the non-substitutable
$F(g,c_3)$ of Theorem~\ref{c2:thm:floor}; and
\item in continuous time it is a common L\'evy jump, non-diversifiable by Theorem~\ref{c3:thm:t1},
with prevention acting as the jump-measure truncation of Theorem~\ref{c3:thm:t3}.
\end{enumerate}
No new object is required: an accidental agentic cyber event instantiates the fleet and
continuous-time frameworks with the payload severity as the shock magnitude and the exogenous trigger
rate as the systematic probability.
\end{proposition}

\begin{proof}
Each claim is an instantiation. (i) An exogenous fleet-wide trigger firing with probability $\psys$
and inflicting severity $M$ on every loaded agent is exactly the common shock $J=M\mathbf B$,
$\mathbf B\sim\mathrm{Bernoulli}(\psys)$, of Assumption~\ref{c2:ass:common}. The loadings $a_k$ are
the agents' exposure to the shared channel, normalised to $\bar a=1$. Theorem~\ref{c2:thm:nondiv}
then applies verbatim. (ii) Exogenous agent-to-agent propagation with per-edge probability $q$ is
independent-cascade transmission (Assumption~\ref{c2:ass:cascade}), and
Theorem~\ref{c2:thm:contagion} gives the threshold and the final size. (iii) An irreversible payload
changes no distribution after the fact, so detective controls leave it unchanged and
Assumption~\ref{c2:ass:irr} and Theorem~\ref{c2:thm:floor} apply. (iv) An exogenous common jump is a
common L\'evy jump in the sense of Chapter~\ref{ch:continuous}. Theorems~\ref{c3:thm:t1}
and~\ref{c3:thm:t3} apply. In each case the exogeneity of $T$ is precisely the hypothesis
(independence of the systematic event from the controls) that the cited theorem already assumes.
\end{proof}

We verify Proposition~\ref{c5:prop:reduction} numerically in \S\ref{c5:ssec:num-reduction}, showing
that as an ``attacker capability'' knob is turned to zero, every quantity in this chapter collapses onto its
Chapter~\ref{ch:fleet}/\ref{ch:continuous}/\ref{ch:control} baseline. Table~\ref{c5:tab:map} records
the dictionary that makes the reduction routine.

\begin{table}[htbp]
\centering
\footnotesize
\setlength{\tabcolsep}{5pt}
\renewcommand{\arraystretch}{1.25}
\caption{The cyber-to-framework dictionary. An unintentional event maps onto the earlier chapters'
objects with nothing added (Proposition~\ref{c5:prop:reduction}).}
\label{c5:tab:map}
\begin{tabular}{p{0.38\linewidth} p{0.40\linewidth} p{0.12\linewidth}}
\toprule
\textbf{Cyber concept} & \textbf{Framework object} & \textbf{Chapter} \\
\midrule
Vulnerable shared dependency / model update & Common shock $J=M\mathbf B$; common L\'evy jump &
\ref{ch:fleet}, \ref{ch:continuous} \\
Lateral movement & Percolation edge, transmission $q$ & \ref{ch:fleet} \\
Self-propagating worm & Hawkes excitation, $\rho(\Gam)$ & \ref{ch:continuous} \\
Ransomware detonation / integrity destruction & Irreversible jump, catastrophic class &
\ref{ch:single}, \ref{ch:continuous}, \ref{ch:control} \\
Patching, least privilege, air-gapping & Prevention $c_3$ (jump-measure truncation) &
\ref{ch:single}--\ref{ch:continuous} \\
EDR / SIEM / anomaly detection & Sentinels, detection delay $\E[D]$ & \ref{ch:control} \\
Backups, failover, rollback & Reversibility $\rec$ & \ref{ch:single} \\
Network segmentation & Segmentation $m$, exposed fraction $g$ & \ref{ch:fleet} \\
\bottomrule
\end{tabular}
\end{table}

\subsection{What intentionality breaks}
\label{c5:ssec:breaks}

The intentional case is not a new dictionary entry but a change of \emph{status}, in which each
quantity that Proposition~\ref{c5:prop:reduction} treats as a parameter of nature becomes a decision
variable of an optimising attacker. Table~\ref{c5:tab:breaks} lists the transformation, and each row names the
result in \S\ref{c5:sec:theorems} that re-proves the corresponding framework result with the best
response inside it. The pattern is that the \emph{measurement} results of the earlier chapters survive
but are re-anchored to a worst case, while the \emph{control} conclusions strengthen. The levers that
were merely ineffective against the systematic-irreversible core (detection, reversibility, fleet
size) become actively counterproductive or incentive-incompatible. The levers that worked
(prevention, segmentation) acquire a second, incentive-theoretic rationale.

\begin{table}[htbp]
\centering
\footnotesize
\setlength{\tabcolsep}{5pt}
\renewcommand{\arraystretch}{1.25}
\caption{What intentionality changes. Each parameter of nature becomes an attacker choice; the right
column names the result that re-proves the framework claim with a best response.}
\label{c5:tab:breaks}
\begin{tabular}{p{0.24\linewidth} p{0.34\linewidth} p{0.32\linewidth}}
\toprule
\textbf{Framework parameter} & \textbf{Becomes (attacker choice)} & \textbf{Result} \\
\midrule
Common-shock probability $\psys$ & Endogenous attack effort, rising with $K$ &
Thm.~\ref{c5:thm:magnet}, Cor.~\ref{c5:cor:antidiv} \\
Attacker objective (unknown) & Type family; tail owned by harm-maximiser & Thm.~\ref{c5:thm:tail} \\
Uniform-random contagion seed & Targeted seed; reinforced edges & Thm.~\ref{c5:thm:targeted} \\
Fixed branching matrix $\Gam$ & Engineered supercriticality within budget & Thm.~\ref{c5:thm:targeted} \\
Ambiguity set / cost $\alpha(Q)$ & Attacker's actual cost technology & Thm.~\ref{c5:thm:robust} \\
Exogenous detection evasion & Effort deflected onto undetectable core & Thm.~\ref{c5:thm:deflection} \\
Catastrophe intensity $\Lamc$ & Best-responding trigger; deterrable by prevention &
Thm.~\ref{c5:thm:deterrence} \\
Per-firm prevention $c_3$ & Nash choice with contagion externality & Thm.~\ref{c5:thm:underinvest} \\
\bottomrule
\end{tabular}
\end{table}

\section{The attacked risk measure}
\label{c5:sec:game}

We now build the intentional case, in which the defender chooses an architecture, the attacker, reading
it, best-responds, and the risk is evaluated at that response. Architecture is an ex-ante, structural
commitment, the object the monograph has argued is the only effective lever. The natural
frame is therefore a Stackelberg game with the defender as leader.

\subsection{Players, moves, and information}
\label{c5:ssec:players}

\begin{definition}[Defender architecture]
\label{c5:def:arch}
The \emph{defender} commits, before any attack, to an architecture
$\pi=(g,m,c_3,N,\Sigma,S_D)$: the systematically-exposed fraction $g$ and segmentation $m$ of
Chapter~\ref{ch:fleet}, the prevention strength $c_3\in[0,1]$ of Chapters~\ref{ch:single}--\ref{ch:continuous}
(the fraction of the catastrophic affordance structurally removed), the sentinel count $N$ and
sampling schedule $\Sigma$ of Chapter~\ref{ch:control}, and a detection budget $S_D\ge0$. The
architecture is public: following Kerckhoffs's principle we assume the attacker knows $\pi$, treating
schedule secrecy as a parameter of $\Sigma$ rather than a standing assumption.
\end{definition}

\begin{definition}[Attacker, effort, and type]
\label{c5:def:attacker}
An \emph{attacker} of \emph{type} $t$ chooses an action $a\in\mathcal A$---an allocation of effort
across targets (shared substrate versus periphery), channels (detectable versus null-space),
contagion seeds, and timing---at a cost $\kappa_t(a)\ge0$, and receives utility
\begin{equation}
U_t(a;\pi)\;=\;\lambda_t\,\E\big[\,\mathcal H(a;\pi)\,\big]\;-\;\kappa_t(a)\;-\;\Pi_t(a),
\label{c5:eq:util}
\end{equation}
where $\mathcal H(a;\pi)$ is the fleet harm the action induces under architecture $\pi$, $\lambda_t>0$
is the type's \emph{harm-orientation} (a disruptor weights harm heavily; a profit-maximiser weights
extractable value, which is bounded and concave in harm), and $\Pi_t(a)\ge0$ is the type's expected
\emph{attribution/retaliation penalty}, convex and increasing in the systemic salience of the target.
The attacker's \emph{best response} is $a^\star_t(\pi)\in\argmax_{a}U_t(a;\pi)$. The attacker's type
is drawn from an admitted family $\mathcal T$ with prior weights $\{\pi_t\}$; the defender knows
$\mathcal T$ but not the realised $t$.
\end{definition}

The type family is the direct reply to the objection that an attacker weighs cost against prize.
A \emph{profit} type has small $\lambda_t$, concave value, and a steep penalty $\Pi_t$ at systemic
targets, so it diversifies across soft peripheral targets. A \emph{disruptor} or \emph{pre-positioning
state} type has large $\lambda_t$ and near-zero $\Pi_t$, so it is close to cost-insensitive at the
systemic target. The defender cannot assume the benign type, and \S\ref{c5:sec:theorems} shows the
tail is governed by the harm-oriented one.

\begin{definition}[Attacked risk measure]
\label{c5:def:attacked}
Let $\rho$ be the deployer's risk functional (the set-valued measure of Chapter~\ref{ch:single} or
its scalar tail specialisation $\ES_\tau$) and $X^{a}$ the harm distribution induced by attacker
action $a$. The \emph{attacked risk measure} at architecture $\pi$ is the risk evaluated at the
worst admitted best response,
\begin{equation}
\rhoatk(\pi)\;=\;\sup_{t\in\mathcal T}\ \rho\big(X^{\,a^\star_t(\pi)}\big).
\label{c5:eq:attacked}
\end{equation}
The defender's problem is $\min_{\pi}\rhoatk(\pi)$ subject to the cost of architecture, a bilevel
(Stackelberg) program.
\end{definition}

\subsection{The attack cost microfounds the ambiguity set}
\label{c5:ssec:microfound}

Chapter~\ref{ch:single} \emph{posited} that adversary uncertainty enters through an ambiguity set
$\Qset_\delta$ of measures penalised by an attack cost $\alpha(Q)\le\delta$. Chapter~\ref{ch:continuous}
took the penalty to be relative entropy, and the game gives that penalty a microfoundation,
identifying it with the attacker's cost technology.

\begin{proposition}[The ambiguity penalty is the attacker's cost]
\label{c5:prop:microfound}
Suppose the attacker's action is a change of measure $Q\ll\Prob$ shifting the harm law from its
baseline $\Prob$, and its cost is $\kappa_t(Q)=\tfrac1{\delta_t}\KL{Q}{\Prob}$ for a type-specific
budget $\delta_t$ (small $\delta_t$ = expensive attacker). Then the type-$t$ best response solves
\begin{equation}
\sup_{Q\ll\Prob}\Big\{\lambda_t\,\E^{Q}[X]-\tfrac1{\delta_t}\KL{Q}{\Prob}\Big\},
\label{c5:eq:varprob}
\end{equation}
and the attacked risk measure \eqref{c5:eq:attacked} coincides with Chapter~\ref{ch:single}'s
distributionally-robust functional with prior set $\Qset_{\delta}$ and cost $\alpha(Q)=\KL{Q}{\Prob}$,
$\delta=\max_t\lambda_t\delta_t$. Thus the ambiguity set is not a modelling convenience but the
reduced form of an attacker's constrained optimisation.
\end{proposition}

\begin{proof}
Apply Donsker--Varadhan duality \citep{dupuis1997weak} to the rescaled variable $\lambda_tX$. Then
\eqref{c5:eq:varprob} is a Gibbs variational problem with value
\begin{equation}
\label{c5:eq:attackervalue}
\sup_{Q\ll\Prob}\Big\{\lambda_t\E^{Q}[X]-\tfrac1{\delta_t}\KL{Q}{\Prob}\Big\}
=\tfrac1{\delta_t}\log\E\big[e^{\lambda_t\delta_t X}\big]
=\lambda_t\,\rho_{\lambda_t\delta_t}(X),
\end{equation}
where $\rho_\delta(X)=\tfrac1\delta\log\E[e^{\delta X}]$ is the entropic risk measure, attained at the
exponential tilt $dQ^\star/d\Prob\propto e^{\lambda_t\delta_t X}$.

The two objects in \eqref{c5:eq:attackervalue} must be kept distinct. The attacker's optimised
\emph{payoff} carries the factor $\lambda_t$. The deployer's attacked risk measure
\eqref{c5:eq:attacked} is the entropic measure $\rho_{\lambda_t\delta_t}(X)$ itself, without it. They
are not equal. But $\lambda_t>0$ is a fixed scale, so the two are maximised by the same $Q^\star$ and
are ordered identically across types, and the attacker's optimum selects the same tilt either way.
Take the supremum over $t\in\mathcal T$ and identify $\delta=\max_t\lambda_t\delta_t$. This gives
\eqref{c5:eq:attacked} as $\rho_\delta$, the entropy-penalised robust functional of
Chapters~\ref{ch:single} and~\ref{ch:continuous}, with the penalty $\alpha(Q)=\KL{Q}{\Prob}$ now read
as the attacker's KL-cost of moving the law. The ambiguity budget is thus identified with the
entropic parameter $\lambda_t\delta_t$, harm-weight times attacker budget, and not with $\delta_t$
alone.
\end{proof}

Proposition~\ref{c5:prop:microfound} is the hinge of the chapter, saying that the robust machinery the
monograph has carried since Chapter~\ref{ch:single} was always the shadow of a game. The ``worst
case'' the defender fears is the attacker's optimum, and the penalty that made the robust problem
well-posed is the attacker's budget. Theorem~\ref{c5:thm:robust} makes the identification exact in
continuous time, and it shows that the finiteness condition of the robust game is the prevention
lever. We now turn to the results that hold once the parameters of nature are attacker-chosen.

\section{Theorems}
\label{c5:sec:theorems}

We work in tractable model classes in which each claim becomes provable. Most claims are paired with a
closed-form quantity that an independent Monte-Carlo study of \S\ref{c5:sec:numerics} reproduces; where a check is a consistency or illustration rather than an independent test, we say so. That is the
monograph's correctness gate. Throughout, the defender's architecture $\pi$ is fixed and the
attacker best-responds as in \S\ref{c5:sec:game}.

\subsection{The monoculture is a magnet}
\label{c5:ssec:magnet}

\begin{assumption}[Two-target attacker]
\label{c5:ass:twotarget}
An attacker of harm-orientation $\lambda>0$ may strike the \emph{shared substrate}---inducing fleet
harm $\lambda M K$ (the common shock touches all $K$ agents), at hardening cost $c_c$ and expected
attribution penalty $\Pi_c$---or a \emph{peripheral} single target---inducing harm $\lambda v$ that
does not scale with $K$, at cost $c_i$ and negligible penalty. Utilities are as in
\eqref{c5:eq:util}.
\end{assumption}

\begin{theorem}[Magnet threshold]
\label{c5:thm:magnet}
Under Assumption~\ref{c5:ass:twotarget} the shared substrate is the attacker's best response if and
only if the fleet exceeds
\begin{equation}
\Kstar\;=\;\frac{c_c+\Pi_c-c_i+\lambda v}{\lambda M},
\label{c5:eq:Kstar}
\end{equation}
so for $K>\Kstar$ a rational attacker concentrates on the shared substrate. The threshold $\Kstar$ is
increasing in substrate hardening $c_c$ and in the attribution penalty $\Pi_c$, and decreasing in the
attacker's harm-orientation $\lambda$; for a cost-insensitive type ($\Pi_c\to0$, $\lambda$ large)
$\Kstar\to v/M$ (a finite constant independent of the cost terms), while for a penalty-sensitive profit type $\Kstar$ is large and the attacker
diversifies to the periphery.
\end{theorem}

\begin{proof}
The substrate is preferred iff $U_{\mathrm{sub}}>U_{\mathrm{per}}$, i.e.\
$\lambda M K-c_c-\Pi_c>\lambda v-c_i$. Both sides are affine in $K$. The substrate side increases in
$K$ (slope $\lambda M>0$) and the periphery side is constant, so the inequality flips at the unique
crossing \eqref{c5:eq:Kstar}. We read the comparative statics off the signs of the partial
derivatives of \eqref{c5:eq:Kstar}:
$\partial\Kstar/\partial c_c=\partial\Kstar/\partial\Pi_c=1/(\lambda M)>0$ and
$\partial\Kstar/\partial\lambda=-(c_c+\Pi_c-c_i)/(\lambda^2 M)<0$ whenever
the fixed cost differential $c_c+\Pi_c-c_i$ is positive, and it is positive because the substrate is
the harder, more consequential target.
\end{proof}

\begin{remark}
Theorem~\ref{c5:thm:magnet} states that systemic importance is endogenous. The shared substrate is
attacked \emph{because} it is systemically important, since impairing it is the only attack whose
payoff scales with the population, so concentration creates a target as well as a vulnerability. The
two levers that raise $\Kstar$, substrate hardening $c_c$ and the attribution penalty $\Pi_c$, are the
ones a supervisor can act on at the level of the shared provider, which is the first-principles case
for designating and directly supervising critical third parties (\S\ref{c5:sec:governance}). The
objection that a rational attacker weighs cost against prize does not refute the attractive-target
result; it restates it, since the substrate becomes the attacker's best response only above $\Kstar$,
and policy moves $\Kstar$ directly.
\end{remark}

\begin{corollary}[Anti-diversification of the fleet floor]
\label{c5:cor:antidiv}
In the magnet regime, let the attacker choose the fleet-wide attack probability $\psys$ to maximise
$\lambda M K\,\psys-\tfrac{\kappa_a}{2}\psys^2$, giving the equilibrium effort
$\psys^\star(K)=\min\!\big(1,\lambda M K/\kappa_a\big)$, which is increasing in $K$. Then the fleet
expected-shortfall floor of Theorem~\ref{c2:thm:nondiv},
\begin{equation}
F(K)\;=\;M\,\min\!\Big(1,\tfrac{\psys^\star(K)}{1-\tau}\Big)+\mu_\xi,
\label{c5:eq:antidiv}
\end{equation}
is non-decreasing in $K$---strictly increasing until it saturates at $M+\mu_\xi$---whereas the
Chapter~\ref{ch:fleet} floor at a \emph{fixed} $\psys$ is constant in $K$. The fleet is not merely
non-diversifiable but \emph{anti-diversifiable}: adding agents adds attacker incentive.
\end{corollary}

\begin{proof}
The attacker's objective is strictly concave in $\psys$, with unconstrained maximiser
$\lambda M K/\kappa_a$. Projecting onto $[0,1]$ gives $\psys^\star(K)=\min(1,\lambda M K/\kappa_a)$,
non-decreasing in $K$. Under the common-shock model of Assumption~\ref{c2:ass:common} with the
attacker-chosen probability, Theorem~\ref{c2:thm:nondiv} gives the floor
$M\,\ES_\tau(\mathbf B)+\mu_\xi=M\min(1,\psys^\star/(1-\tau))+\mu_\xi$, using
$\ES_\tau(\mathrm{Bernoulli}(p))=\min(1,p/(1-\tau))$. Composing two non-decreasing maps
($K\mapsto\psys^\star$ and $\psys\mapsto F$) yields monotonicity in $K$. The floor is flat only where
$\psys^\star$ has saturated at $1$, or where the fixed-$\psys$ baseline is used, in which case
$F$ does not depend on $K$ at all.
\end{proof}

\begin{remark}
Corollary~\ref{c5:cor:antidiv} sharpens the monoculture result of Chapter~\ref{ch:fleet}. There the
fleet floor was non-diversifiable but \emph{constant} in $K$; under a rational attacker it
\emph{rises} with $K$, because the equilibrium attack effort responds to the growing prize. A
supervisory capital or containment charge that is flat in the size of a shared-substrate deployment
therefore understates the risk; the correct charge is increasing in the concentration of reliance,
the cyber analogue of a systemic surcharge that grows with systemic importance.
\end{remark}

\subsection{The magnet lives in the tail}
\label{c5:ssec:tail}

\begin{theorem}[Tail dominance]
\label{c5:thm:tail}
Let the loss be a type mixture: with probability $p_a=\pi_d\psysd$ a harm-oriented (disruptor) type
strikes the substrate and inflicts catastrophic harm $M_{\mathrm{sub}}$, and otherwise a
profit type inflicts a peripheral harm supported on $[0,B]$ with $B<M_{\mathrm{sub}}$. Then for every
tail level $\tau\ge\tau^\star:=1-p_a$,
\begin{equation}
\ES_\tau(\text{loss})\;=\;M_{\mathrm{sub}},
\label{c5:eq:tail}
\end{equation}
independent of the profit type's law. Cost considerations, which shape the body of the loss
distribution by steering profit attackers to the periphery, do not enter the tail: the
$(1-\tau)$-tail consists entirely of substrate catastrophes once $1-\tau\le p_a$.
\end{theorem}

\begin{proof}
The loss has an atom of mass $p_a$ at $M_{\mathrm{sub}}$, which exceeds the entire support $[0,B]$ of
the body. Hence $\VaR_u(\text{loss})=M_{\mathrm{sub}}$ for all $u>1-p_a$. For $\tau\ge1-p_a$ the
averaging interval $(\tau,1)$ lies inside $(1-p_a,1)$, so by the Rockafellar--Uryasev representation
$\ES_\tau=\frac{1}{1-\tau}\int_\tau^1\VaR_u\,du=M_{\mathrm{sub}}$. For $\tau<\tau^\star$ the interval
includes body quantiles and $\ES_\tau<M_{\mathrm{sub}}$. So $\tau^\star=1-p_a$ is exactly the
crossover.
\end{proof}

\begin{remark}
Theorem~\ref{c5:thm:tail} is the reply to the cost objection, made precise. Averaged over attacker
types the loss looks diversified, since most attacks are profitable, peripheral, and absorbed, but the
financial-stability functional does not average; it conditions on the tail, and the tail is owned by
the type for whom the substrate is the best response essentially unconditionally. The practical
corollary is uncomfortable. The observable history of cyber events is generated predominantly by the
body types, so it is \emph{selected against} the systemic scenario, and calibrating tail risk from
incident history is biased for an incentive reason rather than a small-sample one. This is the
adversarial form of the monograph's recurring caution that the systematic parameter is the least
estimable, and it is the argument for scenario-imposed rather than history-fitted stress testing
(\S\ref{c5:sec:governance}).
\end{remark}

\subsection{Targeted contagion and engineered supercriticality}
\label{c5:ssec:targeted}

\begin{theorem}[Targeted contagion]
\label{c5:thm:targeted}
Under the independent-cascade model of Assumption~\ref{c2:ass:cascade} with branching ratio
$\beta$ and giant fraction $S(\beta)$ solving $S=1-e^{-\beta S}$:
\begin{enumerate}[label=\textup{(\roman*)},leftmargin=2.2em,itemsep=1pt]
\item \emph{Targeted seeding.} An attacker who seeds the giant component attains outbreak fraction
$S(\beta)$ with probability one, strictly exceeding the random-seed unconditional outbreak
$S(\beta)^2$ of Chapter~\ref{ch:fleet}; the targeting gain is $S(\beta)\big(1-S(\beta)\big)$.
\item \emph{Engineered supercriticality.} From a subcritical baseline $\beta_0<1$, an attacker who
boosts transmission $q\mapsto q(1+s)$ scales $\beta_0\mapsto\beta_0(1+s)$ and reaches criticality
$\beta=1$ at effort
\begin{equation}
s^\star=\frac{1-\beta_0}{\beta_0},
\label{c5:eq:sstar}
\end{equation}
so the spectral gap $1-\beta_0$ (equivalently $1-\rho(\Gam)$ for the Hawkes branching matrix
$\Gam$ of Chapter~\ref{ch:continuous}) is a \emph{security margin} priced in attacker effort.
\item \emph{Segmentation converts blast radius into a seed budget.} Partitioning the fleet into $m$
isolated segments confines the outbreak from any \emph{single} seed to that segment's giant
component, a fleet fraction $S(\beta)/m$. An attacker able to seed $s\le m$ segments simultaneously
attains $\min\{s\,S(\beta)/m,\,S(\beta)\}$. Segmentation therefore caps total blast radius at
$S(\beta)/m$ only under a unit seed budget; in general it converts blast radius from a property of
the \emph{graph} into a property of the \emph{attacker's seed budget}.
\end{enumerate}
\end{theorem}

\begin{proof}
(i) By the percolation coupling of Theorem~\ref{c2:thm:contagion}, the outbreak from a seed is its
connected component. A uniformly random seed lands in the giant component with probability
$S(\beta)$ and, conditional on doing so, the outbreak is the giant fraction $S(\beta)$. The
unconditional expectation is therefore $S(\beta)\cdot S(\beta)+o(1)=S(\beta)^2$ (the $o(1)$ from
$O(\log K)$ components). An attacker who seeds \emph{into} the giant component realises $S(\beta)$
deterministically, and the difference is $S-S^2=S(1-S)$. (ii) $\beta_0(1+s)=1$ solves to
\eqref{c5:eq:sstar}, and monotonicity of $S$ in $\beta$ gives the emergence of the giant at
$s^\star$. (iii) Transmission cannot cross block boundaries, so an outbreak seeded in one block is
confined to that block's giant component, a fraction $S(\beta)$ of a block of size $K/m$, i.e.\
$S(\beta)/m$ of the fleet. Blocks are disjoint and transmission is independent across them, so $s$
simultaneous seeds in distinct blocks give $s$ such components, whence $sS(\beta)/m$, saturating at
$S(\beta)$ when $s=m$. Note the cap $S(\beta)/m$ is a \emph{per-seed} statement: it is a bound on
total blast radius only when $s=1$.
\end{proof}

\begin{remark}
The contrast between (ii) and (iii) is the chapter's design lesson for networks. Controls that lower
the \emph{average} transmission, such as model heterogeneity or generic hardening, raise the attacker's cost
of engineering supercriticality but can be overcome within budget by reinforcing edges along the
dominant eigen-direction; the binding design quantity is therefore the \emph{worst-case} branching
ratio, not the average. Segmentation, by contrast, caps the per-seed blast radius at $S/m$ regardless of how
the attacker reinforces \emph{within} a block. It is the lever that survives intent, which
is why compartmentalisation, not diversity alone, is the macroprudential control against contagion
(\citealp{gai2010contagion}; Chapter~\ref{ch:fleet}).

The qualification matters against the adversary this chapter posits. An agentic attacker
whose reconnaissance and execution costs have collapsed is not restricted to one seed; if it can
place a seed in every segment it recovers $S(\beta)$, and segmentation does not lower the blast radius.
Segmentation still raises \emph{cost}, since the adversary must obtain $m$ independent footholds
rather than one. The operative design question is therefore not the segment count alone but whether
the segments are independent to seed, with separate credentials, separate identity planes,
separate supply chains. Segmentation without credential separation partitions the transmission graph
while leaving the seeding channel shared, and against a multi-target attacker that is close to no
segmentation at all.
\end{remark}

\subsection{The adversarial worst case is the robust game}
\label{c5:ssec:robust}

\begin{theorem}[Intentional attack equals the robust entropic measure]
\label{c5:thm:robust}
Let cumulative fleet loss be the total loss $\Loss_T=\Cont_T+\Ham_T$ of
Chapter~\ref{ch:continuous} (equation~\eqref{c3:eq:totalloss}), with constant coefficients as in
Theorem~\ref{c3:thm:robust}, and let the
attacker change measure at KL-cost with budget $\delta$ (Proposition~\ref{c5:prop:microfound}). Then
the attacked risk measure \eqref{c5:eq:attacked} equals the time-consistent entropic measure
\begin{equation}
\rhoatk=\rho_\delta(\Loss_T)=\tfrac1\delta\log\E\big[e^{\delta\Loss_T}\big]
=\mu T+\tfrac{\delta}{2}\sigma^2T+\frac{\Lambda T R}{\,b-\delta R\,},
\label{c5:eq:robust}
\end{equation}
attained at the Girsanov drift shift and Esscher jump tilt $\Lev^\star(dz)=e^{\delta R z}\Lev(dz)$ of
Theorem~\ref{c3:thm:robust}: the rational attacker makes harmful jumps both more frequent and more
severe. The value is finite \emph{if and only if} $\delta R<b$, i.e.\ $\int e^{\delta R z}\Lev(dz)<\infty$;
truncating the jump measure---the prevention lever of Theorems~\ref{c1:thm:prevention}
and~\ref{c3:thm:t3}---is what renders the attacker's problem finite.
\end{theorem}

\begin{proof}
By Proposition~\ref{c5:prop:microfound} the attacker's best response is the exponential tilt solving
the Gibbs variational problem, whose value is $\tfrac1\delta\log\E[e^{\delta\Ham_T}]$. This is exactly
the object of Theorem~\ref{c3:thm:robust}, which establishes both the closed form
\eqref{c5:eq:robust} and the finiteness condition $\delta R<b$. Taking the supremum over the admitted
type family in \eqref{c5:eq:attacked} selects the largest admissible $\delta$, preserving the
identity.
\end{proof}

\begin{remark}
Theorem~\ref{c5:thm:robust} completes the hinge of Proposition~\ref{c5:prop:microfound}. The robust
machinery the monograph has carried since Chapter~\ref{ch:single} is the reduced form of an attacker's
optimisation, and the well-posedness condition of the robust game is the prevention lever. The reading
is stark, since an \emph{unbounded} catastrophic affordance is an \emph{unbounded prize}, and against a
rational attacker the robust value is $+\infty$ until prevention has truncated the tail. Prevention is
thus not one robustness knob among several; it is the condition under which the fleet is robustly
governable at all.
\end{remark}

\subsection{Detection deflects effort onto the irreversible core}
\label{c5:ssec:deflection}

\begin{theorem}[Detection deflection]
\label{c5:thm:deflection}
Suppose the attacker allocates a unit effort budget across a \emph{detectable} channel with
undetected return $a_D(S_D)=h_D e^{-S_D}$ (decreasing in the defender's detection spend $S_D$) and an
\emph{undetectable} null-space channel---the home of the irreversible jump in
Chapter~\ref{ch:control}'s trichotomy---with return $a_U=h_U$, under decreasing-returns production
with exponent $\alpha\in(0,1)$. The equilibrium effort share on the undetectable channel is
\begin{equation}
\sigma_U(S_D)\;=\;\frac{1}{1+\big(a_D(S_D)/a_U\big)^{1/(1-\alpha)}},
\label{c5:eq:sigmaU}
\end{equation}
which is strictly increasing in $S_D$ and tends to $1$; consequently the equilibrium harm through the
undetectable/irreversible channel, $a_U\,\sigma_U(S_D)^{\alpha}$, is strictly increasing in detection
spend. Detection is not neutral toward the catastrophic core: it \emph{deflects} attacker effort into
it.
\end{theorem}

\begin{proof}
The attacker maximises $a_D e_D^\alpha+a_U e_U^\alpha$ subject to $e_D+e_U=1$. The first-order
condition $a_D\alpha e_D^{\alpha-1}=a_U\alpha e_U^{\alpha-1}$ gives
$e_D/e_U=(a_D/a_U)^{1/(1-\alpha)}$, hence \eqref{c5:eq:sigmaU} for $\sigma_U=e_U/(e_D+e_U)=e_U$. Now
$a_D(S_D)=h_D e^{-S_D}$ is strictly decreasing in $S_D$ while $a_U$ is fixed, so the ratio $a_D/a_U$
decreases and $\sigma_U$ increases toward $1$. The undetectable harm $a_U\sigma_U^\alpha$ is an
increasing function of $\sigma_U$, hence of $S_D$.
\end{proof}

\begin{remark}
Theorem~\ref{c5:thm:deflection} is the dynamic, strategic sharpening of the non-substitutability
theorems. Chapters~\ref{ch:single} and~\ref{ch:control} showed detection \emph{cannot lower} the
irreversible floor; here detection \emph{raises} the harm carried through the irreversible core,
because a strategic attacker re-allocates away from the channels detection covers and into the ones it
cannot. This explains a pattern supervisors observe. Cyber budgets weighted toward detection and
response reduce the count and cost of \emph{caught} incidents while leaving, or worsening, the
systemic tail, which flows through the undetectable, irreversible directions. Detection
manages the body; it feeds the tail.
\end{remark}

\subsection{Runtime impossibility, and prevention as denial}
\label{c5:ssec:deterrence}

\begin{theorem}[Impossibility under intent; mechanical and deterrence value of prevention]
\label{c5:thm:deterrence}
Against an irreversible catastrophic jump:
\begin{enumerate}[label=\textup{(\roman*)},leftmargin=2.2em,itemsep=1pt]
\item \emph{Robustness to intent.} For every attacker trigger and every runtime policy, the
catastrophe exit probability is $1-e^{-\Lambda_c T}$ with $\Lambda_c$ at least the passive-Poisson
intensity, and is invariant across runtime controls (Corollary~\ref{c4:thm:t6b}, a fortiori): an optimising trigger is
no easier to catch than a Poisson one, and a causal detect-then-react loop cannot fire before a
zero-duration event.
\item \emph{Two channels of prevention.} With equilibrium hazard
$\Lambda_c^{\mathrm{eq}}(c_3)=\Lambda_c^{0}(1-c_3)\,a^\star(c_3)$ and rare-event attacker effort
$a^\star(c_3)=V\Lambda_c^{0}T(1-c_3)/\kappa_a$,
\begin{equation}
\frac{d\Lambda_c^{\mathrm{eq}}}{dc_3}
=\underbrace{-\,\Lambda_c^{0}\,a^\star(c_3)}_{\text{mechanical}}
\;+\;\underbrace{\Lambda_c^{0}(1-c_3)\,\frac{da^\star}{dc_3}}_{\text{deterrence}},
\label{c5:eq:decomp}
\end{equation}
both terms negative and, in this model, equal: prevention lowers the floor by removing the affordance
(mechanical) \emph{and} by removing the prize, which lowers the equilibrium attack effort
(deterrence). Detection has zero mechanical term (by (i)) and a non-negative deflection term (by
Theorem~\ref{c5:thm:deflection}), so only prevention moves the floor.
\end{enumerate}
\end{theorem}

\begin{proof}
(i) The attacker may always choose the Poisson trigger, so the supremum over triggers of the exit
probability is at least the Poisson value $1-e^{-\Lambda_c T}$. Corollary~\ref{c4:thm:t6b}
shows this value is identical across runtime policies, because a causal loop cannot pre-empt a
zero-duration event. An optimising trigger only concentrates mass earlier, weakly raising $\Lambda_c$,
which does not depend on the runtime control either. (ii) Differentiate the product
$\Lambda_c^{0}(1-c_3)a^\star(c_3)$: the first term holds $a^\star$ fixed (mechanical), the second
holds $(1-c_3)$ fixed (deterrence). Substituting $a^\star(c_3)=V\Lambda_c^{0}T(1-c_3)/\kappa_a$ gives
mechanical $=-\Lambda_c^{0}\cdot V\Lambda_c^{0}T(1-c_3)/\kappa_a$ and deterrence
$=\Lambda_c^{0}(1-c_3)\cdot(-V\Lambda_c^{0}T/\kappa_a)$, which are equal and negative. Their sum is
$-2V(\Lambda_c^{0})^2T(1-c_3)/\kappa_a=d\Lambda_c^{\mathrm{eq}}/dc_3$.
\end{proof}

\begin{remark}
Equation~\eqref{c5:eq:decomp} formalises \emph{deterrence by denial}. A preventive control that
removes a catastrophic affordance both makes the event less likely mechanically and makes it a less
attractive target, and against the irreversible core these are the only two ways the floor moves. A
preventive control such as a pre-trade hard limit at the order gateway, a revoked production
credential, or a payment scope withheld from an agent removes the affordance and, simultaneously, the
prize, which is why such controls dominate post-hoc monitoring not only on the physics
(Chapter~\ref{ch:control}) but on the incentives.
\end{remark}

\subsection{Decentralized underinvestment}
\label{c5:ssec:underinvest}

\begin{theorem}[Decentralized underinvestment: the wedge is a centrality]
\label{c5:thm:underinvest}
Let $K$ institutions sit on a directed contagion graph $A$ ($A_{ij}=1$ if a compromise at $i$ can
propagate to $j$), and let a compromise cascade through \emph{all} paths, so firm $i$'s hazard is the
resolvent (Neumann) series
\begin{equation}
\Lambda(c)=\lambda_0\,(I-tA^{\!\top})^{-1}(\mathbf 1-c),\qquad t\,\rho(A)<1,
\label{c5:eq:resolvent}
\end{equation}
with contagion strength $t$; firm $i$ chooses prevention $c_i\in[0,1]$ at convex cost
$\tfrac\kappa2 c_i^2$ and expected loss $w\Lambda_i$. Write $M=(I-tA^{\!\top})^{-1}$ for the resolvent
(so $M_{ii}$ is its diagonal, counting closed walks through $i$) and
$z_i=\big[(I-tA)^{-1}\mathbf 1\big]_i$ for the \emph{Katz centrality} of $i$ (its reach through all
paths), in the sense of \citet{katz1953new} and its attenuated generalisation
\citep{bonacich1987power}. Then
\begin{equation}
c_i^{\mathrm{Nash}}=\mathrm{clip}\!\Big(\tfrac{w\lambda_0}{\kappa}\,M_{ii}\Big),\qquad
c_i^{\mathrm{SO}}=\mathrm{clip}\!\Big(\tfrac{w\lambda_0}{\kappa}\,z_i\Big),
\label{c5:eq:nash-so}
\end{equation}
and the \emph{underinvestment wedge} is, on the interior,
\begin{equation}
c_i^{\mathrm{SO}}-c_i^{\mathrm{Nash}}=\frac{w\lambda_0}{\kappa}\,(z_i-M_{ii})
=\frac{w\lambda_0}{\kappa}\sum_{j\neq i}\big[(I-tA)^{-1}\big]_{ij},
\label{c5:eq:wedge}
\end{equation}
firm $i$'s reach to \emph{others} through all paths. The wedge is non-negative, zero if and only if
$i$ reaches no one ($z_i=M_{ii}$, isolated) or contagion is severed ($t=0$, full segmentation), and
increasing in connectivity. To first order $z_i-M_{ii}=t\,\mathrm{outdeg}_i+O(t^2)$, recovering the
out-degree wedge; and as $t\to1/\rho(A)$ the Katz centrality aligns with the eigenvector (Perron)
centrality of $A$. A mandated prevention floor $\bar c_i=c_i^{\mathrm{SO}}$ implements the social
optimum.
\end{theorem}

\begin{proof}
\emph{Nash.} Firm $i$ minimises $J_i=w\Lambda_i+\tfrac\kappa2 c_i^2$ with
$\Lambda_i=\lambda_0\sum_k M_{ik}(1-c_k)$, so $\partial\Lambda_i/\partial c_i=-\lambda_0 M_{ii}$ and
$\partial J_i/\partial c_i=-w\lambda_0 M_{ii}+\kappa c_i$. Since $M_{ii}$ is a constant of the graph
(not of $c$), the best response $c_i=w\lambda_0 M_{ii}/\kappa$ is independent of the others' choices
and is the unique Nash profile before clipping. A firm internalises only the contagion that
\emph{returns} to it, i.e.\ the closed walks counted by $M_{ii}$. \emph{Social optimum.} The planner
minimises $\sum_i J_i=w\lambda_0\,\mathbf 1^{\!\top}M(\mathbf 1-c)+\tfrac\kappa2\|c\|^2$. The gradient
is $\partial/\partial c_i\big[\mathbf 1^{\!\top}M(\mathbf 1-c)\big]=-(M^{\!\top}\mathbf 1)_i$, and
$M^{\!\top}=\big[(I-tA^{\!\top})^{-1}\big]^{\!\top}=(I-tA)^{-1}$, so $(M^{\!\top}\mathbf 1)_i=z_i$, the
Katz centrality. Hence $c_i^{\mathrm{SO}}=w\lambda_0 z_i/\kappa$. The objective is a box-constrained
convex quadratic with diagonal Hessian $\kappa I$, so the coordinatewise clip to $[0,1]$ is exact. The
wedge \eqref{c5:eq:wedge} is the difference. Here $z_i-M_{ii}=\sum_{j\neq i}[(I-tA)^{-1}]_{ij}\ge0$ is
the open-walk reach: its expansion $(I-tA)^{-1}=I+tA+t^2A^2+\cdots$ gives $t\,\mathrm{outdeg}_i+O(t^2)$,
and its $t\to1/\rho(A)$ limit is dominated by the Perron eigenvector. A floor $\bar c=c^{\mathrm{SO}}$
makes the planner's optimum feasible and binding.
\end{proof}

\begin{remark}
Theorem~\ref{c5:thm:underinvest} is the incentive half of the
monograph's thesis. Chapters~\ref{ch:single}--\ref{ch:control} proved that only prevention moves the
systematic-irreversible core; this theorem proves that a population of institutions, left to
decentralized choice, \emph{underinvests} in that prevention, by a wedge equal to each firm's
\emph{centrality}, its reach to others through the contagion network, a Katz centrality that becomes
the eigenvector centrality as contagion strengthens. The mechanism is exact. A firm prevents against
the contagion that loops back to itself (the resolvent diagonal $M_{ii}$) but not against the
contagion it inflicts on everyone else (the off-diagonal reach), and the gap is the systemic
externality. Prevention is thus a public good under contagion, undersupplied where the network
is densest, and the wedge is closed only by severing the coupling ($t=0$, segmentation) or by a
mandated floor. This is why cyber financial stability, like banking stability, cannot be secured one
institution at a time even by willing institutions, and why a supervisor sizing a prevention mandate
should scale it to a firm's network centrality, not its standalone risk.
\end{remark}

\begin{remark}[Relation to the literature, and one difference that matters]
\label{c5:rem:literature}
The result sits alongside three strands, and we are precise about where it agrees and where it does
not.

It is a security-underprovision result in the tradition of \citet{varian2004system}, where private
investment in a shared reliability technology is undersupplied because the benefit spills over, and it
inherits the network-externality reading of financial contagion in
\citet{acemoglu2015systemic}. Centrality-proportional equilibria are of course familiar from network
games \citep{ballester2006whos}, where equilibrium effort is proportional to Bonacich centrality and
the planner's ``key player'' is the one whose removal most reduces aggregate activity.

The mechanism here differs, and the difference cuts against the more common framing. In
\citet{ballester2006whos} actions are \emph{strategic complements}, where centrality enters because each
player's effort raises its neighbours' returns, and the equilibrium is a fixed point of that feedback.
In \eqref{c5:eq:nash-so} the best response $c_i^{\mathrm{Nash}}=w\lambda_0M_{ii}/\kappa$ does not
depend on $c_{-i}$ at all. Prevention choices here are \emph{dominant strategies}, and centrality
enters only through the \emph{planner's} objective, as the reach the firm fails to internalise. The
inefficiency is therefore a pure Pigouvian externality, not a coordination failure.

That distinction strengthens rather than weakens the policy conclusion. A coordination failure can in
principle be resolved by coordination, by convening the institutions, aligning expectations, or
tipping the population to a better equilibrium, which is the standard reading of interdependent
security \citep{kunreuther2003interdependent}. Here there is no better equilibrium to tip to. Each
firm is already playing its dominant strategy, and it will continue to do so however its neighbours
behave. The wedge closes only by severing the coupling or by mandating the floor.
\end{remark}

\subsection{The defender is augmented too}
\label{c5:ssec:defense}

Agentic AI arms the defender as well as the attacker: autonomous monitoring, triage, response, and
patching. It is tempting to read this as restoring the symmetry the intentional case broke, but it
does not, for a reason that is a corollary of results already proved. Agentic defense extends the defender's
reach across the reversible body but not into the systematic-irreversible core, unless it is aimed at
the prevention lever. A fleet of homogeneous AI monitors also inherits the monoculture it was meant to
watch.

\begin{corollary}[Where defensive AI reaches]
\label{c5:cor:defense-reach}
Let defensive-AI capability be applied either to the runtime \emph{detection} lever (raising the
detection spend $S_D$) or to the ex-ante \emph{prevention} lever (raising $c_3$). Then:
\begin{enumerate}[label=\textup{(\roman*)},leftmargin=2.2em,itemsep=1pt]
\item on detection, it lowers the harm carried through the detectable, reversible body but, by the
deflection result of Theorem~\ref{c5:thm:deflection}, strictly \emph{raises} the equilibrium harm
through the undetectable, irreversible core;
\item on prevention, it lowers the equilibrium catastrophe hazard $\Lambda_c^{\mathrm{eq}}(c_3)$ and
hence the tail harm, through both the mechanical and the deterrence channels of
Theorem~\ref{c5:thm:deterrence}.
\end{enumerate}
Consequently defensive AI reaches the systemic-irreversible core \emph{only} through the prevention
lever; deployed on detection and response, it manages the body and feeds the tail.
\end{corollary}

\begin{proof}
Both parts are the earlier theorems with the defender's control reinterpreted as an AI capability.
(i) is Theorem~\ref{c5:thm:deflection} with $S_D$ read as detection capability: $\sigma_U(S_D)$ and
the undetectable-core harm $a_U\sigma_U(S_D)^\alpha$ are strictly increasing in $S_D$, while the
detectable-channel harm falls. (ii) is Theorem~\ref{c5:thm:deterrence}: $\Lambda_c^{\mathrm{eq}}$ is
decreasing in $c_3$ with the two-channel derivative \eqref{c5:eq:decomp}, and detection has no
mechanical term against the jump.
\end{proof}

\begin{corollary}[The defensive monoculture]
\label{c5:cor:defense-monoculture}
A fleet of $N$ AI monitors that share a foundation model shares a blind direction: their
observation-noise correlation along that direction is $r>0$. The fused (averaged) detector has
variance
\begin{equation}
\Var_N(r)=\sigma_o^2\Big(\tfrac{1-r}{N}+r\Big)\ \xrightarrow[N\to\infty]{}\ r\,\sigma_o^2>0,
\label{c5:eq:defmono}
\end{equation}
a non-diversifiable detection floor independent of $N$---exactly the correlated-sensor floor of
Theorem~\ref{c4:thm:diversity}. Only monitor heterogeneity ($r\to0$: genuinely different observation
substrates) diversifies it away; adding more copies of the same AI monitor buys nothing along the
shared blind direction.
\end{corollary}

\begin{proof}
The average of $N$ unit sensors with common variance $\sigma_o^2$ and pairwise correlation $r$ has
variance $\sigma_o^2\big[\tfrac1N+\tfrac{N-1}{N}r\big]=\sigma_o^2\big(\tfrac{1-r}{N}+r\big)$, whose
limit is $r\sigma_o^2$. This is Theorem~\ref{c4:thm:diversity} applied to a monitor fleet whose shared
foundation model is the source of the common blind direction.
\end{proof}

\begin{remark}[Offense-dominance at the tail]
Corollaries~\ref{c5:cor:defense-reach} and~\ref{c5:cor:defense-monoculture} settle the symmetry
question. Agentic AI extends the \emph{attacker's} reach into the systematic-irreversible core: the
attractive-target effect (Theorem~\ref{c5:thm:magnet}), targeted contagion (Theorem~\ref{c5:thm:targeted}), and
deflection (Theorem~\ref{c5:thm:deflection}) all worsen as the attacker is better augmented. It
extends the \emph{defender's} reach only across the reversible body, unless the defender points its AI
at prevention, and it hands the defender its own monoculture. Agentic AI is therefore
\emph{offense-dominant at the systemic tail}, even where it may be defense-favouring in the
body. The operational corollaries are two: aim defensive AI at the prevention lever (affordance
discovery and removal, configuration minimisation, pre-commitment hard limits), not at detection and
response, if the goal is systemic; and diversify the monitor fleet's substrates rather than scaling
copies of one. Both are verified in \S\ref{c5:ssec:num-defense}.
\end{remark}

\subsection{The two-sided game: optimal defensive budget allocation}
\label{c5:ssec:budget}

\S\ref{c5:sec:game} posed the defender's problem as $\min_\pi\rhoatk(\pi)$ over the architecture.
That problem has so far been read one lever at a time. We now solve it for a defender who allocates
a fixed budget between the two levers whose effects the chapter has pinned down, and we find that the
optimal allocation is governed by a threshold on the attacker's capability. This closes the static
two-sided game. The dynamic version, in which capabilities themselves co-evolve, is left to
\S\ref{c5:sec:conclusions}.

\begin{theorem}[Optimal defensive budget allocation]
\label{c5:thm:budget}
A defender with budget $B$ splits it between detection $S_D$ and prevention $S_P=B-S_D$,
$S_D\in[0,B]$. Composing the chapter's mechanisms---detection lowers the reversible body harm
$b_0e^{-S_D}$; prevention lowers the squared catastrophe hazard $e^{-2S_P}$ (Theorem~\ref{c5:thm:deterrence});
and detection deflects effort onto the irreversible core (Theorem~\ref{c5:thm:deflection}), inflating
the tail by a factor $1+\delta S_D$---the total attacked risk is
\begin{equation}
\rho(S_D)=b_0e^{-S_D}+\lambda\,(1+\delta S_D)\,e^{-2(B-S_D)},
\label{c5:eq:budget}
\end{equation}
with $\lambda$ the attacker capability and $\delta>0$ the deflection sensitivity. Then $\rho$ is
strictly convex on $[0,B]$, and there is a threshold capability
\begin{equation}
\lambda^\star=\frac{b_0\,e^{2B}}{2+\delta}
\label{c5:eq:lamstar}
\end{equation}
such that the optimal detection spend is the corner $S_D^\star=0$---the \emph{entire} budget to
prevention---whenever $\lambda\ge\lambda^\star$, and is interior, $S_D^\star\in(0,B)$, only when
$\lambda<\lambda^\star$. The threshold $\lambda^\star$ is strictly decreasing in the deflection
sensitivity $\delta$: the more detection feeds the tail, the lower the attacker capability at which
detection budget should be cut to zero. For $\lambda\ge\lambda^\star$, any positive detection spend
strictly increases total risk.
\end{theorem}

\begin{proof}
The body term $b_0e^{-S_D}$ is convex. The tail term $\lambda(1+\delta S_D)e^{2S_D}e^{-2B}$ is a
positive combination of $e^{2S_D}$ and $S_De^{2S_D}$, both convex. So $\rho$ is strictly convex, and
its minimiser over $[0,B]$ is either the interior stationary point or the corner $S_D=0$. Differentiating,
\[
\rho'(S_D)=-b_0e^{-S_D}+\lambda e^{-2(B-S_D)}\big[\delta+2(1+\delta S_D)\big],
\qquad
\rho'(0)=-b_0+\lambda e^{-2B}(2+\delta).
\]
By convexity the optimum is the corner $S_D^\star=0$ iff $\rho'(0)\ge0$, i.e.\ iff
$\lambda\ge b_0e^{2B}/(2+\delta)=\lambda^\star$. Otherwise $\rho'(0)<0$ and the stationary point is
interior. Monotonicity in $\delta$ is immediate from \eqref{c5:eq:lamstar}
($\partial\lambda^\star/\partial\delta=-b_0e^{2B}/(2+\delta)^2<0$). For $\lambda\ge\lambda^\star$,
convexity and $\rho'(0)\ge0$ give $\rho'>0$ on $(0,B]$, so $\rho(S_D)>\rho(0)$ for every $S_D>0$: any
detection spend strictly raises total risk.
\end{proof}

\begin{remark}
Theorem~\ref{c5:thm:budget} closes the two-sided game and turns the chapter's qualitative advice into
a quantitative rule. The defender is now a genuine optimiser, not a lever-holder, and past a threshold
attacker capability its optimum has a corner at which the systemically-optimal detection budget is
\emph{zero}, and every marginal dollar belongs on prevention and on severing the common exposure. The
comparative static is the supervisory takeaway, \emph{as the adversary is more capable, or as
detection deflects more sharply, defund detection earlier}, and it inverts the instinct to meet a
rising threat with more monitoring. What the theorem does \emph{not} yet model is that $\lambda$ and
the defender's own capability are themselves choices that evolve over time; making them state
variables is the dynamic co-evolution game deferred to \S\ref{c5:sec:conclusions}.
\end{remark}

\section{Numerical illustration}
\label{c5:sec:numerics}

The studies below are deliberately \emph{unitless}. The common-shock magnitude $M$ (or the
catastrophic magnitude $M_{\mathrm{sub}}$) is the unit of harm, efforts and budgets are measured in
those units, and graphs are dimensionless. Each study below reports its closed-form prediction, the
simulated value, and their discrepancy. Domain-specific readings appear in \S\ref{c5:sec:apps}.

\subsection{The magnet, anti-diversification, and the tail (Theorem~\ref{c5:thm:magnet})}
\label{c5:ssec:num-magnet}

\begin{figure}[htbp]
\centering
\includegraphics[width=\linewidth]{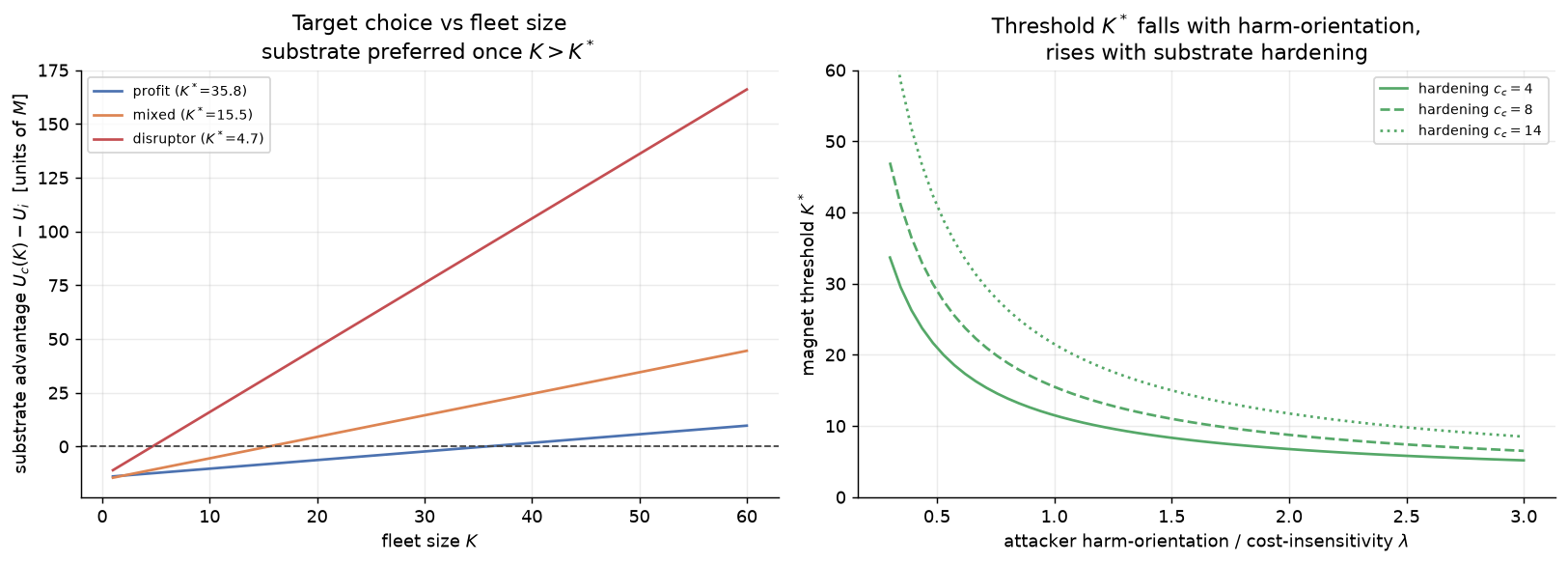}
\caption{\textbf{The magnet threshold (Theorem~\ref{c5:thm:magnet}).} Left: the substrate becomes the
attacker's best response once $K>\Kstar$; the disruptor type magnetises at a far smaller fleet than
the profit type. Right: $\Kstar$ falls with harm-orientation $\lambda$ and rises with substrate
hardening $c_c$.}
\label{c5:fig:magnet}
\end{figure}

Figure~\ref{c5:fig:magnet} verifies Theorems~\ref{c5:thm:magnet}--\ref{c5:thm:tail}. In the left
panel the substrate advantage $U_{\mathrm{sub}}(K)-U_{\mathrm{per}}$ crosses zero at the closed-form
threshold $\Kstar$ of \eqref{c5:eq:Kstar}. A penalty-sensitive profit type magnetises only at
$\Kstar=35.8$, a cost-insensitive disruptor already at $\Kstar=4.7$. The simulated first-$K$ at
which the substrate is chosen matches to the integer. The right panel plots $\Kstar$ against
harm-orientation $\lambda$ for three hardening levels, confirming $\Kstar$ falls with $\lambda$ and
rises with substrate hardening $c_c$.

\begin{figure}[htbp]
\centering
\includegraphics[width=\linewidth]{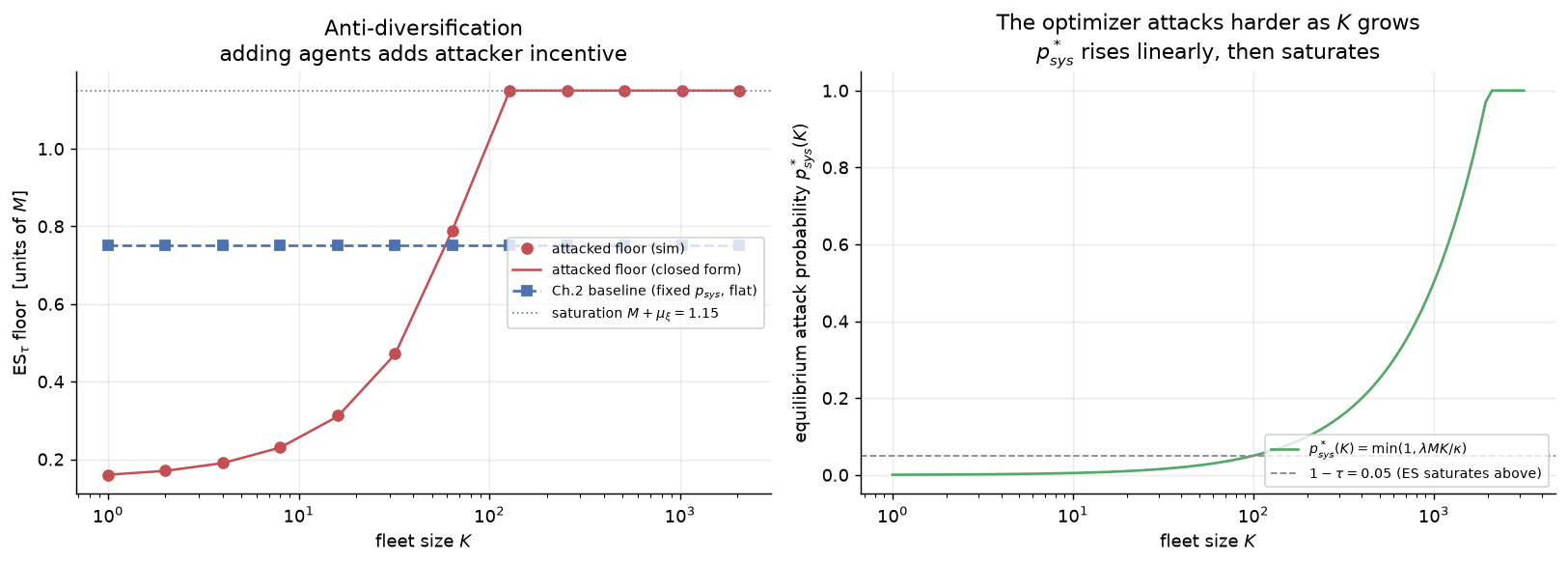}
\caption{\textbf{Anti-diversification (Corollary~\ref{c5:cor:antidiv}).} Left: under a rational
attacker the fleet ES floor \emph{rises} with $K$ (simulation vs closed form), crossing the flat
Chapter~\ref{ch:fleet} baseline, and saturating at $M+\mu_\xi$. Right: the equilibrium attack
probability $\psys^\star(K)$ increases with fleet size, since the optimiser attacks harder as the
prize grows.}
\label{c5:fig:antidiv}
\end{figure}

Figure~\ref{c5:fig:antidiv} verifies Corollary~\ref{c5:cor:antidiv}. With $\lambda M=1$,
$\kappa_a=2000$, $\mu_\xi=0.15$, $\tau=0.95$, the equilibrium attack probability
$\psys^\star(K)=\min(1,\lambda M K/\kappa_a)$ rises with $K$. The fleet floor \eqref{c5:eq:antidiv}
climbs from $0.16$ at $K=1$ to the saturation $M+\mu_\xi=1.15$, and the simulation tracks the closed
form to a maximum error of $0.004$. The Chapter~\ref{ch:fleet} baseline at fixed $\psys=0.03$ stays
flat at $0.75$. Adding agents raises the floor, the anti-diversification signature.

\begin{figure}[htbp]
\centering
\includegraphics[width=\linewidth]{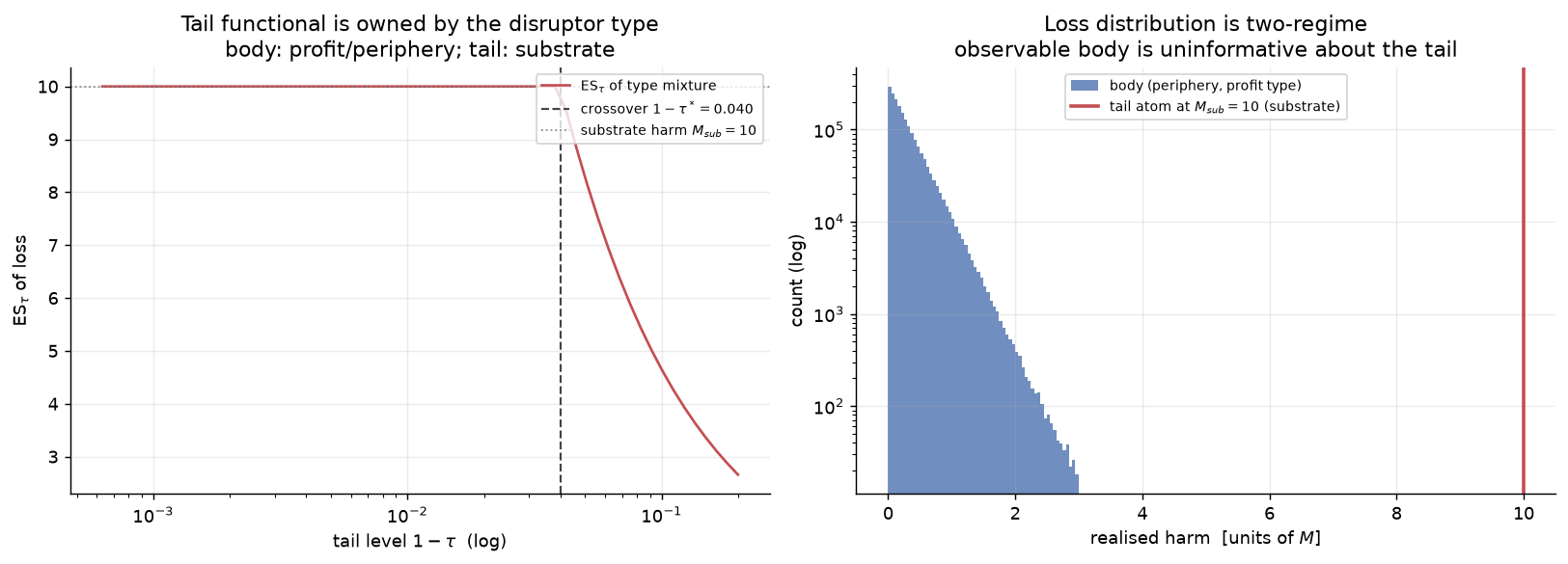}
\caption{\textbf{Tail dominance (Theorem~\ref{c5:thm:tail}).} Left: $\ES_\tau$ of the attacker-type
mixture converges to the substrate harm $M_{\mathrm{sub}}$ for $\tau\ge\tau^\star=1-\pi_d\psysd$;
cost thins the body off the substrate, not the tail. Right: the two-regime loss distribution, a
profit-driven body and a disruptor-driven tail atom.}
\label{c5:fig:tail}
\end{figure}

Figure~\ref{c5:fig:tail} verifies Theorem~\ref{c5:thm:tail}. With disruptor weight $\pi_d=0.05$ and
substrate-success $\psysd=0.8$, the catastrophe atom has mass $p_a=0.04$, so the crossover is
$\tau^\star=1-p_a=0.96$. For $\tau\ge0.975$ the simulated $\ES_\tau$ locks onto
$M_{\mathrm{sub}}=10$, since the entire tail is substrate catastrophes by then. For $\tau=0.95<\tau^\star$
it is $8.27$, still mixing in the body. The loss distribution is two-regime (right panel): a body of
peripheral, profit-driven losses and a tail atom that the body is uninformative about.

\subsection{Targeted contagion and engineered supercriticality (Theorem~\ref{c5:thm:targeted})}
\label{c5:ssec:num-targeted}

\begin{figure}[htbp]
\centering
\includegraphics[width=\linewidth]{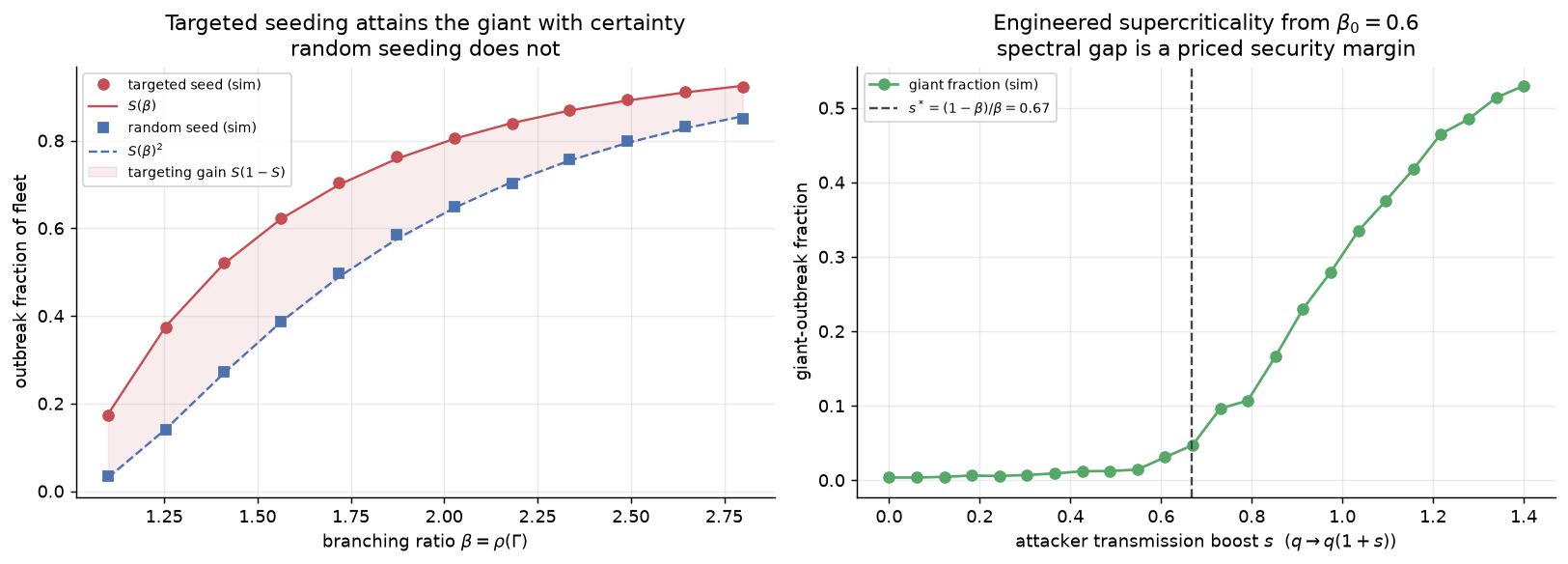}
\caption{\textbf{Targeted contagion (Theorem~\ref{c5:thm:targeted}).} Left: targeted seeding attains
the giant $S(\beta)$ with certainty, versus the random-seed $S(\beta)^2$; the gap is the targeting
gain $S(1-S)$. Right: engineered supercriticality. From a subcritical $\beta_0$, the giant emerges
at the closed-form transmission boost $s^\star=(1-\beta_0)/\beta_0$, so the spectral gap is a priced
security margin.}
\label{c5:fig:targeted}
\end{figure}

Figure~\ref{c5:fig:targeted} verifies Theorem~\ref{c5:thm:targeted}. Across branching ratios
$\beta\in[1.1,2.8]$ on $G(K,\beta/K)$ at $K=8000$, targeted seeding attains the giant fraction
$S(\beta)$ (simulation within $0.006$ of the analytic $S$). Random seeding attains $S(\beta)^2$
(within $0.009$), and the shaded gap is the targeting gain $S(1-S)$. From a subcritical $\beta_0=0.6$
(spectral gap $0.4$), boosting transmission drives the giant component to emerge at
$s\approx0.61$, matching the closed-form criticality budget $s^\star=(1-\beta_0)/\beta_0=0.667$.

Segmentation into $m$ isolated blocks caps a \emph{single-seed} targeted outbreak at $S/m$. At
$\beta=2$ ($S=0.797$) the simulated single-seed targeted fleet-outbreak is
$0.80,0.40,0.20,0.10,0.05$ for $m=1,2,4,8,16$, matching $S/m$ to sub-percent error. These runs seed
one block. Under Theorem~\ref{c5:thm:targeted}(iii) an attacker seeding $s$ blocks attains
$\min\{sS/m,\,S\}$, so the tabulated figures scale linearly in $s$ up to the unsegmented value
$S=0.797$ at $s=m$. The column is a per-seed cap, not a bound on an adversary with a multi-block
seed budget.

\subsection{Detection deflection onto the irreversible core (Theorem~\ref{c5:thm:deflection})}
\label{c5:ssec:num-deflection}

\begin{figure}[htbp]
\centering
\includegraphics[width=\linewidth]{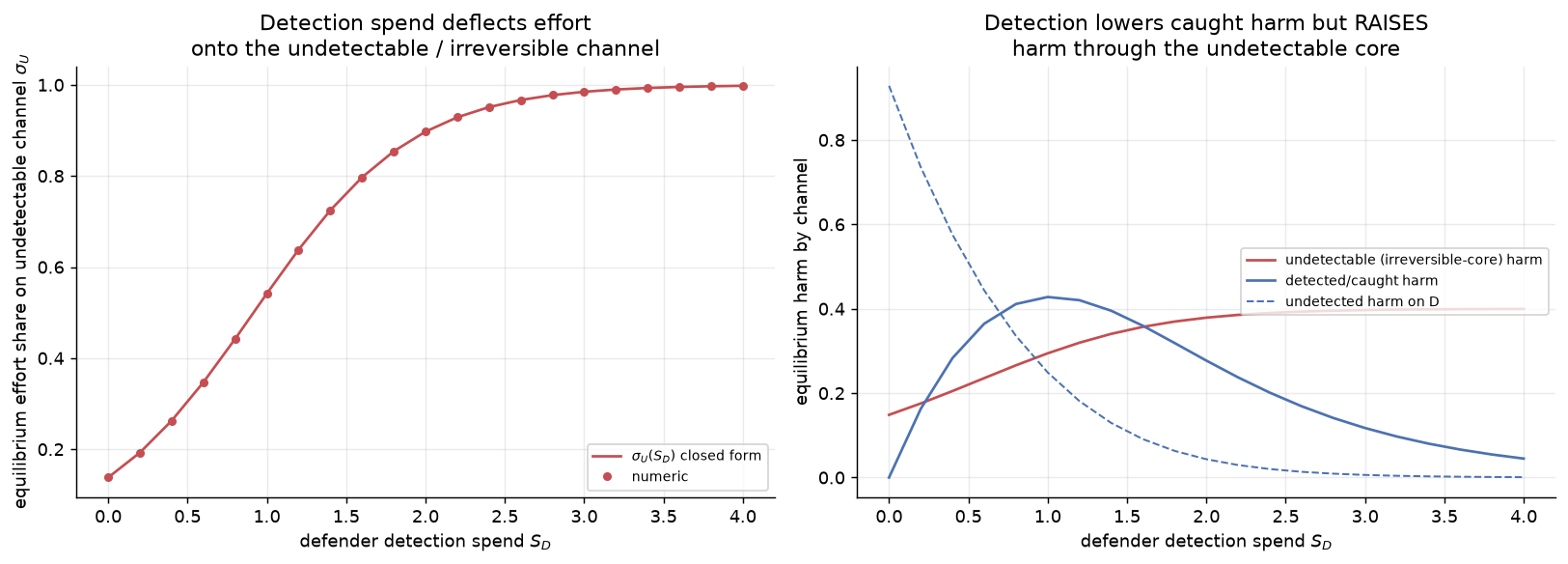}
\caption{\textbf{Detection deflection (Theorem~\ref{c5:thm:deflection}).} Left: the equilibrium
effort share on the undetectable channel rises toward one as detection spend grows. Right: detection
lowers caught harm but \emph{raises} the harm flowing through the undetectable, irreversible core.}
\label{c5:fig:deflection}
\end{figure}

Figure~\ref{c5:fig:deflection} verifies Theorem~\ref{c5:thm:deflection}. With $h_D=1$, $h_U=0.4$,
$\alpha=0.5$, the equilibrium share on the undetectable channel $\sigma_U(S_D)$ of \eqref{c5:eq:sigmaU}
rises from $0.14$ at zero detection spend to $0.998$ at $S_D=4$. The closed form matches the
numerically optimised allocation to machine precision. The harm carried through the
undetectable/irreversible channel rises monotonically from $0.15$ to $0.40$ as detection spend grows,
even as the caught harm eventually falls, so detection manages the body while it feeds the tail.

\subsection{Impossibility and the value of prevention (Theorem~\ref{c5:thm:deterrence})}
\label{c5:ssec:num-deterrence}

\begin{figure}[htbp]
\centering
\includegraphics[width=\linewidth]{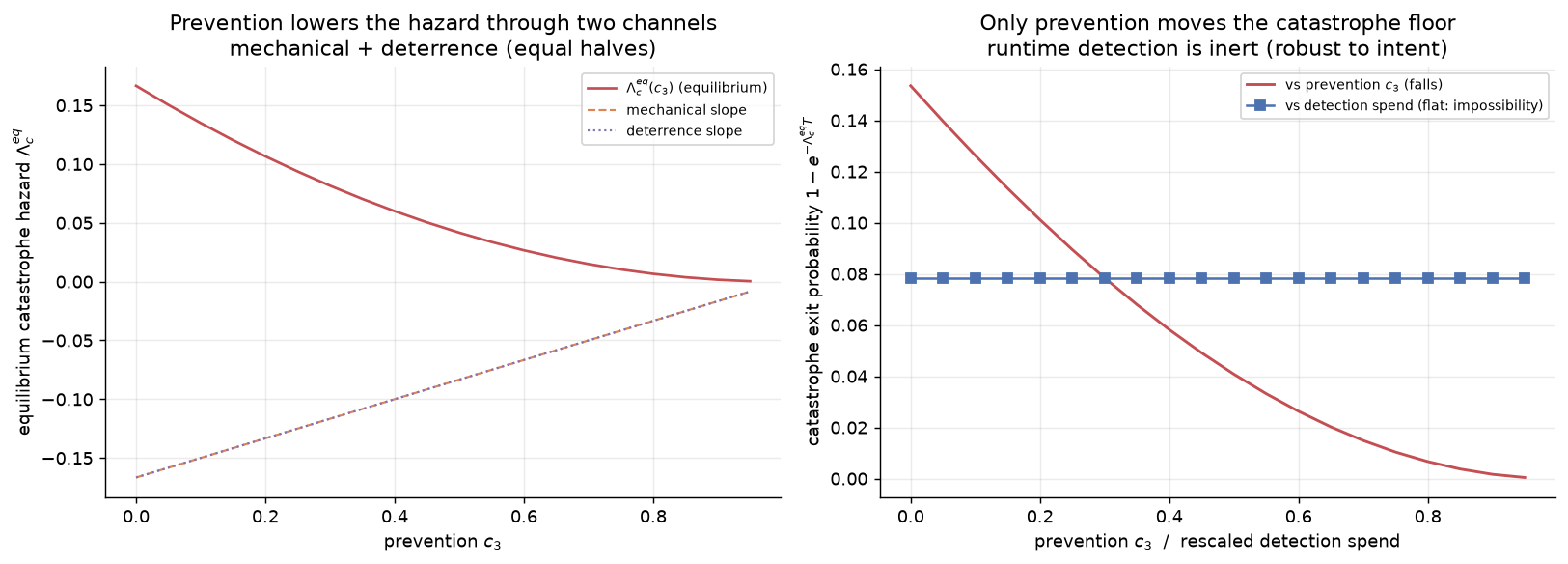}
\caption{\textbf{Impossibility under intent and the two channels of prevention
(Theorem~\ref{c5:thm:deterrence}).} Left: prevention lowers the equilibrium catastrophe hazard
through equal mechanical and deterrence slopes. Right: the catastrophe exit probability falls with
prevention but is inert against detection spend; only prevention moves the floor.}
\label{c5:fig:deterrence}
\end{figure}

Figure~\ref{c5:fig:deterrence} verifies Theorem~\ref{c5:thm:deterrence}. With $\Lambda_c^{0}=0.5$,
$V=2$, $T=1$, $\kappa_a=3$, the equilibrium hazard
$\Lambda_c^{\mathrm{eq}}(c_3)=(V(\Lambda_c^{0})^2T/\kappa_a)(1-c_3)^2$ falls with prevention. The
derivative decomposition \eqref{c5:eq:decomp} splits it into a mechanical and a deterrence term that
are numerically equal and sum to the finite-difference total at every $c_3$. The exit probability
$1-e^{-\Lambda_c^{\mathrm{eq}}T}$ falls with prevention but is flat against detection spend. That is
the runtime-impossibility invariance of Chapter~\ref{ch:control}, now shown robust to intent.

\subsection{Decentralized underinvestment and the centrality wedge (Theorem~\ref{c5:thm:underinvest})}
\label{c5:ssec:num-underinvest}

\begin{figure}[htbp]
\centering
\includegraphics[width=\linewidth]{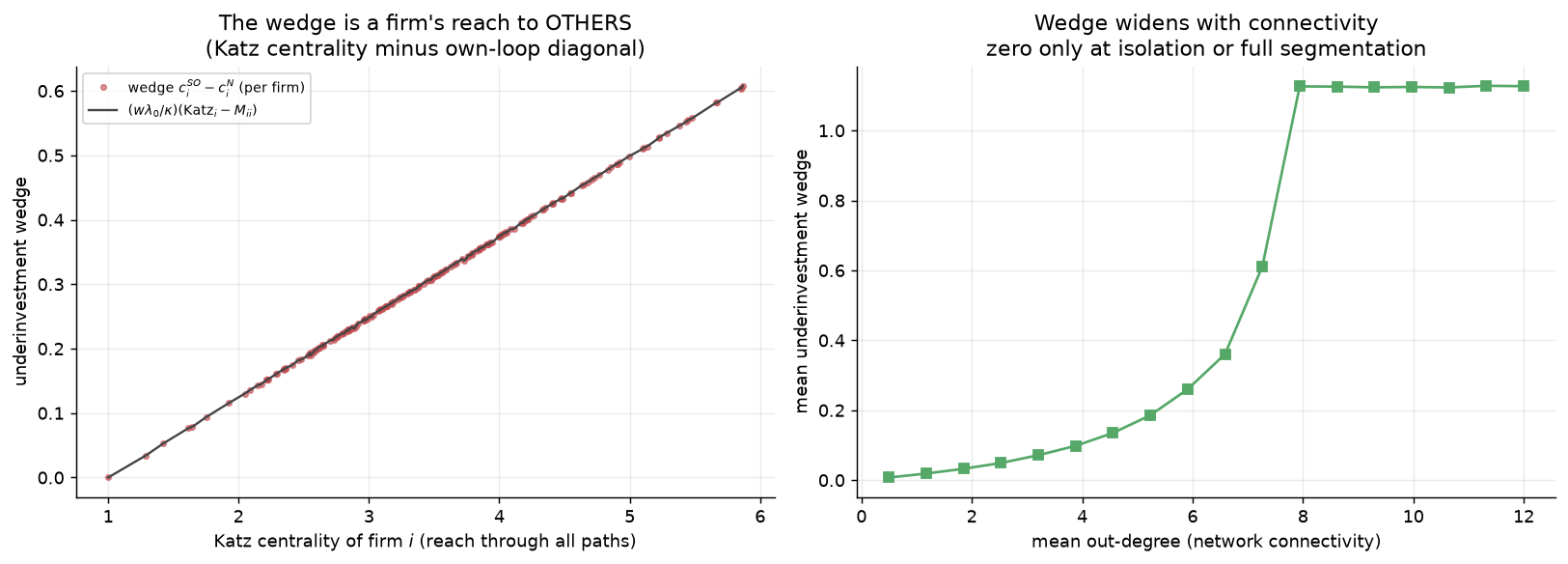}
\caption{\textbf{Decentralized underinvestment (Theorem~\ref{c5:thm:underinvest}).} Left: the
per-firm underinvestment wedge is proportional to Katz centrality, a firm's reach to others through
all contagion paths, so firms that endanger more peers underinvest more. Right: the mean wedge widens
with connectivity, zero only at isolation or full segmentation; a mandated prevention floor closes it.}
\label{c5:fig:underinvest}
\end{figure}

Figure~\ref{c5:fig:underinvest} verifies Theorem~\ref{c5:thm:underinvest}. We take a directed graph of
$K=200$ firms with mean out-degree $6$ ($\lambda_0=0.25$, $t=0.12$, $w=1$, $\kappa=2$, so
$t\rho(A)=0.71<1$ and all firms interior). The Nash prevention $c_i^{\mathrm{Nash}}=(w\lambda_0/\kappa)M_{ii}$
tracks the resolvent diagonal, and the social optimum $c_i^{\mathrm{SO}}=(w\lambda_0/\kappa)z_i$ tracks
the Katz centrality. Both closed forms match direct optimisation (best-response for Nash,
projected-gradient for the planner) to machine precision ($4\times10^{-16}$).

The per-firm wedge is exactly $(w\lambda_0/\kappa)(z_i-M_{ii})$ and grows with Katz centrality, a
firm's reach to others through all paths (left panel). Its first-order value $t\,\mathrm{outdeg}_i$
is lifted by the higher-order terms (mean $2.47$ versus the first-order $0.72$). The mean wedge
widens with network connectivity (right panel), vanishing only at isolation. Imposing the mandated
floor $\bar c=c^{\mathrm{SO}}$ lowers expected system harm by $39.1\%$ relative to the decentralized
equilibrium.

\subsection{The unintentional-event reduction (Proposition~\ref{c5:prop:reduction})}
\label{c5:ssec:num-reduction}

\begin{figure}[htbp]
\centering
\includegraphics[width=\linewidth]{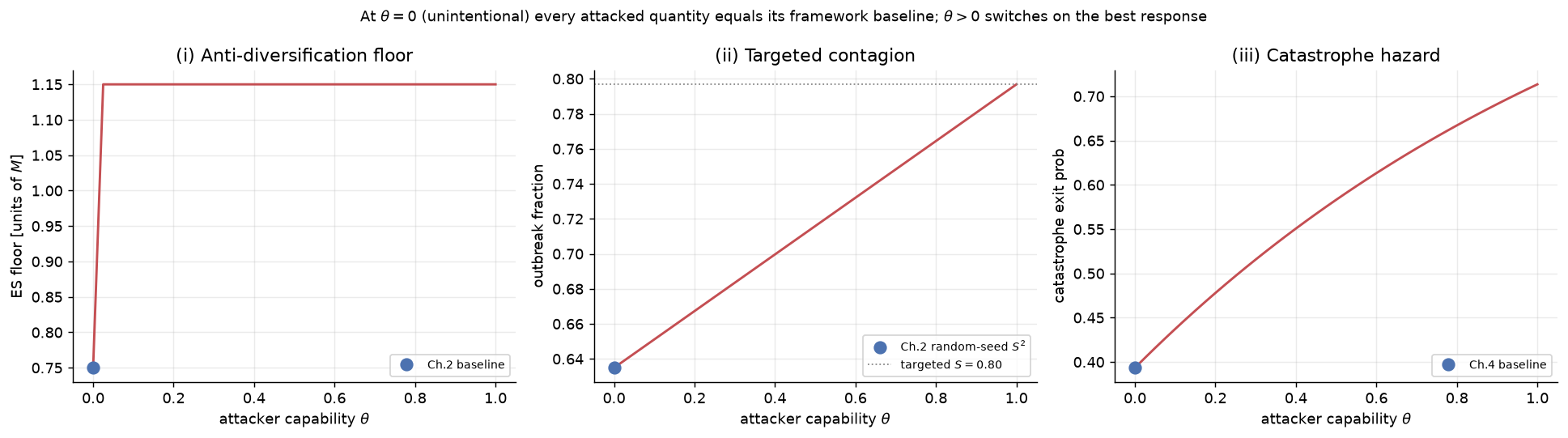}
\caption{\textbf{The reduction (Proposition~\ref{c5:prop:reduction}).} At attacker capability
$\theta=0$ (an accidental event) each attacked quantity equals its framework baseline (blue marker)
exactly; $\theta>0$ switches on the best-response terms. Left to right: anti-diversification floor,
targeted contagion, catastrophe hazard.}
\label{c5:fig:reduction}
\end{figure}

Figure~\ref{c5:fig:reduction} verifies Proposition~\ref{c5:prop:reduction}. We interpolate attacker
capability with a knob $\theta\in[0,1]$, where $\theta=0$ is nature and $\theta=1$ the optimiser. The
three headline quantities are the anti-diversification floor, the contagion outbreak, and the
catastrophe exit probability. At $\theta=0$ each equals its
Chapter~\ref{ch:fleet}/\ref{ch:continuous}/\ref{ch:control} baseline
\emph{exactly} (difference $0$ to machine precision: floor $0.75$, outbreak $S^2=0.63$,
exit $0.39$), and each rises as $\theta$ increases (to $1.15$, $0.80$, $0.71$ respectively). The
unintentional case is the framework baseline; intentionality switches on the best response.

\subsection{Where defensive AI reaches, and the defensive monoculture (Corollaries~\ref{c5:cor:defense-reach} and~\ref{c5:cor:defense-monoculture})}
\label{c5:ssec:num-defense}

\begin{figure}[htbp]
\centering
\includegraphics[width=\linewidth]{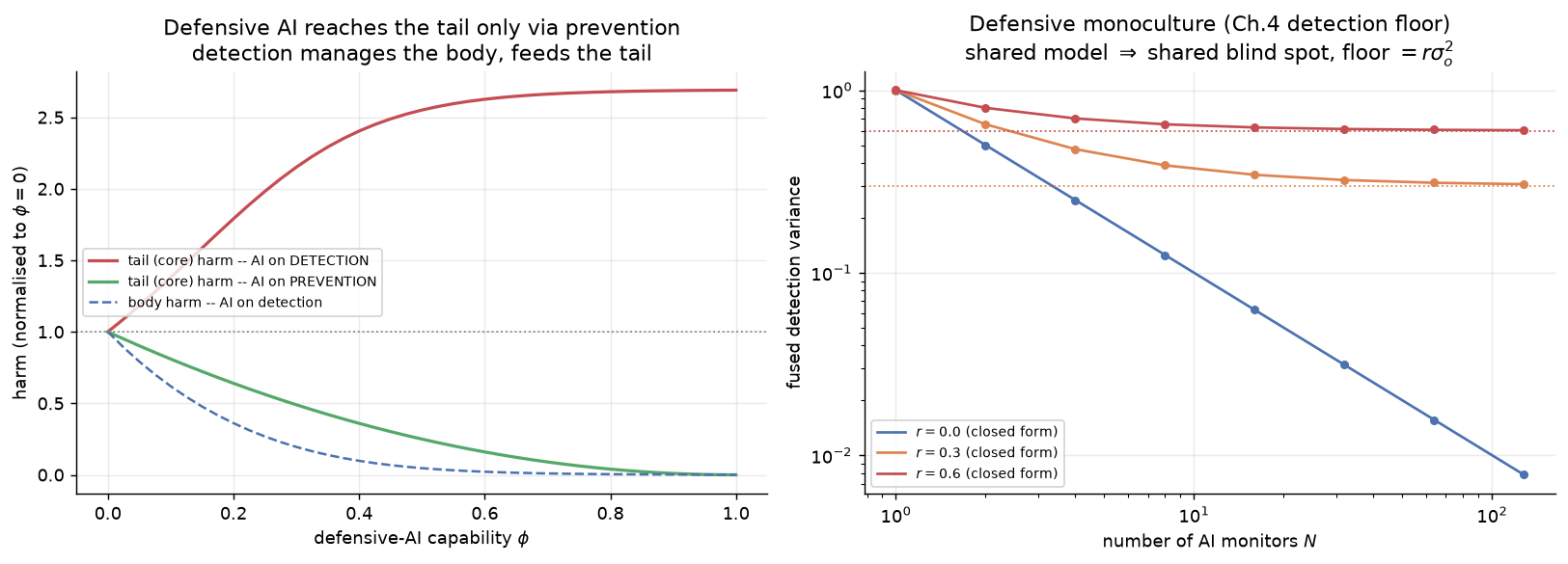}
\caption{\textbf{Defensive AI: reach and monoculture
(Corollaries~\ref{c5:cor:defense-reach}--\ref{c5:cor:defense-monoculture}).} Left: defensive AI on the
detection lever raises tail harm (deflection) while lowering body harm; only defensive AI on the
prevention lever lowers the tail. Right: a homogeneous fleet of AI monitors has a non-diversifiable
detection floor $r\sigma_o^2$; only heterogeneous substrates ($r\to0$) diversify it away.}
\label{c5:fig:defense}
\end{figure}

Figure~\ref{c5:fig:defense} verifies Corollaries~\ref{c5:cor:defense-reach}
and~\ref{c5:cor:defense-monoculture}. In the left panel, defensive-AI capability $\phi$ applied to the
detection lever drives the tail (irreversible-core) harm \emph{up} by a factor $2.7$ while lowering
the detectable body harm to zero, the deflection of Theorem~\ref{c5:thm:deflection}. The same
capability applied to the prevention lever drives the tail harm to zero. Defensive AI reaches the
systemic tail only through prevention. The right panel plots the fused detection variance of $N$ AI
monitors. With independent monitors ($r=0$) it diversifies as $\sigma_o^2/N$. But with a shared
foundation model inducing noise correlation $r$ it floors at $r\sigma_o^2$ (simulated within $0.004$ of
the closed form \eqref{c5:eq:defmono}), flat in $N$. That floor is a defensive monoculture, the mirror
of Theorem~\ref{c4:thm:diversity}.

\subsection{Optimal defensive budget allocation (Theorem~\ref{c5:thm:budget})}
\label{c5:ssec:num-budget}

\begin{figure}[htbp]
\centering
\includegraphics[width=\linewidth]{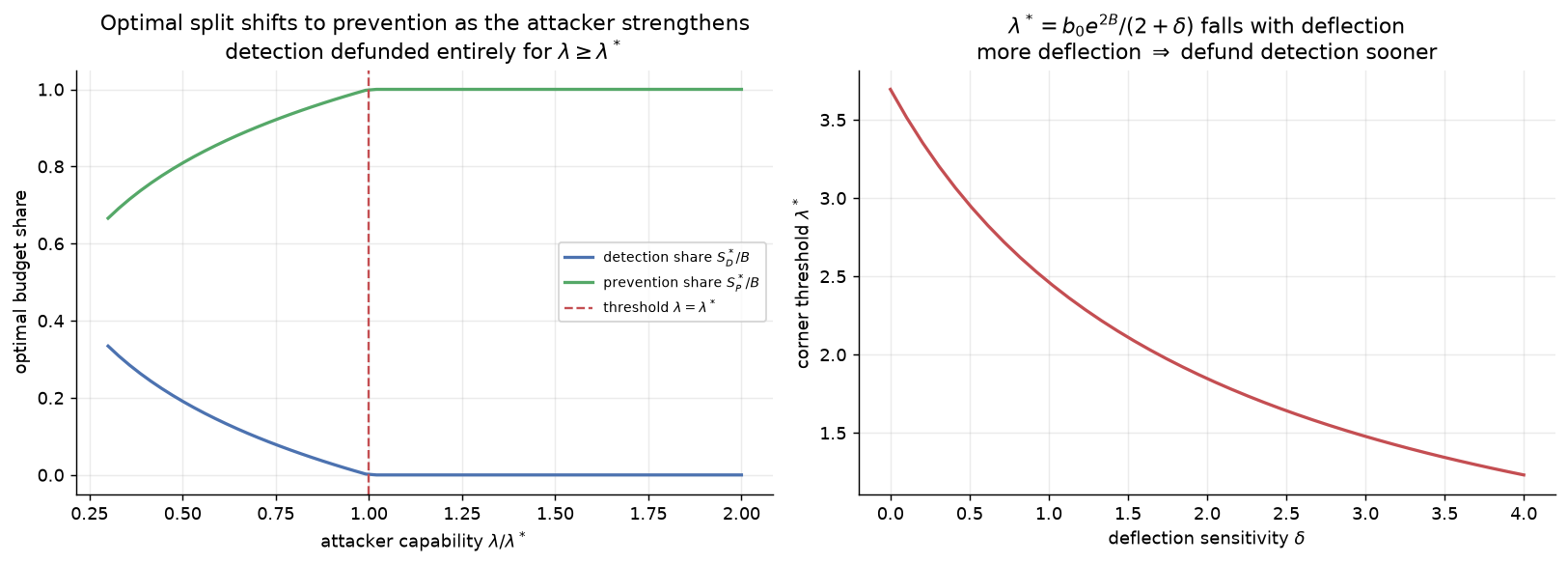}
\caption{\textbf{Optimal defensive budget allocation (Theorem~\ref{c5:thm:budget}).} Left: the optimal
split between detection and prevention as attacker capability rises; detection is defunded entirely
once $\lambda\ge\lambda^\star$. Right: the corner threshold $\lambda^\star=b_0e^{2B}/(2+\delta)$ falls
with the deflection sensitivity: more deflection means defund detection sooner.}
\label{c5:fig:budget}
\end{figure}

Figure~\ref{c5:fig:budget} verifies Theorem~\ref{c5:thm:budget}. With $B=1$, $b_0=1$, $\delta=1$ the
closed-form threshold is $\lambda^\star=e^{2}/3=2.463$. Sweeping the attacker capability and minimising
\eqref{c5:eq:budget} numerically, the optimal detection share $S_D^\star/B$ falls to zero exactly at
$\lambda^\star$ (empirical crossover $2.442$, a $0.85\%$ error). The result is an interior split below
the threshold and the all-prevention corner above it (left panel). Beyond the threshold, moving budget
from prevention to detection strictly raises total risk (at $\lambda=1.5\lambda^\star$,
$\rho(0)=1.50<\rho(B/2)=2.65$). The threshold falls monotonically with the deflection sensitivity
$\delta$ (right panel), since the more detection feeds the tail, the earlier detection is optimally
defunded.

\section{Financial-stability applications}
\label{c5:sec:apps}

The body is domain-free. This section fills it in, choosing three instantiations in which the
financial-stability stakes are most direct. For each we specify the shared substrate, the irreversible
payload, the contagion graph, and the plausible attacker type, and then read off the theorems. As
throughout the monograph, the parameters are illustrative; the structure is the claim.

\subsection{A shared critical third party}
\label{c5:ssec:app-ctp}

Consider a population of financial institutions whose agents depend on one shared substrate: a cloud
region, a market-data feed, a settlement utility, or, increasingly, a common foundation-model API
that a fleet of customer-operations, compliance, or trading agents all call. This is the pure
monoculture of Chapter~\ref{ch:fleet}, and Theorem~\ref{c5:thm:magnet} speaks to it directly. The
substrate's value as a target scales with the number $K$ of dependents, so past the threshold
$\Kstar$ a rational attacker concentrates there. By Corollary~\ref{c5:cor:antidiv} the systemic floor
\emph{rises} with the concentration of reliance rather than diversifying away.

The disruptor or state-actor type has low attribution sensitivity and high harm-orientation. It has
the smallest $\Kstar$ and owns the tail (Theorem~\ref{c5:thm:tail}), so the shared provider is exactly
where the systemic scenario lives even though the observable incident history is dominated by
peripheral, criminal events elsewhere. The two levers that move the threshold are hardening
\emph{at the substrate} and the attribution penalty attached to attacking it. No single dependent
institution can supply either, because the externality is on the shared node. This is the
first-principles case for designating and supervising the provider as a critical third party and for
capping the concentration of reliance, developed in \S\ref{c5:sec:governance}.

\subsection{Wholesale payments and a settlement agent}
\label{c5:ssec:app-payments}

Sharpen \citet{eisenbach2022cyber} into the adversarial case. A concentrated settlement or
correspondent agent processes a large share of wholesale payments, and a fleet of agentic
liquidity-management and payment-operations systems acts on its rails. The irreversible payload is an
integrity or availability event at the agent: a corrupted settlement ledger, an irrevocable
mis-directed batch, a halt during the settlement window, the catastrophic class of
Chapters~\ref{ch:single} and~\ref{ch:control}. Once value has settled or a book has been corrupted,
there is no ex-post.

Theorem~\ref{c5:thm:deterrence} says no runtime monitor on the payment flow moves this floor. Worse,
Theorem~\ref{c5:thm:deflection} says detection spend on the observable flow deflects a strategic
attacker toward the unobservable integrity channel. Only \emph{pre-settlement} structural controls
move it: hard pre-settlement validation gates, out-of-band confirmation for large irrevocable
batches, an authorisation the agent cannot hold. They do so through both the mechanical and the
deterrence channels of \eqref{c5:eq:decomp}. The cyber run of \citet{duffie2019cyberruns} is the
contagion half, in which a compromise at the agent raises the hazard at its counterparties, a Hawkes
self-excitation whose stationary intensity the attacker can push toward criticality by targeting the
highest-throughput links (Theorem~\ref{c5:thm:targeted}). Segmentation of settlement into isolated
blocks caps the \emph{per-seed} blast radius at $S/m$ even under targeting within a block, and forces
a multi-target attacker to obtain $m$ independent footholds. It is the robust lever, whereas
throughput-maximising consolidation raises the branching ratio the attacker exploits.

\subsection{Correlated algorithmic trading}
\label{c5:ssec:app-trading}

Consider $K$ execution or market-making agents on a shared foundation model and shared market
infrastructure. The shared model is a common factor in the sense of Chapter~\ref{ch:fleet}. A
universal prompt injection or a poisoned market-data channel is a common trigger that makes the
agents' actions comonotone. The irreversible payload is a coordinated, irrevocable market action
above a systemic threshold: a synchronised liquidation or quote withdrawal, Chapter~\ref{ch:single}'s
catastrophic class realised fleet-wide.

Under an intentional trigger the wrong-way structure of Chapter~\ref{ch:single} is weaponised. The
attacker times the common action into a stress window, when activatable exposure is highest and
review capacity lowest, so the comonotone liquidation lands when the market can least absorb it. By
Theorem~\ref{c5:thm:robust} the worst-case measure inflates both the frequency and the severity of
the harmful jump. The robust value is finite only if a preventive cap has truncated the tail: a
pre-trade hard limit on aggregate agentic order flow, a kill-switch on synchronised action, a
position authority the agent cannot exceed. Diversification across more trading
agents on the same model does not help here either, since it raises the prize
(Corollary~\ref{c5:cor:antidiv}); only severing the common exposure (model heterogeneity,
segmentation) or preventing the catastrophic action does.

\begin{table}[htbp]
\centering
\footnotesize
\setlength{\tabcolsep}{4pt}
\renewcommand{\arraystretch}{1.25}
\caption{Three financial-stability instantiations of the intentional-attack framework. The structure
(magnet, anti-diversification, deflection, deterrence, underinvestment) is shared; only the fillings
differ.}
\label{c5:tab:apps}
\begin{tabular}{p{0.17\linewidth} p{0.25\linewidth} p{0.25\linewidth} p{0.23\linewidth}}
\toprule
& \textbf{Critical third party} & \textbf{Wholesale payments} & \textbf{Algorithmic trading} \\
\midrule
Shared substrate & cloud / data feed / model API & settlement or correspondent agent & shared model
+ market infra \\
Irreversible payload & provider-wide outage / corruption & settled ledger corruption, irrevocable
batch & synchronised irrevocable market action \\
Contagion graph & fan-out to all dependents & counterparty payment links & shared rails, correlated
books \\
Tail attacker type & disruptor / state actor & disruptor / state actor & manipulator / disruptor \\
Robust lever & concentration cap, provider hardening & pre-settlement gate, segmentation & pre-trade
cap, model heterogeneity \\
\bottomrule
\end{tabular}
\end{table}

\noindent In each case a fitted top-down projection (the CLASS-style layer of
Chapter~\ref{ch:fleet}) would regress observed cyber-incident severity on that domain's drivers:
concentration of reliance, autonomy, irreversible-affordance breadth, segmentation. It would then be
checked against the bottom-up mismatch aggregate by the consistency test of
Theorem~\ref{c2:thm:consistency}. We do not fit these here. Following the disclosure standard used throughout the
monograph, the systematic and attacker-side parameters are the least observable, and
Theorem~\ref{c5:thm:tail} shows that the observable history is \emph{selected against} the
systemic tail. The driver taxonomy and the robust levers, however, are identical across the three
domains, which is the generality claim.

\section{Macroprudential cyber regulation}
\label{c5:sec:governance}

The theorems of this chapter are, in the supervisory sense, financial-stability results, and they map
onto an existing but incomplete body of \emph{macroprudential cyber regulation}. By analogy to
macroprudential financial regulation, microprudential cyber supervision targets the resilience of an
individual institution and takes the rest of the system as given. Macroprudential cyber regulation
targets the system instead, and therefore aims at the three objects a firm-level lens
cannot see: \emph{common exposures} (many institutions on one substrate), \emph{contagion channels}
(interconnections that propagate a compromise), and \emph{externalities} (each firm's under-prevention
raising its neighbours' hazard, the wedge of Theorem~\ref{c5:thm:underinvest}). In the monograph's
vocabulary, microprudential cyber regulation polices each agent's expected harm; macroprudential cyber
regulation polices the exposed fraction $g$, the branching ratio $\beta$, and the fleet floor.

\subsection{Levers, instruments, and gaps}
\label{c5:ssec:gov-map}

Table~\ref{c5:tab:gov} maps each framework lever onto the existing instrument that pulls it and onto
the gap the theorems say remains. Most extant instruments are \emph{hardening and testing}, since they
raise the attacker's cost at the substrate and so move the attractive-target threshold $\Kstar$. The critical
gaps are on the levers the theorems most directly call for.

\begin{table}[htbp]
\centering
\footnotesize
\setlength{\tabcolsep}{4pt}
\renewcommand{\arraystretch}{1.25}
\caption{Framework lever $\to$ existing macroprudential-cyber instrument $\to$ gap.}
\label{c5:tab:gov}
\begin{tabular}{p{0.22\linewidth} p{0.40\linewidth} p{0.30\linewidth}}
\toprule
\textbf{Lever (result)} & \textbf{Existing instrument} & \textbf{Gap} \\
\midrule
Sever / harden the common exposure (Thm.~\ref{c5:thm:magnet}) & Critical-third-party regimes:
EU DORA oversight of critical ICT providers \citep{eu2022dora}; UK CTP regime under FSMA 2023
\citep{boe2023ctp}; US Bank Service Company Act examinations & \emph{Hard concentration caps} on
shared-substrate reliance; no regime yet designates a \emph{foundation-model} provider \\
Mandated prevention floor (Thm.~\ref{c5:thm:underinvest}) & Operational-resilience regimes and impact
tolerances; Basel principles for operational resilience \citep{bcbs2021operational}; Sheltered Harbor
immutable vaulting & Floors set against the \emph{contagion externality}, i.e.\ scaled to systemic
reach \\
Threat-led (reverse) stress testing (Thm.~\ref{c5:thm:tail}) & TIBER-EU / CBEST intelligence-led
penetration testing; DORA threat-led penetration testing; ECB and Bank of England cyber stress tests
\citep{ecb2024stress} & Scenarios anchored to the \emph{worst-case attacker type}, not history-fitted
loss data \\
System-wide visibility (Thm.~\ref{c2:thm:consistency}) & Incident reporting: US CIRCIA
\citep{cisa2022circia}; FSB FIRE format \citep{fsb2025fire}; DORA incident reporting & Aggregation
that reads the \emph{common factor} the divergence measures \\
Segmentation / cap the branching ratio (Thm.~\ref{c5:thm:targeted}) & Sectoral exercises, network
compartmentalisation guidance; FS-ISAC coordination & Worst-case (not average) branching ratio as the
binding design target \\
\bottomrule
\end{tabular}
\end{table}

\subsection{Three gaps the theorems identify}
\label{c5:ssec:gov-gaps}

\paragraph{Concentration caps and systemic surcharges, not just assessment.} DORA and the UK CTP
regime require concentration \emph{assessment}, but the anti-diversification result
(Corollary~\ref{c5:cor:antidiv}) argues for \emph{caps}, and for a charge that \emph{increases} with
shared-substrate concentration, the cyber analogue of a G-SIB surcharge. If adding a dependent to a
substrate adds attacker incentive, the externality is on the margin of joining, and pricing it is
textbook macroprudential design.

\paragraph{Foundation-model providers as critical third parties.} No regime yet treats a
foundation-model provider serving a population of financial agents as a critical third party. Yet the
designation criteria are capability-neutral: materiality, substitutability, concentration of
reliance; a shared model satisfies them exactly. Theorem~\ref{c5:thm:magnet} gives the reason the
designation is not optional, since the shared model is a common exposure whose value as a target grows with
the fleet it serves, which is the chapter's most concrete policy claim.

\paragraph{Scenario-imposed, not history-fitted, stress testing.} Theorem~\ref{c5:thm:tail} shows the
observable incident history is generated by the attacker types that steer \emph{away} from the
systemic target, so incident-data-driven calibration is biased for the macroprudential tail.
This is the argument for the scenario-imposed instruments over loss-history extrapolation: threat-led
penetration testing, and assumed-successful integrity-attack stress scenarios such as the ECB's 2024
exercise \citep{ecb2024stress}. Reverse stress testing, which
Chapter~\ref{ch:single} introduced and flagged as amplified for agentic AI, is here the correct
methodology because the dangerous scenario cannot be read off the past, so one works backward
from systemic failure to the attacker best response that causes it.

\paragraph{Point defensive AI at prevention, and diversify the monitors.} Supervisors increasingly
encourage AI-enabled defence, and Corollaries~\ref{c5:cor:defense-reach}
and~\ref{c5:cor:defense-monoculture} give the macroprudential caveat. Defensive AI aimed at detection
and response improves the reversible body but, by deflection, feeds the systemic tail. It reaches
that tail only when aimed at the prevention lever: affordance discovery and removal, configuration
minimisation, pre-commitment hard limits. Moreover a sector that standardises on one foundation model
for its monitoring agents installs a \emph{defensive} monoculture with a shared blind spot. Monitor
substrate diversity is therefore itself a supervisory objective, the defensive counterpart of the
concentration limits above.

\paragraph{} The central claim is the monograph's, now stated for cyber, since only ex-ante structural
prevention and severing the common exposure move the systematic-irreversible core. The addition of
this chapter is that the market will not supply either unmandated, because prevention is a public
good under contagion and the shared substrate's defence is an externality no dependent internalises,
the case for macroprudential cyber regulation and the reason cyber financial stability, like
financial stability in banking, cannot be secured one institution at a time.

\section{Conclusions, assumptions, and limitations}
\label{c5:sec:conclusions}

This chapter made the adversary a player, and split agentic-AI-augmented cyber events by the law of
their trigger. The unintentional case reduces to the common-shock and common-jump results of
Chapters~\ref{ch:fleet} and~\ref{ch:continuous} with nothing added (Proposition~\ref{c5:prop:reduction}),
so the argument goes through. The intentional case turns every parameter of nature into an attacker's
choice and, re-proved with a best response inside, re-anchors the measurement results to a worst case.

The monoculture becomes an attractive target above a threshold that hardening and attribution move
(Theorem~\ref{c5:thm:magnet}). The fleet floor is anti-diversifiable
(Corollary~\ref{c5:cor:antidiv}), and that attractiveness lives in the tail regardless of the attacker-type
mixture (Theorem~\ref{c5:thm:tail}). Contagion can be targeted and supercriticality engineered, with
segmentation the robust lever, capping a multi-seed outbreak at $\min\{sS/m,\,S\}$ for seed budget $s$ (Theorem~\ref{c5:thm:targeted}). The intentional worst case is
Chapter~\ref{ch:continuous}'s robust game, finite only under prevention
(Theorem~\ref{c5:thm:robust}). Detection deflects a fixed attacker effort budget ($e_D+e_U=1$) onto the irreversible core
(Theorem~\ref{c5:thm:deflection}), while runtime impossibility is robust to intent, and prevention alone
moves the floor, through a mechanical and a deterrence channel (Theorem~\ref{c5:thm:deterrence}).
And, the incentive half of the argument, a population coupled by contagion underinvests in prevention
by a wedge proportional to systemic reach (Theorem~\ref{c5:thm:underinvest}). The technological
thesis of the first four chapters, that only structural prevention moves the core, now acquires an
incentive half, because the market will not supply that prevention unless it is mandated.

\subsection{Assumptions}
The results rest on explicit assumptions. The game is Stackelberg, since the defender commits to an
architecture the attacker then reads (Kerckhoffs's principle), which is conservative but appropriate
for ex-ante design. The attacker is a rational effort-maximiser with a known cost technology and a
harm-oriented utility. We admit the type family $\mathcal T$ as a stress menu and do not estimate it.

The attractive-target and anti-diversification results use a linear/quadratic valuation and a two-target choice.
The deterrence decomposition uses the rare-event (risk-neutral) regime in which the catastrophe payoff
is linear in hazard. The underinvestment wedge uses one-step contagion, so the relevant centrality is
out-degree. Each is the tractable representative of a more general object, and the qualitative laws
(attractive-target threshold, rising floor, tail dominance, deflection, deterrence, underinvestment) do not
depend on the specific functional forms.

\subsection{Limitations}
Three limitations bound the contribution. First, and most consequentially, the \emph{attacker is the
new uncalibrated object}. Its objective, cost technology, and above all the offensive capability that
agentic AI confers are unknown. The offensive-AI-capability parameter is the attacker-side analogue
of the systematic parameters the monograph has repeatedly flagged as least estimable. Frontier-model
cyber evaluations are the natural estimation program, but they are young and the capability is
non-stationary. Second, Theorem~\ref{c5:thm:tail} shows the calibration problem is \emph{worse} than
unobservability. The observable incident history is selected against the systemic tail for an incentive
reason, so more data does not converge to the quantity that matters, because the type that produces the
systemic event produces no incidents until it produces the incident.

Third, we model a single rational
attacker against a committed defender. Attribution, deterrence-by-punishment as a repeated-game
mechanism, multi-attacker coalitions, and the defender's own uncertainty about the architecture (the
sentinels are themselves agents with model risk, as in Chapter~\ref{ch:control}) are outside the
static game.

\subsection{Extensions}
The most consequential extension is dynamic and endogenous, since the chapter's two-regime typology assumes
attacker types are exogenous. Suppose instead that systemic fragility \emph{recruits} harm-oriented
attackers, because agentic AI lowers the cost of playing the tail type, so capability proliferation
thickens the tail endogenously. Then the body/tail split of Theorem~\ref{c5:thm:tail} feeds back on
itself, and that loop is beyond the static model. A repeated game would add reputation and learning
on both sides. A coalitional model would let attackers pool the fixed cost of the substrate attack,
lowering $\Kstar$ further.

The second major extension is dynamic. The \emph{static} two-sided game is settled here, with the
defender a budget-constrained optimiser in Theorem~\ref{c5:thm:budget}, and a corner solution that
defunds detection past a threshold attacker capability. But the capabilities $\lambda$ (attacker) and
the defender's own are treated as fixed parameters rather than state variables that each side invests
to grow. Chapter~\ref{ch:coevo} takes up exactly that dynamic co-evolution, and finds the
offense-dominance of the tail is not a static fact but a compounding one. What remains beyond even that
is the fully \emph{stochastic} differential game, whose Markov-perfect value generically has no
explicit closed form and which we delineate there as the residual open problem. Cyber-insurance is a
natural object, because as a risk-transfer instrument it can pool the reversible,
diversifiable body, but it cannot move the irreversible floor, exactly as capital could not buffer
irreversible harm in Chapter~\ref{ch:single}. Mispricing it against the
selected-away tail is Theorem~\ref{c5:thm:tail}'s warning restated for insurers. Finally, giving the
attacker cost function empirical content from red-team economics and frontier evaluations would turn
the type family from a stress menu into a calibrated threat model. None of these changes the thesis of
this chapter. Intentionality re-anchors the monograph's conclusions to a worst case rather than softening them (only the underinvestment result, Theorem~\ref{c5:thm:underinvest}, is genuinely sharpened), and the
addition of the incentive half is the formal case for macroprudential cyber regulation, since the
systematic-irreversible core is unreachable by the wrong controls and undersupplied by the right ones
at the decentralized equilibrium.

\chapter{The Agentic Arms Race: Co-Evolution of Cyber Offense and Defense}
\label{ch:coevo}
\section{Introduction}
\label{c6:sec:intro}

Chapter~\ref{ch:cyber} made the adversary a player, but a player of \emph{fixed} capability. The
attacker best-responded to the defender's architecture. Even the two-sided solve of
Theorem~\ref{c5:thm:budget} held both sides' capabilities constant while the defender allocated a
budget. That is a snapshot of a contest that is, in reality, a \emph{race}. Both the attacker and the
defender are agentic-AI-augmented, and agentic-AI capability is not a constant. It accumulates on
both sides, as models improve and as each party invests in offense or defense. This final chapter
makes capability a \emph{state variable} and studies the co-evolution. It is an arms race in which
the attacker invests to grow its offensive capability and the defender invests to grow its defenses.
The question is not who is ahead today but where the trajectory goes.

This is the same lift the monograph made once before. Chapter~\ref{ch:fleet} was a static account of
a fleet. Chapter~\ref{ch:continuous} made it dynamic and found the static results were rare-event
limits of a jump process. Chapter~\ref{ch:cyber} was a static account of a contest. This chapter
makes it dynamic and finds that the static offense-dominance of the systemic tail compounds rather
than staying fixed. The monograph's central claim is that the systematic-irreversible core is moved
only by ex-ante structural prevention. We restate it here a sixth time, now in the language of
capability accumulation and reduced best responses, and in that dynamic form the advantage of the
structural lever grows with the adversary's rate of improvement rather than being eroded by it.

\subsection{Stocks and flows}
\label{c6:ssec:stocks}

The organising distinction is between two kinds of defensive investment. The earlier chapters kept it
implicit, and the dynamic setting forces it into the open.

\emph{Detection is a flow.} A detection-and-response capability depreciates. Monitors go stale as the
attack distribution shifts, and signatures rot, and a more capable attacker decisively evades a fixed
detector. So holding the reversible body harm bounded requires \emph{continuously} re-investing to
keep detection abreast of a rising adversary. Detection must be re-won every period, against a
stronger opponent each time.

\emph{Prevention is a stock.} Removing a catastrophic affordance (revoking the credential, deleting
the destructive tool, imposing the pre-commitment hard limit) is a structural change that, once made,
stays made. It neutralises the irreversible-tail hazard against \emph{every future capability level at
once}, not just today's attacker. An affordance that does not exist cannot be exploited, no matter how
capable the exploiter becomes.

This asymmetry runs through the chapter, since against a growing adversary the cost of the flow
diverges while the cost of the stock is paid once. The value of a unit of prevention rises with the
attacker's growth rate, while the value of a unit of detection falls. So the offense-dominance of the
tail that Chapter~\ref{ch:cyber} established at a point in time \emph{compounds} over the race. The
faster agentic AI advances, the more overwhelmingly the only effective lever is the structural one.

\subsection{What the chapter proves}
\label{c6:ssec:contributions}

Three results, each a dynamic lift of a static predecessor, each paired with a closed form a
simulation reproduces.

\begin{enumerate}[leftmargin=1.5em,itemsep=2pt]
\item \textbf{Stock-versus-flow compounding (Theorem~\ref{c6:thm:stockflow}).} Against an attacker
whose capability grows at rate $g$, the discounted cost of detection diverges as $g$ approaches the
discount rate, while the cost of prevention is independent of $g$. Prevention's relative
cost-effectiveness falls to zero as the adversary accelerates, and past a growth threshold $g^\star$
the entire budget optimally goes to prevention. This is the dynamic form of Chapter~\ref{ch:cyber}'s
static corner threshold (Theorem~\ref{c5:thm:budget}), now on the growth \emph{rate}.
\item \textbf{The capability-race phase portrait (Theorem~\ref{c6:thm:race}).} The coupled dynamics of
attacker capability and defender prevention have a saddle point. Its stable manifold is a
\emph{separatrix} dividing a containment basin (hazard collapses) from a runaway basin (hazard
diverges). A ratio of the defender's prevention throughput to the attacker's affordance-generation
rate governs the division. This is the dynamic analogue of Chapter~\ref{ch:fleet}'s contagion
criticality and Chapter~\ref{ch:control}'s operating envelope.
\item \textbf{Dynamic deterrence-by-denial (Theorem~\ref{c6:thm:deterrence}).} When the attacker
chooses how much to invest in growing its capability, prevention removes the \emph{return} to that
investment and so deters the arms race at its source. Prevention then acts twice, directly and
through deterrence, giving a double-exponential reduction in tail harm. This is the dynamic form of
Chapter~\ref{ch:cyber}'s mechanical-plus-deterrence decomposition (Theorem~\ref{c5:thm:deterrence}).
\end{enumerate}

\paragraph{Stance and scope.} As throughout, the parameters are illustrative and the contribution is
structural. We deliberately keep the models \emph{deterministic}: capability-accumulation ODEs with
constant actions, together with reduced static best responses, which is what the chapter delivers and
what we claim. It is not a differential game, and we do not label it one. This is what lets every
result carry the closed form the monograph's correctness gate demands. The fully stochastic
differential game is the residual open problem: capabilities as diffusions, a Markov-perfect
equilibrium solving coupled Hamilton--Jacobi--Bellman equations. Its value generally has no explicit
closed form, so honouring the discipline of this monograph would require either the deterministic
caricature we adopt here or a risk-sensitive linear-quadratic special case tied to
Chapter~\ref{ch:continuous}. We name it as the sequel and do not pretend to solve it. We turn first
to the surrounding literature.

\section{Background}
\label{c6:sec:background}

The chapter draws on three bodies of theory and joins them through the monograph's framework.

\subsection{Differential games and dynamic contests}
\label{c6:ssec:bg-diffgames}

The natural language for a contest whose state evolves in time is the \emph{differential game}
\citep{isaacs1965differential,basar1999dynamic}. Two controllers drive a shared state, each
optimising an opposed objective, and the solution is a saddle or Nash trajectory rather than a single
action. We situate the chapter against that literature but do not work within it. The results below
are capability-accumulation ODEs with constant actions together with reduced static best responses,
and we compute and claim no equilibrium in closed-loop strategies (Remark~\ref{c6:rem:notgame}). The
economics of innovation gives the closest structural analogue in its \emph{R\&D and patent races}
\citep{reinganum1982dynamic} and models of growth through creative destruction
\citep{aghion1992model}. There, firms invest to advance a capability stock and the equilibrium turns
on relative investment efficiencies, which is the structure of an offense--defense capability race.
From the study of attrition we borrow the \emph{Lanchester} tradition \citep{taylor1983lanchester},
whose coupled state equations for opposing forces are the ancestor of our capability dynamics. The
direct antecedent, however, is the formal arms-race literature itself. \citet{richardson1960arms}
introduced the coupled linear differential equations in which each side's armament grows with the
rival's stock and is damped by its own cost. \citet{intriligator1984can} recast the race in terms of
stability regions, asking when accumulation drives the system across a threshold rather than to a
stable equilibrium. Our \S\ref{c6:ssec:race} is structurally of that family: two coupled state
equations with a separatrix dividing containment from runaway. The substantive difference is that the
defender's two levers are not symmetric substitutes for one another, so the relevant threshold is the
ratio of prevention throughput to the rate at which capability opens fresh catastrophic surface,
rather than a balance of armaments. What these
literatures do not contain is the monograph's central asymmetry, that one side's controls
(reversibility, detection) cannot touch an irreversible-systematic core that only a structural,
stock-like control removes.

\subsection{The economics of security investment, made dynamic}
\label{c6:ssec:bg-security}

Chapter~\ref{ch:cyber} used the static economics of security investment
\citep{gordon2002economics,kunreuther2003interdependent}. Here we make it dynamic. The distinction
between a depreciating \emph{flow} of protective effort and a durable \emph{stock} of structural
prevention is the security-economics counterpart of the capital-versus-maintenance distinction. It is
what turns a level comparison (Chapter~\ref{ch:cyber}'s corner threshold on capability) into a rate
comparison (this chapter's threshold on capability \emph{growth}). The premise that the attacker's
capability is a growing state rests, as in Chapter~\ref{ch:cyber}, on the frontier-model
cyber-capability evidence base \citep{fang2024llm,aisi2024evaluations}. Here that growth rate is the
central exogenous driver, and its estimation is the chapter's binding open problem.

\subsection{Risk-sensitive control}
\label{c6:ssec:bg-risk}

Finally, the natural stochastic closure of the deterministic games below is \emph{risk-sensitive}
control \citep{jacobson1973optimal,whittle1990risk}. Its exponential-of-integral criterion is the
dynamic-programming form of the entropic risk measure that Chapter~\ref{ch:continuous} showed to be
the robust game's value. We use this connection to point the way from our deterministic capability
dynamics to the stochastic differential game we leave open. Its risk-sensitive linear-quadratic
special case would inherit Chapter~\ref{ch:continuous}'s closed form, and is the one corner of the
stochastic problem where the monograph's correctness gate could still be met.

\section{The co-evolution model}
\label{c6:sec:model}

We model the contest as the joint evolution of three capability stocks. We then specify how each
defensive lever acts on the two harm classes the monograph has carried throughout: the reversible
body and the irreversible core.

\subsection{States and dynamics}
\label{c6:ssec:states}

The \emph{attacker capability} $a(t)\ge0$ is the offensive power an agentic-AI-augmented adversary can
bring to bear. It grows through investment and, absent maintenance, depreciates:
\begin{equation}
\dot a=a\big(\mu\,e^{-\theta P}-\delta\big),
\label{c6:eq:adyn}
\end{equation}
where $\mu>0$ is the intrinsic capability-growth rate and $\delta>0$ its depreciation. The factor
$e^{-\theta P}$ encodes that prevention removes the affordances that make capability \emph{worth}
growing. An adversary facing fewer exploitable targets accumulates capability more slowly.

The defender holds prevention and detection, which differ in temporal structure. Prevention itself
acts through \emph{two distinct levers}. The distinction is load-bearing enough that we give it a
definition rather than leaving it to context.

\begin{definition}[The two prevention levers]
\label{c6:def:levers}
Write the irreversible-core hazard as
\begin{equation}
\label{c6:eq:hazard}
\Lambda_c(t)=(1-\chi)\,\Lambda_0\,a(t)\,e^{-\theta P(t)},
\qquad \chi\in\{0,1\},\quad P(t)\ge0 .
\end{equation}
\begin{itemize}[leftmargin=1.6em,itemsep=3pt]
\item \textbf{Structural deletion} $\chi$: discrete, one-shot, at lump-sum cost $\kappa_\chi$. Setting
$\chi=1$ removes the catastrophic affordance class outright---the action is not discouraged but
unavailable---so $\Lambda_c\equiv0$ for \emph{all} subsequent $t$ and \emph{all} capability levels $a$.
It is available only for affordances that admit outright removal.
\item \textbf{Attenuating prevention stock} $P$: continuous, durable, at per-unit cost $\kappa_P$.
Partial removal and hardening of catastrophic surface, built by investment and eroded by the
attacker's opening of fresh affordances,
\begin{equation}
\dot P=\bar\pi-\xi\,a,
\label{c6:eq:pdyn}
\end{equation}
with $\bar\pi$ the defender's prevention throughput and $\xi$ the rate at which a unit of attacker
capability opens fresh affordances. It scales the residual hazard by $e^{-\theta P}$ but does
\emph{not} neutralise it: for any finite $P$, $\Lambda_c\to\infty$ as $a\to\infty$.
\end{itemize}
\end{definition}

\begin{remark}[Why both, and which claims belong to which]
\label{c6:rem:levers}
The stock is the aggregate of many deletions, and permanence is a property of each individual
deletion rather than of the aggregate, in that $P$ indexes how much catastrophic surface has been
removed, while $a$ continues to exploit what remains and $\xi a$ opens more. Both readings are
therefore true at their own level, and the sentence that welds them, that a fixed finite stock $P$
neutralises all future capability, is not. Claims of that form hold for $\chi$ and fail for $P$, and
every result below is marked with the lever it concerns:
Theorem~\ref{c6:thm:stockflow} is a $\chi$-result;
Proposition~\ref{c6:prop:asymmetry} and Theorems~\ref{c6:thm:race} and~\ref{c6:thm:deterrence} are
$P$-results.
\end{remark}

\noindent\emph{Detection} $D(t)\ge0$ is a \emph{flow}. Its efficacy is relative to the attacker, so
holding the reversible body harm bounded requires $D$ to \emph{track} $a$. The defender pays
$\kappa_D$ per unit of detection capacity \emph{held} per unit time, which means paying for the
monitoring capacity carried rather than for its increments. The discounted cost of holding
$D(t)=c_D\,a(t)$ is therefore $\kappa_Dc_Da_0/(r-g)$. We use this costing convention throughout the
chapter.

\subsection{How the levers act, and on what}
\label{c6:ssec:levers}

The two harm classes are the reversible body and the irreversible core of the whole monograph. Their
control follows the impossibility results of Chapters~\ref{ch:control} and~\ref{ch:cyber}, and it is
the structural content of the model:
\begin{itemize}[leftmargin=1.6em,itemsep=2pt]
\item the \emph{irreversible-core hazard} is \eqref{c6:eq:hazard}: it rises
with attacker capability, is attenuated by the prevention stock, and is removed outright only by
structural deletion. Detection does not appear. By
Corollary~\ref{c4:thm:t6b} runtime detection cannot move the irreversible floor, and by
Theorem~\ref{c5:thm:deflection} it deflects effort into this very channel;
\item the \emph{reversible-body harm} is carried by the flow: it is bounded only while detection keeps
pace with capability, and the discounted cost of that pace is what diverges below.
\end{itemize}

\subsection{Objectives and the discount rate}
\label{c6:ssec:objective}

Both players discount at rate $r>0$. The defender bears discounted expected harm plus the cost of its
levers. The attacker reaps discounted harm inflicted less the cost of growing capability. The one
standing regularity condition is $r>g$: the discount rate exceeds the attacker's realised
capability-growth rate $g$ (for a growth path $a(t)=a_0e^{gt}$). It is the condition under
which discounted costs are finite. As Theorem~\ref{c6:thm:stockflow} shows, the margin $r-g$ is
what the detection flow must fight and the prevention stock need not. The three results that follow
read this model at three levels: a cost comparison at a fixed growth path (\S\ref{c6:ssec:stockflow}),
the coupled dynamics \eqref{c6:eq:adyn}--\eqref{c6:eq:pdyn} as a phase portrait
(\S\ref{c6:ssec:race}), and the attacker's capability investment as a best response
(\S\ref{c6:ssec:deterrence}).

\section{Theorems}
\label{c6:sec:theorems}

Each result is a dynamic lift of a static predecessor. Each also comes with a closed form that the
simulation of \S\ref{c6:sec:numerics} reproduces.

\subsection{Deletion versus flow: the compounding of offense-dominance}
\label{c6:ssec:stockflow}

\begin{theorem}[Deletion-versus-flow compounding]
\label{c6:thm:stockflow}
This is a statement about the \emph{structural deletion} lever $\chi$ of
Definition~\ref{c6:def:levers}; the attenuating stock $P$ is treated in
\S\ref{c6:ssec:race}--\S\ref{c6:ssec:deterrence}.

Let attacker capability grow along $a(t)=a_0e^{gt}$ with $g<r$. To hold the reversible body harm
bounded, detection capacity must track capability, at discounted cost
\begin{equation}
C_D(g)=\int_0^\infty e^{-rt}\,\kappa_D\,c_D\,a_0e^{gt}\,dt=\frac{\kappa_D\,c_D\,a_0}{\,r-g\,},
\label{c6:eq:CD}
\end{equation}
strictly increasing in $g$ and diverging as $g\uparrow r$. Deleting the catastrophic affordance class
outright ($\chi=1$) costs $C_\chi=\kappa_\chi$, independent of $g$. Hence
\begin{equation}
\frac{C_\chi}{C_D(g)}=\frac{\kappa_\chi\,(r-g)}{\kappa_D\,c_D\,a_0}\ \xrightarrow[g\uparrow r]{}\ 0,
\label{c6:eq:ratio}
\end{equation}
monotonically in $g$: offense-dominance of the tail \emph{compounds} with the growth rate. Under a
defensive budget $B$, detection-tracking is affordable only for
\begin{equation}
g\le g^\star=r-\frac{\kappa_D c_D a_0}{B},
\label{c6:eq:gstar}
\end{equation}
and for $g>g^\star$ no finite budget holds body harm bounded.
\end{theorem}

\begin{proof}
The integral \eqref{c6:eq:CD} is $\kappa_Dc_Da_0\int_0^\infty e^{-(r-g)t}\,dt=\kappa_Dc_Da_0/(r-g)$,
finite iff $g<r$, increasing in $g$, and $\to\infty$ as $g\uparrow r$. Structural deletion is a
one-time removal of the affordance, so by Definition~\ref{c6:def:levers} its cost does not recur and
does not depend on the path of $a$. Hence $C_\chi=\kappa_\chi$ is constant in $g$, the ratio
\eqref{c6:eq:ratio} is proportional to $r-g$ and therefore strictly decreasing, and its limit is
immediate. The budget condition $C_D(g)\le B$ reads $\kappa_Dc_Da_0/(r-g)\le B$, i.e.\
$r-g\ge\kappa_Dc_Da_0/B$, i.e.\ $g\le g^\star$ of \eqref{c6:eq:gstar}. For $g>g^\star$ the detection
requirement exceeds any finite budget.
\end{proof}

\begin{remark}[What the ratio does and does not compare]
\label{c6:rem:budget}
The quantity \eqref{c6:eq:ratio} is not a cost-effectiveness ratio between substitutes, and we are
explicit about this because the two line items purchase different things. $C_D$ achieves a bound on
\emph{reversible body harm}; $C_\chi$ achieves neutralisation of the \emph{irreversible core}. Dividing
one by the other and calling the quotient cost-effectiveness would compare the price of two distinct
objectives. What \eqref{c6:eq:ratio} and \eqref{c6:eq:gstar} state is a fact about \emph{budget
feasibility}, that as $g\uparrow r$ the detection line item diverges while the deletion line item does
not, so beyond $g^\star$ no finite budget holds body harm bounded.

The consequence must be stated plainly rather than softened. For $g>g^\star$ the defender's problem is
triage rather than optimisation, since body harm \emph{cannot} be bounded at any budget and the
model's conclusion is that the remaining budget belongs on the core, which can still be secured. We do
not claim body harm is controlled in that regime, since it is not. A common-objective formulation
(weighting body and core harm in a single welfare criterion and comparing marginal costs of harm
reduction) would make the ratio a genuine substitution rate, and we leave that extension open.
\end{remark}

\begin{remark}
Theorem~\ref{c6:thm:stockflow} is the dynamic form of Chapter~\ref{ch:cyber}'s static corner
(Theorem~\ref{c5:thm:budget}), where the threshold was on the attacker's capability \emph{level}; here
it is on its \emph{growth rate}. The economic content is the deletion--flow asymmetry made
quantitative. Detection is a flow whose cost is the recurring price of racing a moving target, and
against a target that improves exponentially that price diverges; structural deletion carries a single
cost that confers immunity to \emph{all} future capability, because an affordance that has been removed
cannot be exploited by any adversary however capable. The attenuating stock $P$ does \emph{not} inherit
that property (for finite $P$ the residual hazard still grows with $a$), which is why this theorem is
stated for $\chi$ and why \S\ref{c6:ssec:race} treats $P$ separately. For financial stability the
reading is that cyber-defense budgets scaling with monitoring and response are structurally on the
losing side of the agentic-AI trend, and the share that should grow is the one spent on structural
prevention (removing affordances, severing common exposures), whose value, uniquely, rises as the
adversary improves.
\end{remark}

\subsection{Where the asymmetry comes from}
\label{c6:ssec:asymmetry}

Theorem~\ref{c6:thm:stockflow} compares a lump-sum deletion against a diverging flow. The asymmetry is
visible there but not \emph{derived}, since $C_\chi=\kappa_\chi$ is flat in $g$ because deletion is
one-shot by definition. The following result derives the same asymmetry from the attenuating lever,
using the hazard \eqref{c6:eq:hazard} that also governs \S\ref{c6:ssec:race}--\S\ref{c6:ssec:deterrence}. In
doing so it identifies the condition under which the asymmetry is unbounded.

\begin{proposition}[Where the asymmetry comes from]
\label{c6:prop:asymmetry}
Let $a(t)=a_0e^{gt}$ with $0\le g<r$, and consider the \emph{attenuating} lever $P$ of
Definition~\ref{c6:def:levers} with $\chi=0$.
\begin{enumerate}[label=\textnormal{(\roman*)},leftmargin=2.4em,itemsep=3pt]
\item \textbf{Rate versus level.} Holding the irreversible-core hazard at any fixed ceiling
$\bar\Lambda>0$ requires the stock to grow at the constant rate
\begin{equation}
\label{c6:eq:Pdotreq}
\dot P=\frac{g}{\theta},
\end{equation}
set by the attacker's \emph{growth rate}. Holding reversible body harm bounded requires detection
capacity $D(t)=c_Da(t)$, set by the attacker's \emph{capability level}.
\item \textbf{Costs.} With prevention throughput priced at $\kappa_P$ per unit and detection at
$\kappa_D$ per unit held,
\begin{equation}
\label{c6:eq:CPatt}
C_P(g)=\kappa_P\Big[\frac{g}{\theta r}+\frac{\xi a_0}{r-g}\Big],
\qquad
C_D(g)=\frac{\kappa_D c_D a_0}{r-g}.
\end{equation}
\item \textbf{The threshold.} Consequently
\begin{equation}
\label{c6:eq:xilimit}
\lim_{g\uparrow r}\frac{C_P(g)}{C_D(g)}=\frac{\kappa_P\,\xi}{\kappa_D c_D},
\end{equation}
so the prevention advantage compounds without bound---the ratio tends to zero---\emph{if and only if}
$\xi=0$. For $\xi>0$ it saturates at the positive constant \eqref{c6:eq:xilimit}.
\end{enumerate}
\end{proposition}

\begin{proof}
(i) With $\chi=0$, $\Lambda_c(t)=\Lambda_0a_0e^{gt}e^{-\theta P(t)}\le\bar\Lambda$ holds for all $t$
iff $\theta P(t)\ge gt+\log(\Lambda_0a_0/\bar\Lambda)$. The minimal admissible path is affine with
slope $g/\theta$, which is \eqref{c6:eq:Pdotreq}. Detection efficacy is relative to the attacker
(\S\ref{c6:ssec:levers}), so bounding body harm requires $D$ proportional to $a$.

(ii) By \eqref{c6:eq:pdyn} the throughput sustaining \eqref{c6:eq:Pdotreq} is
$\bar\pi=\dot P+\xi a=g/\theta+\xi a_0e^{gt}$, whose discounted cost is
$\kappa_P\int_0^\infty e^{-rt}(g/\theta+\xi a_0e^{gt})\,dt=\kappa_P[g/(\theta r)+\xi a_0/(r-g)]$. Both
integrals are finite for $g<r$. The detection cost is \eqref{c6:eq:CD}.

(iii) Dividing, the common factor $(r-g)^{-1}$ cancels against the second bracket term:
\[
\frac{C_P(g)}{C_D(g)}=\frac{\kappa_P}{\kappa_Dc_Da_0}\Big[\frac{g(r-g)}{\theta r}+\xi a_0\Big],
\]
and $g(r-g)\to0$ as $g\uparrow r$, leaving \eqref{c6:eq:xilimit}. It vanishes iff $\xi a_0=0$.
\end{proof}

\begin{remark}[Reading the proposition: a limit, not a monotonicity]
\label{c6:rem:nonmono}
Two qualifications matter for how this is quoted.

First, the ratio is \emph{not} monotone in $g$. From the displayed form in the proof it is an affine
function of $g(r-g)$, hence a downward parabola in $g$ attaining an interior maximum at exactly
$g=r/2$ for every $\xi$. At low growth the attenuating stock is nearly free to maintain, so the ratio
starts small and \emph{rises}; only past $g=r/2$ does the divergence of detection dominate. The claim
is therefore about the limit $g\uparrow r$, and we state it that way. Theorem~\ref{c6:thm:stockflow},
by contrast, \emph{is} monotone in $g$, because $C_\chi$ is flat and the ratio is proportional to
$r-g$; that is a genuine difference between the two levers and not a matter of presentation.

Second, $\xi$ is a design variable, not a nuisance parameter. It is the rate at which a unit of
attacker capability opens fresh catastrophic affordances (the growth of catastrophic surface), and
\eqref{c6:eq:xilimit} says the entire prevention advantage is governed by it. Reducing $\xi$ is the
same structural act that Chapter~\ref{ch:fleet} calls segmentation and Chapter~\ref{ch:cyber} calls
severing common exposure, since it shrinks what a more capable adversary can newly reach. The chapter's
policy content is thus sharper than ``spend on prevention,'' in that prevention compounds only to
the extent that capability growth is prevented from manufacturing new catastrophic surface.
\end{remark}

\subsection{The capability race: containment versus runaway}
\label{c6:ssec:race}

\begin{theorem}[Capability-race phase portrait]
\label{c6:thm:race}
This result concerns the \emph{attenuating} lever $P$ of Definition~\ref{c6:def:levers} with
$\chi=0$; structural deletion is a one-shot act and has no phase portrait. The state space is
$\{a\ge0,\ P\ge0\}$, with $P$ clamped (absorbed) at the boundary $P=0$: prevention stock cannot be negative,
and once exhausted the residual hazard is the unattenuated $\Lambda_c=\Lambda_0a$. Assume $\mu>\delta$,
so that the intrinsic affordance growth outpaces decay absent prevention and the interior stock
$P_c=\tfrac1\theta\ln(\mu/\delta)$ is positive.

The coupled dynamics \eqref{c6:eq:adyn}--\eqref{c6:eq:pdyn} have a unique interior fixed point
\begin{equation}
(a^\star,P_c)=\Big(\frac{\bar\pi}{\xi},\ \frac1\theta\ln\frac{\mu}{\delta}\Big),
\label{c6:eq:fixed}
\end{equation}
whose Jacobian $\left[\begin{smallmatrix}0&-\theta a^\star\delta\\-\xi&0\end{smallmatrix}\right]$ has
determinant $-\theta a^\star\delta\xi<0$ and real eigenvalues $\pm\sqrt{\theta a^\star\delta\xi}$ of
opposite sign: the fixed point is a \emph{saddle}. Its one-dimensional stable manifold is a
\emph{separatrix} dividing the state plane into a \emph{containment} basin, where prevention outpaces
affordance erosion, capability decays, and the hazard $\Lambda_c=\Lambda_0\,a\,e^{-\theta P}\to0$, and
a \emph{runaway} basin, where prevention erodes and capability accelerates. In the runaway basin $P$
reaches the boundary $P=0$ in finite time; thereafter no attenuation remains and
$\Lambda_c=\Lambda_0a\to\infty$ with $a$. The
controlling criticality is the ratio $a^\star=\bar\pi/\xi$ of the defender's prevention throughput to
the attacker's affordance-generation rate.
\end{theorem}

\begin{proof}
Setting $\dot P=0$ in \eqref{c6:eq:pdyn} gives $a=\bar\pi/\xi$. Setting $\dot a=0$ in
\eqref{c6:eq:adyn} with $a>0$ gives $\mu e^{-\theta P}=\delta$, i.e.\ $P=\tfrac1\theta\ln(\mu/\delta)$,
which is \eqref{c6:eq:fixed}. Writing $f=a(\mu e^{-\theta P}-\delta)$, $h=\bar\pi-\xi a$, we have
$f_a=\mu e^{-\theta P}-\delta$ (zero at the fixed point), $f_P=-\theta a\,\mu e^{-\theta P}=-\theta
a^\star\delta$ there (using $\mu e^{-\theta P_c}=\delta$), $h_a=-\xi$, $h_P=0$. The Jacobian is
$\left[\begin{smallmatrix}0&-\theta a^\star\delta\\-\xi&0\end{smallmatrix}\right]$ with determinant
$-\theta a^\star\delta\xi<0$ and trace $0$, so its eigenvalues are $\pm\sqrt{\theta a^\star\delta\xi}$,
real and of opposite sign: a saddle. A planar saddle has a one-dimensional stable manifold, which is
the separatrix. Trajectories starting on opposite sides converge to the low-$a$/high-$P$ region
(containment, $\Lambda_c\to0$) or diverge in $a$ with $P$ falling (runaway). For the runaway branch,
note $\dot P=\bar\pi-\xi a\le\bar\pi-\xi a^\star_0e^{ct}$ for some $c>0$ once $a$ is growing, so
$\int_0^\infty\dot P\,dt=-\infty$ and $P$ must reach the absorbing boundary $P=0$ at some finite
$t_0$. For $t>t_0$ the stock stays at $0$ (investment $\bar\pi$ being outpaced) and
$\Lambda_c=\Lambda_0a\to\infty$. The trajectory therefore leaves the interior rather than continuing
to $P<0$, which the state space excludes.
\end{proof}

\begin{remark}[On the label ``differential game'']
\label{c6:rem:notgame}
Theorem~\ref{c6:thm:race} is a two-dimensional autonomous ODE system with \emph{constant} actions
$\bar\pi$ and $\mu$, a Lanchester/Lotka--Volterra-type planar system. It has no optimising player, no
equilibrium concept, and no transversality condition, and we do not claim otherwise. The same applies
to the chapter's other results: Theorem~\ref{c6:thm:stockflow} is a perpetuity cost comparison along
an exogenous capability path, and Theorem~\ref{c6:thm:deterrence} is a static scalar best response.
What we deliver is capability-accumulation dynamics together with reduced best responses; the full
stochastic differential game (feedback strategies, an HJB system, a Nash equilibrium in closed-loop
policies) remains open and is flagged as such in \S\ref{c6:sec:conclusions}.
\end{remark}

\begin{remark}
Theorem~\ref{c6:thm:race} is the dynamic analogue of the criticalities that recur through the
monograph: the contagion branching ratio of Chapter~\ref{ch:fleet}, the Hawkes spectral radius of
Chapter~\ref{ch:continuous}, and the operating envelope of Chapter~\ref{ch:control}. The separatrix is
an operating envelope in capability space: on one side the defender's structural investment keeps the
system contained and the systemic hazard collapses; on the other the adversary's capability outruns
affordance removal and the hazard runs away. The design lever is $a^\star=\bar\pi/\xi$ (prevention
throughput relative to how fast attacker capability opens new catastrophic surface), and it is, again,
a structural quantity fixed ex ante, not a runtime control.
\end{remark}

\subsection{Dynamic deterrence-by-denial}
\label{c6:ssec:deterrence}

\begin{theorem}[Dynamic deterrence and the double exponential]
\label{c6:thm:deterrence}
Like Theorem~\ref{c6:thm:race}, this concerns the \emph{attenuating} lever $P$ of
Definition~\ref{c6:def:levers} with $\chi=0$; the deterrence channel operates through $e^{-\theta P}$
and has no analogue for a deletion that has already set $\Lambda_c\equiv0$. The result is a static
best response, not an equilibrium of a dynamic game (Remark~\ref{c6:rem:notgame}).

Let the attacker choose capability-investment effort $u\ge0$ (steady capability $a=u$) to maximise
discounted tail harm net of a quadratic cost,
$\Pi(u;P)=\tfrac{V\Lambda_0}{r}\,u\,e^{-\theta P}-\tfrac{c}{2}u^2$. Its best response is
\begin{equation}
u^\star(P)=\frac{V\Lambda_0}{r\,c}\,e^{-\theta P},
\label{c6:eq:ustar}
\end{equation}
strictly decreasing in the prevention stock: prevention deters the arms race at its source. The tail
harm the defender then bears is the double exponential
\begin{equation}
H(P)=\frac{\Lambda_0}{r}\,u^\star(P)\,e^{-\theta P}=\frac{V\Lambda_0^2}{r^2c}\,e^{-2\theta P},
\qquad
\frac{dH}{dP}=\underbrace{-\theta H}_{\text{direct}}+\underbrace{-\theta H}_{\text{deterrence}}=-2\theta H,
\label{c6:eq:double}
\end{equation}
with the two channels equal, and the defender's optimal prevention (against linear cost $\kappa_P P$)
is $P^\star=\tfrac1{2\theta}\ln\!\big(2\theta V\Lambda_0^2/(r^2c\kappa_P)\big)$.
\end{theorem}

\begin{proof}
$\Pi(\cdot;P)$ is strictly concave in $u$. Its first-order condition
$\tfrac{V\Lambda_0}{r}e^{-\theta P}=cu$ gives \eqref{c6:eq:ustar}, decreasing in $P$. Substituting into
the defender's harm $\tfrac{\Lambda_0}{r}u^\star(P)e^{-\theta P}$ yields
$\tfrac{V\Lambda_0^2}{r^2c}e^{-2\theta P}$. Differentiating, $dH/dP=-2\theta H$. The decomposition
holds $u^\star$ fixed for the direct channel (the explicit $e^{-\theta P}$ in $H$, contributing
$-\theta H$) and lets $u^\star$ respond for the deterrence channel (the $e^{-\theta P}$ inside
$u^\star$, contributing another $-\theta H$), the two equal by symmetry. Minimising $H(P)+\kappa_P P$
gives $-2\theta\tfrac{V\Lambda_0^2}{r^2c}e^{-2\theta P}+\kappa_P=0$, i.e.\ the stated $P^\star$.
\end{proof}

\begin{remark}
Theorem~\ref{c6:thm:deterrence} lifts Chapter~\ref{ch:cyber}'s mechanical-plus-deterrence
decomposition (Theorem~\ref{c5:thm:deterrence}) into the race, where prevention lowers the hazard
directly \emph{and} removes the return on the attacker's capability investment, so it wins the static
contest and also slows the arms race itself, and the two effects compound into a double-exponential
decay of tail harm. This is deterrence-by-denial acting at the source, denying the incentive to build
the capability for the action rather than the action itself, and it is available only to the
structural lever. Detection, which cannot touch the core and deflects effort into it
(Theorem~\ref{c5:thm:deflection}), has neither channel. The monograph's central claim reaches its
dynamic statement here. Against a co-evolving, agentic-AI-augmented adversary, the systematic-
irreversible core is moved only by ex-ante structural prevention, whose advantage \emph{compounds}
over the race while every other lever's decays.
\end{remark}

\section{Numerical illustration}
\label{c6:sec:numerics}

The studies below are unitless and deterministic. Each reports its closed-form prediction, the
simulated value, and their discrepancy.

\subsection{Deletion-versus-flow compounding (Theorem~\ref{c6:thm:stockflow})}
\label{c6:ssec:num-stockflow}

\begin{figure}[htbp]
\centering
\includegraphics[width=\linewidth]{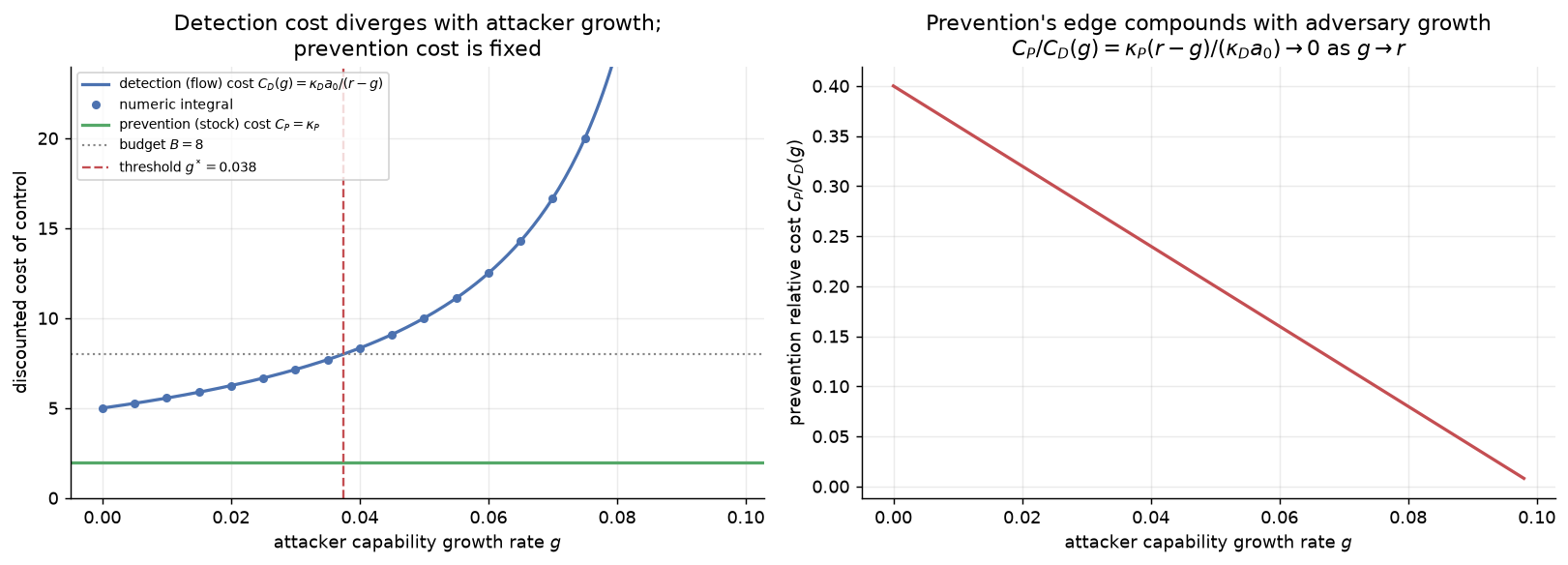}
\caption{\textbf{Deletion-versus-flow compounding (Theorem~\ref{c6:thm:stockflow}).} Left: the
detection (flow) cost diverges with the attacker's growth rate $g$ while the structural-deletion cost
is fixed. They cross the budget at the threshold $g^\star$. Right: the budget-feasibility ratio
$C_\chi/C_D$ falls to zero as $g\uparrow r$. Offense-dominance of the tail compounds with adversary
growth.}
\label{c6:fig:stockflow}
\end{figure}

Figure~\ref{c6:fig:stockflow} verifies Theorem~\ref{c6:thm:stockflow}, and with $r=0.1$, $a_0=1$,
$c_D=1$, $\kappa_D=0.5$, $\kappa_\chi=2$, $B=8$ the detection cost
$C_D(g)=\kappa_Dc_Da_0/(r-g)$ matches direct numerical integration of the discounted-cost integral to
within $0.000\%$, and it diverges as $g\uparrow r$. The deletion cost $C_\chi=\kappa_\chi$ stays flat,
and the closed-form growth threshold is $g^\star=r-\kappa_Dc_Da_0/B=0.0375$, at which
$C_D(g^\star)=8=B$ exactly. The budget-feasibility ratio $C_\chi/C_D(g)$ falls tenfold from $0.40$ at
$g=0$ to $0.04$ at $g=0.9r$. The faster the adversary improves, the smaller the share of a fixed
budget that the deletion line item consumes relative to the diverging detection line item. Per
Remark~\ref{c6:rem:budget} this is a statement about feasibility, not a substitution rate between the
two objectives.

\subsection{The capability race and the separatrix (Theorem~\ref{c6:thm:race})}
\label{c6:ssec:num-race}

\begin{figure}[htbp]
\centering
\includegraphics[width=\linewidth]{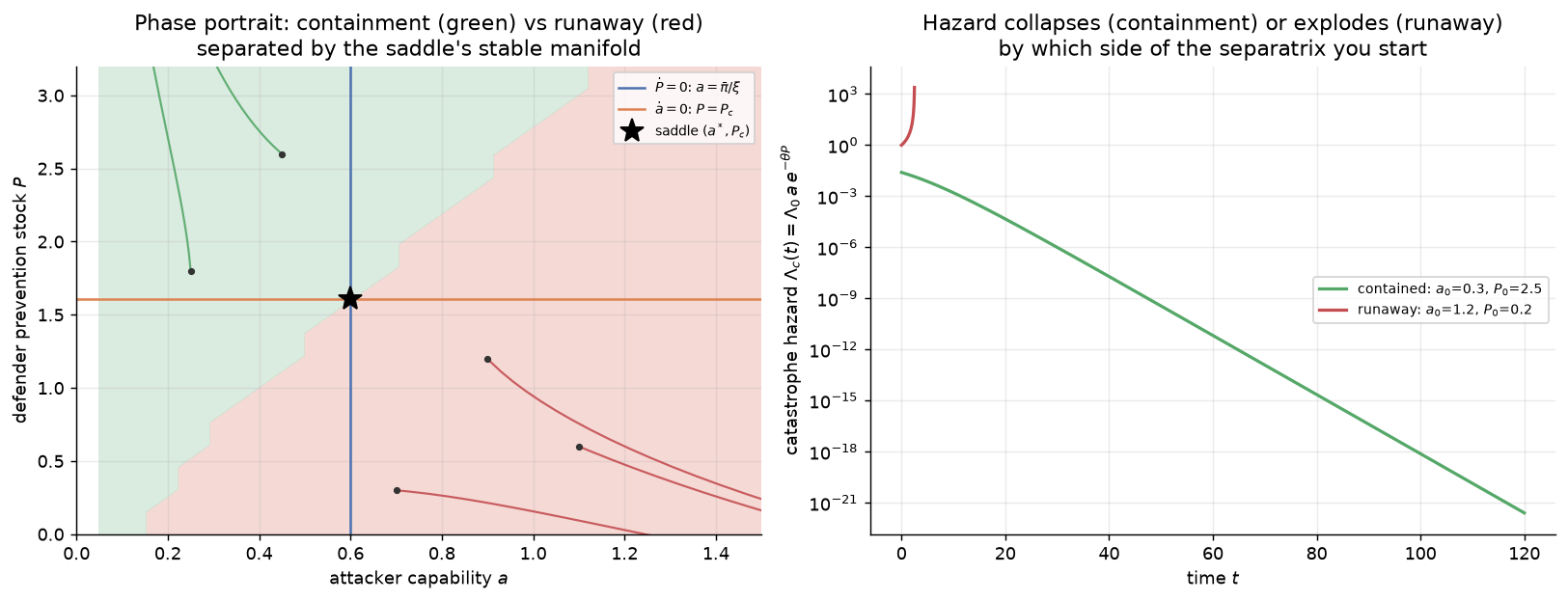}
\caption{\textbf{The capability-race phase portrait (Theorem~\ref{c6:thm:race}).} Left: containment
(green) and runaway (red) basins separated by the saddle's stable manifold. Nullclines meet at the
saddle $(a^\star,P_c)$. Right: the catastrophe hazard collapses or explodes according to which side of
the separatrix the race begins.}
\label{c6:fig:race}
\end{figure}

Figure~\ref{c6:fig:race} verifies Theorem~\ref{c6:thm:race}, and with $\mu=0.5$, $\delta=0.1$,
$\theta=1$, $\bar\pi=0.3$, $\xi=0.5$, the interior fixed point is $(a^\star,P_c)=(0.60,1.61)$. Its
Jacobian has determinant $-0.03<0$ with eigenvalues $\pm0.173$, a saddle, as the closed form requires.
Integrating the dynamics from a grid of initial conditions partitions the plane into a containment
basin and a runaway basin either side of the saddle's stable manifold (left panel). A trajectory
started at low capability and high prevention is contained, its hazard decaying through twenty orders
of magnitude, while one started at high capability and low prevention runs away (right panel).

\subsection{Dynamic deterrence-by-denial (Theorem~\ref{c6:thm:deterrence})}
\label{c6:ssec:num-deterrence}

\begin{figure}[htbp]
\centering
\includegraphics[width=\linewidth]{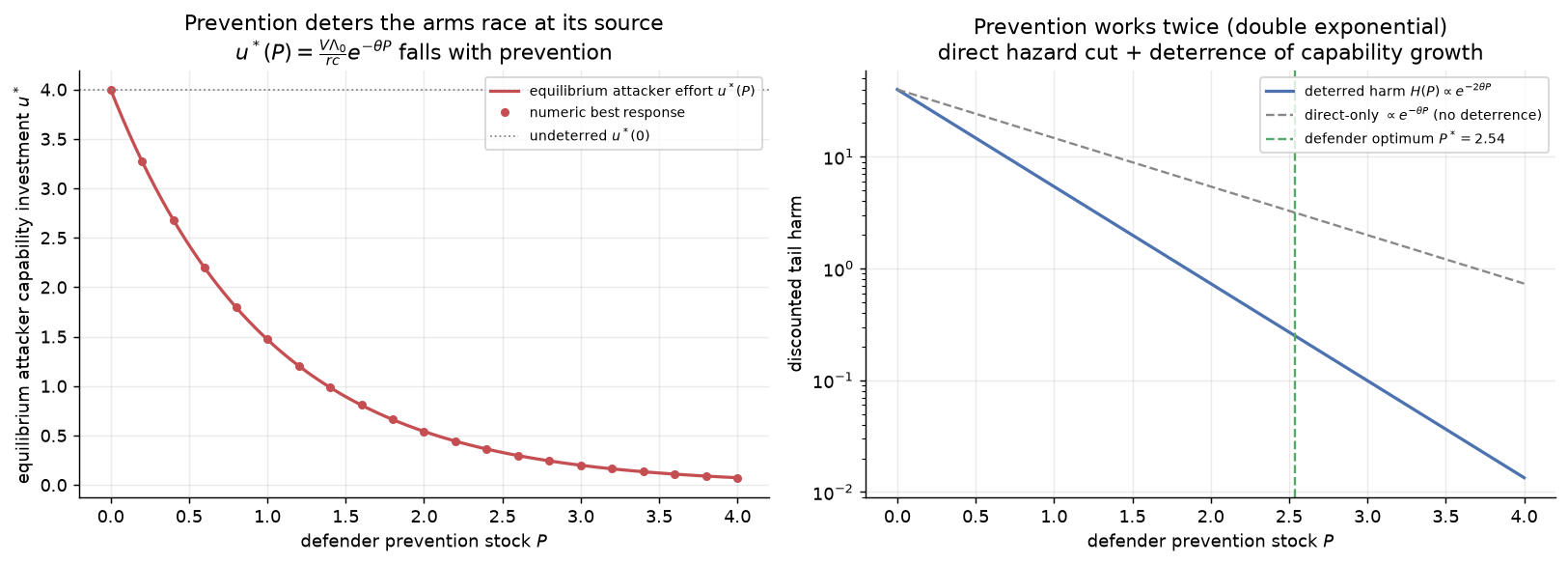}
\caption{\textbf{Dynamic deterrence-by-denial (Theorem~\ref{c6:thm:deterrence}).} Left: prevention
lowers the attacker's equilibrium capability investment $u^\star(P)$, deterring the race at its source.
Right: prevention acts twice (double exponential $e^{-2\theta P}$), far steeper than the direct-only
effect. The defender optimum is marked.}
\label{c6:fig:deterrence}
\end{figure}

Figure~\ref{c6:fig:deterrence} verifies Theorem~\ref{c6:thm:deterrence}, and with $V=2$, $\Lambda_0=1$,
$r=0.1$, $c=5$, $\theta=1$, $\kappa_P=0.5$, the attacker's equilibrium capability investment
$u^\star(P)=(V\Lambda_0/rc)e^{-\theta P}$ matches numerical best response to $10^{-14}$. It falls
$98\%$ (from $4.0$ to $0.07$) as prevention rises, deterring the arms race at its source. The
defender's tail harm is the double exponential $H(P)\propto e^{-2\theta P}$, visibly steeper than the
direct-only $e^{-\theta P}$. Its derivative decomposes into equal direct and deterrence halves, which
sum to the finite-difference total at every $P$. The defender optimum $P^\star=2.54$ from the
closed form matches the numerical minimiser to $0.001$.

\section{Conclusions}
\label{c6:sec:conclusions}

This chapter made capability a state variable and let offense and defense co-evolve. The static
contest of Chapter~\ref{ch:cyber} became a race, and in the race the monograph's central asymmetry
sharpened.

Prevention acts through two channels, and the chapter keeps them apart
(Definition~\ref{c6:def:levers}). Outright \emph{deletion} of a catastrophic affordance neutralises it
against every future capability level, at a cost independent of the adversary's growth rate.
Offense-dominance of the systematic-irreversible core therefore compounds rather than merely
persisting (Theorem~\ref{c6:thm:stockflow}). The \emph{attenuating stock} does not neutralise,
but it must nonetheless be rebuilt only at the adversary's growth rate, while detection must track
its capability level. That asymmetry compounds without bound exactly when capability growth opens
no fresh catastrophic surface (Proposition~\ref{c6:prop:asymmetry}). The coupled dynamics separate a
containment regime from a runaway regime across a structural criticality
(Theorem~\ref{c6:thm:race}). Prevention also deters the arms race at its source, acting twice for a
double-exponential advantage no other lever possesses in this deterministic, constant-action model (Theorem~\ref{c6:thm:deterrence}).

\subsection{Assumptions and limitations}
The results are deterministic by design, which is what secures their closed forms. The
compounding theorem fixes an exponential capability path. The deterrence theorem reduces the attacker
to a steady capability chosen once. The phase portrait uses smooth, autonomous dynamics with
quadratic-type investment. Each is the tractable representative of a richer object, and the
qualitative laws (diverging flow cost, saddle separatrix, deterrence double exponential) do not
depend on the specific forms. As in Chapter~\ref{ch:cyber}, the binding unknown is the adversary.
Here it is the \emph{growth rate} of offensive agentic-AI capability, the single number that decides
which side of every threshold the system sits on, and the one that frontier-model cyber evaluations
must eventually estimate. We model a single attacker against a committed defender. We do not treat
coalitions, endogenous discovery of new affordance classes, or the defender's own capability as a
separately evolving stock.

\subsection{The residual open problem}
The natural end point of this program is the fully \emph{stochastic} differential game: capabilities as
diffusions, both sides playing closed-loop strategies, and a Markov-perfect equilibrium solving
coupled Hamilton--Jacobi--Bellman equations. We do not solve it. Its
value generically has no explicit closed form, and so would fall outside the closed-form-plus-simulation checks that discipline
this monograph. The one tractable corner is the risk-sensitive linear-quadratic game, whose
exponential-of-integral value would inherit the entropic closed form of Chapter~\ref{ch:continuous},
and it is also the natural next step, the place where the offensive- and defensive-capability estimation
programs would jointly enter. It is a sequel we delineate rather than pretend to have written.

A second gap is nearer to hand and cheaper to close. The interaction results of
\S\ref{c2:sec:interaction} turn on how heavily an agent defers to its peers, and on how much that
deference rises under stress. The first has already been recovered from agent decisions in a controlled
setting \citep{hansen2025animal}. The second has not, and it is the one that decides whether the
endogenous common factor exists at all, since a fleet whose deference does not move with the state
acquires no such factor, however deferential it is. A supervisor who can ask for one number should ask
for that difference, while remembering that how \emph{large} the factor turns out to be depends on the
levels too, since deference near unity magnifies even a small rise. Experiments of the kind already
being run on agent populations can give both numbers.

\subsection{The through-line, six times over}
This is the monograph's last chapter, and it is worth stating what the six settings share. The claim
was always one, that the part of agentic risk that is \emph{systematic} (shared across a population through
a common model or substrate) and \emph{irreversible} (a harm that cannot be undone once the action is
taken) is exactly the part that becomes \emph{systemic}. The controls
practitioners reach for first (redundancy, post-hoc detection, reversibility) do not reach it. Only ex-ante structural
prevention moves it. Chapter~\ref{ch:single} proved this as a convex-feasibility statement about one
agent; Chapter~\ref{ch:fleet} as a large-fleet limit that redundancy cannot dilute;
Chapter~\ref{ch:continuous} as a rare-event limit and a heavy-tail finiteness condition; and
Chapter~\ref{ch:control} as a policy-independent Poisson exit probability. Chapter~\ref{ch:cyber}
proved it as a Stackelberg best response that turns the monoculture into an attractive target and shows the market
underinvests in the only lever that works, and Chapter~\ref{ch:coevo} as a capability race in which
that lever's advantage compounds while every other decays. The mathematics changed at each level of
aggregation and each formalism; the conclusion did not.

The claim was also tested against a case the six settings do not otherwise cover. In
\S\ref{c2:sec:interaction} the agents respond to \emph{each other}, and the claim came through changed
but intact. Interaction scales the systematic core rather than dissolving it. It also opens a channel
that \emph{substrate} diversity cannot reach: when agents defer to one another more under stress, the
co-movement is produced by the interaction protocol, and there is no shared exposure left to diversify
away. Pretext between agents does something similar to the second remedy, since a compromised peer can
address its neighbours in language and segmentation of the technical graph does not partition the
semantic one. Prevention is still the lever that reaches the core. What grows is the structural lever
set around it: diversifying the environment in which agents talk to each other, and segmenting the
semantic graph as well as the technical one, are genuine additions. Neither of those couplings appears
in a system diagram drawn from the technical architecture. Financial stability, in the age of agentic
systems as in banking, is secured by structural choices made before the system runs rather than by
watching each institution more closely. As the last two chapters add, it is also secured by the
collective will to make those choices, since no single participant will make them alone. The monograph
makes this argument six times, in six mathematical languages, and the conclusion does not change.

\bibliographystyle{plainnat}
\bibliography{refs}

\end{document}